\documentclass[11pt,a4paper]{article}
\usepackage{indentfirst}
\usepackage{comment}
\usepackage{setspace}
\usepackage[font=small,  margin=2cm]{caption}
\usepackage{amsthm,amsmath,amssymb}
\usepackage{natbib}
\usepackage{multirow}
\usepackage[pdftex]{graphicx}
\usepackage{subfigure}
\usepackage{makecell}
\usepackage{booktabs}
\usepackage{array}
\usepackage{fullpage}
\usepackage{url}
\usepackage{algorithm}
\usepackage{algpseudocode}
\usepackage{bm}
\usepackage{smile}
\usepackage{mathtools}
\usepackage{wrapfig}
\usepackage{lipsum}
\usepackage{mathrsfs}
\usepackage{dsfont}
\usepackage{rotating} 
\usepackage{color}
\usepackage{natbib}
\usepackage{multirow}
\usepackage{apalike}
\usepackage[normalem]{ulem}
\usepackage{appendix}

\def\P{{\mathbb P}}
\def\E{{\mathbb E}}

\def\supp{\mathop{\text{supp}\kern.2ex}}
\def\argmin{\mathop{\text{\rm arg\,min}}}
\def\argmax{\mathop{\text{\rm arg\,max}}}

\def\supp{\mathop{\text{supp}}}

\DeclarePairedDelimiter\floor{\lfloor}{\rfloor}

\usepackage[usenames,dvipsnames,svgnames,table]{xcolor}
\usepackage[colorlinks=true,
            linkcolor=blue,
            urlcolor=blue,
            citecolor=blue]{hyperref}

\numberwithin{equation}{section}
\numberwithin{theorem}{section}
\numberwithin{corollary}{section}
\numberwithin{asmp}{section}
\numberwithin{definition}{section}

\renewcommand{\baselinestretch}{1.5}
\begin{document}

\title{Change Point Detection for High-dimensional Linear Models: A General Tail-adaptive Approach}

\author{
Bin Liu\footnotemark[1],
Zhengling Qi\footnotemark[2]
Xinsheng Zhang\footnotemark[1],
Yufeng Liu\footnotemark[3]
}
\renewcommand{\thefootnote}{\fnsymbol{footnote}}
 
\footnotetext[1]{Department of Statistics and Data Science, School of Management at Fudan University, Shanghai, China; E-mail:{\tt liubin0145@gmail.com; xszhang@fudan.edu.cn }}
\footnotetext[2]{Department of Decision Sciences, George Washington University, U.S.A; E-mail:{\tt qizhengling321@gmail.com }}
\footnotetext[3]{Department of Statistics and Operations Research, Department of Genetics, and Department of Biostatistics, Carolina Center for Genome Sciences, Linberger Comprehensive Cancer Center, University of North Carolina at Chapel Hill, U.S.A; E-mail:{\tt yfliu@email.unc.edu }}

\maketitle\vspace{-0.4in}
\begin{abstract}
We propose a novel approach for detecting change points in high-dimensional linear regression models. Unlike previous research that relied on strict Gaussian/sub-Gaussian error assumptions and had prior knowledge of change points, we propose a tail-adaptive method for  change point detection and estimation. We use a weighted combination of composite quantile and least squared losses to build a new loss function, allowing us to leverage information from both conditional means and quantiles. For change point testing, we develop a family of individual testing statistics with different weights to account for unknown tail structures. These individual tests are further aggregated to construct a powerful tail-adaptive test for sparse regression coefficient changes. For change point estimation, we propose a family of argmax-based individual estimators. We provide theoretical justifications for the validity of these tests and change point estimators. Additionally, we introduce a new algorithm for detecting multiple change points in a tail-adaptive manner using the wild binary segmentation. Extensive numerical results show the effectiveness of our method. Lastly, an R package called ``TailAdaptiveCpt" is developed to implement our algorithms. 
\end{abstract}


\section{Introduction}		
With the advances of  data collection and storage capacity, large scale/high-dimensional data are ubiquitous in many scientific fields ranging from genomics, finance, to social science. Due to the complex data generation mechanism, the heterogeneity, also known as the structural break, has become a common phenomenon for high-dimensional data, where the underlying model of data generation changes and the identically distributed assumption may not hold anymore. Change point analysis is a powerful tool for handling structural changes since the seminal work by \cite{page1955control}. It received considerable attentions in recent years and has a lot of real applications in various fields including genomics \citep{Liu2019Unified}, social science \citep{Roy2017Change}, 
and even for the recent COVID-19 pandemic studies \citep{jiang2021modelling}. Motivated by this, in this paper, we study change point testing and estimation for  high-dimensional linear regression models.

Specifically, suppose we have $n$ independent but (time) ordered realizations $\{(Y_i,\bX_i),i=1,\ldots,n\}$ with $\bX_i=(X_{i1},\ldots,X_{ip})^\top$. For each time point $i$, consider the following regression model:
\begin{equation}\label{equation: linear models}
	Y_i=\bX_i^\top \bbeta_i+\epsilon_i,
\end{equation} 
where $Y_i\in\RR$ is the response variable, $\bbeta_i=(\beta_{i1},\ldots,\beta_{ip})^\top$ is the regression coefficient vector for the $i$-th observation, and $\epsilon_i$ is the error term. For the above model, our first question is whether there is a change point. This can be formulated as the following hypothesis testing problem:
\begin{equation}\label{hypothesis: H0}
	\begin{array}{ll}
		\Hb_0:\bbeta_1=\cdots=\bbeta_n~~\text{v.s.}~
		\Hb_1: \bbeta_1=\cdots=\bbeta_{k_1}\neq \bbeta_{k_1+1}=\cdots=\bbeta_{n},
	\end{array}
\end{equation}
where $k_1$ is the possible but unknown change point location. 
According to (\ref{hypothesis: H0}), the linear regression structure between $Y$ and $\bX$ remains homogeneous if $\Hb_0$ holds, and otherwise there is a change point $k_1$ that divides the data into two segments with different regression coefficients, say $\bbeta^{(1)}$ and $\bbeta^{(2)}$. Our second goal of this paper is to identify the change point location if we reject $\Hb_{0}$ in (\ref{hypothesis: H0}). 
Note that the above two goals are referred as change point testing and estimation, respectively. In the practical use, both testing and estimation are important since practitioners typically have no prior knowledge about either the existence of a change point or its location. Therefore, it is very useful to consider simultaneous change point detection and estimation. {Furthermore, the tail structure of the error $\epsilon_i$ in Model \eqref{equation: linear models} is typically unknown, which could significantly affect the performance of the change point detection and estimation. In the existing literature, the performance guarantee of most methods on change point estimation relies on the assumption that the error $\epsilon_i$ follows a Gaussian/sub-Gaussian distribution. Such an assumption could be violated in practice when the data distribution is heavy-tailed or contaminated by outliers. While some robust methods can address these issues, they may lose efficiency when errors are indeed sub-Gaussian distributed.  It is also very difficult to estimate the tail structures and construct a corresponding change point method based on that.   Hence, it is of great interest to construct a tail-adaptive change point detection and estimation method for high-dimensional linear models. }

\subsection{Contribution}

{Motivated by our previous discussion, in this paper, under the high-dimensional setup with $p\gg n$, we propose a tail-adaptive procedure for simultaneous change point testing and estimation in linear regression models. The proposed method relies on a new loss function  in our change point estimation procedure, which is a weighted combination between the composite quantile loss proposed in \cite{zou2008composite} and the least squared loss with the weight $\alpha\in[0,1]$ for balancing the efficiency and robustness. Thanks to this new loss function with different $\alpha$, we are able to borrow information related to the possible change point from both the conditional mean and quantiles in Model \eqref{equation: linear models}. Therefore, besides controlling the type I error to any desirable level when $\Hb_{0}$ holds, the proposed method  simultaneously enjoys high power and accuracy for change point testing and identification across various underlying error distributions including both lighted and heavy-tailed errors when there exists a change point. 
	By combining our single change point estimation method with the wild binary segmentation (WBS) technique \citep{fryzlewicz2014wild}, we also generalize our method for detecting multiple change points in Model \eqref{equation: linear models}.

	{In terms of our theoretical contribution, for each given  $\alpha$, a novel score-based $\RR^p$-dimensional individual CUSUM process $	\{{\bC}_\alpha(t),t\in[0,1]\}$ is proposed.  Based on this, we construct a family of individual-based testing statistics $\{T_{\alpha},\alpha\in[0,1]\}$
		via aggregating $\bC_\alpha(t)$ using the $\ell_2$-norm of its first $s_0$ largest order statistics, known as the $(s_0,2)$-norm proposed in \cite{Zhou2017An}. 	
		A high-dimensional  bootstrap procedure is introduced to approximate $T_{\alpha}$'s limiting null distributions. The proposed bootstrap method only requires mild conditions on the covariance structures of $\bX$ and the underlying error distribution $\epsilon$,  and is free of tuning parameters and computationally efficient.  Furthermore, combining the corresponding individual tests in $\{T_{\alpha},\alpha\in[0,1]\}$,  we  construct a tail-adaptive test statistic $T_{\rm ad}$ by taking the minimum $P$-values of $\{T_{\alpha},\alpha\in[0,1]\}$. The proposed tail-adaptive method $T_{\rm ad}$ chooses the best individual test according to the data and thus enjoys simultaneous high power across various tail structures.  Theoretically, we adopt a low-cost bootstrap method for approximating the limiting distribution of $T_{\rm ad}$.  In terms of size and power,  for both individual and tail-adaptive tests, we prove that the corresponding test can control the type I error for any given significance level if $\Hb_0$ holds, and reject the null hypothesis with probability tending to one otherwise.  
	}

	As for the change point estimation, once $\Hb_0$ is rejected by our test, based on each individual test statistic, we can estimate its location via taking argmax with respect  to different candidate locations $t\in(0,1)$ for the $(s_0,2)$-norm aggregated process $\{\|\bC_\alpha(t)\|_{(s_0,2)},t\in[0,1]\}$. Under some regular conditions, for each individual based estimator $\{\hat{t}_{\alpha},\alpha\in[0,1]\}$, we can show that the estimation error  is rate optimal up to a $\log(pn)$ factor. Hence, the proposed individual estimators for the change point location allow the signal jump size scale well with  $(n,p)$ and are consistent.

	\subsection{Related literature}
	
	For the  low dimensional setting with a fixed $p$ and $p<n$, change point analysis for linear regression models has been well-studied. Specifically, \cite{quandt1958tests} considered testing (\ref{hypothesis: H0}) for a simple regression model with $p=2$. Other extensions include the  maximum likelihood ratio tests \citep{horvath1995detecting}, partial sums of regression residuals \citep{bai1998estimating},  and so on.  Other related  methods include \cite{qu2008testing,zhang2014testing,oka2011estimating} and among  others. 
	
	Due to the curse of dimensionality, on the other hand, only a few papers studied high-dimensional change point analysis, which mainly focused on  the change point estimation. See \cite{lee2016LASSO,JMLR:v20:18-460,lee2018oracle,leonardi2016computationally,JMLR:v22:19-531,WLLZ2022}. However, none of the aforementioned papers handle hypothesis testing, which is the prerequisite for the change point detection. Furthermore, most existing literature requires strong conditions on the underlying errors $\epsilon_i$ for deriving the desirable theoretical properties, which may not be applicable when the data are heavy-tailed or contaminated by outliers. One possible solution is to use the robust method such as median regression in \cite{lee2018oracle} for change point estimation. As discussed in \cite{zou2008composite,2014A}, when making statistical inference for homogeneous linear models, the asymptotic relative efficiency of median regression-based estimators compared to least squared-based is only about 64\%  in both low  and high dimensions. In addition, inference based on quantile regression can have arbitrarily  small relative efficiency compared to the least squared based regression. Our proposed tail-adaptive method is the one that can perform simultaneous  change point testing and estimation for high-dimensional linear regression models with different distributions. In addition to controlling the type I error to any desirable level, the proposed method enjoys simultaneously high power and accuracy for the change point testing and identification  across various underlying error distributions when there exists a change point.

	The rest of this paper is organized as follows. In Section \ref{section: methodology}, we introduce our new tail-adaptive methodology for detecting a single change point as well as multiple change points. In Section \ref{sec: theory}, we derive the theoretical results in terms of size and power as well as the change point estimation.  In Sections  \ref{section: simulation studies} and \ref{sec: real analysis}, we provide extensive numerica studies and an real data analysis to the S\&P100 data set.  The concluding remarks are provided in Section \ref{section: summary}. Detailed proofs and the full numerical results as well as an application to the S\&P 100 dataset are given in the online supplementary materials.

	{\bf{Notations:}} For $\bv=(v_1,\ldots,v_p)^\top \in \mathbb{R}^p$,  we define its $\ell_p$ norm as $\|\bv\|_p=(\sum_{j=1}^d|v_j|^p)^{1/p}$ for $1\leq p\leq \infty$. For $p=\infty$, define $\|\bv\|_\infty=\max_{1\leq j\leq d}|v_j|$. 
	For any set $\cS$, denote its cardinality by $|\cS|$. For two real numbered sequences $a_n$ and $b_n$, we set $a_n=O(b_n)$ if there exits a constant $C$ such that $|a_n|\leq C|b_n|$ for a sufficiently large $n$; $a_n=o(b_n)$ if $a_n/b_n\rightarrow0$ as $n\rightarrow\infty$; $a_n\asymp b_n$ if there exists constants $c$ and $C$ such that $c|b_n|\leq|a_n|\leq C|b_n|$ for a sufficiently large $n$. For a sequence of random variables (r.v.s) $\{\xi_1,\xi_2,\ldots\}$, we set $\xi_n\xrightarrow{\P} \xi$ if $\xi_n$ converges to $\xi$ in probability as $n\rightarrow\infty$. We also denote $\xi_n=o_p(1)$ if $\xi_n\xrightarrow{\P} 0$. 
	We define $\floor{x}$ as the largest integer less than or equal to $x$ for $x\geq 0$. Denote $(\mathcal{X},\cY)=\{(\bX_1,Y_1),\ldots,(\bX_n,Y_n)\}$.
\section{Methodology}\label{section: methodology}

\subsection{Single change point detection}\label{sec: method of single cpt}
In this section, we introduce our new methodology for Problem \eqref{hypothesis: H0}.  We first focus on detecting a single change point in Model \eqref{equation: linear models} with
\begin{equation}\label{equation: single cpt model}
	Y_i=\bX_i^\top\bbeta^{(1)}\mathbf{1}\{i\leq k_1\}+\bX_i^\top\bbeta^{(2)}\mathbf{1}\{i> k_1\}+\epsilon_i,~\text{for}~i=1,\ldots,n.
\end{equation}
In this paper, we assume $k_1=\floor{nt_1}$ for some constant $t_1\in(0,1)$. Note that $t_1$ is called the relative change point location. We assume  the change point does not occur at the begining or end of data observations. Specifically, suppose there exists a constant $q_0\in(0,0.5)$ such that $q_0\leq t_1\leq 1-q_0$ holds, which is a common assumption in the literature \citep[e.g.,][]{dette2018relevant,Jirak2015Uniform}. For Model \eqref{equation: single cpt model}, the conditional mean of $Y_i$  is:
\begin{equation}\label{model: mean}
	\E[Y_i\, | \, \bX_i]=\bX_i^\top\bbeta^{(1)}\mathbf{1}\{i\leq k_1\}+\bX_i^\top\bbeta^{(2)}\mathbf{1}\{i> k_1\}.
\end{equation}
Moreover, let $0<\tau_1<\ldots<\tau_K<1$ be $K$ candidate quantile indices. For each $\tau_k\in(0,1)$, let $b_k^{(0)}:=\inf\{t:\P(\epsilon\leq t)\geq \tau_k\}$ be the $\tau_k$-th theoretical quantile for the generic error term $\epsilon$ in Model \eqref{equation: single cpt model}. Then, conditional on $\bX_i$, the $\tau_k$-th quantile for $Y_i$ becomes:
\begin{equation}\label{model: quantile}
	Q_{\tau_k}(Y_i|\bX_i)=b_k^{(0)}+\bX_i^\top\bbeta^{(1)}\mathbf{1}\{i\leq k_1\}+\bX_i^\top\bbeta^{(2)}\mathbf{1}\{i> k_1\}, k=1,\ldots,K,
\end{equation}
where $Q_{\tau_k}(Y_i|\bX_i):=\inf\{t:\P(Y_i\leq t|\bX_i)\geq \tau_k\}$.  
Hence,  if there exists a change point in Model \eqref{equation: single cpt model}, both the conditional mean and the conditional quantile change after the change point.  This suggests that  we can make change point inference for $\bbeta^{(1)}$ and $\bbeta^{(2)}$ using either (\ref{model: mean}) or (\ref{model: quantile}).  To propose our new testing statistic, we first introduce the following weighted composite loss function. In particular, 
let $\alpha\in[0,1]$ be some candidate weight.  Define the weighted composite loss function as:
\begin{equation}\label{equation: weighted loss function}
	\ell_\alpha(\bx,y;\tilde{\btau},\bb,\bbeta):=(1-\alpha)\dfrac{1}{K}\sum_{k=1}^{K}\rho_{\tau_k}(y-b_k-\bx^\top\bbeta)+ \dfrac{\alpha}{2}(y-\bx^\top\bbeta)^2,
\end{equation}
where $\rho_{\tau}(t):=t(\tau-\mathbf{1}\{t\leq 0\})$ is the check loss function \citep{koenker1978regression},  $\tilde{\btau}:=(\tau_1,\ldots,\tau_K)^\top$ are user-specified $K$ quantile  levels, and $\bb=(b_1,\ldots,b_K)^\top\in \RR^K$ and $\bbeta=(\beta_1,\ldots,\beta_p)^\top\in \RR^p$. Note that we can regard $\ell_\alpha(\bx,y;\tilde{\btau},\bb,\bbeta)$ as a weighted loss function between the composite quantile loss and the squared error loss. For example, for $\alpha=1$, it reduces to the ordinary least squared-based loss function with $\ell_1(\bx,y)=(y-\bx^\top\bbeta)^2/2$. When $\alpha=0$, it is the composite quantile loss function $\ell_0(\bx,y)=\sum_{k=1}^{K}\rho_{\tau_k}(y-b_k-\bx^\top\bbeta)/K$ proposed in \cite{zou2008composite}. It is known that the least squared loss-based method has the best statistical  efficiency when errors follow Gaussian distributions and the composite quantile loss is more robust when the error distribution is heavy-tailed or contaminated  by outliers. As discussed before, in practice, it is  challenging to obtain the tail structure of the error distribution and construct a corresponding change point testing method based on the error structure. Hence, we propose a weighted loss function by borrowing the information related to the possible change point from both the conditional mean and quantiles. We use the weight $\alpha$ to balance the efficiency and robustness. 

Our new testing statistic is based on a novel high-dimensional weighted score-based CUSUM process of the weighted composite loss function. In particular, 
for the composite loss function  $\ell_\alpha(\bx,y;\tilde{\btau},\bb,\bbeta)$,  define its score (subgradient) function ${\partial \ell_\alpha(\bx,y;\tilde{\btau},\bb,\bbeta) }/{\partial \bbeta}$ with respect to $\bbeta$  as:  
\begin{equation}\label{equation: score loss}
	\begin{array}{ll}
		\bZ_{\alpha}(\bx,y;\tilde{\btau},\bb,\bbeta):=\Big[\dfrac{1-\alpha}{K}\sum\limits_{k=1}^{K}\bx\big(\mathbf{1}\{y-b_k-\bx^\top\bbeta\leq 0\}-\tau_k\big)\Big]-\alpha\big[\bx (y-\bx^\top\bbeta)\big].
	\end{array}
\end{equation}
Using $	\bZ_{\alpha}(\bx,y;\tilde{\btau},\bb,\bbeta)$, for each $\alpha\in[0,1]$ and $t\in(0,1)$, we first define the oracle score-based CUSUM as follows:
\begin{equation}\label{equation: score cusum}
	\tilde{\bC}_\alpha(t;\tilde{\btau},\bb,\bbeta)
	=\dfrac{1}{\sqrt{n}\sigma(\alpha,\tilde{\btau})}\big(\sum\limits_{i=1}^{\lfloor nt\rfloor}\bZ_{\alpha}(\bX_i,Y_i;\tilde{\btau},\bb,\bbeta)-\dfrac{\lfloor nt \rfloor }{n}\sum\limits_{i=1}^{n}\bZ_{\alpha}(\bX_i,Y_i;\tilde{\btau},\bb,\bbeta)\big),
\end{equation}
where $\sigma^2(\alpha,\tilde{\btau}):=\text{Var}[(1-\alpha)e_i(\tilde{\btau})-\alpha\epsilon_i)]$ with  	$e_i(\tilde{\btau}):=K^{-1}\sum\limits_{k=1}^{K}(\mathbf{1}\{\epsilon_i\leq b_{k}^{(0)}\}-\tau_k)$. 
Note that we call $\tilde{\bC}_\alpha(t;\tilde{\btau},\bb,\bbeta)$ oracle  since we assume $\sigma^2(\alpha,\tilde{\btau})$ is known. In Section \ref{sec: method of variance estimation}, we will give the explicit form of $\sigma^2(\alpha,\tilde{\btau})$ under various combinations of $\alpha$ and $\tilde{\btau}$ and  introduce its consistent estimator under both $\Hb_0$ and $\Hb_1$. {In the following, to motivate our test statistics, we study the behaviors of $\tilde{\bC}_\alpha(t;\tilde{\btau},\bb,\bbeta)$  under $\Hb_0$ and $\Hb_1$ respectively.}
First, under  $\Hb_0$, if we replace $\bbeta=\bbeta^{(0)}$ and $\bb=\bb^{(0)}$ in	(\ref{equation: score cusum}), the  score based CUSUM becomes
\begin{equation*}
	\begin{array}{ll}
		\tilde{\bC}_\alpha(t;\tilde{\btau},\bb^{(0)},\bbeta^{(0)})
		=\dfrac{1}{\sqrt{n}\sigma(\alpha,\tilde{\btau})}\Big(\sum\limits_{i=1}^{\lfloor nt\rfloor}\bZ_{\alpha}(\bX_i,Y_i;\tilde{\btau},\bb^{(0)},\bbeta^{(0)})-\dfrac{\lfloor nt \rfloor }{n}\sum\limits_{i=1}^{n}\bZ_{\alpha}(\bX_i,Y_i;\tilde{\btau},\bb^{(0)},\bbeta^{(0)})\Big).
	\end{array}
\end{equation*}
By noting that under $\Hb_0$, we have $Y_i=\bX_i^\top\bbeta^{(0)}+\epsilon_i$, the above CUSUM   reduces to the following $\RR^p$--dimensional random noise based CUSUM:
\begin{equation}\label{equation: random noise based CUSUM}
	\begin{array}{ll}
		\tilde{\bC}_\alpha(t;\tilde{\btau},\bb^{(0)},\bbeta^{(0)})
		=\dfrac{1}{\sqrt{n}\sigma(\alpha,\tilde{\btau})}\Big(\sum\limits_{i=1}^{\floor{nt}}\bX_i((1-\alpha) e_i(\tilde{\btau})-\alpha\epsilon_i)-\dfrac{\floor{nt}}{n}\sum\limits_{i=1}^n\bX_i((1-\alpha) e_i(\tilde{\btau})-\alpha\epsilon_i)\Big),\\
	\end{array}
\end{equation}
{whose asymptotic distribution can be easily characterized.}
Since both $\bb^{(0)}$ and $\bbeta^{(0)}$ are unknown, we need some proper estimators  that can approximate them  well under $\Hb_0$.  In this paper, for each $\alpha\in[0,1]$, we obtain the estimators by solving the following penalized optimization problem:
\begin{equation}\label{equation: lasso estimator}
	(\hat{\bb}_\alpha,\hat{\bbeta}_\alpha)=\argmin_{\bb\in\RR^K,\atop\bbeta\in\RR^p}\Big[(1-\alpha)\dfrac{1}{n}\sum_{i=1}^{n}\dfrac{1}{K}\sum_{k=1}^{K}\rho_{\tau_k}(Y_i-b_i-\bX_i^\top\bbeta)+ \dfrac{\alpha}{2n}\sum_{i=1}^{n}(Y_i-\bX_i^\top \bbeta)^2+\lambda_{\alpha}\big\|\bbeta\big\|_1\Big],
\end{equation}
where $\hat{\bb}_\alpha:=(\hat{b}_{\alpha,1},\cdots,\hat{b}_{\alpha,K})^\top$, $\hat{\bbeta}_{\alpha}:=(\hat{\beta}_{\alpha,1},\ldots,\hat{\beta}_{\alpha,p})^\top$, and $\lambda_{\alpha}$ is the non-negative tuning parameter controlling the overall sparsity of $\hat{\bbeta}_{\alpha}$.
Note that the above estimators are obtained using all observations $(\mathcal{X},\cY)$.  After obtaining $(\hat{\bb}_\alpha^\top,\hat{\bbeta}_\alpha^\top)$, we plug them into the score function 	in (\ref{equation: score cusum}) and obtain the $\RR^p$--dimensional oracle score based CUSUM statistic as follows:
\begin{equation}\label{equation: score CUSUM}
	\tilde{\bC}_\alpha(t;\tilde{\btau},\hat{\bb}_\alpha,
	\hat{\bbeta}_\alpha)=\big(\tilde{C}_{\alpha,1}(t;\tilde{\btau},\hat{\bb}_\alpha,
	\hat{\bbeta}_\alpha),\ldots,\tilde{C}_{\alpha,p}(t;\tilde{\btau},\hat{\bb}_\alpha,
	\hat{\bbeta}_\alpha)\big)^\top.
\end{equation}
Using $(\hat{\bb}_\alpha,\hat{\bbeta}_\alpha)$, we can prove that under $\Hb_0$, for each $\alpha\in[0,1]$,  (\ref{equation: score CUSUM}) can approximate the random-noise based CUSUM process in (\ref{equation: random noise based CUSUM}) under some proper norm {aggregations}. {Next, we investigate the behavior of (\ref{equation: score CUSUM}) under $\Hb_1$. Observe that the score based CUSUM  has the following decomposition:}
\begin{equation}
	\tilde{\bC}_\alpha(t;\tilde{\btau},\hat{\bb}_\alpha,\hat{\bbeta}_\alpha)={{\tilde{\bC}_\alpha(t;\tilde{\btau},\bb^{(0)},\bbeta^{(0)})}}+{{\bdelta_{\alpha}(t)}}+{{\bR_{\alpha}(t;\hat{\bb}_\alpha,\hat{\bbeta}_\alpha)}},
\end{equation}
where $\tilde{\bC}_\alpha(t;\tilde{\btau},\bb^{(0)},\bbeta^{(0)})$ is the random noise based CUSUM process defined in (\ref{equation: random noise based CUSUM}), $\bR_{\alpha}(t;\hat{\bb}_\alpha,\hat{\bbeta}_\alpha)$ is some random bias which has a very complicated form but can be controlled properly under $\Hb_1$, and $\bdelta_{\alpha}(t)$ is the signal jump function. More specifically, let
\begin{equation}\label{equation: theoretical SNR}
	SNR(\alpha,\tilde{\btau}):=\dfrac{(1-\alpha) \big(\dfrac{1}{K}\sum\limits_{k=1}^{K}f_{\epsilon}(b_{k}^{(0)})\big)+\alpha}{\sigma(\alpha,\tilde{\btau})},
\end{equation}
where $f_{\epsilon}(t)$ is the probability density function of $\epsilon$, and define the signal jump function
\begin{equation}
	\bDelta(t;\bbeta^{(1)},\bbeta^{(2)}):=	\left\{
	\begin{array}{ll}
		\dfrac{{\lfloor nt \rfloor (n-\floor {nt_1})}}{n^{3/2}}\bSigma\big(\bbeta^{(1)}-\bbeta^{(2)}\big),~\text{if}~t\leq t_1,\\
		\dfrac{{\lfloor nt_1 \rfloor (n-\floor {nt})}}{n^{3/2}}\bSigma\big(\bbeta^{(1)}-\bbeta^{(2)}\big),~\text{if}~t> t_1.
	\end{array} \right.
\end{equation}
Then, the signal jump $\bdelta_{\alpha}(t)$ can be explicitly represented as the products of $SNR(\alpha,\tilde{\btau})$ and $\bDelta(t;\bbeta^{(1)},\bbeta^{(2)})$, which has the following explicit form:
\begin{equation}\label{equation: decomposition of the signal function}
	\bdelta_{\alpha}(t):=SNR(\alpha,\tilde{\btau})\times \bDelta(t;\bbeta^{(1)},\bbeta^{(2)}).
\end{equation}
By (\ref{equation: decomposition of the signal function}), we can see that $\bdelta_{\alpha}(t)$ can be decomposed into a  loss-function-dependent part  $SNR(\alpha,\tilde{\btau})$  and a change-point-model-dependent part  \\
$\bDelta(t;\bbeta^{(1)},\bbeta^{(2)})$. More specifically, the first term $SNR(\alpha,\tilde{\btau})$ (short for the signal-to-noise-ratio) is only related to the parameters $\alpha,K,\bb^{(0)}$ as well as $\sigma(\alpha,\tilde{\btau})$, resulting from a user specified weighted loss function defined in (\ref{equation: weighted loss function}). In contrast, the second term $\{\bDelta(t;\bbeta^{(1)},\bbeta^{(2)}),t\in[0,1]\}$ is only related to   Model \eqref{equation: single cpt model}, which is based on parameters $t_1$, $\bSigma$, $\bbeta^{(1)}$, and $\bbeta^{(2)}$ and is independent of the loss function. Moreover, for any weighted composite loss function, the process $\{\bDelta(t;\bbeta^{(1)},\bbeta^{(2)}),t\in[0,1]\}$ has the following properties: First, under $\Hb_1$, the non-zero elements of $\bDelta(t;\bbeta^{(1)},\bbeta^{(2)})$ are at most $(s^{(1)}+s^{(2)})$-sparse since we require sparse regression coefficients in the model; Second, we can see that $\|\bDelta(t;\bbeta^{(1)},\bbeta^{(2)})\|$ with $t\in[q_0,1-q_0]$ obtains its maximum value at the true change point location $t_1$, where $\|\cdot\|$ denotes some proper norm such as $\|\cdot\|_{\infty}$. Hence, in theory, the signal jump function $\bdelta_{\alpha}(t)$ also achieves its maximum value at  $t_1$ under some proper norm. This is the key reason why using the score based CUSUM can correctly localize the true change point if $\bbeta^{(1)}-\bbeta^{(2)}$ is big enough. More importantly, for a given  underlying error distribution $\epsilon$ in Model \eqref{equation: single cpt model}, we can use $SNR(\alpha,\tilde{\btau})$ to further amplify the magnitude of $\bdelta_{\alpha}(t)$ via choosing a proper combination of $\alpha$ and $\tilde{\btau}$. In particular, recall $\sigma^2(\alpha,\tilde{\btau}):=\text{Var}[(1-\alpha)e_i(\tilde{\btau})-\alpha\epsilon_i)]$. Then, we have
\begin{equation}\label{equation: orcale sigma square}
	\sigma^2(\alpha,\tilde{\btau})=(1-\alpha)^2\text{Var}[e_i(\tilde{\btau})]+\alpha^2\sigma^2-2\alpha(1-\alpha)\text{Cov}(e_i(\tilde{\btau}),\epsilon_i),
\end{equation}
where $\sigma^2:=\text{Var}(\epsilon)$. Using (\ref{equation: theoretical SNR}) and (\ref{equation: orcale sigma square}), $SNR(\alpha,\tilde{\btau})$ can be further simplified under some specific $\alpha$. For example, if $\alpha=1$, then $SNR(\alpha,\tilde{\btau})=1/\sigma$; If $\alpha=0$, then
\begin{equation*}
	SNR(\alpha,\tilde{\btau})=\dfrac{\sum\limits_{k=1}^{K}f_{\epsilon}(b_{k}^{(0)})}{\sqrt{\sum_{k_1=1}^{K}\sum_{k_2=1}^{K}\gamma_{k_1k_2}}}
\end{equation*}
with $\gamma_{k_1k_2}:=\min(\tau_{k_1},\tau_{k_2})-\tau_{k_1}\tau_{k_2}$; If we choose $\alpha\in(0,1)$,  $K=1$ and $\tilde{\btau}=\tau$ for some $\tau\in(0,1)$. Then we have
\begin{equation}
	SNR(\alpha,\tilde{\btau})=\dfrac{(1-\alpha)f_{\epsilon}(b_{\tau}^{(0)})+\alpha}{[(1-\alpha)^2\tau(1-\tau)+\alpha^2\sigma^2-2\alpha(1-\alpha)\text{Cov}(e({\tau}),\epsilon)]^{1/2}}.
\end{equation}
Hence, for any underlying error distribution $\epsilon$ in Model \eqref{equation: single cpt model}, it is possible for us to choose a proper $\alpha$ and $\tilde{\btau}$ that makes $SNR(\alpha,\tilde{\btau})$ as large as possible for that distribution. See Figure \ref{figure: snr} for a direct illustration.


\begin{figure}[!h]
	\begin{center}
		\includegraphics[width=12cm]{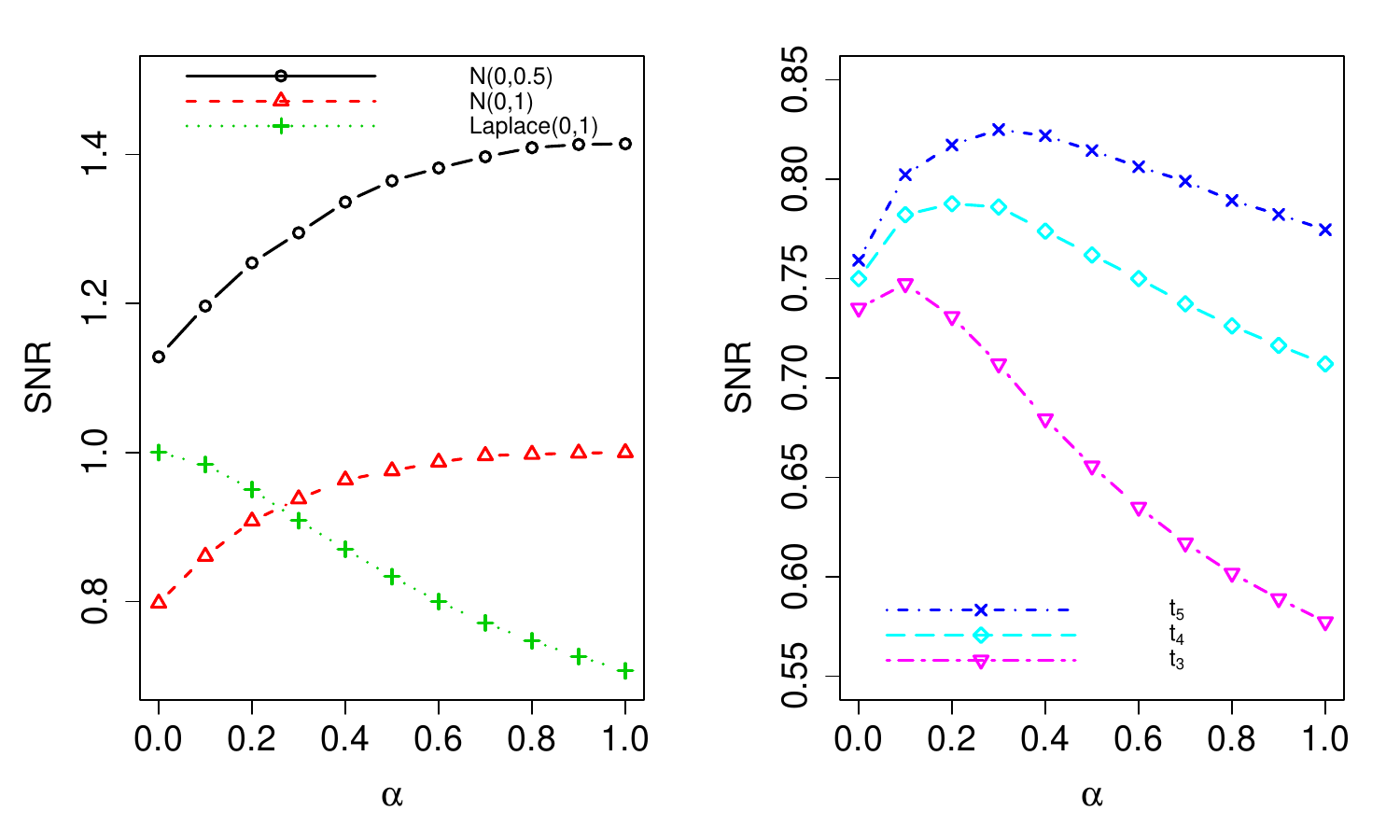}
		\caption{$SNR(\alpha,\tilde{\btau})$ under various errors with different weights $\alpha\in\{0,0.1,\ldots,0.9,1\}$ for the weighted loss with $\tilde{\tau}=0.5$ and $K=1$.}
		\label{figure: snr}
	\end{center}
\end{figure}
For change point detection, a natural  question is how to aggregate the $\RR^p$--dimensional CUSUM process $\tilde{\bC}_\alpha(t;\tilde{\btau},\hat{\bb}_\alpha,\hat{\bbeta}_\alpha)$. Note that for high-dimensional sparse linear models, there are at most $s=s^{(1)}+s^{(2)}$ coordinates in $\bbeta^{(1)}-\bbeta^{(2)}$ that can have a change point, which can be much smaller than the data dimension $p$, although we allow $s$ to grow with the sample size $n$. Motivated by this, in this paper, we aggregate the first $s_0$ largest statistics of  $\tilde{\bC}_\alpha(t;\tilde{\btau},\hat{\bb}_\alpha,
\hat{\bbeta}_\alpha)$. To that end, we introduce the $(s_0,2)$-norm as follows. Let $\bv=(v_1,\ldots,v_p)\in \RR^p$. For any $1\leq s_0\leq p$, define  $\big\|\bv\big\|_{(s_0,2)}=(\sum_{j=1}^{s_0}|v_{(j)}|^2)^{1/2}$,
where $|v_{(1)}\geq |v_{(2)}|\cdots\geq |v_{(p)}|$ are the order statistics of $\bv$. By definition, we can see that  $\|\bv\|_{(s_0,2)}$ is the $L_2$-norm for the first $s_0$ largest order statistics of $(|v_1|,\ldots,|v_p|)^\top$, which can be regarded as an adjusted $L_2$-norm in high dimensions.  Note that the $(s_0,2)$-norm is a special case of the $(s_0, \tilde p)$-norm proposed in \cite{Zhou2017An} by setting $\tilde p=2$. We also remark that taking the first $s_0$ largest order statistics can account for the sparsity structure of $\bbeta^{(1)}-\bbeta^{(2)}$. 
Using the $(s_0,2)$-norm with a user-specified $s_0$ and  known variance $\sigma^2(\alpha,\tilde{\tau})$, we introduce the oracle individual testing statistic with respect to a given $\alpha\in[0,1]$ as

\begin{equation*}
	\tilde{T}_{\alpha}=\max_{q_0\leq t\leq 1-q_0}\Big\|\tilde{\bC}_\alpha(t;\tilde{\btau},\hat{\bb}_\alpha,\hat{\bbeta}_{a})\Big\|_{(s_0,2)},~~\text{with}~~\alpha\in[0,1].
\end{equation*}

By construction, $\tilde{T}_{\alpha}$ can capture the tail structure  of the underlying regression errors by choosing a special $\alpha$ and $\tilde{\btau}$. Specifically, for $\alpha=1$, it equals to the least square loss-based method. In this case, since $\tilde{T}_{\alpha}$ only uses the moment information of the errors, it is powerful for detecting a change point with light-tailed errors such as Gaussian or sub-Gaussian distributions. For $\alpha=0$, $\tilde{T}_{\alpha}$ reduces to the composite quantile loss-based method, which only uses the information of ranks or quantiles. In this case, $\tilde{T}_{\alpha}$ is more robust to data with heavy tails such as Cauchy distributions. As a special case of $\alpha=0$, if we further choose $\tilde{\tau}=0.5$ and $K=1$, our testing statistic reduces to the median regression-based method. Moreover, if we choose a proper non-trivial weight $\alpha$, $\tilde{T}_{\alpha}$ enjoys satisfactory power performance for data with a moderate magnitude of tails such as the  Student's $t_v$ or Laplace distributions. Hence, our proposed individual testing statistics can adequately capture the tail structures of the data. 

Another distinguishing feature for using $\tilde{T}_{\alpha}$ is that, we can establish a general and flexible framework for aggregating the score based CUSUM for high-dimensional sparse linear models. Instead of taking the $\ell_\infty$-norm  as most papers adopted for making statistical inference of high-dimensional linear models \citep[e.g.,][]{xia2018two-sample}, we choose to aggregate them via using the $\ell_2$-norm of the first $s_0$ largest order statistics. Under this framework,  the $\ell_\infty$-norm is a special case by taking $s_0=1$. 


\subsection{Variance estimation under $\Hb_0$ and $\Hb_1$}\label{sec: method of variance estimation}
Note that 	$\tilde{T}_{\alpha}$ is constructed using a known variance $\sigma^2(\alpha,\tilde{\btau})$ which is defined in (\ref{equation: orcale sigma square}). In practice, however, $\sigma^2(\alpha,\tilde{\btau})$ is typically  unknown. Hence, to yield a  powerful testing method, a consistent variance estimation is needed especially under the alternative hypothesis. For high-dimensional change point analysis, the main difficulty comes from the unknown change point $t_1$. To overcome this problem, we propose a weighted variance estimation. In particular, for each fixed $\alpha\in[0,1]$ and $t\in(0,1)$, define the score based CUSUM statistic without standardization as follows:
\begin{equation}\label{equation: score cusum without sd}
	\breve{\bC}_\alpha(t;\tilde{\btau},\hat{\bb}_{\alpha},\hat{\bbeta}_\alpha)=\dfrac{1}{\sqrt{n}}\big(\sum\limits_{i=1}^{\lfloor nt\rfloor}\bZ_{\alpha}(\bX_i,Y_i;\tilde{\btau},\hat{\bb}_\alpha,\hat{\bbeta}_\alpha)-\dfrac{\lfloor nt \rfloor }{n}\sum\limits_{i=1}^{n}\bZ_{\alpha}(\bX_i,Y_i;\tilde{\btau},\hat{\bb}_\alpha,\hat{\bbeta}_\alpha)\big).
\end{equation}
For each $\alpha\in[0,1]$, we obtain the individual based change point estimator: 
\begin{equation}\label{equation: individual single cpt estimator}
	\hat{t}_{\alpha}=\argmax_{q_0\leq t\leq 1-q_0}\big\|\breve{\bC}_{\alpha}(t;\tilde{\btau},\hat{\bb}_\alpha,
	\hat{\bbeta}_\alpha)\big\|_{(s_0,2)}.
\end{equation}
In Theorem \ref{theorem: cpt estimation results}, we  prove that under some regular conditions, if $\Hb_1$ holds, $\hat{t}_\alpha$ is a consistent estimator for $t_1$, e.g. $|n\hat{t}_{\alpha}-nt_1|=o_p(n)$. 
This result enables us to propose a variance estimator which is consistent under both $\Hb_0$ and $\Hb_1$. Specifically, let $h\in(0,1)$ be a user specified constant, and define the samples $n_-=\{i: i\leq nh\hat{t}_{\alpha}\}$ and $n_+=\{i:   \hat{t}_{\alpha}n+(1-h)(1-\hat{t}_{\alpha})n   \leq i\leq n \}$. Let $((\hat{\bb}^{(1)}_\alpha)^\top,(\hat{\bbeta}^{(1)}_\alpha)^\top)$ and $((\hat{\bb}^{(2)}_\alpha)^\top,(\hat{\bbeta}^{(2)}_\alpha)^\top)$ be the estimators using the samples in $n_-$ and $n_+$.
For each $\alpha$, we  calculate the regression residuals:
\begin{equation}\label{equation: regression residuals}
	\hat{\epsilon}_i=	(Y_i-\bX_i^\top\hat{\bbeta}_{\alpha}^{(1)})\mathbf{1}\{i \in n_-\}+(Y_i-\bX_i^\top\hat{\bbeta}_{\alpha}^{(2)})\mathbf{1}\{i\in n_+\}, ~\text{for}~ i\in n_-\cup n_+.
\end{equation}
Moreover, compute
$\hat{e}_i(\tilde{\btau})=K^{-1}\sum\limits_{k=1}^{K}\hat{e}_i(\tau_k)$ with $\hat{e}_i(\tau_k)$ defined as 
\begin{equation}\label{equation: regression quantile residuals}
	\hat{e}_i(\tau_k):=(\mathbf{1}\{\hat{\epsilon}_i\leq \hat{b}_{\alpha,k}^{(1)}\}-\tau_k)\mathbf{1}\{i \in n_-\}+(\mathbf{1}\{\hat{\epsilon}_i\leq \hat{b}_{\alpha,k}^{(2)}\}-\tau_k)\mathbf{1}\{i\in n_+\}.
\end{equation}
Lastly, based on $\hat{\epsilon}_i$ and $\hat{e}_i(\tilde{\btau})$, we construct a weighted estimator for 	$\sigma^2(\alpha,\tilde{\btau})$:
\begin{equation}\label{equation: variance estimator}
	\hat{\sigma}^2(\alpha,\tilde{\btau})=\hat{t}_{\alpha}\times\hat{\sigma}_-^2(\alpha,\tilde{\btau})+(1-\hat{t}_{\alpha})\times\hat{\sigma}_+^2(\alpha,\tilde{\btau}),
\end{equation}
where:
\begin{equation*}
	\begin{array}{ll}
		\hat{\sigma}_-^2(\alpha,\tilde{\btau}):=\dfrac{1}{|n_-|}\sum_{i\in n_-}\big[(1-\alpha)\hat{e}_i(\tilde{\btau})-\alpha\hat{\epsilon}_i\big]^2,\\\hat{\sigma}_+^2(\alpha,\tilde{\btau}):=\dfrac{1}{|n_+|}\sum_{i\in n_+}\big[(1-\alpha)\hat{e}_i(\tilde{\btau})-\alpha\hat{\epsilon}_i\big]^2.
	\end{array}
\end{equation*}
For the above variance estimation, we can prove that $|\hat{\sigma}^2(\alpha,\tilde{\btau})-{\sigma}^2(\alpha,\tilde{\btau})|=o_p(1)$ under either $\Hb_0$ or $\Hb_1$. As a result, the proposed variance estimator $\hat{\sigma}^2(\alpha,\tilde{\btau})$ avoids the problem of non-monotonic power performance as discussed in \cite{shao2010testing}, which is a serious issue in change point analysis. Hence, we replace ${\sigma}(\alpha,\tilde{\btau})$ in (\ref{equation: score CUSUM}) by $\hat{\sigma}(\alpha,\tilde{\btau})$ and define the  data-driven score-based CUSUM process 
\begin{equation}\label{equation: data-driven score CUSUM}
	\begin{array}{ll}
		{\bC}_\alpha(t;\tilde{\btau},\hat{\bb}_\alpha,
		\hat{\bbeta}_\alpha)
		=\dfrac{1}{\sqrt{n}\widehat \sigma(\alpha,\tilde{\btau})}\big(\sum\limits_{i=1}^{\lfloor nt\rfloor}\bZ_{\alpha}(\bX_i,Y_i;\tilde{\btau},\hat{\bb}_\alpha,\hat{\bbeta}_\alpha)-\dfrac{\lfloor nt \rfloor }{n}\sum\limits_{i=1}^{n}\bZ_{\alpha}(\bX_i,Y_i;\tilde{\btau},\hat{\bb}_\alpha,\hat{\bbeta}_\alpha)\big). 
	\end{array}
\end{equation}
For a user-specified $s_0\in\{1,\ldots,p\}$ and any $\alpha\in[0,1]$, we define the final individual-based testing statistic as follows:
\begin{equation}\label{equation: final individual test statistic}
	{T}_{\alpha}=\max_{q_0\leq t\leq 1-q_0}\Big\|{\bC}_\alpha(t;\tilde{\btau},\hat{\bb}_\alpha,\hat{\bbeta}_{a})\Big\|_{(s_0,2)},~~\text{with}~~\alpha\in[0,1].
\end{equation}
In what follows, we use $\{{T}_{\alpha},\alpha\in[0,1]\}$ as our individual  testing statistics. 
\subsection{Bootstrap approximation for the individual testing statistic }\label{sec: method of bootstrap}
In high dimensions, it is very difficult to obtain the limiting null distribution of $T_{\alpha}$. To overcome this problem, we propose a novel bootstrap procedure. In particular, suppose we implement the bootstrap procedure for $B$ times. Then, for each $b$-th bootstrap with $b=1,\ldots,B$, we generate i.i.d. random variables  $e_1^b,\ldots,e_n^b$  with  $e_i^b\sim N(0,1)$. Let $e_i^{b}(\tilde{\btau})=K^{-1}\sum\limits_{k=1}^{K}e_i^{b}(\tau_k)$ with $e_i^{b}(\tau_k):=\mathbf{1}\{\epsilon_i^b\leq \Phi^{-1}(\tau_k)\}-\tau_k$, where $\Phi(x)$ is the CDF for the standard normal distribution.
Then,  for each individual-based testing statistic $T_{\alpha}$, with a user specified $s_0$, we define its $b$-th bootstrap sample-based score CUSUM process as:

\begin{equation}\label{equation: bootstrap score CUSUM}
	{\bC}^b_\alpha(t;\tilde{\btau})
	=\dfrac{1}{\sqrt{n}v(\alpha,\tilde{\btau})}\big(\sum\limits_{i=1}^{\floor{nt}}\bX_i((1-\alpha) e^{b}_i(\tilde{\btau})-\alpha e^b_i)-\dfrac{\floor{nt}}{n}\sum\limits_{i=1}^n\bX_i((1-\alpha) e^{b}_i(\tilde{\btau})-\alpha e^b_i)\big),
\end{equation}
where $v^2(\alpha,\tilde{\btau})$ is the corresponding variance for the bootstrap samples with
\begin{equation}\label{equation: oracle variance bootstrap}
	v^2(\alpha,\tilde{\btau}):=(1-\alpha)^2\text{Var}[e_i^{b}(\tilde{\btau})]+\alpha^2-2\alpha(1-\alpha)\text{Cov}(e_i^{b}(\tilde{\btau}),e^b_i).
\end{equation}
Note that for bootstrap, the calculation or estimation of $v^2(\alpha,\tilde{\btau})$  is not a difficult task since we use $N(0,1)$ as the error term. For example, when $\tilde{\tau}=0.5$, it has an explicit form of 
\begin{equation*}
	v^2(\alpha,\tilde{\btau})=(1-\alpha)^2\sum\limits_{k_1=1}^{K}\sum\limits_{k_2=1}^{K}\gamma_{k_1k_2}+\alpha^2-\alpha(1-\alpha)\sqrt{\dfrac{2}{\pi}}.
\end{equation*}
Hence, for simplicity, we directly use the oracle variance $v^2(\alpha,\tilde{\btau})$ in (\ref{equation: bootstrap score CUSUM}). Using ${\bC}^b_\alpha(t;\tilde{\btau})$ and for a user specified $s_0$, we define the $b$-th bootstrap version of the individual-based testing statistic $T_{\alpha}$ as
\begin{equation}\label{equation: bootstrap individual test}
	T_\alpha^b=\max_{q_0\leq t\leq 1-q_0}\Big\|\bC^b_{\alpha}(t;\tilde{\btau})\Big\|_{(s_0,2)},~~\text{with}~~\alpha\in[0,1].
\end{equation}
Let $\gamma\in(0,0.5)$ be the significance level.
For each individual-based  testing statistic $T_\alpha$, let $F_{\alpha}=\P(T_{\alpha}\leq t)$ be its theoretical CDF and $P_\alpha=1-F_{\alpha}(T_{\alpha})$ be its theoretical $p$-value. Using the bootstrap samples $\{T_{\alpha}^1,\ldots,T_{\alpha}^B\}$, we estimate $P_\alpha$ by 
{
	
	\begin{equation}\label{equation: p-value for individual test}
		\hat{P}_{\alpha}=\dfrac{\sum_{b=1}^B\mathbf{1}\{T_{\alpha}^b> T_{\alpha}|\cX,\cY\}}{B+1},~~\text{with}~~\alpha\in[0,1].
	\end{equation}
}
Given the significance level $\gamma$, we can construct the individual test as
\begin{equation}\label{equation: final individual test}
	\Psi_{\gamma,\alpha}=\mathbf{1}\{	\hat{P}_{\alpha}\leq \gamma\}, ~~\text{with}~~\alpha\in[0,1].
\end{equation}
For each $T_{\alpha}$, we reject $\Hb_0$ if and only if $\Psi_{\gamma,\alpha}=1$. Note that  the above bootstrap procedure is  easy to implement since it does not require any model fitting such as obtaining the LASSO estimators which  is required by the data-based testing statistic $T_{\alpha}$.  
\subsection{Constructing the tail-adaptive testing procedure }\label{sec: method of tail-adaptive}

In Sections \ref{sec: method of single cpt} -- \ref{sec: method of bootstrap}, we propose a family of  individual-based testing statistics $\{T_{\alpha},\alpha\in[0,1]\}$ and introduce a bootstrap-based procedure for approximating their theoretical $p$-values. As discussed in Sections \ref{sec: method of single cpt} and seen from Figure \ref{figure: snr}, $T_{\alpha}$ with different $\alpha$'s can have various power performance  for a given underlying error distribution. For example, $T_{\alpha}$ with a larger $\alpha$ (e.g. $\alpha=1$) is more sensitive to change points with  light-tailed error distributions by using more moment information. In contrast, $T_{\alpha}$ with a smaller $\alpha$ (e.g. $\alpha=0,0.1$) is more powerful for data with heavy tails such as Student's $t_v$ or even Cauchy distribution. In general, as shown in Figure \ref{figure: snr}, a properly chosen $\alpha$ can give the most satisfactory power performance for data with a particular magnitude of tails. In practice, however, the tail structures of data are typically unknown. Hence, it is desirable to construct a tail-adaptive method which is simultaneously powerful under various tail structures of data. One candidate method is to find $\alpha^*$ which maximizes the theoretical $SNR(\alpha,\tilde{\btau})$, i.e. $\alpha^*=\argmax_{\alpha}SNR(\alpha,\tilde{\btau})$, and constructs a corresponding individual testing statistic $T_{\alpha^*}$. Note that in theory, calculating $SNR(\alpha,\tilde{\btau})$ needs to know  $\sigma(\alpha,\tilde{\btau})$ and $\{f_{\epsilon}(b_{k}^{(0)}),k=1,\ldots,K\}$, which could be difficult to estimate  especially under the high-dimensional change point model. Instead, we construct our tail-adaptive method by combining all candidate individual tests for yielding a powerful one. In particular, as a small $p$-value leads to rejection of $\Hb_0$, for the individual tests $T_{\alpha}$ with $\alpha\in[0,1]$, we construct the tail-adaptive testing statistic as their minimum $p$-value, which is defined as follows:
\begin{equation}\label{equation: tail-adaptive test statistic}
	T_{\rm ad}=\min_{\alpha\in\cA}	\hat{P}_{\alpha},
\end{equation}
where $\hat{P}_{\alpha}$ is  defined in (\ref{equation: p-value for individual test}), and $\cA$ is a candidate subset of $\alpha$. 

\begin{remark}
	{	It's worth noting that for the construction of $T_{\rm ad}$, the issue of selecting $K$ in the composite quantile regression is typically left to the user. Based on our extensive numerical studies, choosing $K=1$ and $\tau=0.5$ has  satisfactory performance across data with various tail structures.}
\end{remark}

In this paper, we require $|\cA|$ to be  finite, which is a theoretical  requirement. Note that our tail-adaptive method is flexible and user-friendly.  In practice, if the users have some prior knowledge about the tails of errors, we can choose $\cA$ accordingly.  For example, we can choose $\mathcal{A}=\{0.9,1\}$ for light-tailed errors, and $\mathcal{A}=\{0\}$ for extreme heavy tails such as Cauchy distributions. However, if the tail structure is unknown, we can choose $\cA$ consisting both small and large values of $\alpha\in[0,1]$. For example, according to our theoretical analysis of $SNR(\alpha,\tilde{\btau})$, we find that $SNR(\alpha,\tilde{\btau})$ tends to be maximized near the boundary of $[0,1]$. Hence, we recommend to use $\cA=\{0,0.1,0.5,0.9,1\}$ in real applications, which is shown by our numerical studies to enjoy stable size performance as well as high powers across various error distributions. 
Let $F_{\rm ad}(x)$  be its theoretical distribution function. Note that  $F_{\rm ad}(x)$ is unknown. Hence, we can not use $T_{\rm ad}$ directly for  Problem \eqref{hypothesis: H0}. To approximate its theoretical $p$-value, we adopt the low-cost bootstrap method proposed by \cite{Zhou2017An}, which is also used in \cite{Liu2019Unified}. 
Let $\hat{P}_{\rm ad}$ be an estimation for the theoretical $p$-value of $T_{\rm ad}$
using the low-cost bootstrap. Given the significance level $\gamma\in(0,0.5)$, define the final  tail-adaptive test:
\begin{equation}\label{statistic: adaptive tests}
	\begin{array}{lc}
		\Psi_{\gamma,\rm ad}=\mathbf{1}\{\hat{P}_{\rm ad}\leq \gamma  \}.
	\end{array}
\end{equation}
For the tail-adaptive testing procedure, given  $\gamma$, we reject $\Hb_0$ if $\Psi_{\gamma,\rm ad}=1$. 

\section{Theoretical results}\label{sec: theory}
In this section, we give   some theoretical results. In Section \ref{sec: basic assumptions}, we provide some basic model assumptions. In Sections \ref{sec: Theoretical results of the individual test statistics} and  \ref{sec: Theoretical results of the adaptive test statistics}, we discuss the theoretical properties of the individual  and tail-adaptive methods.


\subsection{Basic assumptions}\label{sec: basic assumptions}

We introduce some basic assumptions for deriving our main theorems. Before that, we introduce some notations. Let $ e_i(\tilde{\btau}):=K^{-1}\sum\limits_{k=1}^{K}\big(\mathbf{1}\{\epsilon_i\leq b^{(0)}_k\}-\tau_k\big):=K^{-1}\sum\limits_{k=1}^{K}e_i(\tau_k)$.  We set $\mathcal{V}_{s_0}:=\{\bv\in \mathbb{S}^{p}: \|\bv\|_0\leq s_0\}$, where $\mathbb{S}^{p}:=\{\bv\in\RR^p: \|\bv\|=1\}$. For each $\alpha\in[0,1]$, we introduce $\uwave{\bbeta}^*=((\bbeta^*)^\top,(\bb^*)^\top)^\top\in \RR^{p+K}$ with $\bbeta^*\in \RR^p$,  $\bb^*=(b_1^*,\ldots,b_K^*)^\top\in \RR^K$, where 
\begin{equation}
	\uwave{\bbeta}^*:=\argmin_{\bbeta\in \RR^p,\bb\in \RR^K}\E \Big[(1-\alpha)\dfrac{1}{n}\sum_{i=1}^{n}\dfrac{1}{K}\sum_{k=1}^{K}\rho_{\tau_k}(Y_i-b_i-\bX_i^\top\bbeta)+ \dfrac{\alpha}{2n}\sum_{i=1}^{n}(Y_i-\bX_i^\top \bbeta)^2\Big].
\end{equation}
Note that by definition, we can regard $\uwave{\bbeta}^*$ as the true parameters under the population level with pooled samples. We can prove that under $\Hb_0$, $\uwave{\bbeta}^*=((\bbeta^{(0)})^\top,(\bb^{(0)})^\top)^\top$ with $\bb^{(0)}=(b_{1}^{(0)},\ldots,b_{K}^{(0)})^\top$. Under $\Hb_1$, $\uwave{\bbeta}^*$ is  generally a weighted combination of the parameters before the change point and those after the change point. For example, when $\alpha=1$, it has the explicit form of $ \uwave{\bbeta}^*=((t_1\bbeta^{(1)}+t_2\bbeta^{(2)})^\top,(\bb^{(0)})^\top)^\top$. More discussions about $\uwave{\bbeta}^*$ under our weighted composite loss function are provided in the appendix. With the above notations, we are ready to introduce our assumptions as follows:\\
{\bf{Assumption A (Design matrix):}} The design matrix $\Xb$ has i.i.d rows $\{\bX_i\}_{i=1}^n$. 
(A.1) Assume that there are positive constants $\kappa_1$ and $\kappa_2$ such that $\lambda_{\rm min}(\bSigma)\geq\kappa_1>0$ and $\lambda_{\rm max}(\bSigma)\leq \kappa_2<\infty$ hold. (A.2) There exists some constant $M\geq 1$ such that $\max_{1\leq i\leq n}\max_{1\leq j\leq p}|X_{ij}|\leq M$ almost surely {for every $n$ and $p$.}\\
{\bf{Assumption B} (Error distribution):} The error terms $\{\epsilon_i\}_{i=1}^n$ are i.i.d. with mean zero and finite variance $\sigma_{\epsilon}^2$. There exist positive constants 
$c_{\epsilon}$ and $C_{\epsilon}$ such that $c^2_\epsilon\leq \text{Var}(\epsilon_i)\leq C^2_{\epsilon}$ hold. In addition, $\epsilon_i$ is independent with $\bX_i$ for $i=1,\ldots,n$.\\
{\bf{Assumption C (Moment constraints)}: }
(C.1) There exists some constant $b>0$ such that $\E(\bv^\top\bX_i\epsilon_i)^{2}\geq b$  and  $\E(\bv^\top\bX_ie_i(\tilde{\btau}))^{2}\geq b$,
for $\bv\in \cV_{s_0}$ and all $i=1,\ldots,n$. Moreover, assume that $\inf_{i,j}\E[X^2_{ij}]\geq b$ holds. 
(C.2) There exists a constant $K>0$ such that  $\E|\epsilon_i|^{2+\ell}\leq K^{\ell}$, for $\ell=1,2$.\\
\
{\bf{Assumption D (Underlying distribution):}} The  distribution function $\epsilon$ has a continuously  differentiable density function $f_{\epsilon}(t)$ whose derivative is denoted by $f'_{\epsilon}(t)$. Furthermore, suppose there exist some constants $C_{+}$, $C_{-}$ and $C'_{+}$ such that
\begin{equation*}
	\begin{array}{l}
		\text{(D.1)} \sup\limits_{t\in\RR}f_{\epsilon}(t)\leq C_{+}; 
		\text{(D.2)}  \inf\limits_{j=1,2}\inf\limits_{1\leq k\leq K} f_{\epsilon}(\bx^\top(\bbeta^*-\bbeta^{(j)})+b_k^*)\geq C_-;\\
		\text{(D.3)} \sup\limits_{t\in\RR}|f'_{\epsilon}(t)|\leq C'_{+}.
	\end{array}
\end{equation*}
{\bf{Assumption E (Parameter space):}} \\
(E.1) We require $s_0^3\log(pn)=O(n^{\xi_1})$ for some $0<\xi_1<1/7$ and $s_0^4\log(pn)=O(n^{\xi_2})$ for some $0<\xi_2<\frac{1}{6}$. \\
(E.2) Assume that $\dfrac{s_0^2s^3\log^3(pn)}{n}\rightarrow0$ as $(n,p)\rightarrow \infty$, where $s$ is the overall sparsity of $\bbeta^{(1)}$ and $\bbeta^{(2)}$.\\ (E.3) We require $\max (\|\bbeta^{(1)}\|_{\infty}, \|\bbeta^{(2)}\|_{\infty} )<C_{\bbeta}$ for some $C_{\bbeta}>0$. Moreover, we require $\|\bbeta^{(2)}-\bbeta^{(1)}\|_1\leq C_{\bDelta}$ for some constant $C_{\bDelta}>0$.\\
(E.4) For the tuning parameters $\lambda_{\alpha}$ in (\ref{equation: lasso estimator}), we require $\lambda_{\alpha}=C_{\lambda}\sqrt{\log(pn)/n}$ for some $C_\lambda>0$.

Assumption A gives some conditions for the design matrix, requiring  $\bX$ has a non-degenerate covariance matrix $\bSigma$ in terms of its eigenvalues. This is important for deriving the  high-dimensional LASSO property with $\alpha\in[0,1]$ under both $\Hb_0$ and $\Hb_1$.   Assumption B mainly requires the underlying error term $\epsilon_i$ has non-degenerate variance. Assumption C imposes some restrictions on the moments of the error terms as well as the design matrix. In particular, Assumption C.1 requires that $\bv^\top\bX\epsilon$, $\bv^\top\bX e(\tilde{\btau})$, as well as $X_{ij}$ have non-degenerate variances. Moreover, Assumption C.2 requires that the errors have at most  fourth moments, which is much weaker than the commonly used Gaussian or sub-Gaussian assumptions. Both Assumptions C.1 and C.2 are basic moment conditions for  bootstrap approximations for the individual-based tests. See Lemma C.6 in the proof.  Assumptions D.1 - D.3 are some regular conditions for the underlying distribution of the errors, requiring $\epsilon$ has a bounded density function as well as bounded derivatives. Assumption D.2 also requires the density function at $\bx^\top(\bbeta^*-\bbeta^{(j)})+b_k^*$ to be strictly bounded away from zero.  Lastly, Assumption E imposes some conditions for the parameter spaces in terms of $(s_0,n,p,s,\bbeta^{(1)},\bbeta^{(2)})$. Specifically, Assumption E.1 scales the relationship between $s_0$, $n$, and $p$, which allows $s_0$ can grow with the sample size $n$. {This condition is mainly used to establish the high-dimensional Gaussian approximation for our individual tests.}   Assumption E.2 also gives some restrictions on $(s_0,s,n,p)$. Note that both Assumptions E.1 and E.2 allow the data dimension $p$ to be much larger than the sample size $n$  as long as the required conditions hold. Assumption E.3 requires that the regression coefficients as well as signal jump  in terms of its $\ell_1$-norm are bounded. Assumption E.4 imposes the regularization parameter $\lambda_\alpha=O(\sqrt{\log(pn)/n})$, which is important for deriving the desired error bound for the LASSO estimators under both $\Hb_0$ and $\Hb_1$ using our weighted composite loss function. See Lemmas C.9 - C.11 in the proof.

\begin{remark}
	Assumption C.2 with the finite  fourth moment is mainly for the individual test with $\alpha=1$, while Assumption D is for that with $\alpha=0$. Note that Assumption D only imposes some conditions on the density functions of the errors instead of the moments, which can be statisfied for the errors with heavy tails. Hence, in both cases, our proposed individual-based change point method extends the high-dimensional linear models with sub-Gaussian distributed errors to those with only finite moments or without any moments, covering a wide range of errors with different tails. 
\end{remark}

\subsection{Theoretical results of the individual-based testing statistics}\label{sec: Theoretical results of the individual test statistics}
\subsubsection{Validity of controlling the testing size}
Before giving the size results, we first consider the variance estimation. Recall $\sigma^2(\alpha,\tilde{\btau})$ in (\ref{equation: orcale sigma square}) and $\hat{\sigma}^2(\alpha,\tilde{\btau})$ in (\ref{equation: variance estimator}).
Theorem \ref{theorem: variance estimator under H0} shows that the pooled weighted variance estimator $\hat{\sigma}^2(\alpha,\tilde{\btau})$ is consistent under $\Hb_0$, which is crucial for deriving the Gaussian approximation results as shown in Theorem \ref{theorem: size control for individual test} and shows that our testing method has satisfactory size performance.
\begin{theorem}\label{theorem: variance estimator under H0}
	For $\alpha=1$,	suppose  Assumptions A, B, C, E hold. For $\alpha=0$,	suppose {Assumptions A,  C.1, D, E} hold. For $\alpha\in(0,1)$, 	suppose { Assumptions A - E} hold.
	Let $r_{\alpha}(n)=\sqrt{s\log(pn)/n}$ if $\alpha=1$ and $r_{\alpha}(n)=s\sqrt{\dfrac{\log(pn)}{n}}\vee s^{\frac{1}{2}}(\dfrac{\log(pn)}{n})^{\frac{3}{8}}$ if $\alpha\in[0,1)$.
	Under $\Hb_0$, for  $\alpha\in[0,1]$, we have 
	\begin{equation*}\label{equation: variance estimation  under H0}
		|\hat{\sigma}^2(\alpha,\tilde{\btau})-{\sigma}^2(\alpha,\tilde{\btau})|=O_p(r_{\alpha}(n)).
	\end{equation*}
\end{theorem}
Based on Theorem \ref{theorem: variance estimator under H0} as well as some other regularity conditions, the following Theorem \ref{theorem: size control for individual test} justifies the validity of our bootstrap-based procedure. 
\begin{theorem}\label{theorem: size control for individual test}
	Suppose the assumptions in Theorem \ref{theorem: variance estimator under H0} hold.
	Then, under $\Hb_0$, for the  individual test with $\alpha\in[0,1]$, we have
	\begin{equation}\label{equation: size individual test }
		\sup_{z\in (0,\infty)}\big|\P(T_{\alpha}\leq z)-\P(T_{\alpha}^{b}\leq z|\mathcal{X},\cY)\big|=o_p(1), ~\text{as}~  n,p\rightarrow\infty.
	\end{equation}
\end{theorem}
\noindent Theorem \ref{theorem: size control for individual test} demonstrates that we can uniformly approximate the distribution of $T_{\alpha}$ by that of $T_{\alpha}^b$. The following Corollary further shows that our proposed new test $\Psi_{\gamma,\alpha}$ can control the Type I error asymptotically for any given  significant level $\gamma\in(0,1)$.
\begin{corollary}\label{corollary: size}
	Suppose assumptions in Theorem \ref{theorem: size control for individual test} hold. Under $\Hb_0$, we have
	\begin{equation*}
		\P(\Psi_{\gamma,\alpha}=1)\rightarrow\gamma, ~ \text{as} ~n,p,B\rightarrow\infty.
	\end{equation*}
\end{corollary}	
\subsubsection{Change point estimation}
We next consider the performance of the individual test under $\Hb_1$. We first give some theoretical results on the change point estimation. To that end, some additional assumptions are needed. Recall $\Pi=\{j: \beta^{(1)}_j\neq \beta^{(2)}_j\}$ as the set with change points. For $j\in\{1,\ldots,p\}$, define the signal jump $\bDelta=(\Delta_1,\ldots,\Delta_p)^\top$ with $\Delta_j:=\beta^{(1)}_j-\beta^{(2)}_j$. Let $\Delta_{\min}=\min_{j\in\Pi}|\Delta_j|$ and $\Delta_{\max}=\max_{j\in \Pi}|\Delta_j|$.  We now introduce the following {{Assumption F}}.\\
{\bf{Assumption F}}. There exist constants $\underline{c}>0$ and $\overline{C}>0$ such that 
\begin{equation}\label{inequality: ratio of maximum and minimum signal}
	0<\underline{c}\leq \liminf\limits_{p\rightarrow\infty} \dfrac{\Delta_{\min}}{\Delta_{\max}} \leq\limsup\limits_{p\rightarrow\infty}\dfrac{\Delta_{\max}}{\Delta_{\min}}\leq \overline{C}<\infty.
\end{equation}
Note that Assumption F is only a technical condition requiring that $\Delta_{\min}$ and $\Delta_{\max}$ are of the same order.
Theorem \ref{theorem: cpt estimation results} provides a non-asymptotic estimation error bound of  the argmax-based individual  estimators. 
\begin{theorem}\label{theorem: cpt estimation results}
	Suppose $\|\bDelta\|_{(s_0,2)}\gg \sqrt{\log(pn)/n}$  and  {Assumption F} hold.  
	For $\alpha=1$, suppose  { Assumptions A, B, C, E.2 - E.4 } as well as $n^{1/4}=o(s)$ hold;
	For $\alpha=0$, suppose  { Assumptions A,  C.1, D, E.2 - E.4} as well as
		$	\lim\limits_{n,p\rightarrow\infty}s_0^{1/2}s^2\sqrt{\log(p)/n}\|\bDelta\|_{(s_0,2)}=0$
	hold;
	For $\alpha\in(0,1)$, suppose { Assumptions A, B, C, D, E.2 - E.4 } as well as $n^{1/4}=o(s)$ and  $\lim\limits_{n,p\rightarrow\infty}s_0^{1/2}s^2\sqrt{\log(pn)/n}\|\bDelta\|_{(s_0,2)}=0$ hold.
Then,  under $\Hb_1$,  for each $\alpha\in[0,1]$, with probability tending to one, we have 	
\begin{equation}\label{inequality: estimation error bound of cpt}
	\big|\hat{t}_{\alpha}-t_1\big|\leq C^*(s_0,\tilde{\btau},\alpha)\dfrac{\log(pn)}{nSNR^2(\alpha,\tilde{\btau})\|\bSigma\bDelta\|^2_{(s_0,2)}},
\end{equation}
where $C^*(s_0,\tilde{\btau},\alpha)>0$ is some  constant only depending on $s_0, \tilde{\btau}$ and $\alpha$.	
\end{theorem}
\begin{remark}
Theorem \ref{theorem: cpt estimation results} shows that our individual estimators are consistent under the condition $\|\bDelta\|_{(s_0,2)}\gg \sqrt{\log(pn)/n}$. Moreover,
according to \cite{rinaldo2021localizing}, for  high-dimensional linear models, under Assumption F, if $\|\bDelta\|_{\infty}\gg 1/\sqrt{n} $, any change point estimator $\hat{t}$ has  an estimation lower bound $|\hat{t}-t_1|\geq c_*\dfrac{1}{n\|\bDelta\|^2_{\infty}}$, for some constant $c_*>0$. Hence, considering (\ref{inequality: ratio of maximum and minimum signal}) and (\ref{inequality: estimation error bound of cpt}), with a fixed $s_0$, Theorem  \ref{theorem: cpt estimation results} demonstrates that our individual-based estimators are rate optimal up to a $\log(pn)$ factor.
\end{remark}



\subsubsection{Power performance}
We discuss the power properties of the individual tests.  Note that for the change point problem, variance estimation under the alternative is a difficult but important task. As pointed out in \cite{shao2010testing}, due to the unknown change point, any improper estimation may lead to nonmonotonic power performance. This distinguishes the change point problem substantially from one-sample or two-sample 
tests where homogenous data are used to construct consistent variance estimation. Hence, for yielding a powerful change point test, we need to guarantee a consistent variance estimation.
Theorem \ref{theorem: variance estimator under H1} shows that the  pooled weighted variance estimation is consistent under $\Hb_1$. This guarantees that our proposed testing method has reasonable power performance. 
\begin{theorem}\label{theorem: variance estimator under H1}
Suppose the assumptions in Theorem \ref{theorem: cpt estimation results} hold. 
Let $r_{\alpha}(n)=\sqrt{s\log(pn)/n}$ if $\alpha=1$ and $r_{\alpha}(n)=s\sqrt{\dfrac{\log(pn)}{n}}\vee s^{\frac{1}{2}}(\dfrac{\log(pn)}{n})^{\frac{3}{8}}$ if $\alpha\in[0,1)$.
Under $\Hb_1$, for each $\alpha\in[0,1]$, we have
\begin{equation*}\label{equation: variance estimation  under H1}
	|\hat{\sigma}^2(\alpha,\tilde{\btau})-{\sigma}^2(\alpha,\tilde{\btau})|=O_p(r_{\alpha}(n)).
\end{equation*}
\end{theorem}

Using the consistent variance estimation, we are able to discuss the power properties of the individual tests. 
Define the oracle signal to noise ratio vector $\bD=(D_1,\ldots,D_{p})^\top$ with
\begin{equation}\label{equation: signal to noise ration}
D_j:=\left\{\begin{array}{ll}
	0,& \text{for}~~ j\in \Pi^c\\
	SNR(\alpha,\tilde{\btau})\times\Big|{t_1(1-t_1)\big(\bSigma(\bbeta^{(1)}-\bbeta^{(2)})\big)_j}\Big|,& \text{for}~~j\in \Pi,
\end{array}\right.
\end{equation}  
where $SNR(\alpha,\tilde{\btau})$ is defined in (\ref{equation: theoretical SNR}).
Theorem \ref{theorem: power control for individual test} stated below  shows that we can reject the null hypothesis  with probability tending to $1$.
\begin{theorem}\label{theorem: power control for individual test} 
Let $\epsilon_n:=O(s^{1/2}_0s\sqrt{{\log (pn)}/{n}}) \vee O({s_0^{1/2}s^2\sqrt{{\log (pn)}/{n}}\|\bDelta\|_{(s_0,2)}})$. For each $\alpha\in[0,1]$, assume the following conditions hold: When $\alpha=1$, suppose that { Assumptions A, B, C, E.2 - E.4 } hold; When $\alpha=0$, suppose that  {Assumptions A,  C.1, D, E.2 - E.4} as well as
\begin{equation}\label{euqation: condion for alpha=0}			\lim\limits_{n,p\rightarrow\infty}s_0^{1/2}s^2\sqrt{\log(p)/n}\|\bDelta\|_{(s_0,2)}=0
\end{equation}
hold; When $\alpha\in(0,1)$, suppose that {Assumptions A - D, E.2 - E.4 } as well as  (\ref{euqation: condion for alpha=0}) hold. Under $\Hb_1$, if $\bD$ in (\ref{equation: signal to noise ration}) satisfies
\begin{equation}\label{inequality: theoretical signal strengh}
	\sqrt{n}\times \|\bD\|_{(s_0,2)}\geq \dfrac{C(\tilde{\btau},\alpha)}{1-\epsilon_n}s^{1/2}_0\big(\sqrt{\log(pn)}+\sqrt{\log(1/\gamma)}\big),
\end{equation}
then we have
\begin{equation*}
	\P(\Phi_{\gamma,\alpha}=1)\rightarrow 1, \text{as}~n,p,B\rightarrow\infty,
\end{equation*}
where $C(\tilde{\btau},\alpha)$ is some  positive constant only depending on $\tilde{\btau}$ and $\alpha$.
\end{theorem}

Theorem \ref{theorem: power control for individual test}  demonstrates that with probability tending to one, our proposed individual test with $\alpha\in[0,1]$ can detect the existence of a change point for high-dimensional linear models as long as the corresponding signal to noise ratio satisfies (\ref{inequality: theoretical signal strengh}). Combining (\ref{equation: signal to noise ration}) and (\ref{inequality: theoretical signal strengh}), for each individual test, we note that with a larger signal jump and a closer change point location $t_1$ to the middle of data observations, it is more likely to trigger a rejection of  the null hypothesis. More importantly, considering $\epsilon_{n}=o(1)$, Theorem \ref{theorem: power control for individual test} illustrates that for consistently detecting a change point, we require the signal to noise ratio vector to be at least an order of  $\|\bD\|_{(s_0,2)}\asymp s_0^{1/2}\sqrt{\log(pn)/n}$, which is particularly interesting to further discuss under several special cases. For example, if we choose $s_0=1$ and $\alpha=1$, our proposed individual test reduces to the least squared loss based testing statistic with the $\ell_\infty$-norm aggregation. In this case, we require $\|\bD\|_{\infty}\asymp \sqrt{\log(pn)/n}$ for detecting a change point. If we choose $\alpha=0$ with the composite quantile loss, the test is still consistent as long as $\|\bD\|_{\infty}\asymp \sqrt{\log(pn)/n}$. Note that the latter one is of special interest for the robust change point detection. Hence, our theorem provides the unified  condition for detecting a change point under a general framework, which may be of independent interest.  Moreover, Theorem \ref{theorem: power control for individual test} reveals that for detecting a change point, our individual-based method with $\alpha\in[0,1]$ can account for the tails of the data. For Model \eqref{equation: single cpt model} with a fixed signal jump $\bDelta$ and a change point location $t_1$, considering (\ref{equation: signal to noise ration}) and (\ref{inequality: theoretical signal strengh}), the individual test $T_{\alpha}$ is more powerful with a larger $SNR(\alpha, \tilde{\btau})$. 

\subsection{Theoretical results of the tail-adaptive testing statistics}\label{sec: Theoretical results of the adaptive test statistics}

In this section, we  discuss the size and power properties of the tail-adaptive test $\Psi_{\gamma,\rm ad}$ defined in (\ref{statistic: adaptive tests}). To present the theorems,  we need additional notations. Let
$F_{T_\alpha}(x):=\P(T_{\alpha}\leq x)$ be the CDF of $T_{\alpha}$. Then $\hat{P}_{\alpha}$  in (\ref{equation: p-value for individual test}) approximates the following individual tests' theoretical $P$-values defined as $P_{\alpha}:=1-F_{T_{\alpha}}(T_{\alpha})$. Hence, based on the above  theoretical $P$-values,
we can define the oracle tail-adaptive testing statistic
$\tilde{T}_{\rm ad}=\min_{\alpha\in\cA}P_{\alpha}.$
Let $\tilde{F}_{T,\rm ad}(x):=\P(\tilde{T}_{\rm ad}\leq x)$  be the CDF of $\tilde{T}_{\rm ad}$. Then we can also define the theoretical tail-adaptive test's $P$-value as
$\tilde{P}_{\rm ad}:=\tilde{F}_{T,\rm ad}(\tilde{T}_{\rm ad})$.
Recall $\hat{P}_{\rm ad}$ be the low cost bootstrap $P$-value for $\tilde{P}_{\rm ad}$.  In what follows, we show that   $\hat{P}_{\rm ad}$ converges to $\tilde{P}_{\rm ad}$  in probability as $n,p,B\rightarrow\infty$. 

We introduce Assumption $\mathbf{E.1'}$ to describe the scaling relationships among $n$, $p$, and $s_0$.
Let $\bG_{i}=(G_{i1},\ldots,G_{ip})^\top$ with $\bG_i\sim N(\mathbf{0},\bSigma)$  being i.i.d. Gaussian random vectors, where $\bSigma:=\text{Cov}(\bX_1)$.  Define 
\begin{equation*}
\bC^{\bG}(t)=\dfrac{1}{\sqrt{n}}\big(\sum_{i=1}^{\floor{nt}}\bG_i-\dfrac{\floor{nt}}{n}\sum_{i=1}^n\bG_i\big) ~~\text{and}~~T^{\bG}=\max_{q_0\leq t\leq 1-q_0}\| \bC^{\bG}(t)\|_{(s_0,2)}.
\end{equation*}
As shown in the proof of Theorem \ref{theorem: size control for individual test}, 
we use  $T^{\bG}=\max\limits_{q_0\leq t\leq 1-q_0}\big\|\bC^{\bG}(t)\big\|_{(s_0,2)}$ to approximate $T_{\alpha}$. For $T^{\bG}$, let $f_{T^{\bG}}(x)$ and $c_{T^{\bG}}(\gamma)$ be the probability density function (pdf), and the $\gamma$-quantile of $T^{\bG}$, respectively. We then define $h(\epsilon)$ as $h(\epsilon)=\max_{x\in I(\epsilon)}f_{T^{\bG}}^{-1}(x)$,
where $I(\epsilon):=[c_{T^{\bG}}(\epsilon),c_{T^{\bG}}(1-\epsilon)]$.

With the above definitions, we now introduce Assumption $\mathbf{E.1'}$:
\begin{description}
\item [(E.1)$'$] For any $0<\epsilon< 1$, we require 
$ h^{0.6}(\epsilon)s_0^3\log(pn)=o(n^{1/10})$.
\end{description}
Note that Assumption $\mathbf{E.1'}$ is more stringent than Assumption $\bf{E.1}$. The intuition of Assumption $\mathbf{E.1'}$ is that, we construct our tail-adaptive testing statistic by taking the minimum $P$-values of the individual tests. For analyzing the combinational tests, we need not only the uniform convergence of the distribution functions, but also the uniform convergence of their quantiles on $[\epsilon,1-\epsilon]$ for any $0<\epsilon<1$.

The following Theorem \ref{theorem: adaptive size}  justifies the validity of the low-cost bootstrap procedure in Section \ref{sec: method of tail-adaptive}.
\begin{theorem}\label{theorem: adaptive size}
For $T_{\rm ad}$, suppose  {Assumptions A - D, E.1$'$, E.2 - E.4 } hold. Under $\Hb_0$, we have
\begin{equation*}\label{equation: adaptive size1}
	\begin{array}{lc}
		\P(\Psi_{\gamma,\rm ad}=1)\rightarrow \gamma,&\text{and}~~
		\hat{P}_{\rm ad}-\tilde{P}_{\rm ad}\xrightarrow{\P}0,~\text{as}~n,p,B\rightarrow\infty.
	\end{array}
\end{equation*}
\end{theorem}
We now discuss the power.  Theorem \ref{theorem: adaptive power} shows that under some regularity conditions, our tail-adaptive test has its power converging to one.
\begin{theorem}\label{theorem: adaptive power}
Let $\epsilon_n:=O(s^{1/2}_0s\sqrt{{\log (pn)}/{n}}) \vee O({s_0^{1/2}s^2\sqrt{{\log (pn)}/{n}}\|\bDelta\|_{(s_0,2)}})$.
Suppose    {Assumptions A - D, E.2 - E.4 } as well as   $\lim\limits_{n,p\rightarrow\infty}s_0^{1/2}s^2\sqrt{\log(p)/n}\|\bDelta\|_{(s_0,2)}=0$ hold. 
If $\Hb_1$ holds with 
\begin{equation}\label{inequality: theoretical signal strengh2}
	\sqrt{n}\times \|\bD\|_{(s_0,2)}\geq \dfrac{C(\tilde{\btau},\alpha)}{1-\epsilon_n}s^{1/2}_0\big(\sqrt{\log(pn)}+\sqrt{\log(|\cA|/\gamma)}\big),
\end{equation}
then for $T_{\rm ad}$, we have
\begin{equation*}\label{equation: adaptive power }
	\begin{array}{lc}
		\P(\Psi_{\gamma,\rm ad}=1)\rightarrow 1 ~\text{as}~n,p,B\rightarrow\infty,
	\end{array}	
\end{equation*}
where $C(\tilde{\btau},\alpha)$ is some positive constant only depending on $\tilde{\btau}$ and $\alpha$.	
\end{theorem}

Note that based on the theoretical results obtained in Section \ref{sec: Theoretical results of the individual test statistics}, Theorems \ref{theorem: adaptive size} and \ref{theorem: adaptive power} can be proved using some modifications of the proofs of Theorems 3.5 and 3.7  in \cite{Zhou2017An}. Hence, we omit the detailed proofs for brevity. Lastly, recall the tail-adaptive based change point estimator $\hat{t}_{\rm ad} = \widehat{t}_{\widehat{\alpha}}$ with $\widehat{\alpha} = \argmin_{\alpha \in \mathcal{A}} \hat{P}_{\alpha}$.  According to Theorem \ref{theorem: cpt estimation results}, the tail-adaptive estimator is also consistent. 

\section{Numerical experiments}\label{section: simulation studies}
In this section, we investigate the numerical performance of our proposed method and compare with the existing techniques in terms of change point detection and identification. In Sections \ref{sec: single testing} - \ref{sec: empirical powers}, we consider single change point testing and estimation. In Section \ref{sec: multiple detection}, we investigate multiple change points detection.
\subsection{Single change point testing}\label{sec: single testing}
We consider the performace of singe change point testing  for the following model:
\begin{equation}\label{equation: single cpt model simulation}
	Y_i=\bX_i^\top\bbeta^{(1)}\mathbf{1}\{i\leq k_1\}+\bX_i^\top\bbeta^{(2)}\mathbf{1}\{i> k_1\}+\epsilon_i,~~i=1,\ldots,n.
\end{equation}
where $k_1=\floor {nt_1}$. To show the broad applicability of our method, we generate data from various model settings. Specifically, for the design matrix $\Xb$, we generate $\bX_i$ (i.i.d) from $N(\mathbf{0},\bSigma)$ under two different models:
\begin{itemize}
	\item \textbf{Model~1:} We generate $\bX_i$ with banded $\bSigma$. Specifically, we set $\bSigma=\bSigma'$, where $\bSigma'=(\sigma'_{ij})\in \mathbb{R}^{p\times p}$ with $\sigma'_{ij}=0.8^{|i-j|}$ for $1\leq i,j\leq p$. 
	\item \textbf{Model~2:} We generate $\bX_i$ with blocked $\bSigma$. Specifically, we set $\bSigma=\bSigma^\star$, where $\bSigma^\star=(\sigma^*_{ij})\in \mathbb{R}^{p\times p}$  with $\sigma^\star_{ii}\overset{\rm i.i.d.}{\sim} {\rm U}(1,2)$,  $\sigma^\star_{ij}=0.6$ for
	$5(k-1)+1\le i\ne j\le 5k$ ($k=1,\ldots,\lfloor p/5\rfloor$), and $\sigma^\star_{ij} = 0$ otherwise.  
\end{itemize}
Moreover, to show the tail-adaptivity of our new testing method, we generate the error term $\epsilon_i$ from  various types of distributions including both lighted-tailed and heavy-tailed distributions. In particular, we generate $\epsilon_i$ from the Gaussian distribution $N(0,1)$  and the Student's $t_v$ distribution with a degree of freedom $v\in \{1,2,3,4\}$. Note that $t_v$ with $v=2$ and $v=1$ corresponds to the error without second moments and first moments, respectively. For the regression coefficient $\bbeta^{(1)}$, for each replication, we generate $\bbeta^{(1)}=(1,1,1,1,1,0,\cdots,0)^\top\in\RR^p$. In other words, only the first five elements in $\bbeta^{(1)}$ are non-zero with magitudes of ones, which are called the active set. Under $\Hb_0$, we set $\bbeta^{(2)}=\bbeta^{(1)}:=\bbeta^{(0)}$. Under $\Hb_1$,  we set $\bbeta^{(2)}=\bbeta^{(1)}+\bdelta$, where $\bdelta=(\delta_1,\ldots,\delta_p)^\top\in\RR^p$ is the signal jump with 
\begin{equation*}
	\delta_s=
	\left\{\begin{array}{ll}
		c\sqrt{\log(p)/n},&\text{for}~s\in\{1,2,3,4,5\},\\
		0,&\text{for}~s\in\{6,\ldots,p\}.
	\end{array}\right.
\end{equation*}
In other words, we add a signal jump with a magnitude of $c\sqrt{\log(p)/n}$ on the first five elements of $\bbeta^{(1)}$. To avoid the trivial power performance (too low or high powers), we set $c=1$ and $c=1.5$ for the normal and the Student's $t$  distributions, respectively. 

Throughout the simulations, we fix the sample size  at $n=200$ and  the dimension at $p=400$. The number of bootstrap replications is $B=200$. Without additional specifications, all numerical results are based on 1000 replications. In addition, we consider the $L_1-L_2$ composite loss  by setting $\tilde{\tau}=0.5$ and $K=1$ in (\ref{equation: weighted loss function}), which is of special interest in high -dimensional data analysis. 
Note that our proposed method  involves  the  optimization problem in (\ref{equation: lasso estimator}). 
We use the coordinate descent algorithm for obtaining the corresponding LASSO estimators. As for  the tuning parameters $\lambda_{\alpha}$, for $\alpha=1$, we use the cross-validation technique to select the "best" $\lambda_1$; for $\alpha=0$, we adopt the method recommended in \cite{belloni2011} (see Section 2.3 therein) to set $\lambda_0$; for $\alpha \in(0,1)$, we use an idea of weighted combination and let $\lambda_{\alpha}=(1-\alpha)\lambda_0+\alpha\lambda_1$. 
\subsection{Empirical sizes}\label{sec: empirical sizes}
\begin{table}[!hpt]
	\small
	\caption{Empirical sizes of the individual and tail-adaptive tests for {\bf{Models 1-2}} with banded and blocked covariance matrices for $s_0\in\{1,3,5,7\}$. The results are based on 1000 replications with $B=200$ for each replication.}\label{table: sizes of individual band}
	\vspace{-0.3cm}
	\addtolength{\tabcolsep}{1.5pt}
	\begin{center}
		\begin{tabular}{cccc cccc}
			\toprule[2pt]
			\multicolumn{8}{c}{\bf{Empirical sizes for Model 1 with $p=400$}}\\
			\bf{Dist}	&$\bf{s_0}$&$\bf{\balpha=0}$&$\bf{\balpha=0.1}$&$\bf{\balpha=0.5}$&$\bf{\balpha=0.9}$&$\bf{\balpha=1}$&\bf{Adaptive}\\
			\hline
			$N(0,1)$	&	$s_0=1$	&	0.062 	&	0.041 	&	0.036 	&	0.033 	&	0.038 	&	0.040 
			\\
			&	$s_0=3$	&	0.052 	&	0.056 	&	0.041 	&	0.032 	&	0.034 	&	0.045 
			\\
			&	$s_0=5$	&	0.051 	&	0.053 	&	0.040 	&	0.032 	&	0.027 	&	0.040 
			\\
			&	$s_0=7$	&	0.050 	&	0.048 	&	0.041 	&	0.035 	&	0.027 	&	0.046 
			\\
			\cline{2-8}
			$t_4$	&	$s_0=1$	&	0.049 	&	0.056 	&	0.052 	&	0.048 	&	0.040 	&	0.062\\ 
			&	$s_0=3$	&	0.058 	&	0.057 	&	0.052 	&	0.050 	&	0.040 	&	0.051 
			\\
			&	$s_0=5$	&	0.049 	&	0.035 	&	0.041 	&	0.038 	&	0.035 	&	0.041 
			\\
			&	$s_0=7$	&	0.064 	&	0.043 	&	0.045 	&	0.050 	&	0.048 	&	0.048 
			\\
			\cline{2-8}
			$t_3$	&	$s_0=1$	&	0.058 	&	0.052 	&	0.046 	&	0.038 	&	0.048 	&	0.064\\ 
			&	$s_0=3$	&	0.053 	&	0.053 	&	0.045 	&	0.050 	&	0.052 	&	0.058 
			\\
			&	$s_0=5$	&	0.062 	&	0.060 	&	0.055 	&	0.053 	&	0.063 	&	0.074 
			\\
			&	$s_0=7$	&	0.051 	&	0.051 	&	0.053 	&	0.053 	&	0.055 	&	0.066 \\
			
			\hline
			\multicolumn{8}{c}{\bf{Empirical sizes Model~2 with $p=400$}}\\
			\bf{Dist}	&$\bf{s_0}$&$\bf{\balpha=0}$&$\bf{\balpha=0.1}$&$\bf{\balpha=0.5}$&$\bf{\balpha=0.9}$&$\bf{\balpha=1}$&\bf{Adaptive}\\
			\hline
			\cline{2-8}
			$N(0,1)$	&	$s_0=1$	&	0.052 	&	0.049 	&	0.043 	&	0.030 	&	0.028 	&	0.042 
			\\
			&	$s_0=3$	&	0.068 	&	0.062 	&	0.042 	&	0.025 	&	0.025 	&	0.048 
			\\
			&	$s_0=5$	&	0.068 	&	0.058 	&	0.028 	&	0.018 	&	0.017 	&	0.043 
			\\
			&	$s_0=7$	&	0.043 	&	0.043 	&	0.022 	&	0.015 	&	0.011 	&	0.031 
			\\
			\cline{2-8}
			\cline{2-8}
			$t_4$	&	$s_0=1$	&	0.059 	&	0.050 	&	0.047 	&	0.041 	&	0.040 	&	0.053\\ 
			&	$s_0=3$	&	0.048 	&	0.046 	&	0.044 	&	0.041 	&	0.036 	&	0.044 
			\\
			&	$s_0=5$	&	0.073 	&	0.059 	&	0.034 	&	0.036 	&	0.042 	&	0.060 
			\\
			&	$s_0=7$	&	0.064 	&	0.051 	&	0.030 	&	0.038 	&	0.038 	&	0.055 
			\\
			\cline{2-8}
			$t_3$	&	$s_0=1$	&	0.059 	&	0.063 	&	0.044 	&	0.036 	&	0.041 	&	0.058\\ 
			&	$s_0=3$	&	0.070 	&	0.055 	&	0.042 	&	0.041 	&	0.043 	&	0.057 
			\\
			&	$s_0=5$	&	0.052 	&	0.055 	&	0.047 	&	0.042 	&	0.042 	&	0.049 
			\\
			&	$s_0=7$	&	0.056 	&	0.048 	&	0.044 	&	0.042 	&	0.035 	&	0.054 \\
			\bottomrule[2pt]
		\end{tabular}
	\end{center}
\end{table}

\begin{table}[!hpt]
	\small
	\caption{Empirical sizes of the individual and tail-adaptive tests for {\bf{Models 1-2}} for the error term being Student's $t_2$ and $t_1$ distributed. The results are based on 1000 replications with $B=200$ for each replication.}\label{table: size t2+t1}
	\vspace{-0.1cm}
	\addtolength{\tabcolsep}{1pt}
	\begin{center}
		\begin{tabular}{ccccc cccc }
			\toprule[2pt]
			\multicolumn{9}{c}{\bf{Empirical sizes for Mode~1 with heavy tails} }\\
			$\bf{p}$&\bf{Dist}&$\bf{s_0}$&$\bf{\balpha=0}$&$\bf{\balpha=0.1}$&$\bf{\balpha=0.5}$&$\bf{\balpha=0.9}$&$\bf{\balpha=1}$&\bf{Adaptive}\\
			\hline
			$400$	&	$t_2$	&	$s_0=1$	&	0.041 	&	0.057 	&	0.079 	&	0.087 	&	0.085 	&	0.077 
			\\
			&	$t_2$	&	$s_0=3$	&	0.057 	&	0.060 	&	0.092 	&	0.088 	&	0.090 	&	0.090 
			\\
			&	$t_2$	&	$s_0=5$	&	0.068 	&	0.074 	&	0.110 	&	0.107 	&	0.129 	&	0.128 
			\\
			&	$t_2$	&	$s_0=7$	&	0.062 	&	0.065 	&	0.116 	&	0.115 	&	0.113 	&	0.126 \\
			\cline{2-9}
			\cline{2-9}
			$400$	&	$t_1$	&	$s_0=1$	&	0.057 	&	0.207 	&	0.222 	&	0.217 	&	0.217 	&	0.192 
			\\
			&	$t_1$	&	$s_0=3$	&	0.043 	&	0.228 	&	0.244 	&	0.240 	&	0.232 	&	0.208 
			\\
			&	$t_1$	&	$s_0=5$	&	0.058 	&	0.299 	&	0.315 	&	0.310 	&	0.300 	&	0.266 
			\\
			&	$t_1$	&	$s_0=7$	&	0.057 	&	0.276 	&	0.306 	&	0.308 	&	0.300 	&	0.267 \\
			\hline
			\multicolumn{9}{c}{\bf{Empirical sizes  for Mode~2 with heavy tails} }\\
			$\bf{p}$&\bf{Dist}&$\bf{s_0}$&$\bf{\balpha=0}$&$\bf{\balpha=0.1}$&$\bf{\balpha=0.5}$&$\bf{\balpha=0.9}$&$\bf{\balpha=1}$&\bf{Adaptive}\\
			\hline
			$400$	&	$t_2$	&	$s_0=1$	&	0.066 	&	0.082 	&	0.095 	&	0.096 	&	0.094 	&	0.106 
			\\
			&	$t_2$	&	$s_0=3$	&	0.072 	&	0.086 	&	0.115 	&	0.113 	&	0.118 	&	0.127 
			\\
			&	$t_2$	&	$s_0=5$	&	0.058 	&	0.074 	&	0.138 	&	0.151 	&	0.152 	&	0.147 
			\\
			&	$t_2$	&	$s_0=7$	&	0.056 	&	0.085 	&	0.113 	&	0.127 	&	0.118 	&	0.136 \\
			\cline{2-9}
			$400$	&	$t_1$	&	$s_0=1$	&	0.070 	&	0.220 	&	0.249 	&	0.251 	&	0.247 	&	0.223 
			\\
			&	$t_1$	&	$s_0=3$	&	0.046 	&	0.402 	&	0.440 	&	0.432 	&	0.430 	&	0.406 
			\\
			&	$t_1$	&	$s_0=5$	&	0.055 	&	0.455 	&	0.500 	&	0.487 	&	0.489 	&	0.462 
			\\
			&	$t_1$	&	$s_0=7$	&	0.057 	&	0.479 	&	0.526 	&	0.511 	&	0.501 	&	0.488 \\
			\bottomrule[2pt]
		\end{tabular}
	\end{center}
\end{table}

We consider the size performance with a significance level $\gamma=5\%$. Tables \ref{table: sizes of individual band} provides the size results for the individual tests $T_{\alpha}$ with $\alpha\in\cA=\{0,0.1,0.5,0.9,1\}$ and the tail-adaptive test $T_{\rm ad}$ under \textbf{Models 1} and \textbf{2} with various error distributions. Note that the construction of our testing statistic involves a selection of $s_0\in\{1,\ldots,p\}$. To show the effect of different $s_0$, we consider various $s_0\in\{1,3,5,7\}$. Note that $s_0=1$ corresponds to the $\ell_\infty$-norm based individual test and $s_0=5$ corresponds to the test that aggregates the active set of variables in $\bbeta^{(0)}$. As shown in Table \ref{table: sizes of individual band}, for a given $s_0$, our individual test $T_{\alpha}$ and tail-adaptive test $T_{\rm ad}$ can have a size that is very close to the nominal level.
This strongly suggests that our bootstrap-based procedure in Algorithms 1 and 2 can approximate the theoretical distributions very well. Interestingly, it can be seen that under a specific error distribution,
the individual test $T_{\alpha}$ may have different size performance  in the sense that the corresponding size can be slightly above or below the nominal level. In contrast, after the combination, the size of the tail-adaptive test $T_{\rm ad}$ is near the nominal level as compared to its individual test. This indicates that in practice, the tail-adaptive test is more reliable in terms of size control.

Table \ref{table: size t2+t1} provides additional size performance under Student's $t_2$ and $t_1$ distributions. Note that these two distributions are known as seriously heavy-tailed. It is also well known that controlling the size for these two distributions is a challenging task, especially for high-dimensional change point analysis. As can be seen from Table \ref{table: size t2+t1}, in these cases, the individual test $T_{\alpha}$ except $\alpha=0$ suffers from serious size distortion. In particular, as $\alpha$ increases from 0.1 to 1, it is more difficult to control the size. Moreover, when the error is Cauchy distributed, the size is completely out of control for $\alpha\in\{0.1,\ldots,1\}$. As a result, the corresponding tail-adaptive method becomes oversized. As an exception, we can see that the individual test $T_{\alpha}$ with $\alpha=0$ enjoys satisfactory size performance for both $t_2$ and $t_1$ distributions. A reasonable explanation is that for $\alpha=0$, our individual test reduces to the median regression based method which does not  require any moment constraints on the error terms. Hence, our proposed individual test with $\alpha=0$ contributes to the literature for handling the extremely  heavy-tailed case. In practice, if the practitioners strongly believe the data are seriously heavy-tailed, we can just set $\cA=\{0\}$.

\subsection{Empirical powers}\label{sec: empirical powers}

We next consider the power performance, where various error distributions, data dimensions as well as change point locations are investigated. The results are summarized in Tables \ref{table: power model1} and \ref{table: power model2}. Note that according to our model setups, there are five coordinates in $\bbeta^{(1)}$ having a change point. 
It can be seen that for light-tailed error distributions such as  $N(0,1)$, the individual tests with $\alpha=0.5,0.9,1$ have the best power performance and those with $\alpha=0$ have the worst performance. This indicates that for a light-tailed error distribution, using median regression can lose power efficiency, and using the moment information with a larger weight $\alpha$ can increase the signal to noise ratio. Interestingly, in this empirical study,  the individual test with $\alpha=0.5$ generally has slightly higher powers than that with $\alpha=1$, even though the latter one is expected to have the best power performance (see Figure \ref{figure: snr}). As for the tail-adaptive test, in the light-tailed case, it has very close powers to the best individual tests. We next turn to the heavy-tailed case,  where the individual tests have power performance that is very different from the light-tailed case. Specifically, for $t_3$ distributions, the individual test with $\alpha=0$ and $\alpha=0.1$ have higher powers than the remaining ones. This indicates that for data with heavy tails, it is beneficial to use more rank information instead of using only moments. More specifically, we see that $T_{\alpha}$ with $\alpha=0.1$ has the highest powers and that with $\alpha=1$ has the lowest powers. This result is consistent with the theoretical SNR in Figure \ref{figure: snr}. In this case, using a non-trivial  weight  ($\alpha=0.1$) can significantly enhance the power efficiency via increasing the SNR. As for the tail-adaptive method, it still has very close powers to the best individual test, i.e. $\alpha=0.1$ when the data are heavy-tailed. In addition to $N(0,1)$ and $t_3$ distributions, we can observe  that for $t_4$ distributions, even though the individual tests may present various power performances, the tail-adaptive method consistently has powers close to that of the corresponding best individual test. The above results suggest that our proposed tail-adaptive method can sufficiently account for the unknown tail-structures, and enjoy satisfactory power performance under various data generating mechanisms. Lastly, we remark that when the change point location gets closer to the boundary of data observation, e.g. from $t_1=0.5$ to $t_1=0.3$,  it becomes more difficult to detect a change point, {which is also consistent with our theoretical result.}

Next, we consider the effect of different $s_0$ on the power performance.  We find that for any given $s_0$, the performance of the individual and the tail-adaptive tests are similar to our above findings. This suggests that the tail-adaptivity of our testing method is robust to the choice of $s_0$. Moreover, for each case with a specific error distribution and data dimension,  both the individual and tail-adaptive tests with $s_0=3,5,7$ have higher powers than  those  with $s_0=1$. More specifically, tests with $s_0=5$ generally have the best performance and those with $s_0=3$ and $s_0=7$ have close powers to $s_0=5$. This indicates that for high -dimensional sparse linear models, instead of using the $\ell_\infty$-norm,  it is more efficient to detect a change point  via aggregating the CUSUM statistics using the first $s_0>1$ order statistics.  

\begin{table}[!hpt]
	\scriptsize
	\caption{Empirical powers  of the individual and tail-adaptive tests for {\bf{Model 1}} with banded covariance matrix under various distributions with $s_0\in\{1,3,5,7\}$ and $t_1\in\{0.3,0.5\}$. The dimension is  $p=400$. The results are based on 1000 replications with $B=200$ for each replication.}\label{table: power model1}
	\vspace{-0.3cm}
	\addtolength{\tabcolsep}{2pt}
	\begin{center}
		\begin{tabular}{ccccc cccc}
			\toprule[2pt]
			
			\multicolumn{9}{c}{\bf{Empirical powers  for N(0,1)} }\\
			\bf{	Dist}&$\bf{t_1}$&$\bf{s_0}$&$\bf{\balpha=0}$&$\bf{\balpha=0.1}$&$\bf{\balpha=0.5}$&$\bf{\balpha=0.9}$&$\bf{\balpha=1}$&\bf{Adaptive}\\
			\hline
			&		&$s_0=1$		&	0.482 	&	0.612 	&	0.768 	&	0.732 	&	0.722 	&	0.733 
			\\
			$N(0,1)$	&	0.5	&	$s_0=3$	&	0.529 	&	0.655 	&	0.787 	&	0.759 	&	0.739 	&	0.759 
			\\
			&		&	$s_0=5$	&	0.546 	&	0.641 	&	0.783 	&	0.759 	&	0.749 	&	0.760 
			\\
			&		&	$s_0=7$	&	0.525 	&	0.634 	&	0.802 	&	0.778 	&	0.773 	&	0.765 
			\\
			&		&	$s_0=1$	&	0.295 	&	0.398 	&	0.546 	&	0.506 	&	0.489 	&	0.495 
			\\
			$N(0,1)$	&	0.3	&	$s_0=3$	&	0.318 	&	0.415 	&	0.573 	&	0.534 	&	0.516 	&	0.518 
			\\
			&		&	$s_0=5$	&	0.286 	&	0.418 	&	0.568 	&	0.543 	&	0.514 	&	0.505 
			\\
			&		&	$s_0=7$	&	0.315 	&	0.418 	&	0.560 	&	0.522 	&	0.505 	&	0.522 \\
			
			\hline
			
			\multicolumn{9}{c}{\bf{Empirical powers  for Student's $t_4$} }\\
			\bf{	Dist}&$\bf{t_1}$&$\bf{s_0}$&$\bf{\balpha=0}$&$\bf{\balpha=0.1}$&$\bf{\balpha=0.5}$&$\bf{\balpha=0.9}$&$\bf{\balpha=1}$&\bf{Adaptive}\\
			\hline
			&		&	$s_0=1$	&	0.792 	&	0.880 	&	0.835 	&	0.769 	&	0.756 	&	0.873 
			\\
			$t_4$	&	0.5	&	$s_0=3$	&	0.836 	&	0.903 	&	0.852 	&	0.795 	&	0.782 	&	0.904\\ 
			&		&$s_0=5$		&	0.847 	&	0.914 	&	0.884 	&	0.813 	&	0.787 	&	0.915 
			\\
			&		&	$s_0=7$	&	0.831 	&	0.895 	&	0.861 	&	0.813 	&	0.807 	&	0.896 
			\\
			&		&	$s_0=1$	&	0.595 	&	0.724 	&	0.687 	&	0.588 	&	0.555 	&	0.722 
			\\
			$t_4$	&	0.3	&	$s_0=3$	&	0.644 	&	0.762 	&	0.737 	&	0.627 	&	0.591 	&	0.765 
			\\
			&		&$s_0=5$		&	0.646 	&	0.773 	&	0.745 	&	0.618 	&	0.606 	&	0.765 
			\\
			&		&	$s_0=7$	&	0.603 	&	0.765 	&	0.712 	&	0.582 	&	0.564 	&	0.743 \\
			\hline
			\multicolumn{9}{c}{\bf{Empirical powers  for Student's $t_3$} }\\
			\bf{	Dist}&$\bf{t_1}$&$\bf{s_0}$&$\bf{\balpha=0}$&$\bf{\balpha=0.1}$&$\bf{\balpha=0.5}$&$\bf{\balpha=0.9}$&$\bf{\balpha=1}$&\bf{Adaptive}\\
			\hline
			&		&	$s_0=1$	&	0.773 	&	0.819 	&	0.663 	&	0.580 	&	0.572 	&	0.802 
			\\
			$	t_3	$&	0.5	&	$s_0=3$	&	0.777 	&	0.826 	&	0.693 	&	0.612 	&	0.583 	&	0.827\\ 
			&		&	$s_0=5$	&	0.791 	&	0.840 	&	0.685 	&	0.604 	&	0.594 	&	0.822 
			\\
			&		&	$s_0=7$	&	0.789 	&	0.847 	&	0.713 	&	0.623 	&	0.602 	&	0.829 
			\\
			\cline{2-9}
			&		&	$s_0=1$	&	0.554 	&	0.629 	&	0.475 	&	0.380 	&	0.362 	&	0.599 
			\\
			$t_3$	&	0.3	&	$s_0=3$	&	0.599 	&	0.692 	&	0.527 	&	0.422 	&	0.403 	&	0.656 
			\\
			&		&$s_0=5$		&	0.587 	&	0.697 	&	0.525 	&	0.404 	&	0.390 	&	0.640 
			\\
			&		&	$s_0=7$	&	0.549 	&	0.650 	&	0.487 	&	0.362 	&	0.348 	&	0.613 \\
			\bottomrule[2pt]
		\end{tabular}
	\end{center}
\end{table}

\begin{table}[!hpt]
	\scriptsize
	\caption{Empirical powers  of the individual and data-adaptive tests for {\bf{Model 2}} with blocked covariance matrix under various distributions with $s_0\in\{1,3,5,7\}$ and $t_1\in\{0.3,0.5\}$. The dimension $p$ is  400. The results are based on 1000 replications with $B=200$ for each replication.}\label{table: power model2}
	\vspace{-0.1cm}
	\addtolength{\tabcolsep}{2pt}
	\begin{center}
		\begin{tabular}{ccccc cccc}
			\toprule[2pt]
			\multicolumn{9}{c}{\bf{Empirical powers  for N(0,1)} }\\
			\bf{	Dist}&$\bf{t_1}$&$\bf{s_0}$&$\bf{\balpha=0}$&$\bf{\balpha=0.1}$&$\bf{\balpha=0.5}$&$\bf{\balpha=0.9}$&$\bf{\balpha=1}$&\bf{Adaptive}\\
			\hline
			&		&	$s_0=1$	&	0.328 	&	0.441 	&	0.651 	&	0.626 	&	0.626 	&	0.594 
			\\
			$N(0,1)$	&	0.5	&	$s_0=3$	&	0.428 	&	0.547 	&	0.733 	&	0.714 	&	0.704 	&	0.695 
			\\
			&		&	$s_0=5$	&	0.462 	&	0.585 	&	0.761 	&	0.712 	&	0.702 	&	0.714 
			\\
			&		&	$s_0=7$	&	0.476 	&	0.591 	&	0.760 	&	0.718 	&	0.703 	&	0.712 
			\\
			&		&	$s_0=1$	&	0.175 	&	0.232 	&	0.361 	&	0.326 	&	0.321 	&	0.301 
			\\
			$N(0,1)$	&	0.3	&	$s_0=3$	&	0.245 	&	0.334 	&	0.486 	&	0.453 	&	0.437 	&	0.458 
			\\
			&		&	$s_0=5$	&	0.244 	&	0.356 	&	0.483 	&	0.428 	&	0.412 	&	0.428 
			\\
			&		&	$s_0=7$	&	0.246 	&	0.338 	&	0.470 	&	0.409 	&	0.389 	&	0.419 \\
			
			\hline
			\multicolumn{9}{c}{\bf{Empirical powers  for Student's $t_4$} }\\
			\bf{	Dist}&$\bf{t_1}$&$\bf{s_0}$&$\bf{\balpha=0}$&$\bf{\balpha=0.1}$&$\bf{\balpha=0.5}$&$\bf{\balpha=0.9}$&$\bf{\balpha=1}$&\bf{Adaptive}\\
			\hline
			&		&	$s_0=1$	&	0.618 	&	0.722 	&	0.749 	&	0.654 	&	0.659 	&	0.742 
			\\
			$t_4$	&	0.5	&	$s_0=3$	&	0.791 	&	0.862 	&	0.849 	&	0.784 	&	0.763 	&	0.873\\ 
			&		&	$s_0=5$	&	0.780 	&	0.866 	&	0.855 	&	0.778 	&	0.769 	&	0.874 
			\\
			&		&	$s_0=7$	&	0.802 	&	0.879 	&	0.868 	&	0.806 	&	0.782 	&	0.889 
			\\
			&		&	$s_0=1$	&	0.398 	&	0.518 	&	0.547 	&	0.427 	&	0.404 	&	0.511 
			\\
			$t_4$	&	0.3	&	$s_0=3$	&	0.511 	&	0.661 	&	0.645 	&	0.535 	&	0.514 	&	0.665\\ 
			&		&	$s_0=5$	&	0.522 	&	0.663 	&	0.631 	&	0.505 	&	0.483 	&	0.651 
			\\
			&		&	$s_0=7$	&	0.531 	&	0.661 	&	0.637 	&	0.509 	&	0.482 	&	0.655 \\
			\hline
			\multicolumn{9}{c}{\bf{Empirical powers  for Student's $t_3$} }\\
			\bf{	Dist}&$\bf{t_1}$&$\bf{s_0}$&$\bf{\balpha=0}$&$\bf{\balpha=0.1}$&$\bf{\balpha=0.5}$&$\bf{\balpha=0.9}$&$\bf{\balpha=1}$&\bf{Adaptive}\\
			\hline
			&		&	$s_0=1$	&	0.571 	&	0.640 	&	0.571 	&	0.460 	&	0.455 	&	0.621 
			\\
			$t_3$	&	0.5	&	$s_0=3$	&	0.726 	&	0.790 	&	0.683 	&	0.588 	&	0.579 	&	0.782\\ 
			&		&	$s_0=5$	&	0.761 	&	0.794 	&	0.674 	&	0.583 	&	0.576 	&	0.808 
			\\
			&		&	$s_0=7$	&	0.753 	&	0.803 	&	0.717 	&	0.619 	&	0.600 	&	0.807 
			\\
			\cline{2-9}
			&		&	$s_0=1$	&	0.359 	&	0.455 	&	0.360 	&	0.293 	&	0.261 	&	0.422 
			\\
			$	t_3	$&	0.3	&	$s_0=3$	&	0.470 	&	0.574 	&	0.485 	&	0.370 	&	0.349 	&	0.560\\ 
			&		&	$s_0=5$	&	0.490 	&	0.581 	&	0.451 	&	0.339 	&	0.328 	&	0.567 
			\\
			&		&	$s_0=7$	&	0.498 	&	0.601 	&	0.440 	&	0.330 	&	0.308 	&	0.581 \\
			\bottomrule[2pt]
		\end{tabular}
	\end{center}
\end{table}

\subsection{{The choices of $\cA$ and $s_0$}}

\begin{table}[!h]
	\caption{{Empirical powers of the tail-adaptive tests for \textbf{Model 1} with banded covariance matrix under various choices of $s_0$ and $\cA$. The dimension is $p=400$. The results are based on 1000 replications with $B=200$ for each replication}.}
	\label{table: s choice}
	\addtolength{\tabcolsep}{-3pt}
	\begin{tabular}{lcccccccc cccccccc}
		\toprule[2pt]
		& \multicolumn{7}{c}{$N(0,1)$} &  & \multicolumn{7}{c}{$t_3$} \\
		\cline{1-16}
		$s_0$	& $\cA_1$ & $\cA_2$ & $\cA_3$  & $\cA_4$  & $\cA_5$  & $\cA_6$  & $\cA_7$   && $\cA_1$  & $\cA_2$  & $\cA_3$ & $\cA_4$  & $\cA_5$  & $\cA_6$  & $\cA_7$   \\ 
		$1$		&0.456 	&	0.746 	&	0.725 	&	0.546 	&	0.682 	&	0.698 	&	0.700 	&&	0.760 	&	0.634 	&	0.524 	&	0.804 	&	0.486 	&	0.786 	&	0.748 \\
		$2$	&0.479 	&	0.800 	&	0.748 	&	0.584 	&	0.718 	&	0.706 	&	0.708 	&&	0.792 	&	0.666 	&	0.544 	&	0.828 	&	0.520 	&	0.786 	&	0.778 \\
		$4$	&	0.521 	&	0.773 	&	0.752 	&	0.608 	&	0.738 	&	0.704 	&	0.662 	&&	0.766 	&	0.728 	&	0.584 	&	0.840 	&	0.534 	&	0.812 	&	0.788 \\
		$[\log(p)]$	&	0.498 	&	0.780 	&	0.746 	&	0.586 	&	0.682 	&	0.742 	&	0.682 	&&	0.822 	&	0.676 	&	0.560 	&	0.820 	&	0.538 	&	0.814 	&	0.758 \\
		$8$		&	0.488 	&	0.798 	&	0.724 	&	0.586 	&	0.650 	&	0.710 	&	0.678 	&&	0.788 	&	0.688 	&	0.592 	&	0.860 	&	0.536 	&	0.826 	&	0.762 \\
		$16$		&	0.442 	&	0.738 	&	0.716 	&	0.554 	&	0.680 	&	0.714 	&	0.626 	&&	0.722 	&	0.616 	&	0.480 	&	0.806 	&	0.516 	&	0.750 	&	0.676 \\
		$32$		&	0.426 	&	0.692 	&	0.644 	&	0.470 	&	0.590 	&	0.586 	&	0.526 	&&	0.646 	&	0.616 	&	0.416 	&	0.720 	&	0.448 	&	0.736 	&	0.608 \\
		$64$		&	0.346 	&	0.586 	&	0.512 	&	0.398 	&	0.476 	&	0.524 	&	0.508 	&&	0.584 	&	0.516 	&	0.350 	&	0.630 	&	0.384 	&	0.574 	&	0.508 \\
		$128$		&	0.264 	&	0.556 	&	0.436 	&	0.320 	&	0.426 	&	0.396 	&	0.354 	&&	0.468 	&	0.476 	&	0.322 	&	0.550 	&	0.336 	&	0.540 	&	0.396 \\
		\bottomrule[2pt]
	\end{tabular}
\end{table}
Note that our approach involves the selection of the candidate set 
$\cA$ and the parameter $s_0$,  both of which can be regarded as tuning parameters. Intuitively, $\cA$ determines the weight  between the quantile loss and the least squared losses, whereas $s_0$ dictates how much information on changepoints among regression coefficient components should be integrated into the CUSUM statistic. Therefore, we conducted numerical simulations to investigate how different choices of  $\cA$ and $s_0$  affect the efficacy of changepoint detection. We selected seven different subsets for $\cA$ including $\cA_1=\{0\}$, $\cA_2=\{0.5\}$, $\cA_3=\{1\}$, $\cA_4=\{0,0.1\}$, $\cA_5=\{0.9,1\}$, $\cA_6=\{0,0.1,0.5,0.9,1\}$ and $\cA_7=\{0,0.1,0.2,0.3,0.4,0.5,0.6,0.7,0.8,0.9,1\}$. Additionally, we selected different values for $s_0$,  including $s_0=2^0, 2^1, \ldots, 2^{\lfloor\log_2(p)+1\rfloor}$. Table \ref{table: s choice} displays the performance of the adaptive changepoint detection method under $N(0,1)$ and $t_3$ distributions for these selections of $\cA$ and $s_0$. The model settings are the same as in Section S4.3. We observed that for any given $\cA$, when $s_0$ increases from small to large, the efficacy of the adaptive detection method initially increases and then decreases, indicating that as $s_0$
increases, the statistic extracts more changepoint information from the regression components, enhancing the efficacy of changepoint detection. However, once $s_0$ becomes larger, additional noise accumulates, leading to a decrease in the detection efficacy. Considering the sparsity assumptions for regression coefficients  and the requirements of Gaussian approximation theory,
which requires  $s_0^3\log(pn)=O(n^{\xi_1})$ for some $0<\xi_1<1/7$ and $s_0^4\log(pn)=O(n^{\xi_2})$ for some $0<\xi_2<\frac{1}{6}$. We recommend the use of $s_0=\floor{\log(p)}$ in practice.

Regarding the selection of $\cA$, we noted that for data with light-tailed distributions, sets with larger values such as $\cA_2,\cA_3,\cA_5,\cA_6,\cA_7$
exhibit higher efficacy. Conversely, for data with heavy-tailed distributions, sets with smaller values such as $\cA_1,\cA_4,\cA_6,\cA_7$ perform satisfactorily. Therefore, if the tail structure of the data is unknown in practical applications, we might consider a candidate set that includes both larger and smaller values. Interestingly, we found that adding too many weights, such as in $\cA_7$ does not yield much additional benefit. Considering the balance between detection efficacy and computational efficiency, we recommend $\cA=\{0,0.1,0.5,0.9,1\}$ for practical use.

\subsection{Multiple change points detection}\label{sec: multiple detection}

\begin{table}[H]
	\centering
	\caption{Multiple change points estimation results with $(n,p)=(1000,100)$. The results are based on 100 replications with $B=100$ for each replication.}\label{table: multiple cpt}
	\begin{tabular}{@{}lcccccccc@{}}
		\toprule[2pt]
		& \multicolumn{2}{c}{N(0,1) ($c=3$)} & \multicolumn{2}{c}{N(0,1) ($ c=6$)} & \multicolumn{2}{c}{t3 ($c=4$)} & \multicolumn{2}{c}{t3 ($ c=6$)} \\ 
		\cmidrule(r){2-3} \cmidrule(r){4-5} \cmidrule(r){6-7} \cmidrule(r){8-9}
		Methods	& Haus & (Sd) & Haus & (Sd) & Haus & (Sd) & Haus & (Sd) \\ 
		\midrule
		$\alpha=0$ (BS)      & 0.186 & (0.1310) & 0.028 & (0.0374) & 0.144 & (0.1300)  & 0.040 & (0.0603)  \\
		$\alpha=0.1$ (BS)    & 0.138 & (0.1286) & 0.033 & (0.0442)  & 0.111 & (0.1229)  & 0.032 & (0.0580)  \\
		$\alpha=0.5$ (BS)   & 0.072 & (0.1018) & 0.028 & (0.0432)  & 0.145 & (0.1353)  & 0.051 & (0.0740)  \\
		$\alpha=0.9$ (BS)   & 0.088 & (0.1092) & 0.030 & (0.0443)  & 0.180 & (0.1371)  & 0.063 & (0.0877)  \\
		$\alpha=1$ (BS)    & 0.095 & (0.1146) & 0.024 & (0.0394)  & 0.187 & (0.1405)  & 0.065 & (0.0963)  \\
		Adaptive (BS) & 0.092 & (0.1085) & 0.032 &(0.0428)  & 0.113 & (0.1260)  & 0.039 & (0.0638)  \\
		\hline
		$\alpha=0$ (WBS)     & 0.087 & (0.0926) & 0.018 & (0.0381)  & 0.046 & (0.0690)  & 0.018 & (0.0381)  \\
		$\alpha=0.1$ (WBS)   & 0.060 & (0.0846) & 0.012 & (0.0277)  & 0.049 & (0.0710)  & 0.012 & (0.0277)  \\
		$\alpha=0.5$ (WBS)   & 0.036 & (0.0599) & 0.012 & (0.0279)  & 0.077 & (0.0908)  & 0.012 & (0.0279)  \\
		$\alpha=0.9$ (WBS)  & 0.041 & (0.0662) & 0.011 & (0.0216)  & 0.095 & (0.0999)  & 0.011 & (0.0216)  \\
		$\alpha=1$ (WBS)    & 0.043 & (0.0659) & 0.012 & (0.0216)  & 0.101 & (0.1014)  & 0.012 & (0.0216)  \\
		Adaptive (WBS) & 0.031 & (0.0478) & 0.014 & (0.0292)  & 0.033 & (0.0511)  & 0.014 & (0.0292)  \\
		\hline
		VPWBS       & 0.135 & (0.1004) & 0.038 & (0.0474)  & 0.138 & (0.0782)  & 0.085 & (0.0636)  \\
		DPDU        & 0.082 & (0.1097) & 0.009 & (0.0087)  & 0.118 & (0.0830)  & 0.045 & (0.0598)  \\
		\bottomrule[2pt]
	\end{tabular}
\end{table}

In this section, we consider the performance of multiple change point detection and compare our method with the existing techniques. In this numerical study, we set $n=1000$ and $p=100$ with three change points ($m=3$) at $k_1=300$, $k_2=500$, and $k_3=700$, respectively. The above three change points divide the data into four segments with piecewise constant regression coefficients $\bbeta^{(1)}$, $\bbeta^{(2)}$, $\bbeta^{(3)}$ and $\bbeta^{(4)}$ as follows:
\begin{equation*}\label{equation: linear model with multiple change point}
	\left\{\begin{array}{ll}
		Y_i=\bX_i^\top\bbeta^{(1)}+\epsilon_i, & \text{for}~i=1,\ldots, k_1 ,\\
		Y_i=\bX_i^\top\bbeta^{(2)}+\epsilon_i,& \text{for}~i=k_1+1,\ldots, k_2,\\
		Y_i=\bX_i^\top\bbeta^{(3)}+\epsilon_i,& \text{for}~i=k_{2}+1,\ldots, k_3,\\
		Y_i=\bX_i^\top\bbeta^{(4)}+\epsilon_i,& \text{for}~i=k_3+1,\ldots, n.
	\end{array}\right.
\end{equation*}
The covariates $\bX_i$ are generated from $N(\mathbf{0},\bSigma )$ with $\bSigma$ being banded which is introduced in Model 1. For each replication,  we first randomly select five covariates (denoted by $\cS_1$) from $\{1,\ldots,10\}$. For generating $\bbeta^{(1)}$, we set $\beta^{(1)}_{s}=1$ if 
$s\in \cS_1$ and $\beta^{(1)}_{s}=0$ if $s\notin \cS_1$. For $\bbeta^{(2)}$, we set $\beta^{(2)}_{s}=\beta^{(1)}_{s}+c\sqrt{\log(p)/n}$ if  $s\in \cS_1$ and $\beta^{(2)}_{s}=0$ if $s\notin \cS_1$. Then, we set $\bbeta^{(3)}=\bbeta^{(1)}$ and $\bbeta^{(4)}=\bbeta^{(2)}$. {We compare our proposed method with the  
	Variance-Projected Wild Binary Segmentation (VPWBS) method in \cite{JMLR:v22:19-531} and the dynamic programming with dynamic update method in \cite{xu2022change}. As compared in \cite{JMLR:v22:19-531}, VPWBS has better performance than  the binary segmentation based technique in \cite{leonardi2016computationally} and the sparse graphical LASSO based method in \cite{zhang2015change-point}. Hence, we don't compare with  \cite{leonardi2016computationally} and \cite{zhang2015change-point}. For VPWBS, we used the R codes published by the authors on GitHub (\textcolor{blue}{https://github.com/darenwang/VPBS}) and employed a cross-validation method to select tuning parameters for estimating changepoints. For DPDU, we utilized the DPDU.regression.R function from the  R package named ``changepoints" to estimate multiple changepoints.}
{As for our methods, we combine the individual and tail-adaptive procedures with the Binary Segmentation and Wild Binary Segmentation techniqus. For WBS, we use Algorithm \ref{alg:WBS} with parameters as $\gamma=0.05$, $s_0=5$, $q_0=0.1$, $B=100$, $V=150$, and $v_0=0.1$. 
	In this numerical study, we set the replication number as 100.}

To evaluate the performance in identifying the change point, we use  the scaled Hausdorff distance  to evaluate the performance in change point estimation, which is defined as:
\begin{equation*}
	d(\cS_1,\cS_2)=\dfrac{\max(\max_{s_1\in\cS_1}\min_{s_2\in\cS_2} |s_1-s_2|,\max_{s_2\in\cS_2}\min_{s_1\in\cS_1} |s_1-s_2|  )}{1000},
\end{equation*}
where $\cS_1=\{300,500,7000,1000\}$ are the true change points and $\cS_2$ are the estimated change points.
Note that scaled Hausdorff distance is a number between 0 and 1, and a smaller one indicates better change point estimation. Table \ref{table: multiple cpt} provides the results for $N(0,1)$ and $t_3$ distributions with various signal strength $c\in\{3,4,6\}$. For light-tailed error distributions,  the individual methods with a larger $\alpha$ generally have better performance than those with a smaller one for identifying the change point number and locations. This can be seen by  smaller Hausdorff. 
On the contrary, in the heavy-tailed case, the individual methods with a smaller $\alpha$ are more preferred. As for the tail-adaptive method, it has comparable performance to the best individual one under both light and heavy-tailed errors. 

{Additionally, we note that for both individual  and  tail-adaptive testing methods,
	those based on Wild Binary Segmentation (WBS) generally outperform those based on Binary Segmentation (BS). Therefore, we recommend combining our proposed method with the WBS algorithm for multi-changepoint estimation in practical applications. For VPWBS and DPDU, these methods show satisfactory performance under light-tailed distributions such as the normal distribution. Particularly, DPDU, which employs a dynamic programming algorithm for multiple changepoint estimation, achieves the lowest estimation errors when data follow a normal distribution with strong signals. Our adaptive method performs comparably to these two methods under light-tailed distributions. However, their detection capabilities decrease when the data follow heavy-tailed distributions, such as the Student's $t_3$ distribution. This indicates that methods based on the least squared loss are not robust for heavy-tailed data.}

{Lastly, we report the computational complexity of the algorithm. For our individual and tail-adaptive testing methods, when combined with the WBS algorithm, the complexity is $O(M\text{Lasso}(n,p))$ and $O(M|\cA|\text{Lasso}(n,p))$, respectively, where $M$ represents the number of small intervals in WBS, and $\text{Lasso}(n,p)$ denotes the computational cost for calculating lasso with sample size $n$ and data dimension $p$. For the DPDU algorithm, it uses a backward iterative dynamic programming approach, and its complexity is  $O(n^2p^2+n^2\text{Lasso(p)})$. Figure \ref{figure: time} shows the computation times of our method and the DPDU algorithm under various $n\in\{200,300,400,500\}$ and $p\in\{200,300,400,500\}$, where the model setup is as in Section S2.3. We set the number of intervals in WBS to $\log^2(n)$. We can observe that the computational costs of both our method and the DPDU method increase with $n$ and $p$. Our individual testing method has comparable  computation time  to that of the DPDU. The computation cost for the tail-adaptive testing method is the highest. This is not surprising, as we aim to construct a testing method that is adaptive to the tail structure of the error terms. As a cost, we need to calculate  lasso estimates with different weights $\alpha$ to obtain the best individual testing method.}

\begin{figure}[!h]
	\begin{center}
		\begin{tabular}{ll}
			\includegraphics[width=8cm]{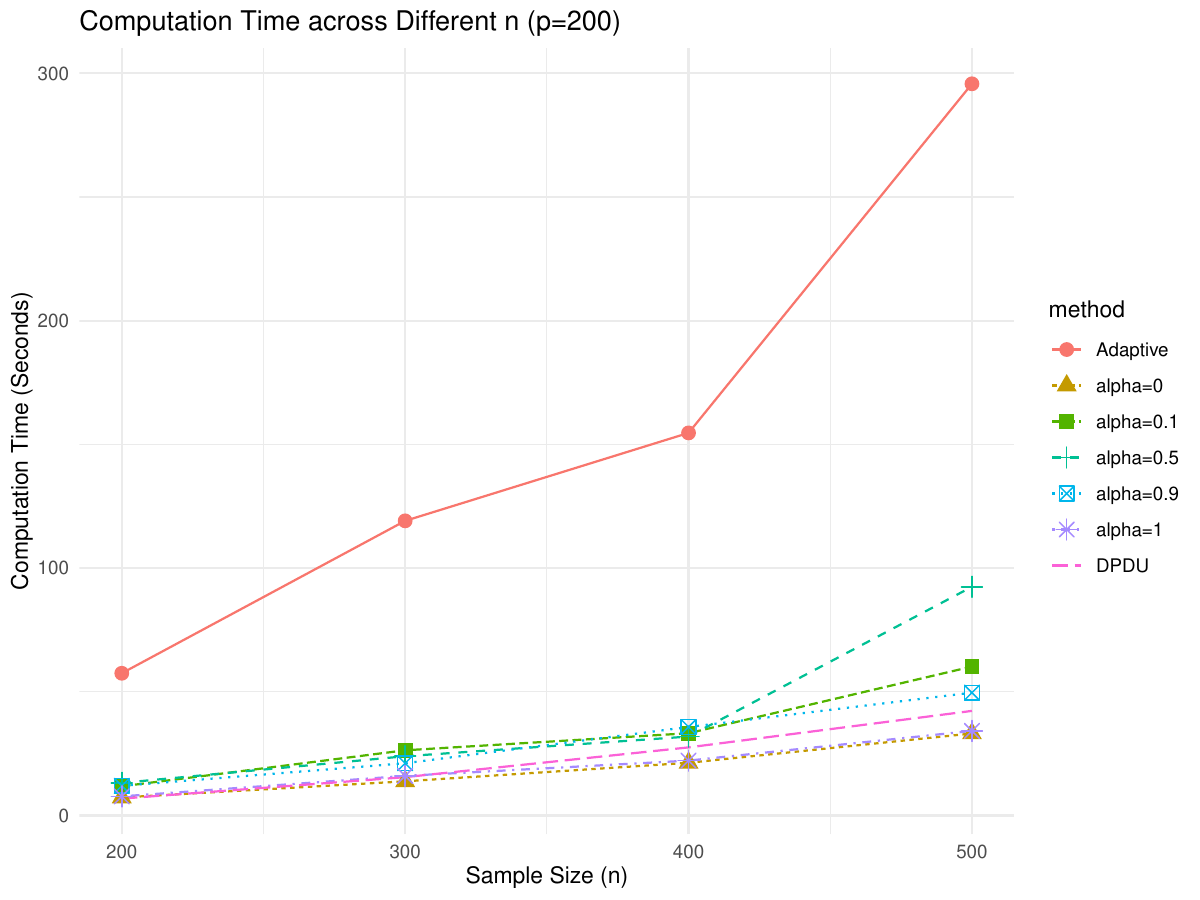}&	\includegraphics[width=8cm]{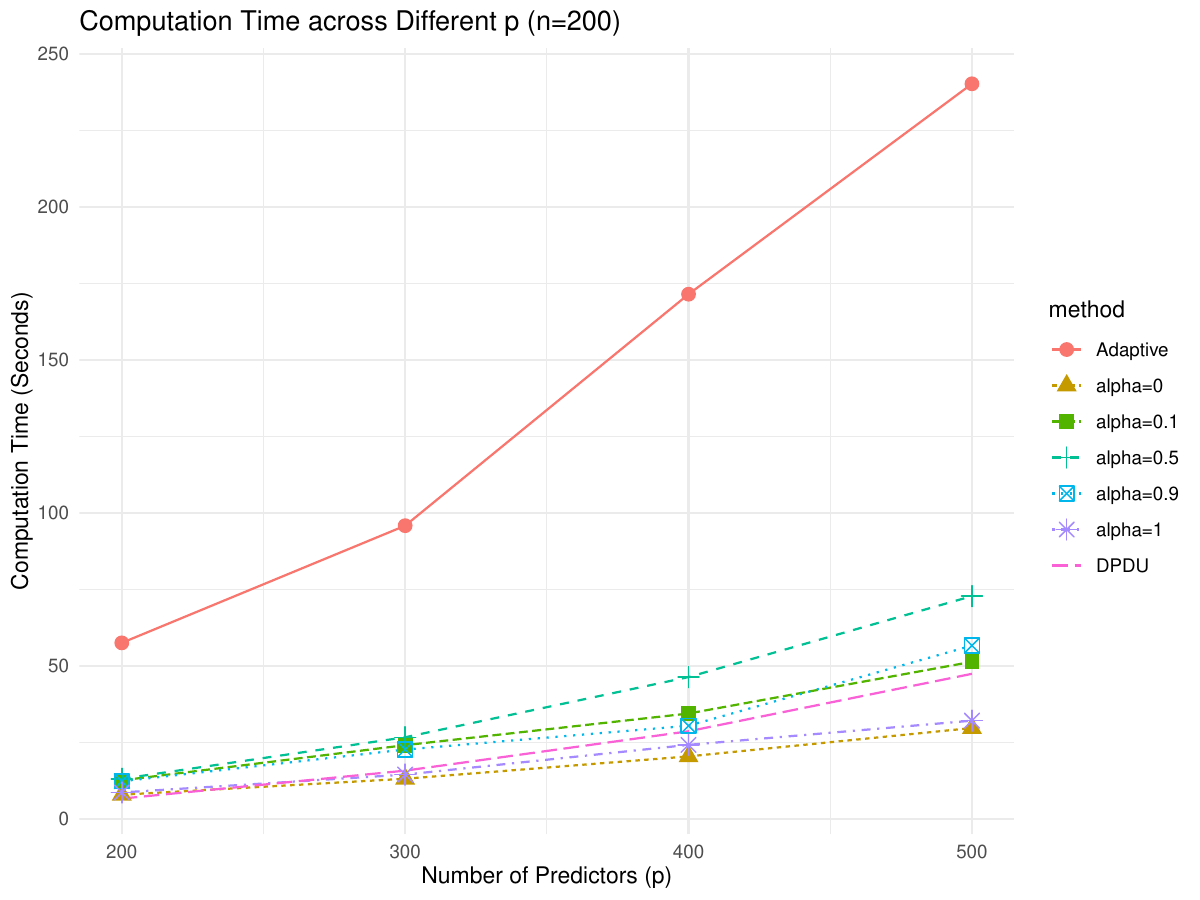}\\
		\end{tabular}
	\end{center}
	\caption{Computational time for our proposed method and the DPDU algorithm with $n\in\{200,300,400,500\}$ and $p\in\{200,300,400,500\}$.
	}\label{figure: time} 
\end{figure}

\newpage
\section{An application to the S\&P 100 dataset }\label{sec: real analysis}

In this section, we apply our proposed methods to the S\&P 100 dataset to find multiple change points. We obtain the S$\&$P 100 index as well as the associated stocks from Yahoo! Finance (https://finance.yahoo.com/) including the largest and most established 100 companies in the S$\&$P 100. For this dataset, we collect the daily prices of 76 stocks that have remained in the S\&P 100 index consistently from January 3, 2007 to December 30, 2011. This covers the recent financial crisis beginning in 2008 and some other important events, resulting in a  sample size $n=1259$. 

In financial marketing, it is of great interest to predict the S\&P 100 index since it reveals the direction of the entire financial system. To this end, we use the daily prices of the 76 stocks to predict the S\&P100 index. Specifically, let $Y_t\in \RR^1$ be the S\&P 100 index for the $t$-th day and $\bX_t \in \RR^{76\times 2}$ be the stock prices with  lag-1 and lag-3  for the $t$-th day. Our goal is to predict $Y_t$ using $\bX_t$ under the high dimensional linear regression models and  detect multiple change points for the linear relationships between the S\&P100 index and the 76 stocks' prices. Note that we have differenced the data to remove the temporal trend. It is well known that the financial data are typically heavy-tailed and we have no prior-knowledge about the tail structure of the data. Hence, for this real data analysis, it seems very suitable to use our proposed tail-adaptive method.  We combine our proposed tail-adaptive test with the WBS method (\cite{fryzlewicz2014wild}) to detect multiple change points, which is 
demonstrated in Algorithm \ref{alg:WBS}. To implement this algorithm, we set $\cA=\{0,0.1,0.5,0.9,1\}$, $s_0=5$, $B=100$, and $V=500$ (number of random intervals). Moreover, we consider the $L_1-L_2$ weighted loss by setting $\tilde{\btau}=0.5$ in (\ref{equation: weighted loss function}). The data are scaled to have mean zeros and variance ones before the change point detection. There are 14 change points detected which are reported in Table \ref{table: realdata}. 

To further justify the meaningful findings of our proposed new methods, we refer to the T-bills and ED (TED) spread, which is short for the difference between the 3-month of London Inter-Bank Offer Rate (LIBOR), and the 3-month short-term U.S. government debt (T-bills). It is well-known that TED spread is an indicator of perceived risk in the general economy and an increased TED spread during the financial crisis reflects an increase in credit risk. Figure \ref{figure: Ted} shows the plot of TED where the red dotted lines correspond to the estimated change points. We can see that during the financial crisis from 2007 to 2009, the TED spread has experienced very dramatic fluctuations and the estimated change points can capture some big changes in the TED spread. In addition, the S\&P 100 index obtains its highest level during the financial crisis in October 2007 and then has a huge drop. Our method identifies October 29, 2007 as a change point. Moreover, the third detected change-point is January 10th, 2008. The National Bureau of Economic Research (NBER) identifies December of 2007 as the beginning of the great recession which is captured by our method. In addition, it is well known that affected by the 2008 financial crisis, Europe experienced a debt crisis from 2009 to 2012, with the Greek government debt crisis in October, 2009 serving as the starting point. Our method identifies October 5, 2009 as a change point after which S$\&$P 100 index began to experience a significant decline. Moreover, it is known that countries such as Italy and Spain were facing severe debt issues in July, 2011, raising fears about the stability of the Eurozone and the potential impact on global financial markets. As a result,  there exists another huge drop for the S$\&$P 100 index in July 26, 2011, which can be successfully detected by our method.

\begin{figure}[!h]
	\begin{center}
		\includegraphics[width=12cm]{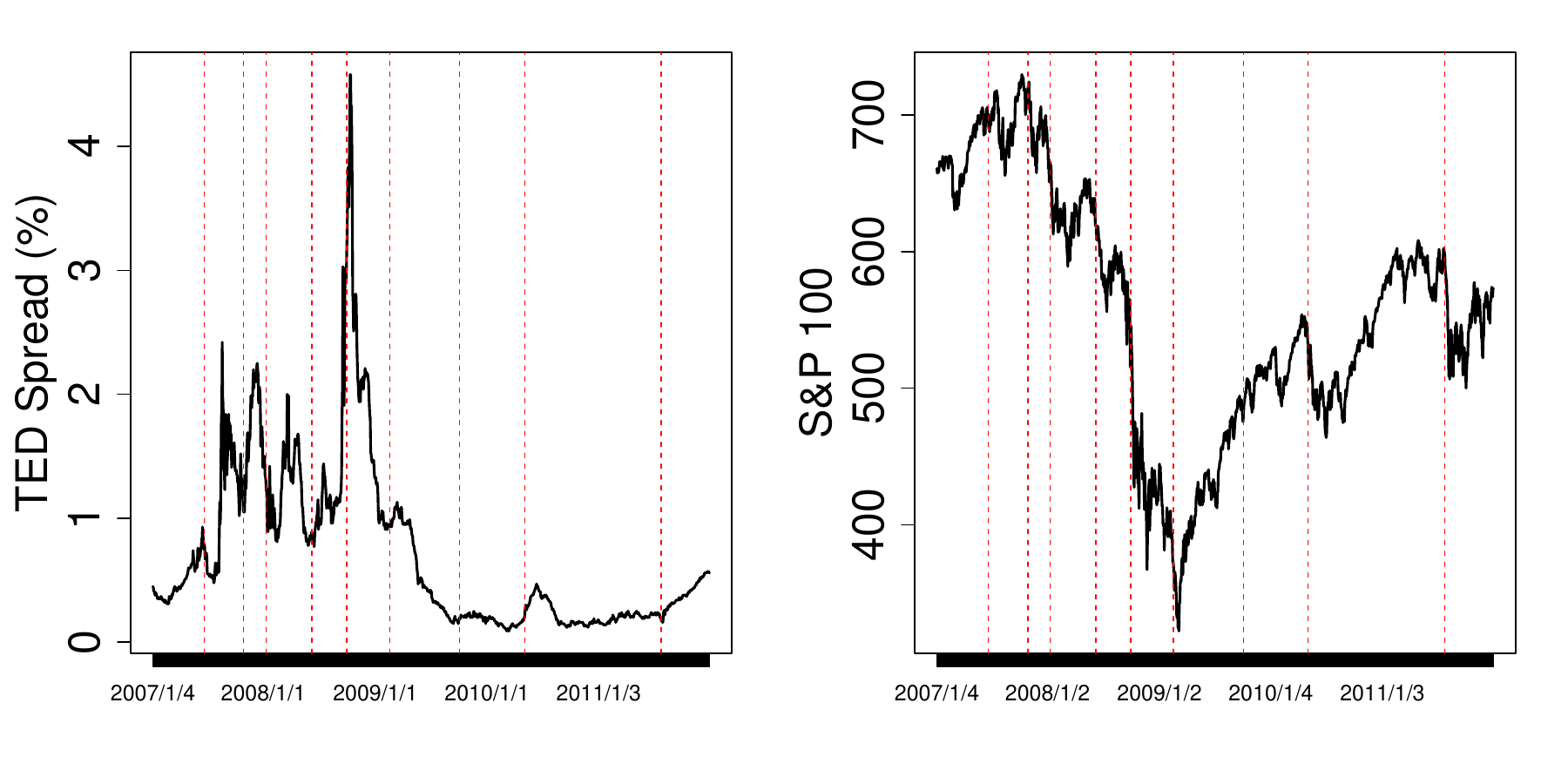}
		\vspace{-0.6cm}
		\caption{Plots of the Ted spread (left) and the S\&P 100 index (right) with the estimated change-points (vertical lines) marked by $\#$ in Table \ref{table: realdata}. }
		\label{figure: Ted}
	\end{center}
	\vspace{-0.3in}
\end{figure}

\begin{table}[!h]
	\caption{Multiple change poins detection for the S\&P 100 dataset.}\label{table: realdata}
	\vspace{-0.8cm}
	\addtolength{\tabcolsep}{-1pt}
	\begin{center}
		\begin{tabular}{cccccccc} 
			\toprule[2pt]
			&&Change points &Date&Events&&&   \\
			\hline
			&&117&2007/06/21& TED Spread$\#$&&&  \\
			&	&207&2007/10/29&TED Spread$\#$&&&  \\
			&	&257&2008/01/10&Global Financial Crisis (TED Spread)$\#$&&&  \\
			&	&360&2008/06/09&TED Spread$\#$&&&  \\
			&	&439&2008/09/30&TED Spread$\#$&&&  \\					
			&	&535&2009/02/18&Nadir of the crisis$\#$&&&  \\
			&	&632&2009/07/08&&&&  \\
			&	&694&2009/10/05&Greek debt crisis$\#$&&&  \\
			&	&840&2010/05/05&Global stock markets fell due to fears of&&&  \\
			&	&&&	 contagion of the European sovereign debt crisis$\#$ &&&  \\
			&	&890&2010/07/16&&&&  \\
			&	&992&2010/12/09&&&&  \\
			&	&1074&2011/04/07&&&&  \\
			&	&1149&2011/07/26&Spread of the European debt crisis to Spain and Italy$\#$&&&  \\
			&	&1199&2011/10/05&&&&   \\
			\hline	
			\toprule[2pt]   
		\end{tabular}
	\end{center}
\end{table}

\section{Summary}\label{section: summary}

In this article, we propose a general tail-adaptive approach for simultaneous change point testing and estimation for high-dimensional linear regression models. The method is based on the observation that both the conditional mean and quantile change if the regression coefficients have a change point. Built on a weighted composite loss, we propose a family of individual testing statistics with different weights to account for the unknown tail structures. Then, we combine the individual tests to construct a tail-adaptive method, which is powerful against sparse alternatives under various tail structures. In theory, with  mild conditions on the regression covariates and errors, we show the optimality of our methods theoretically in terms of size, power and change point estimation. In the presence of multiple change points, we combine our tail-adaptive approach with the WBS technique to detect multiple change points. With extensive numerical studies, our proposed methods  have better performance than the existing methods under various model setups.



		\bibliographystyle{ims}
		\bibliography{RefDatBas_Tail_Reg}

\renewcommand{\baselinestretch}{1}
\setcounter{footnote}{0}
\clearpage
\setcounter{page}{1}
\title{
	\begin{center}
		\Large Supplementary Materials to
		``Change Point Detection for High-dimensional Linear Models:A General Tail-adaptive Approach''
	\end{center}
}
\date{}
\begin{center}
\author{
	Bin Liu\footnotemark[1],
	Zhengling Qi\footnotemark[2]
	Xinsheng Zhang\footnotemark[1],
	Yufeng Liu\footnotemark[3]
}
\renewcommand{\thefootnote}{\fnsymbol{footnote}}

\footnotetext[1]{Department of Statistics and Data Science, School of Management at Fudan University, Shanghai, China; E-mail:{\tt liubin0145@gmail.com; xszhang@fudan.edu.cn }}
\footnotetext[2]{Department of Decision Sciences, George Washington University, U.S.A; E-mail:{\tt qizhengling321@gmail.com }}
\footnotetext[3]{Department of Statistics and Operations Research, Department of Genetics, and Department of Biostatistics, Carolina Center for Genome Sciences, Linberger Comprehensive Cancer Center, University of North Carolina at Chapel Hill, U.S.A; E-mail:{\tt yfliu@email.unc.edu }}
\end{center}
\maketitle
\appendix

This document provides detailed proofs of the main theoretical results as well as full numerical studies. 
In Section \ref{sec: method of multiple cpt}, we demonstrate how to combine our proposed tail-adaptive methods with the wild binary segmentation technique to detect multiple change points. In Section \ref{section: some notations}, we introduce some additional notations.   In Section \ref{section: useful lemmas}, some useful lemmas are provided. In Section \ref{sec:proof of main results}, we give the detailed proofs of theoretical results in the main paper. In Section \ref{sec: proofs of lemmas used in the main theory}, we provide the proof of lemmas used in Section \ref{sec:proof of main results}. In  Sections \ref{sec: proof of useful lemmas}  and \ref{sec: additional lemmas}, we prove the useful lemmas in Section \ref{section: useful lemmas} as well as some additional lemmas. 

\section{Extensions to multiple change points detection }\label{sec: method of multiple cpt}
\begin{algorithm}[!h]
	\caption{: A WBS-typed tail-adaptive test for multiple change point detection}\label{alg:WBS}
	\begin{description}
		\item[Input:] Given the data $(\mathcal{X},\cY)=\{(\bX_1,Y_1),\ldots,(\bX_n,Y_n)\}$, set the values for $\tilde{\btau}$, the significance level $\gamma$,  $s_0$, $q_0$, the bootstrap replication number $B$, the candidate subset $\cA\subset[0,1]$, and a set of random intervals $\{(s_\nu, e_\nu\}_{\nu = 1}^V$ with thresholds $v_0$ and $v_1$. Initialize an empty set $\cC$.
		\item[Step~1:] For each $\nu = 1, \cdots, V$, compute $\hat{P}_{\rm ad}(s_\nu, e_\nu)$ following Section \ref{sec: method of tail-adaptive}. 
		\item[Step~2:] Perform the following function with $S=q_0$ and $E=1-q_0$.
		\item[Function(S, E):] $S$ and $E$ are the starting and ending points for the change point detection.
		\begin{enumerate}
			\item[(a)] RETURN  if $E-S \leq v_1$.
			\item[(b)] Define $\cM = \{1 \leq \nu \leq V \, | \, [s_\nu, e_\nu] \subset [E, S]\}$.
			\item[(c)] Compute the test statistics as $\overline{P}_{\rm ad} = \underset{\nu \in \cM, v_0 \leq e_{\nu} - s_\nu}{\min} \hat{P}_{\rm ad}(s_\nu, e_\nu)$ and the corresponding optimal solution $\nu^\ast$.
			\item[(d)] If $\overline{P}_{\rm ad} \geq \gamma / V$, RETURN. Otherwise, add the corresponding change point estimator $\hat t_{\nu^\ast}$ to $\cC$, and perform Function(S, $\nu^\ast$) and Function($\nu^\ast$, E).
		\end{enumerate}
		\item[Output:] The set of multiple change points $\cC$.
	\end{description}
\end{algorithm}

In practical applications, it may exist multiple change points in describing the relationship between $\bX$ and $Y$. Therefore, it is essential to perform estimation of multiple change points if $\Hb_0$ is rejected by our powerful tail-adaptive test. In this section, we extend our single change point detection method by the idea of WBS proposed in \cite{fryzlewicz2014wild}  to estimate the locations of all possible multiple change points.

Consider a single change point detection task in any interval $[s, e]$, where $0 \leq q_0 \leq s < e \leq 1-q_0$. Following Section \ref{sec: method of tail-adaptive}, we can compute the corresponding adaptive test statistics as $\hat{P}_{\rm ad}(s, e)$ using the subset of our data, i.e., $\{ \bX_{\floor{ns}}, \bX_{\floor{ns} + 1}, \cdots, \bX_{\floor{ne}} \}$ and $\{Y_{\floor{ns}}, Y_{\floor{ns} + 1}, \cdots, Y_{\floor{ne}}\}$. Following the idea of WBS, we first independently generate a series of random intervals by the uniform distribution. Denote the number of these random intervals as $V$. For each random interval $[s_\nu, e_\nu]$ among $\nu = 1, 2, \cdots, V$, we compute $\hat{P}_{\rm ad}(s_\nu, e_\nu)$ as long as $0 \leq q_0 \leq s_\nu < e_\nu \leq 1-q_0$ and $e_\nu - s_\nu \geq v_0$, where $v_0$ is the minimum length for implementing Section \ref{sec: method of tail-adaptive}. The threshold $v_0$ is used to reduce the variability of our algorithm for multiple change point detection. Based on the test statistics computed from the random intervals, we consider the final test statistics as $\overline{P}_{\rm ad} = \underset{1 \leq \nu \leq V, v_0 \leq e_{\nu} - s_\nu}{\min} \hat{P}_{\rm ad}(s_\nu, e_\nu)$, based on which we make decisions if there exists at least one change point among these intervals. We stop the algorithm if $\overline{P}_{\rm ad} \geq \bar c$, otherwise we report the change point estimation in $[s_{\nu^\ast}, e_{\nu^\ast}]$, where $\nu^\ast \in \underset{1 \leq \nu \leq V, v_0 \leq e_{\nu} - s_\nu}{\argmin} \hat{P}_{\rm ad}(s_\nu, e_\nu)$, and continue our algorithm. Given the first change point estimator denoted by $\hat t_{\nu^\ast}$, we split our data into two folds, i.e., before and after the estimated change point. Then we apply the previous procedure on each fold of the data using the same set of the random intervals as long as it satisfies the constraints. We repeat this step until the algorithm stops returning the change point estimation. For each step, we choose $\bar c = \gamma / V$, where $\gamma$ is the significance level used in each single change point detection algorithm.  While we do not have the theoretical guarantee of using $\bar c$ in the proposed algorithm for controlling the size, the selection of this constant is based on the idea of Bonferroni correction, which is conservative. The numerical experiments in the appendix demonstrate the superiority of our proposed method in detecting multiple change points. Nevertheless, it is interesting to study the asymptotic property of $\overline{P}_{\rm ad}$, which we leave  for the future work. The  full algorithm of the multiple change point detection can be found in Algorithm \ref{alg:WBS}.

\section{Some notations}\label{section: some notations}
Before the proofs, we give some notations. Under $\Hb_0$, we set $\bbeta^{(0)}:=\bbeta^{(1)}=\bbeta^{(2)}$ and $s^{(0)}:=s^{(1)}=s^{(2)}$.  Under $\Hb_1$, For the  regression vectors $\bbeta^{(1)}$ and $\bbeta^{(2)}$, define $\cS^{(1)}=\{1\leq j\leq p: \beta_j^{(1)}\neq 0\}$ and
$\cS^{(2)}=\{1\leq j\leq p: \beta_j^{(2)}\neq 0\}$
as the active sets of variables. Denote $s^{(1)}=|\cS^{(1)}| $  and $s^{(2)}=|\cS^{(2)}| $ as the cardinalities of $\cS^{(1)}$ and $\cS^{(2)}$, respectively. We set $\cS=\cS^{(1)}\cup \cS^{(2)}$ and $s=|\cS|$.
For a vector $\bv\in \RR^p$, we denote $J(\bv)=\{1\leq j\leq p: v_j\neq 0\}$ as the set of non-zero elements of $\bv$ and set $\cM(\bv):=|J(\bv)|$ as the number of non-zero elements of $\bv$. For a set $J$ and $\bv\in \RR^p$, denote $\bv_J$ as the vector in $\RR^p$ that has the same coordinates as $\bv$ on $J$ and zero coordinates on the complement $J^c$ of $J$. For any vector $\bx\in \RR^p$ and a matrix $\Ab\in \RR^{p\times p}$, define $\|\bx\|^2_{\Ab}=\bx^\top\Ab\bx$.
Denote $\cX=\{\Xb,\bY\}$. 
We use $C_1, C_2,\ldots$ to denote constants that may vary from line to line. We use  w.p.a.1 for the abbreviation of with probability approaching to one. For $\beta>0$, we define the function $\psi_{\beta}: [0,\infty)\rightarrow [0,\infty)$ as $\psi_{\beta}(x):=\exp(x^\beta)-1$. Then, for any random variable $X$, we define 
\begin{equation*}
	\|X\|_{\psi_{\beta}}:=\inf\big\{C>0: \E\psi_{\beta}(|X|/C)|)\leq 1\big\}.
\end{equation*}
For any $0\leq s<t\leq 1$, we denote
\begin{equation}\label{equation: big sigma}
	\hat{\bSigma}(s:t)=\dfrac{1}{\lfloor nt \rfloor -\lfloor ns \rfloor+1 }\sum_{i=\floor{ns}+1}^{\floor{nt}}\bX_i\bX_i^\top.
\end{equation}

\section{Useful lemmas}\label{section: useful lemmas}
\begin{lemma}[Lemma E.1 in \cite{chernozhukov2017central}]\label{lemma: maximum inequality}
	Let $\bX_1,\ldots,\bX_n\in \RR^{p}$ with $\bX_i=(X_{i1},\ldots,X_{ip})^\top$ be independet and centered random vectors. Define $Z=\max_{1\leq j\leq p}|\sum_{i=1}^n X_{ij}|$, $M=\max_{1\leq i\leq n}\max_{1\leq j\leq p}|X_{ij}|$ and $\sigma^2=\max_{j}\sum_{i}\E[X_{ij}^2]$. Then,
	\begin{equation*}
		\E [Z]\leq C(\sigma\sqrt{\log p}+\sqrt{\E[M^2]}\log p),
	\end{equation*}
	where $C$ is some universal constant. 
\end{lemma}	

\begin{lemma}[Lemma E.2 in \cite{chernozhukov2017central}]\label{lemma: concentration inequality for maximum} (a)Assume the setting of Lemma \ref{lemma: maximum inequality} holds. For every $\eta>0,\beta\in(0,1]$ and $t>0$, we have
	\begin{equation*}
		\P(Z\geq (1+\eta)\E [Z]+t)\leq \exp(-\dfrac{t^2}{3\sigma^2})+ 3\exp\Big(-\big(\dfrac{t}{K\|M\|_{\psi_{\beta}}}\big)^\beta\Big),
	\end{equation*}
	where $K=K(\eta,\beta)$ is a constant only depending on $\eta$ and $\beta$. 
	\\(b) Assume the setting of Lemma \ref{lemma: maximum inequality} holds. For every $\eta>0,s\geq 1$ and $t>0$, we have
	\begin{equation*}
		\P(Z\geq (1+\eta)\E [Z]+t)\leq \exp(-\dfrac{t^2}{3\sigma^2})+ K'\dfrac{\E [M^s]}{t^s},
	\end{equation*}
	where $K'=K(\eta,s)$ is a constant only depending on $\eta$ and $s$. 
\end{lemma}

\begin{lemma}[Hoeffding's inequality]\label{lemma: hoeffding's inequality}
	Suppose $X_1,\ldots,X_n\in \RR^1$ be independent random variables with $|X_i|\leq K $ for some $K>0$. Let $\bar{X}$ be the sample mean. Then, for any $x>0$, we have
	\begin{equation}
		\P(|\bar{X}-\E \bar{X}|\geq x)\leq 2\exp\Big(-\dfrac{nx^2}{2K^2}\Big).
	\end{equation}
\end{lemma}
\begin{lemma}[Nazarovs inequality in \cite{nazarov2003maximal}] \label{lemma:anti consentration inequality}
	Let $\bW=(W_1,W_2,\cdots,W_d)^\top \in \mathbb{R}^{p}$ be centered Gaussian random vector with  $\inf_{1\leq k\leq p}\E(W_k)^2\geq b>0$. Then for
	any $\bx\in \mathbb{R}^p$ and $a>0$, we have
	\begin{equation*}
		\P(\bW\leq \bx+a)-\P(\bW\leq \bx)\leq Ca\sqrt{\log p},
	\end{equation*}
	where $C$ is a constant only depending on $b$.
\end{lemma}

Before introducing Lemma \ref{lemma: convex approximation}, we need some definitions for an $m$-generated convex set $A^{m}$. We say a set $A^m$ is $m$-generated if it is generated by intersecting $m$ half spaces. In other words, the set $A^m$ is a convex polytope with at
most $m$ facets. Moreover, for any $\epsilon>0$ and an $m$-generated convex set $A^{m}$, we define
\begin{equation}\label{equation: Ame}
	A^{m,\epsilon}=\bigcap_{\bv\in \mathcal{V}(A^{m})}\{\bw \in \mathbb{R}^{d}: \bw^\top \bv\leq S_{A^m}(\bv)+\epsilon\},
\end{equation}
where $\mathcal{V}(A^{m})$ consists $m$ unit vectors that are outward normal to the facets of $A^m$, and  $S_{A^m}(\bv)$ is the support function for $A^{m}$ (see \cite{chernozhukov2017central}).

Let $\bx=(x_1,\ldots,x_p)\in \RR^p$. For any $1\leq s_0\leq p$, define $\|\bx\|_{(s_0,2)}=(\sum_{j=1}^{s_0}|x_{(j)}|^2)^{1/2}$, where $|x_{(1)}\geq |x_{(2)}|\cdots\geq |x_{(p)}|$ be the order statistics of $\bx$. 
The following lemma shows that the set $V_{(s_0,2)}^{z,p}:=\{\bx\in \mathbb{R}^{p}: \|\bx\|_{(s_0,2)}\leq z\}$ can be approximated by $m$ generated convex set. 
\begin{lemma}[\cite{Zhou2017An}]\label{lemma: convex approximation}
	Let $\mathcal{E}^{R,p}=\{\bx\in \mathbb{R}^{p}: \|\bx\|\leq R\}$ and $V_{(s_0,2)}^{z,p}=\{\bx\in \mathbb{R}^{p}: \|\bx\|_{(s_0,2)}\leq z\}$. For any $\gamma>e/4\sqrt{2}$, there is a $m$-generated convex set $A^{m}\in \RR^{p}$ and a constant $\epsilon_{\gamma}$ such that for any $0<\epsilon<\epsilon_{\gamma}$, we have
	\begin{equation*}
		A^{m}\subset \mathcal{E}^{R,p}\cap V_{(s_0,2)}^{z,p}\subset A^{m,R\epsilon} ~~\text{and}~~ m\leq p^{s_0}\Big(\frac{\gamma}{\sqrt{\epsilon}}\ln(\dfrac{1}{\epsilon})\Big)^{s_0^2}.
	\end{equation*}
\end{lemma}
The following Lemma \ref{lemma: key lemma for gaussian approximation} shows the Gaussian approximation theory for the testing statistic, which is very important for the size control. To show that, we need some notations and assumptions. In particular, let $\bZ_1,\ldots,\bZ_n\sim(\mathbf{0},\bSigma)$ be independent and centered random vectors in $\mathbb{R}^{p}$ with $\bZ_i=(Z_{i1},\ldots, Z_{ip})^\top$ for $i=1,\ldots,n$. Let $\bG_1,\ldots,\bG_n$ be independent centered Gaussian random vectors in $\mathbb{R}^{p}$ such that each $\bG_i$ has the same covariance matrix as $\bZ_i$. Let $\mathcal{V}_{s_0}:=\{\bv\in \mathbb{S}^{q-1}: \|\bv\|_0\leq s_0\}$, where $\mathbb{S}^{q-1}:=\{\bv\in\RR^p: \|\bv\|=1\}$. We require the following conditions:
\begin{description}
	\item[\bf (M1)] There is a constant $b>0$ such that $\inf_{\bv\in \mathcal{V}_{s_0} }\dfrac{1}{n}\sum\limits_{i=1}^n\E(\bv^\top\bZ_i)^{2}\geq b$  for  $i=1,\ldots,n$. 
	\item[\bf (M2)] There exists some constant $K>0$ such that $\max\limits_{1\leq j
		\leq p} \dfrac{1}{n}\sum\limits_{i=1}^{n}\E|Z_{ij}|^{2+\ell}\leq K^{\ell}$ for $\ell=1,2$.
	\item[\bf (M3)] There exists a constant $K>0$ and $q>0$ such that $\E ((\max_{1\leq j \leq p } |Z_{ij}|/K)^q)\leq 2$ holds for all $i=1,\ldots,n$.
\end{description}
\begin{lemma}\label{lemma: key lemma for gaussian approximation}
	Assume that that $s_0^3K^{2/7}\log(pn)=O(n^{\xi_1})$ for some $0<\xi_1<1/7$ and $s_0^4K^{2/3}\log(pn)=O(n^{\xi_2})$ for some $0<\xi_2<\frac{1}{3}(1-2/q)$. Let
	\begin{equation}\label{equation: SZ+SG}
		\bS^{\bZ}(k)=\dfrac{1}{\sqrt{n}}\Big(\sum_{i=1}^{k}\bZ_i-\dfrac{k}{n}\sum_{i=1}^n\bZ_i\Big),~~	\bS^{\bG}(k)=\dfrac{1}{\sqrt{n}}\Big(\sum_{i=1}^{k}\bG_i-\dfrac{k}{n}\sum_{i=1}^n\bG_i\Big),
	\end{equation}
	be the partial sum processes for $(\bZ_i)_{i\geq 1}$ and $(\bG_i)_{i\geq 1}$, respectively.
	If $\bZ_1,\ldots,\bZ_n$ satisfy $\bf (M1)$, $\bf (M2)$ and $\bf (M3)$, then there is a constant $\zeta_0>0$ such that
	\begin{equation}\label{equation: key lemma for gaussian approximation1}
		\sup_{z\in(0,\infty)} \big|\P(\max_{k_0\leq k\leq n-k_0}\|\bS^{\bZ}(k)\|_{(s_0,2)}\leq z\big)-\P(\sup_{k_0\leq k\leq n-k_0}\|\bS^{\bG}(k)\|_{(s_0,2)}\leq z\big)\big|\leq Cn^{-\zeta_0},
	\end{equation}
	where $C$ is a constant only depending on $b,q$, $K$ and $k_0:=\floor {nq_0}$ for some $0<q_0<0.5.$
\end{lemma}

The following Lemmas \ref{lemma: exponential inequality for partial sum process} and Lemma \ref{lemma: concentration for covariance} present the orders for the partial sum process of $\{\bX\epsilon\}_{i=1}^n$ as well as the $\ell_\infty$-norm based  uniform large deviation bound for $\hat{\bSigma}(0:t)$ and $\hat{\bSigma}(t:1)$ , which will be frequently used throughout the proofs.
\begin{lemma}\label{lemma: exponential inequality for partial sum process}
	Let $\bX_1,\ldots,\bX_n$ be independent centered random vectors in $\mathbb{R}^{p}$ and $\epsilon_1,\ldots,\epsilon_n$ be independent centered random vectors in $\mathbb{R}^{1}$. Suppose further that $\{\bX_i\}_{i=1}^n$ and $\{\epsilon_i\}_{i=1}^n$ satisfy {\bf{Assumptions A -- C}} in the main paper. Then, for any sequence $a_n\in(0,1)$ and $b_n\in(0,1)$ satisfying  $\floor{n a_n}\rightarrow \infty$ and  $\floor{n b_n}\rightarrow \infty$ as $n\rightarrow \infty$,  we have
	
	\begin{equation}
		\begin{array}{ll}
			\max\limits_{t\in[a_n,1-b_n]}\max\limits_{1\leq j\leq p}\Big|\dfrac{1}{\sqrt{n}}\Big( \sum\limits_{i=1}^{\lfloor nt \rfloor}X_{ij}\epsilon_i-\dfrac{\lfloor nt \rfloor}{n} \sum\limits_{i=1}^{n}X_{ij}\epsilon_i\Big)\Big|\\
			\quad\quad	=\max\limits_{t\in[a_n,1-b_n]}\Big\|\dfrac{1}{\sqrt{n}}\Big( \sum\limits_{i=1}^{\lfloor nt \rfloor}\bX_{i}\epsilon_i-\dfrac{\lfloor nt \rfloor}{n} \sum\limits_{i=1}^{n}\bX_{i}\epsilon_i\Big)\Big\|_{\infty}\\
			\quad\quad= O_p( M\sqrt{\log(p(n-\underline{k}_n-\overline{k}_n))}),
		\end{array}
	\end{equation}
	where $\underline{k}_n:=\floor{na_n}$ and $\overline{k}_n:=\floor{nb_n}$. Moreover, we can also have the following results:
	\begin{equation}
		\begin{array}{ll}
			\max\limits_{t\in[a_n,1-b_n]}\max\limits_{1\leq j\leq p}\Big|\dfrac{1}{{\floor{nt}}} \sum\limits_{i=1}^{\lfloor nt \rfloor}X_{ij}\epsilon_i\Big|
			=\max\limits_{t\in[a_n,1-b_n]}\Big\|\dfrac{1}{{\floor{nt}}} \sum\limits_{i=1}^{\lfloor nt \rfloor}\bX_{i}\epsilon_i\Big\|_{\infty}\\
			\quad =O_p\Big( M\sqrt{\dfrac{\log(pn)}{\underline{k}_n}}\max\big\{1,n^{1/4}\sqrt{\dfrac{\log(pn)}{\underline{k}_n}}\big\}\Big).
		\end{array}
	\end{equation}
\end{lemma}	

\begin{lemma}\label{lemma: concentration for covariance}
	Let $\bX_1,\ldots,\bX_n$ be independent centered random vectors in $\mathbb{R}^{p}$ satisfying {\bf{ Assumption A}}. Let $\bSigma=\text{Cov}(\bX_1)$. Recall $\hat{\bSigma}(s:t)$ defined in (\ref{equation: big sigma}). Then, for any sequence $a_n\in(0,1)$ and $b_n\in(0,1)$ satisfying  $\floor{n a_n}\rightarrow \infty$ and  $\floor{n b_n}\rightarrow \infty$ as $n\rightarrow \infty$, with probability at least $1-(np)^{-C_1}$, we have:
	\begin{equation*}
		\begin{array}{ll}
			\max\limits_{a_n\leq t \leq 1-b_n}\|\hat{\bSigma}(0:t)-\bSigma\|_{\infty}\leq C_2M^2\sqrt{\dfrac{\log(pn)}{\floor{na_n}}},\\
			\max\limits_{a_n\leq t \leq 1-b_n}\|\hat{\bSigma}(t:1)-\bSigma\|_{\infty}\leq C_3M^2\sqrt{\dfrac{\log(pn)}{\floor{nb_n}}}.\\
		\end{array}
	\end{equation*}
	Moreover, if we take $a_n=b_n=q_0\in(0,0.5)$, we have 
	\begin{equation*}
		\begin{array}{ll}
			\max\limits_{q_0\leq t \leq 1-q_0}\|\hat{\bSigma}(0:t)-\bSigma\|_{\infty}\leq C_4M^2\sqrt{\dfrac{\log(pn)}{n}},\\
			\max\limits_{q_0\leq t \leq 1-q_0}\|\hat{\bSigma}(t:1)-\bSigma\|_{\infty}\leq C_5M^2\sqrt{\dfrac{\log(pn)}{n}},\\
		\end{array}
	\end{equation*}
	where $C_1,\ldots,C_5$ are some universal constants.
	Note that Lemma \ref{lemma: concentration for covariance} is a direct consequence of Lemma \ref{lemma: hoeffding's inequality}. The proof is ommitted.
\end{lemma}	

Note that for proving our results, we need some theoretical analysis for the lasso estimator defined in (\ref{equation: lasso estimator}). The following Lemmas \ref{lemma: basic inequality for lasso with alpha=1} - \ref{lemma: basic inequality for lasso with alpha}  show the lasso property for $\alpha\in[0,1]$ under both $\Hb_0$ and $\Hb_1$, which is very important for deriving the theoretical results for the individual test. Before presenting the details, for each $\alpha\in[0,1]$, we introduce $\uwave{\bbeta}^*=((\bbeta^*)^\top,(\bb^*)^\top)^\top\in \RR^{p+K}$ with $\bb^*=(b_1^*,\ldots,b_K^*)^\top\in \RR^K$, where 
\begin{equation}\label{equation: parameters under population level}
	\uwave{\bbeta}^*:=\argmin_{\bbeta\in \RR^p,\bb\in \RR^K}\E \Big[(1-\alpha)\dfrac{1}{n}\sum_{i=1}^{n}\dfrac{1}{K}\sum_{k=1}^{K}\rho_{\tau_k}(Y_i-b_i-\bX_i^\top\bbeta)+ \dfrac{\alpha}{2n}\sum_{i=1}^{n}(Y_i-\bX_i^\top \bbeta)^2\Big].
\end{equation}
Note that by definition, we can regard $\uwave{\bbeta}^*$ as the true parameters under the population level. In this paper, we assume $\bbeta^*$ enjoys some sparsity property in the sense that $\cM(\bbeta^*)=O(s)$. Moreover, the properties of $\uwave{\bbeta}^*$ are discussed in Sections \ref{sec: Proof of lasso property with alpha=1} - \ref{sec: Proof of lasso property with alpha}, respectively.  

The following Lemma \ref{lemma: basic inequality for lasso with alpha=1} shows the lasso property with $\alpha=1$. The proof of Lemma \ref{lemma: basic inequality for lasso with alpha=1} is given in Section \ref{sec: Proof of lasso property with alpha=1}.
\begin{lemma}[Lasso property with $\alpha=1$]\label{lemma: basic inequality for lasso with alpha=1}
	Let $\hat{\bbeta}$ be the lasso estimator with $\alpha=1$ defined in (\ref{equation: lasso estimator}). Let $\lambda_\alpha=C_{\lambda}M^2\sqrt{{\log(pn)}/{n}}$ for some big enough constant $C_\lambda>0$.
	Assume {\bf{Assumptions A, B, C.2, E.2 - E.3}} hold.  Then, with probability tending to one, we have 
	\begin{equation}\label{inequality: lasso estimator inequality}
		\begin{array}{ll}
			\dfrac{1}{2n}\big\|\Xb(\hat{\bbeta}-\bbeta^*)\big\|^2\leq C_2\lambda^2s,~~\|\hat{\bbeta}-\bbeta^*\|_{q}\leq C_3\lambda s^{1/q}, ~\text{and}~~\cM(\hat{\bbeta})\leq C_4s, ~~\text{for}~q=1,2.
		\end{array}
	\end{equation} 
\end{lemma}	
The following Lemma \ref{lemma: basic inequality for lasso with alpha=0} shows the lasso property with $\alpha=0$ under both $\Hb_0$ and $\Hb_1$. The proof of Lemma \ref{lemma: basic inequality for lasso with alpha=0} is given in Section \ref{sec: Proof of lasso property with alpha=0}.
\begin{lemma}[Lasso property with $\alpha=0$]\label{lemma: basic inequality for lasso with alpha=0}
	Let $\hat{\bbeta}$ be the lasso estimator with $\alpha=0$ defined in (\ref{equation: lasso estimator}). Let $\lambda=C_{\lambda}M\sqrt{{\log(pn)}/{n}}$ for some big enough constant $C_{\lambda}>0$. 
	Assume {\bf{Assumptions A, D, E.2 - E.3}} hold. Then, with probability tending to one, we have 
	\begin{equation}\label{inequality: lasso estimator inequality with alpha=0}
		\begin{array}{ll}
			\dfrac{1}{n}\big\|\Xb(\hat{\bbeta}-\bbeta^*)\big\|^2\leq C_2\lambda^2s,~~\|\uwave{\hat{\bbeta}}-\uwave{\bbeta}^*\|_{q}\leq C_3\lambda s^{1/q}, ~\text{and}~~\cM(\hat{\bbeta})\leq C_4s, ~~\text{for}~q=1,2.
		\end{array}
	\end{equation} 
\end{lemma}	
The following Lemma \ref{lemma: basic inequality for lasso with alpha} show the lasso property with $\alpha\in(0,1)$ under both $\Hb_0$ and $\Hb_1$. The proof of Lemma \ref{lemma: basic inequality for lasso with alpha} is given in Section \ref{sec: Proof of lasso property with alpha}.
\begin{lemma}[Lasso property with $\alpha\in(0,1)$]\label{lemma: basic inequality for lasso with alpha}
	Let $\hat{\bbeta}$ be the lasso estimator with $\alpha\in(0,1)$ defined in (\ref{equation: lasso estimator}). Let $\lambda=C_{\lambda}M^2\sqrt{{\log(pn)}/{n}}$ for some big enough constant $C_{\lambda}>0$. Assume {\bf{Assumptions A, B, C.2, D, E.2 - E.3}} hold. Then, with probability tending to one, we have 
	\begin{equation}\label{inequality: lasso estimator inequality with alpha}
		\begin{array}{ll}
			\dfrac{1}{n}\big\|\Xb(\hat{\bbeta}-\bbeta^*)\big\|^2\leq C_2\lambda^2s,~~\|\uwave{\hat{\bbeta}}-\uwave{\bbeta}^*\|_{q}\leq C_3\lambda s^{1/q}, ~\text{and}~~\cM(\hat{\bbeta})\leq C_4s,~~\text{for}~q=1,2.
		\end{array}
	\end{equation} 
\end{lemma}

\section{Proof of main results}\label{sec:proof of main results}

\subsection{Proof of Theorem \ref{theorem: variance estimator under H0}}\label{sec: proof of variance estimation under H0}
In this section, we prove the variance estimation results under $\Hb_0$, which are given in  Sections \ref{sec: variance with alpha=1 H0}
- \ref{sec: variance with alpha=0 H0}, respectively. For simplicity, we  omit the subscript $\alpha$ whenever needed.
\subsubsection{Proof of Theorem \ref{theorem: variance estimator under H0} with $\alpha=1$}\label{sec: variance with alpha=1 H0}
Note that for $\alpha=1$, the variance estimators $\hat{\sigma}_-^2(1,\tilde{\btau})$ and $\hat{\sigma}_+^2(1,\tilde{\btau})$ reduce to
\begin{equation*}
	\hat{\sigma}_-^2(1,\tilde{\btau}):=\dfrac{1}{|n_-|}\sum_{i\in n_-}\big[\hat{\epsilon}_i\big]^2,~~\hat{\sigma}_+^2(1,\tilde{\btau}):=\dfrac{1}{|n_+|}\sum_{i\in n_+}\big[\hat{\epsilon}_i\big]^2,
\end{equation*}
where $\hat{\epsilon}_i$ is defined in (\ref{equation: regression residuals}). Moreover, under $\Hb_0$, the change point estimator $\hat{t}_1$ can be an arbitrary number which satisfies $\hat{t}_1\in[q_0,1-q_0]$. We aim to prove both $\hat{\sigma}_-^2(1,\tilde{\btau})$ and $\hat{\sigma}_+^2(1,\tilde{\btau})$ are consistent. We first consider $\hat{\sigma}_-^2(1,\tilde{\btau})$.
In fact, by the definition of $Y_i=\epsilon_i+\bX_i^\top\bbeta^{(0)}$, we have:
\begin{equation*}
	\begin{array}{ll}
		\hat{\sigma}_-^2(1,\tilde{\btau})\\
		=\underbrace{\dfrac{1}{|n_-|}\sum_{i\in n_-}\epsilon_i^2}_{I}+\underbrace{(\hat{\bbeta}^{(1)}-\bbeta^{(0)})^\top\dfrac{1}{|n_-|}\sum_{i\in n_-}\bX_i\bX_i^\top(\hat{\bbeta}^{(1)}-\bbeta^{(0)})}_{II}\\
		\qquad\qquad+\underbrace{\dfrac{2}{|n_-|}\sum_{i\in n_-}\epsilon_i\bX_i^\top(\bbeta^{(0)}-\hat{\bbeta}^{(1)})}_{III}.
	\end{array}
\end{equation*}
For $I$, by Assumption C.2 and according to the law of large numbers, we have $I-\sigma^2=O_p(\dfrac{1}{\sqrt{n}})$. For $II$, similar to the proof of Lemma \ref{lemma: basic inequality for lasso with alpha=1}, under $\Hb_0$ and {\bf{Assumptions A, B, C.2, E.2 - E.4}},  it is not hard to prove: $II=O_p(s\dfrac{\log(pn)}{n})$. For $III$, using the Cauchy-Swartz inequality, we have:
\begin{equation*}
	III\leq 2\sqrt{\dfrac{1}{|n_-|}\sum_{i\in n_-}\epsilon_i^2}\times \sqrt{(\hat{\bbeta}^{(1)}-\bbeta^{(0)})^\top\dfrac{1}{|n_-|}\sum_{i\in n_-}\bX_i\bX_i^\top(\hat{\bbeta}^{(1)}-\bbeta^{(0)})}=O_p(\sqrt{s\dfrac{\log(pn)}{n}}).
\end{equation*}
Combining the above results, by Assumption E.2, we have:
\begin{equation*}
	\hat{\sigma}_-^2(1,\tilde{\btau})-\sigma^2=O_{p}(\sqrt{s\dfrac{\log(pn)}{n}}). 
\end{equation*}
With a similar analysis, we can prove that the same bound applies to $\hat{\sigma}_+^2(1,\tilde{\btau})-\sigma^2$, which yields:
\begin{equation*}
	|\hat{\sigma}^2(1,\tilde{\btau})-\sigma^2|=|\hat{t}_{1}\times(\hat{\sigma}_-^2(1,\tilde{\btau})-\sigma^2)+(1-\hat{t}_{1})\times(\hat{\sigma}_+^2(1,\tilde{\btau})-\sigma^2)|=O_{p}(\sqrt{s\dfrac{\log(pn)}{n}}). 
\end{equation*}
\subsubsection{Proof of Theorem \ref{theorem: variance estimator under H0} with $\alpha=0$}\label{sec: variance with alpha=0 H0}
Note that  for $\alpha=0$, the true variance has the following  explicit form:
\begin{equation*}
	{\sigma}^2(0,\tilde{\btau}):=\text{Var}[e_i(\tilde{\btau})]=\dfrac{1}{K^2}\sum_{k_1=1}^{K}\sum_{k_2=1}^{K}\gamma_{k_1k_2},~~\text{with}~~ \gamma_{k_1k_2}:=\min(\tau_{k_1},\tau_{k_2})-\tau_{k_1}\tau_{k_2}.
\end{equation*}
In this case, the variance estimators $\hat{\sigma}_-^2(0,\tilde{\btau})$ and $\hat{\sigma}_+^2(0,\tilde{\btau})$ reduce to
\begin{equation*}
	\hat{\sigma}_-^2(0,\tilde{\btau}):=\dfrac{1}{|n_-|}\sum_{i\in n_-}\big[\hat{e}_i(\tilde{\btau})\big]^2,~~\hat{\sigma}_+^2(0,\tilde{\btau}):=\dfrac{1}{|n_+|}\sum_{i\in n_+}\big[\hat{e}_i(\tilde{\btau})\big]^2,
\end{equation*}
where  $\hat{e}_i(\tilde{\btau})=K^{-1}\sum\limits_{k=1}^{K}\hat{e}_i(\tau_k)$ with $\hat{e}_i(\tau_k)$ being defined in (\ref{equation: regression quantile residuals}). Let $\uwave{\bbeta}^{(0)}:=((\bbeta^{(0)})^\top,$\\$(\bb^{(0)})^\top)^\top$ be the true parameters under $\Hb_0$ and $\hat{\uwave{\bbeta}}^{(1)}:=(\hat{\bbeta}^{(1)})^\top,\hat{\bb}^{(1)})^\top)^\top$ and $\hat{\uwave{\bbeta}}^{(2)}:=(\hat{\bbeta}^{(2)})^\top,\hat{\bb}^{(1)})^\top)^\top$ be the estimators using samples in $n_-$ and $n_+$, respectively.  Similar to the proof of Lemma \ref{lemma: basic inequality for lasso with alpha=0}, under $\Hb_0$ and {\bf{Assumptions A, D, E.2 - E.4}}, it is not hard to prove that:
\begin{equation}\label{equation: bound of lasso variance with alpha=0}
	\|\hat{\uwave{\bbeta}}^{(1)}-\uwave{\bbeta}^{(0)}\|_1=O_p(s\sqrt{\dfrac{\log(pn)}{n})},~~\|\hat{\uwave{\bbeta}}^{(2)}-\uwave{\bbeta}^{(0)}\|_1=O_p(s\sqrt{\dfrac{\log(pn)}{n}}).
\end{equation}
We first prove the consistency of $\hat{\sigma}_-^2(0,\tilde{\btau})$.
For $\hat{e}_{i}(\tilde{\btau})$ with $i\in n_-$, it has  the following decomposition: 
\begin{equation}
	\hat{e}_{i}(\tilde{\btau})=e_{i}(\tilde{\btau})+\E[\hat{e}_i(\tilde{\btau})-e_i(\tilde{\btau})]+\underbrace{\{\hat{e}_i(\tilde{\btau})-e_i(\tilde{\btau})-\E[\hat{e}_i(\tilde{\btau})-e_i(\tilde{\btau})]\}}_{V_i(\tilde{\btau})},
\end{equation}
where 
\begin{equation}\label{equation: decomposition of ei}
	\begin{array}{ll}
		e_i(\tilde{\btau}):=\dfrac{1}{K}\sum\limits_{k=1}^{K}e_i(\tau_k),~\text{with}~e_i(\tau_k)=\mathbf{1}\{\epsilon_i\leq b^{(0)}_{k}\}-\tau_k,\\
		\E[\hat{e}_i(\tilde{\btau})-e_i(\tilde{\btau})]:=\dfrac{1}{K}\sum\limits_{k=1}^{K}\E\big[\hat{e}_i({\tau_k})-e_i({\tau_k})\big],&\\
		V_i(\tilde{\btau}):=\dfrac{1}{K}\sum\limits_{k=1}^{K}V_{i}(\tau_k),&
	\end{array}
\end{equation}
and 
\begin{equation*}
	\begin{array}{ll}
		V_{i}(\tau_k)=[\mathbf{1}\{Y_i-\hat{b}^{(1)}_k-\bX_i^\top\hat{\bbeta}^{(1)}\leq 0\}-\mathbf{1}\{\epsilon_i\leq b^{(0)}_{k}\}]\\
		\qquad\qquad-\E[\mathbf{1}\{Y_i-\hat{b}^{(1)}_k-\bX_i^\top\hat{\bbeta}^{(1)}\leq 0\}-\mathbf{1}\{\epsilon_i\leq b^{(0)}_{k}\}].
	\end{array}
\end{equation*}
By Taylor's expansion, for $\E[\hat{e}_i(\tilde{\btau})-e_i(\tilde{\btau})]$, we can further decompose into two terms:
\begin{equation}\label{equation: taylor expansion for composite quantile variance H0}
	\begin{array}{ll}	
		\E[\hat{e}_i(\tilde{\btau})-e_i(\tilde{\btau})]
		&=\dfrac{1}{K}\sum\limits_{k=1}^{K}\E\big[\hat{e}_i({\tau_k})-e_i({\tau_k})\big]=\underbrace{\dfrac{1}{K}\sum\limits_{k=1}^{K}M_i^{(1)}(\tau_{k})}_{M_i^{(1)}(\tilde{\btau})}+\underbrace{\dfrac{1}{K}\sum\limits_{k=1}^{K}M_i^{(2)}(\tau_{k})}_{M_i^{(2)}(\tilde{\btau})},
	\end{array}
\end{equation}
where
\begin{equation}
	\begin{array}{ll}
		M_i^{(1)}(\tau_{k})
		:=f_\epsilon(b_{k}^{(0)})\big(\hat{b}^{(1)}_k-b_{k}^{(0)}+\bX_i^\top(\hat{\bbeta}^{(1)}-\bbeta^{(0)})\big),\\
		M_i^{(2)}(\tau_{k}):=\dfrac{1}{2}f'_{\epsilon}(\xi_{ik})\big(\hat{b}^{(1)}_k-b_{k}^{(0)}+\bX_i^\top(\hat{\bbeta}^{(1)}-\bbeta^{(0)})\big)^2,
	\end{array}
\end{equation}
with $\xi_{ik}$ being some constant that between ${b}_{k}^{(0)}$ and $\hat{b}^{(1)}_k+\bX_i^\top(\hat{\bbeta}^{(1)}-\bbeta^{(0)})$. Hence, based on the above decomposition, for $\hat{\sigma}_-^2(0,\tilde{\btau})-{\sigma}^2(0,\tilde{\btau})$, it can be  decomposed into ten terms:
\begin{equation*}
	\hat{\sigma}_-^2(0,\tilde{\btau})-{\sigma}^2(0,\tilde{\btau})=A_1+\cdots+A_{10},
\end{equation*}
where $A_1+\cdots+A_{10}$ are defined as:
\begin{equation*}
	\begin{array}{ll}
		A_1:= \dfrac{1}{|n_-|}\sum\limits_{i\in n_-}\big[{e}_i(\tilde{\btau})\big]^2-{\sigma}^2(0,\tilde{\btau}),~~
		A_2:= \dfrac{1}{|n_-|}\sum\limits_{i\in n_-}\big[  M_i^{(1)}(\tilde{\btau}) \big]^2\\
		A_3=\dfrac{1}{|n_-|}\sum\limits_{i\in n_-}\big[  M_i^{(2)}(\tilde{\btau}) \big]^2,~~
		A_4=\dfrac{1}{|n_-|}\sum\limits_{i\in n_-}\big[  V_i(\tilde{\btau}) \big]^2\\
		A_5:= \dfrac{2}{|n_-|}\sum\limits_{i\in n_-}\big[{e}_i(\tilde{\btau})M_i^{(1)}(\tilde{\btau}) \big],~~A_6:= \dfrac{2}{|n_-|}\sum\limits_{i\in n_-}\big[{e}_i(\tilde{\btau})M_i^{(2)}(\tilde{\btau}) \big]\\
		A_7:= \dfrac{2}{|n_-|}\sum\limits_{i\in n_-}\big[{e}_i(\tilde{\btau})V_i(\tilde{\btau}) \big],~~A_8:= \dfrac{2}{|n_-|}\sum\limits_{i\in n_-}\big[M_i^{(1)}(\tilde{\btau})M_i^{(2)}(\tilde{\btau}) \big]\\
		A_9:= \dfrac{2}{|n_-|}\sum\limits_{i\in n_-}\big[M_i^{(1)}(\tilde{\btau})V_{i}(\tilde{\btau}) \big],~~A_{10}:= \dfrac{2}{|n_-|}\sum\limits_{i\in n_-}\big[M_i^{(2)}(\tilde{\btau})V_{i}(\tilde{\btau}) \big].
	\end{array}
\end{equation*}
Next, we consider the above ten terms, respectively. For $A_1$, by the law of large numbers, we have $A_1=O_p(n^{-1/2})$. For $A_2$ and $A_3$, by Assumption A.2 and the bounds in (\ref{equation: bound of lasso variance with alpha=0}), we can prove that 
\begin{equation*}
	|A_2|=O_p(\|\hat{\uwave{\bbeta}}^{(1)}-\uwave{\bbeta}^{(0)}\|^2_1),~~ |A_3|=O_p(\|\hat{\uwave{\bbeta}}^{(1)}-\uwave{\bbeta}^{(0)}\|^4_1).
\end{equation*}
For $A_4$, similar to the proof in Lemma \ref{lemma: upper bound for the empirical process of alpha=0 under H0} but using very tedious modifications, we can prove
\begin{equation*}
	|A_4|=O_p\Big(\|\hat{\uwave{\bbeta}}^{(1)}-\uwave{\bbeta}^{(0)}\|_1^{1/2} \sqrt{s\dfrac{\log(pn)}{n}}  \Big)=O_p\Big(s(\dfrac{\log(pn)}{n})^{\frac{3}{4}}\Big).
\end{equation*}
For $A_5 - A_{10}$, using the obtained bounds and  the Cauchy-Swartz inequality, we have:
\begin{equation*}
	\begin{array}{ll}
		|A_{5}|=O_p(\|\hat{\uwave{\bbeta}}^{(1)}-\uwave{\bbeta}^{(0)}\|_1),~~|A_{6}|=O_p(\|\hat{\uwave{\bbeta}}^{(1)}-\uwave{\bbeta}^{(0)}\|^2_1),\\
		|A_{7}|=O_p\Big(\|\hat{\uwave{\bbeta}}^{(1)}-\uwave{\bbeta}^{(0)}\|^{1/4}_1 \big(s\dfrac{\log(pn)}{n}\big)^{1/4}\Big)=O_p\Big(s^{\frac{1}{2}}(\dfrac{\log(pn)}{n})^{\frac{3}{8}}\Big),\\
		|A_{8}|=O_p(\|\hat{\uwave{\bbeta}}^{(1)}-\uwave{\bbeta}^{(0)}\|^3_1),~~|A_{9}|=O_p\Big(\|\hat{\uwave{\bbeta}}^{(1)}-\uwave{\bbeta}^{(0)}\|^{5/4}_1 \big(s\dfrac{\log(pn)}{n}\big)^{1/4}\Big),\\
		|A_{10}|=O_p\Big(\|\hat{\uwave{\bbeta}}^{(1)}-\uwave{\bbeta}^{(0)}\|^{9/4}_1 \big(s\dfrac{\log(pn)}{n}\big)^{1/4}\Big).\\
	\end{array}
\end{equation*}
By Assumption E.2, we can see that $|A_{5}|$ and $|A_{7}|$ dominate the other terms. Hence, we have:
\begin{equation*}
	\hat{\sigma}_-^2(0,\tilde{\btau})-{\sigma}^2(0,\tilde{\btau})=O_p\Big( s\sqrt{\dfrac{\log(pn)}{n}}\vee s^{\frac{1}{2}}(\dfrac{\log(pn)}{n})^{\frac{3}{8}}\Big).
\end{equation*}
With a similar analysis, we can prove  the same bound applies to $\hat{\sigma}_+^2(0,\tilde{\btau})-{\sigma}^2(0,\tilde{\btau})$, which yields:
\begin{equation*}
	\begin{array}{ll}
		|\hat{\sigma}^2(0,\tilde{\btau})-{\sigma}^2(0,\tilde{\btau})|\\
		=|\hat{t}_{0}\times(\hat{\sigma}_-^2(0,\tilde{\btau})-{\sigma}^2(0,\tilde{\btau}))+(1-\hat{t}_{0})\times(\hat{\sigma}_+^2(0,\tilde{\btau})-{\sigma}^2(0,\tilde{\btau}))|\\
		=O_p\Big( s\sqrt{\dfrac{\log(pn)}{n}}\vee s^{\frac{1}{2}}(\dfrac{\log(pn)}{n})^{\frac{3}{8}}\Big).
	\end{array}
\end{equation*}
\subsubsection{Proof of Theorem \ref{theorem: variance estimator under H0} with $\alpha\in(0,1)$}\label{sec: variance with alpha H0}
Note that  for $\alpha\in(0,1)$, the true variance has the following  explicit form:
\begin{equation*}
	\sigma^2(\alpha,\tilde{\btau})=(1-\alpha)^2\E[e^2_i(\tilde{\btau})]+\alpha^2\sigma^2-2\alpha(1-\alpha)\E [e_i(\tilde{\btau})\epsilon_i].
\end{equation*}
In this case, the variance estimators $\hat{\sigma}_-^2(\alpha,\tilde{\btau})$ and $\hat{\sigma}_+^2(\alpha,\tilde{\btau})$ reduce to
\begin{equation*}
	\hat{\sigma}_-^2(\alpha,\tilde{\btau}):=\dfrac{1}{|n_-|}\sum_{i\in n_-}\big[(1-\alpha)\hat{e}_i(\tilde{\btau})-\alpha\hat{\epsilon}_i\big]^2,~~\hat{\sigma}_+^2(\alpha,\tilde{\btau}):=\dfrac{1}{|n_+|}\sum_{i\in n_+}\big[(1-\alpha)\hat{e}_i(\tilde{\btau})-\alpha\hat{\epsilon}_i\big]^2,
\end{equation*}
where  $\hat{\epsilon}_i$ is defined in (\ref{equation: regression residuals}) and $\hat{e}_i(\tilde{\btau})=K^{-1}\sum\limits_{k=1}^{K}\hat{e}_i(\tau_k)$ with $\hat{e}_i(\tau_k)$ being defined in (\ref{equation: regression quantile residuals}). Recall  $\uwave{\bbeta}^{(0)}:=((\bbeta^{(0)})^\top,(\bb^{(0)})^\top)^\top$ are the true parameters under $\Hb_0$, and $\hat{\uwave{\bbeta}}^{(1)}:=(\hat{\bbeta}^{(1)})^\top,\hat{\bb}^{(1)})^\top)^\top$ as well as  $\hat{\uwave{\bbeta}}^{(2)}:=(\hat{\bbeta}^{(2)})^\top,\hat{\bb}^{(1)})^\top)^\top$ are the estimators using samples in $n_-$ and $n_+$, respectively.  Similar to the proof of Lemma \ref{lemma: basic inequality for lasso with alpha}, under $\Hb_0$ and {\bf{Assumptions A, B, C.2, D, E.2 - E.4}} , it is not hard to prove that:
\begin{equation}\label{equation: bound of lasso variance with alpha}
	\begin{array}{ll}
		\|\hat{\uwave{\bbeta}}^{(1)}-\uwave{\bbeta}^{(0)}\|_1=O_p(s\sqrt{\dfrac{\log(pn)}{n})},~~\|\hat{\uwave{\bbeta}}^{(2)}-\uwave{\bbeta}^{(0)}\|_1=O_p(s\sqrt{\dfrac{\log(pn)}{n}}),\\
		(\hat{\bbeta}^{(1)}-\bbeta^{(0)})^\top\dfrac{1}{|n_-|}\sum\limits_{i\in n_-}\bX_i\bX_i^\top(\hat{\bbeta}^{(1)}-\bbeta^{(0)})=O_p(s\dfrac{\log(pn)}{n}),\\
		(\hat{\bbeta}^{(2)}-\bbeta^{(0)})^\top\dfrac{1}{|n_+|}\sum\limits_{i\in n_+}\bX_i\bX_i^\top(\hat{\bbeta}^{(2)}-\bbeta^{(0)})=O_p(s\dfrac{\log(pn)}{n}).
	\end{array}
\end{equation}
We first prove the consistency of $\hat{\sigma}_-^2(\alpha,\tilde{\btau})$. For $\hat{\sigma}_-^2(\alpha,\tilde{\btau})-{\sigma}^2(\alpha,\tilde{\btau})$, it can be  decomposed into three terms:
\begin{equation}\label{equation: sigma-1 =A+B+C}
	\begin{array}{ll}
		\hat{\sigma}_-^2(\alpha,\tilde{\btau})-{\sigma}^2(\alpha,\tilde{\btau})\\
		=(1-\alpha)^2\underbrace{\dfrac{1}{|n_-|}\sum\limits_{i\in n_-}(\hat{e}_i(\tilde{\btau})^2-\E[e_i^2(\tilde{\btau})])}_{A}+\alpha^2\underbrace{\dfrac{1}{|n_-|}\sum\limits_{i\in n_-}(\hat{\epsilon}_i^2-\sigma^2)}_{B}\\-2\alpha(1-\alpha)\underbrace{\dfrac{1}{|n_-|}\sum\limits_{i\in n_-}(\hat{e}_i(\tilde{\btau})\hat{\epsilon}_i-\E[e_i(\tilde{\btau})\epsilon_i])}_{C}.
	\end{array}
\end{equation}
Next, we consider the three terms $A$, $B$, and $C$, respectively. For $B$, using the bounds obtained in Section \ref{sec: variance with alpha=1 H0}, we have: $B=O_{p}(\sqrt{s\dfrac{\log(pn)}{n}}) $. For $A$, using the bounds in 
Section \ref{sec: variance with alpha=0 H0}, we have:
\begin{equation*}
	A=O_p\Big( s\sqrt{\dfrac{\log(pn)}{n}}\vee s^{\frac{1}{2}}(\dfrac{\log(pn)}{n})^{\frac{3}{8}}\Big).
\end{equation*}
Next, we consider $C$. By the decomposition of $e_i(\tilde{\btau})$ in (\ref{equation: decomposition of ei}) and the fact that $\hat{\epsilon_i}=\epsilon_i+\hat{\epsilon_i}-\epsilon_i$, we can decompose $C$ into eight terms:
\begin{equation*}
	\dfrac{1}{|n_-|}\sum\limits_{i\in n_-}(\hat{e}_i(\tilde{\btau})\hat{\epsilon}_i-\E[e_i(\tilde{\btau})\epsilon_i])=C_1+\cdots+C_8,
\end{equation*}
where 
\begin{equation*}
	\begin{array}{ll}
		C_1=\dfrac{1}{|n_-|}\sum\limits_{i\in n_-}({e}_i(\tilde{\btau}){\epsilon}_i-\E[e_i(\tilde{\btau})\epsilon_i])~~C_2=\dfrac{1}{|n_-|}\sum\limits_{i\in n_-}M_i^{(1)}(\tilde{\btau}){\epsilon}_i,\\
		C_3=\dfrac{1}{|n_-|}\sum\limits_{i\in n_-}M_i^{(2)}(\tilde{\btau}){\epsilon}_i,~~ C_4=\dfrac{1}{|n_-|}\sum\limits_{i\in n_-}V_i(\tilde{\btau}){\epsilon}_i,\\
		C_5=\dfrac{1}{|n_-|}\sum\limits_{i\in n_-}e_i(\tilde{\btau})(\hat{\epsilon_i}-\epsilon_i),~~C_6=\dfrac{1}{|n_-|}\sum\limits_{i\in n_-}M^{(1)}_i(\tilde{\btau})(\hat{\epsilon_i}-\epsilon_i),\\
		C_7=\dfrac{1}{|n_-|}\sum\limits_{i\in n_-}M_i^{(2)}(\tilde{\btau})(\hat{\epsilon_i}-\epsilon_i),~~C_8=\dfrac{1}{|n_-|}\sum\limits_{i\in n_-}V_i(\tilde{\btau})(\hat{\epsilon_i}-\epsilon_i).
	\end{array}
\end{equation*}
For $C_1$, by the law of large numbers, we have $C_1=O_p(n^{-1/2})$. Note that using the bounds in (\ref{equation: bound of lasso variance with alpha}), with a very similar proof techniques as in Sections \ref{sec: variance with alpha=1 H0} and \ref{sec: variance with alpha=0 H0}, we can prove:
\begin{equation*}
	\begin{array}{cc}
		\dfrac{1}{|n_-|}\sum\limits_{i\in n_-}\big[  M_i^{(1)}(\tilde{\btau}) \big]^2=O_p(\|\hat{\uwave{\bbeta}}^{(1)}-\uwave{\bbeta}^{(0)}\|^2_1),~~ \dfrac{1}{|n_-|}\sum\limits_{i\in n_-}\big[  M_i^{(2)}(\tilde{\btau}) \big]^2=O_p(\|\hat{\uwave{\bbeta}}^{(1)}-\uwave{\bbeta}^{(0)}\|^4_1),\\
		\dfrac{1}{|n_-|}\sum\limits_{i\in n_-}\big[  V_i(\tilde{\btau}) \big]^2=O_p\Big(\|\hat{\uwave{\bbeta}}^{(1)}-\uwave{\bbeta}^{(0)}\|_1^{1/2} \sqrt{s\dfrac{\log(pn)}{n}}  \Big)=O_p\Big(s(\dfrac{\log(pn)}{n})^{\frac{3}{4}}\Big),\\
		\dfrac{1}{|n_-|}\sum\limits_{i\in n_-}\big[  \hat{\epsilon}_i-\epsilon_i \big]^2=O_p(s\dfrac{\log(pn)}{n}).
	\end{array}
\end{equation*}
Hence, for $C_2 - C_8$ using the above bounds and the Cauchy-Swartz inequality, it is not hard to see that $C_2$ and $C_4$ dominate the other terms. Specifically, we have:
\begin{equation*}
	|C_2|\leq \sqrt{\dfrac{1}{|n_-|}\sum_{i\in n_-}\epsilon_i^2}\times\sqrt{\dfrac{1}{|n_-|}\sum\limits_{i\in n_-}\big[  M_i^{(1)}(\tilde{\btau}) \big]^2}=O_p\Big(s\sqrt{\dfrac{\log(pn)}{n}}\Big),
\end{equation*}
and
\begin{equation*}
	|C_4|\leq \sqrt{\dfrac{1}{|n_-|}\sum_{i\in n_-}\epsilon_i^2}\times\sqrt{\dfrac{1}{|n_-|}\sum\limits_{i\in n_-}\big[  V_i(\tilde{\btau}) \big]^2}=O_p\Big(s^{\frac{1}{2}}(\dfrac{\log(pn)}{n})^{\frac{3}{8}}\Big),
\end{equation*}
which implies
\begin{equation*}
	C=O_p\Big( s\sqrt{\dfrac{\log(pn)}{n}}\vee s^{\frac{1}{2}}(\dfrac{\log(pn)}{n})^{\frac{3}{8}}\Big).
\end{equation*}
Lastlt, combining (\ref{equation: sigma-1 =A+B+C}) and the obtained upper bounds for $A, B$ and $C$, we have 
\begin{equation*}
	\hat{\sigma}_-^2(\alpha,\tilde{\btau})-{\sigma}^2(\alpha,\tilde{\btau})=O_p\Big( s\sqrt{\dfrac{\log(pn)}{n}}\vee s^{\frac{1}{2}}(\dfrac{\log(pn)}{n})^{\frac{3}{8}}\Big).
\end{equation*}
With a similar analysis, we can prove  the same bound applies to $\hat{\sigma}_+^2(\alpha,\tilde{\btau})-{\sigma}^2(\alpha,\tilde{\btau})$, which yields:
\begin{equation*}
	\begin{array}{ll}
		|\hat{\sigma}^2(\alpha,\tilde{\btau})-{\sigma}^2(\alpha,\tilde{\btau})|
		=O_p\Big( s\sqrt{\dfrac{\log(pn)}{n}}\vee s^{\frac{1}{2}}(\dfrac{\log(pn)}{n})^{\frac{3}{8}}\Big).
	\end{array}
\end{equation*}
\subsection{Proof of Theorem \ref{theorem: size control for individual test} }
In this section, we prove the Gaussian approximation results under $\Hb_0$, which are given in Sections \ref{sec: size control for alpha=1} - \ref{sec: size control for alpha}, respectively. For simplicity, we will omit the subscript $\alpha$ whenever  needed.
\subsubsection{Gaussian approximation for $\alpha=1$}\label{sec: size control for alpha=1}
\begin{proof}
	In this section, we give the size results for $\alpha=1$. Note that in this case, our individual based test statistic $T_1$ reduces to the least squared based score typed test statistic. Let  $\bz(\bx,y;\bbeta)=\bx(y-\bx^\top\bbeta):=-S_{1}(y,\bx;\tilde{\btau},\bb,\bbeta)$ be the negative score for the $\ell_2$-loss. Let $\bZ_i(\bX_i,Y_i;\bbeta)=\bX_i(Y_i-\bX_i^\top\bbeta)$ be the sample version. 
	In this section, we aim to prove:
	\begin{equation}\label{equation: proof of gaussian approximation}
		\sup_{z\in (0,\infty)}\big|\P(T_{1}\leq z)-\P(T_{1}^{b}\leq z|\mathcal{X})\big|=o_p(1), ~\text{as}~  n,p\rightarrow\infty.
	\end{equation}
	The proof proceeds into three steps.\\
	{\bf{Step~1: Decomposition of $T_1$}}.
	Note that for $\alpha=1$, the score based CUSUM process reduces to:
	\begin{equation}\label{equation: C1 for alpha=0}
		\bC_1(t)=\dfrac{1}{\sqrt{n}\hat{\sigma}(1,\tilde{\btau})}\big(\sum\limits_{i=1}^{\lfloor nt\rfloor}\bZ_i(\bX_i,Y_i;\hat{\bbeta})-\dfrac{\lfloor nt \rfloor }{n}\sum\limits_{i=1}^{n}\bZ_i(\bX_i,Y_i;\hat{\bbeta})\big),
	\end{equation}
	where $\hat{\bbeta}$ is the lasso estimator defined in (\ref{equation: lasso estimator}) and $\hat{\sigma}(\alpha,\tilde{\btau})$ is the variance estimator defined in (\ref{equation: variance estimator}).
	By definition, we have $T_1=\max\limits_{q_0\leq t\leq 1-q_0}\|\bC_1(t)\|_{(s_0,2)}$. Replacing $\hat{\bbeta}$ by $\bbeta^{(0)}$ in $\bC_1(t)$, we have:
	\begin{equation*}
		\bC_1(t)=\bC^{I}_1(t)+\bC^{II}_1(t),
	\end{equation*}
	where $\bC^{I}_{1}(t)$ and $\bC^{I}_{I1}(t)$ are defined as
	\begin{equation}\label{equation: C11+C12}
		\begin{array}{cc}
			\bC^{I}_{1}(t)=\dfrac{1}{\sqrt{n}\hat{\sigma}(1,\tilde{\btau})}\big(\sum\limits_{i=1}^{\floor{nt}}\bX_i\epsilon_i-\dfrac{\floor{nt}}{n}\sum\limits_{i=1}^n\bX_i\epsilon_i\big),\\
			\bC^{II}_{1}(t)=\dfrac{1}{\sqrt{n}\hat{\sigma}(1,\tilde{\btau})}\big(\sum\limits_{i=1}^{\floor{nt}}\bX_i\bX_i^\top(\bbeta^{(0)}-\hat{\bbeta})-\dfrac{\floor{nt}}{n}\sum\limits_{i=1}^n\bX_i\bX_i^\top(\bbeta^{(0)}-\hat{\bbeta})\big).
		\end{array}
	\end{equation}
	Note that we can regard $\bC^{I}_{1}(t)$ as the leading term of $\bC_1(t)$ and $\bC^{II}_{1}(t)$ as the residual term. Moreover, replacing $\hat{\sigma}(1,\tilde{\btau})$ by $\sigma^2:=\text{Var}(\epsilon)$ in $\bC^{I}_{1}(t)$, we can define the oracle leading term as:
	\begin{equation}
		\tilde{\bC}^{I}_{1}(t)=\dfrac{1}{\sqrt{n}\sigma}\big(\sum\limits_{i=1}^{\floor{nt}}\bX_i\epsilon_i-\dfrac{\floor{nt}}{n}\sum\limits_{i=1}^n\bX_i\epsilon_i\big).
	\end{equation}
	The following Lemma  \ref{lemma: negligible for alpha=1} 
	shows that we can approximate $T_1$ by $\tilde{\bC}^{I}_{1}(t)$ in terms of the $(s_0,2)$-norm. The proof of Lemma  \ref{lemma: negligible for alpha=1} is provided in Section \ref{sec: Proof of negligible alpha=1}.
	
	\begin{lemma}\label{lemma: negligible for alpha=1}
		Assume {\bf{Asssumptions A, B, C.2, E.2-E.4}} hold. Under $\Hb_0$, we have
		\begin{equation}\label{equation: Lemma B.1 alpha=1}
			\P\big(\max_{q_0\leq t\leq 1-q_0} \big\|\bC_1(t)-\tilde{\bC}_1^{I}(t)\big\|_{(s_0,2)}\geq \epsilon \big)=o(1),
		\end{equation}
		where ${\epsilon:=Cs_0^{1/2}sM^2{\log(p)}/{\sqrt{n}}}$ for some big enough universal constant $C>0$.
	\end{lemma}
	\noindent{\bf{Step~2: Gaussian approximation for the oracle leading term}}. By Lemma \ref{lemma: negligible for alpha=1}, we only need to consider Gaussian approximation for the process $\{\tilde{\bC}_1^{I}(t), q_0\leq t\leq 1-q_0\}$. Recall the bootstrap based CUSUM process for $\alpha=1$ as:
	\begin{equation}\label{equation: bootstrap based CUSUM alpha=1}
		{\bC}^{b}_{1}(t)=\dfrac{1}{\sqrt{n}}\big(\sum\limits_{i=1}^{\floor{nt}}\bX_ie^b_i-\dfrac{\floor{nt}}{n}\sum\limits_{i=1}^n\bX_ie^b_i\big),
	\end{equation}
	where $e_i^b\sim N(0,1)$. By definition, the bootstrap based testing statistic is $$T_{1}^b=\max_{q_0\leq t\leq 1-q_0}\|{\bC}^{b}_{1}(t)\|_{(s_0,2)}.$$ Let  $\bZ_i=(Z_{i1},\ldots,Z_{ip})^\top$ with $Z_{ij}:=X_{ij}\epsilon_i/\sigma$ and $\bG_{i}=(G_{i1},\ldots,G_{ip})^\top$ with $\bG_i\sim N(\mathbf{0},\bSigma)$ where $\bSigma=\text{Cov}(\bX_1)$.
	It is easy to see that $\bG_i$ has the same covariance matrix as $\bZ_i$. Define 
	\begin{equation*}
		\bC_1^{\bG}(t)=\dfrac{1}{\sqrt{n}}\big(\sum_{i=1}^{\floor{nt}}\bG_i-\dfrac{\floor{nt}}{n}\sum_{i=1}^n\bG_i\big), ~~\text{and}~~T_{1}^{\bG}=\max_{q_0\leq t\leq 1-q_0}\| \bC_1^{\bG}(t)\|_{(s_0,2)}.
	\end{equation*}
	By {\bf{Assumptions A, C, E.1}}, we can verify that the Conditions {(\bf{M1}) - (\bf{M3})} in Lemma \ref{lemma: key lemma for gaussian approximation} hold. Hence, by Lemma \ref{lemma: key lemma for gaussian approximation}, we can prove
	\begin{equation}\label{equation: proof of gaussian approximation2}
		\sup_{z\in (0,\infty)}\big|\P(\max_{q_0\leq t\leq 1-q_0}\|\tilde{\bC}_{1}^{I}\|_{(s_0,2)}\leq z)-\P(T_{1}^{\bG}\leq z)\big|\leq n^{-\xi_0}, ~\text{for~some}~  \xi_0>0.
	\end{equation}
	Next, we aim to approximate $T_1^b|\cX$ by $T_{1}^{\bG}$. The result is based on the following Lemma \ref{lemma: Gaussian approximation 2}. 
	\begin{lemma}\label{lemma: Gaussian approximation 2}
		Suppose {\bf{Assumptions A, E.1}} are satisfied. Then, under $\Hb_0$, we have
		\begin{equation*}\label{equation: gaussian approximation 2}
			\sup_{z\in(0,\infty)} \big|\P(\max_{q_0\leq t\leq 1-q_0}\| \bC_1^{\bG}(t)\|_{(s_0,2)}> z\big)-\P(\max_{q_0\leq t\leq 1-q_0}\|\bC_1^b(t)\|_{(s_0,2)}> z|\mathcal{X}\big)\big|=o_p(1).
		\end{equation*}
	\end{lemma}
	Hence, based on Lemma \ref{lemma: Gaussian approximation 2}, we show that the two Gaussian processes $\bC_1^{\bG}(t)$ and $\bC^b(t)|\mathcal{X}$ with $q_0\leq t\leq 1-q_0$ can be uniformly close to each other with the $(s_0,2)$-norm. The proof of Lemma \ref{lemma: Gaussian approximation 2} is provided in Section \ref{sec: proof of Gaussian approximation 2 with alpha=1}.\\
	{\bf{Step~3: Combining the previous results}}.  In this step, we aim to combine the previous two steps for proving:
	\begin{equation}\label{equation: proof of gaussian approximation3}
		\sup_{z\in (0,\infty)}\big|\P(T_{1}\leq z)-\P(T_{1}^{b}\leq z|\mathcal{X})\big|=o_p(1), ~\text{as}~  n,p\rightarrow\infty.
	\end{equation}
	In particular, we need to obtain the upper and lower bounds of $\rho_0$, where 
	\begin{equation}
		\rho_0:=\P(T_{1}> z)-\P(T_{1}^{b}> z|\mathcal{X}).
	\end{equation}
	We first consider the upper bound. Note that $T_{1}=\max\limits_{t\in[q_0,1-q_0]}\|\bC_1(t)\|_{(s_0,2)}$. By plugging $\tilde{\bC}_1^{I}(t)$ in $T_{1}$ and using 
	the triangle inequality of $\|\cdot\|_{(s_0,2)}$, we have
	\begin{equation}\label{step4: inequality1}
		\P(T_{1}> z)\leq \P(\max_{t\in[q_0,1-q_0]}\|\tilde{\bC}_1^{I}(t)\|_{(s_0,2)}>z-\epsilon)+\rho_1,
	\end{equation}
	where $\rho_1:=\P(\max\limits_{t\in[q_0,1-q_0]}\|\bC_1(t)-\tilde{\bC}_1^{ I}(t)\|_{(s_0,2)}>\epsilon)$. By Lemma \ref{lemma: negligible for alpha=1}, we have $\rho_1=o(1)$. For $\P(\max_{q_0\leq t\leq 1-q_0} \|\tilde{\bC}_1^{\text{I}}(t)\|_{(s_0,2)}\geq z-\epsilon)$, by the triangle inequality, we have
	\begin{equation}\label{step4: inequality2}
		\P\big(\max\limits_{q_0\leq t\leq 1-q_0} \big\|\tilde{\bC}_1^{\text{I}}(t)\big\|_{(s_0,2)}\geq z-\epsilon \big)\leq \P\big(\max\limits_{q_0\leq t\leq 1-q_0} \big\|\bC_1^{\bG}(t)\big\|_{(s_0,2)}\geq z-\epsilon \big)+\rho_2,
	\end{equation}
	where
	\begin{equation*}
		\rho_2=\max_{x>0}\big|\P(\max_{q_0\leq t\leq 1-q_0}\|\bC_{1}^{\bG}(t)\|_{(s_0,2)}> x\big)-\P(\max_{q_0\leq t\leq 1-q_0}\|\tilde{\bC}^{\text{I}}(t)\|_{(s_0,2)}> x\big)\big|.
	\end{equation*}
	By Lemma \ref{lemma: key lemma for gaussian approximation}, we have $\rho_2\leq Cn^{-\zeta_0}$. Therefore, by (\ref{step4: inequality1}) and (\ref{step4: inequality2}), we have proved that
	\begin{equation}\label{step4: inequality3}
		\P\big(\max\limits_{q_0\leq t\leq 1-q_0} \big\|\bC_1(t)\big\|_{(s_0,2)}\geq z \big)\leq\underbrace{\P\big(\max\limits_{q_0\leq t\leq 1-q_0} \big\|\bC_{1}^{\bG}(t)\big\|_{(s_0,2)}\geq z-\epsilon \big)}_{\rho_3}+o(1). 
	\end{equation}
	We next consider $\rho_3$. We decompose $\rho_3$ as $\rho_3=\rho_4+\rho_5$, where $\rho_4$ and $\rho_5 $ are defined as
	\begin{equation*}
		\rho_4=\P\big(z-\epsilon\leq \max\limits_{q_0\leq t\leq 1-q_0} \big\|\bC_{1}^{\bG}(t)\big\|_{(s_0,2)}\leq z\big),~~\rho_5=\P(\max\limits_{q_0\leq t\leq 1-q_0} \|\bC_{1}^{\bG}(t)\|_{(s_0,2)}\geq z).
	\end{equation*}
	By Lemmas \ref{lemma:anti consentration inequality} and  \ref{lemma: convex approximation}, we can show that $\rho_4=o(1)$. For $\rho_5$, we have
	\begin{equation}\label{step4: inequality4}
		\P\big(\max\limits_{q_0\leq t\leq 1-q_0} \big\|\bC_{1}^{\bG}(t)\big\|_{(s_0,2)}\geq z \big)\leq \P\big(\max\limits_{q_0\leq t\leq 1-q_0} \big\|\bC_1^b(t)\big\|_{(s_0,2)}\geq z|\mathcal{X}) \big)+\rho_6,
	\end{equation}
	where $$\rho_6=\sup_{z\in(0,\infty)} |\P(\max\limits_{q_0\leq t\leq 1-q_0}\|\bC_{1}^{\bG}(t)\|_{(s_0,2)}> z)-\P(\max\limits_{q_0\leq t\leq 1-q_0}\|\bC_1^b(t)\|_{(s_0,2)}> z|\mathcal{X})|.$$ By Lemma \ref{lemma: Gaussian approximation 2}, we have $\rho_6=o_p(1)$. Therefore, by (\ref{step4: inequality1}) -- (\ref{step4: inequality4}), we have proved
	\begin{equation*}
		\P(T_{1}\geq z)-\P(T_{1}^{b}\geq z|\mathcal{X})=o_p(1),
	\end{equation*}
	uniformly for $z>0$. Similarly, we can obtain the lower bound and prove that
	\begin{equation*}
		\sup_{z\in (0,\infty)}\big|\P(T_{1}\geq z)-\P(T_{1}^{b}\geq z|\mathcal{X})\big|=o_p(1),
	\end{equation*}
	which finishes the proof of Theorem \ref{theorem: size control for individual test} for the individual test with $\alpha=1$.
\end{proof}

\subsubsection{Gaussian approximation for $\alpha=0$}
\begin{proof}
	In this section, we give the size results for $\alpha=0$. Recall $0<\tau_1<\cdots<\tau_K<1$ are user-specified $K$ quantile levels.  Let 
	$\tilde{\btau}:=(\tau_1,\ldots,\tau_K)^\top$ and $\bb=(b_1,\ldots,b_K)^\top$. Note that in this case, our individual based test statistic $T_0$ reduces to composite quantile loss based score typed test statistic. Define the score function as:
	\begin{equation}
		z(\bx,y;\tilde{\btau},\bb,\bbeta):=\dfrac{1}{K}\sum_{k=1}^{K}\bx \big(\mathbf{1}\{y-b_k-\bx^\top\bbeta\leq 0\}-\tau_k\big),
	\end{equation}
	and $Z(\bX_i,Y_i;\tilde{\btau},\bb,\bbeta)$ as its sample version. For $\alpha=0$, we aim to prove:
	\begin{equation}\label{equation: proof of gaussian approximation for alpha=0}
		\sup_{z\in (0,\infty)}\big|\P(T_{0}\leq z)-\P(T_{0}^{b}\leq z|\mathcal{X})\big|=o_p(1), ~\text{as}~  n,p\rightarrow\infty.
	\end{equation}
	The proof proceeds into three steps.\\
	Note that for $\alpha=0$, the score based CUSUM process reduces to:
	\begin{equation}\label{equation: C1 for alpha=00}
		\bC_0(t)=\dfrac{1}{\sqrt{n}\hat{\sigma}(\alpha,\tilde{\btau})}\big(\sum\limits_{i=1}^{\lfloor nt\rfloor}\bZ(\bX_i,Y_i;\tilde{\btau},\hat{\bb},\hat{\bbeta})-\dfrac{\lfloor nt \rfloor }{n}\sum\limits_{i=1}^{n}\bZ(\bX_i,Y_i;\tilde{\btau},\hat{\bb},\hat{\bbeta})\big),
	\end{equation}
	where $\hat{\bb}$ and $\hat{\bbeta}$ are the lasso estimators defined in (\ref{equation: lasso estimator}), and $\hat{\sigma}(\alpha,\tilde{\btau})$ is the variance estimator defined in (\ref{equation: variance estimator}). By definition  of $T_0$, we have $T_0=\max\limits_{q_0\leq t\leq 1-q_0}\|\bC_0(t)\|_{(s_0,2)}$. Before the proof, we need some notations. Let $\bDelta={\bbeta}-\bbeta^{(0)}\in \RR^p$, $\bdelta={\bb}-\bb^{(0)}\in \RR^K$, $\delta_k=b_k-b_k^{(0)}\in \RR^1$, $\uwave{\bDelta_k}=(\bDelta^\top,\delta_k)^\top\in \RR^{p+1}$, and $\uwave{\bDelta}=(\bDelta^\top,\bdelta^\top)^\top\in \RR^{p+K}$. Accordingly, we define $\hat{\bDelta}$, $\hat{\bdelta}$, $\hat{\delta}_k$, $\uwave{\hat{\bDelta}_k}$, $\uwave{\hat{\bDelta}}$ by using the corresponding estimators.
	Moreover, we define $\uwave{\bX_i}=(\bX_i^\top,1)^\top\in \RR^{p+1}$ or $\uwave{\bX_i}=(\bX_i^\top,\mathbf{1}_K)^\top\in \RR^{p+K}$ whenever it is used, where $\mathbf{1}_K$ is an $\RR^{K}$ dimensional vector with elements being 1s. By the definition of $Y_i=\bX_i^\top \bbeta^{(0)}+\epsilon_i$, we have $Y_i\leq b_k+\bX_i^\top\bbeta$ which is equal to $\epsilon_i\leq \uwave{\bX_i}^\top\uwave{\bDelta_k}+b_k^{(0)}$. Hence, by replacing $\hat{\bbeta}$ by $\bbeta^{(0)}$ and $\hat{\bb}$ by $\bb^{(0)}$ in $\bC_0(t)$, we have the following decomposition:
	\begin{equation}\label{equation: C0t}
		\bC_0(t)=\bC^{I}_0(t)+\bC^{II}_0(t),
	\end{equation}
	where $\bC^{I}_{0}(t)$ and $\bC^{II}_{0}(t)$ are defined as
	\begin{equation}\label{equation: C11+C12 alpha=0}
		\begin{array}{cc}
			\bC^{I}_{0}(t)=\dfrac{1}{\sqrt{n}\hat{\sigma}(\alpha,\tilde{\btau})}\big(\sum\limits_{i=1}^{\floor{nt}}\bX_ie_i(\tilde{\btau})-\dfrac{\floor{nt}}{n}\sum\limits_{i=1}^n\bX_ie_i(\tilde{\btau})\big),\\
			\bC^{II}_{0}(t)=\dfrac{1}{\sqrt{n}\hat{\sigma}(\alpha,\tilde{\btau})}\big(\sum\limits_{i=1}^{\floor{nt}}\dfrac{1}{K}\sum\limits_{k=1}^K\bX_i\big(\mathbf{1}\{\epsilon_i\leq \uwave{\bX_i}^\top\uwave{\hat{\bDelta}_k}+b_k^{(0)}\}-\mathbf{1}\{\epsilon_i\leq b_k^{(0)} \}\big)\\
			\qquad\qquad\qquad\qquad\qquad-\dfrac{\floor{nt}}{n}\sum\limits_{i=1}^n\dfrac{1}{K}\sum\limits_{k=1}^K\bX_i\big(\mathbf{1}\{\epsilon_i\leq \uwave{\bX_i}^\top\uwave{\hat{\bDelta}_k}+b_k^{(0)}\}-\mathbf{1}\{\epsilon_i\leq b_k^{(0)} \}\big)\big),
		\end{array}
	\end{equation}
	where $ e_i(\tilde{\btau}):=\dfrac{1}{K}\sum\limits_{k=1}^{K}\big(\mathbf{1}\{\epsilon_i\leq b^{(0)}_k\}-\tau_k\big):=\dfrac{1}{K}\sum\limits_{k=1}^{K}e_i(\tau_k)$ be a random sample satisfying
	\begin{equation}
		\E[e_i(\tilde{\btau})]=0~~\text{and}~~\text{Var}[e_i(\tilde{\btau})]=\dfrac{1}{K^2}\sum_{k_1=1}^{K}\sum_{k_2=1}^{K}\gamma_{k_1k_2}
	\end{equation}
	with $\gamma_{k_1k_2}:=\min(\tau_{k_1},\tau_{k_2})-\tau_{k_1}\tau_{k_2}$  for $\tau_{k_1},\tau_{k_2}\in(0,1)$. Under this decomposition,  we can regard $\bC^{I}_{0}(t)$ as the leading term of $\bC_0(t)$ and $\bC^{II}_{0}(t)$ as the residual term. Moreover, replacing $\hat{\sigma}(\alpha,\tilde{\btau})$ by $\sigma^2:=\text{Var}( e_i(\tilde{\btau}))$ in $\bC^{I}_{0}(t)$, we can define the oracle leading term as:
	\begin{equation}\label{equation: tilde-C0-I}
		\tilde{\bC}^{I}_{0}(t)=\dfrac{1}{\sqrt{n}\sigma}\big(\sum\limits_{i=1}^{\floor{nt}}\bX_ie_i(\tilde{\btau})-\dfrac{\floor{nt}}{n}\sum\limits_{i=1}^n\bX_ie_i(\tilde{\btau})\big).
	\end{equation}
	The following Lemma  \ref{lemma: negligible for alpha=0} shows that we can approximate $T_0$ by $\tilde{\bC}^{0}_{1}(t)$ in terms of $(s_0,2)$-norm. The proof of Lemma  \ref{lemma: negligible for alpha=0} is provided in Section \ref{sec: Proof of negligible alpha=0}.
	
	\begin{lemma}\label{lemma: negligible for alpha=0}
		Assume {\bf{Assumptions A, D, E.2-E.4}} hold. Under $\Hb_0$, we have
		\begin{equation}\label{equation: Lemma B.1}
			\P\big(\max_{q_0\leq t\leq 1-q_0} \big\|\bC_0(t)-\tilde{\bC}_0^{I}(t)\big\|_{(s_0,2)}\geq \epsilon \big)=o(1),
		\end{equation}
		where ${\epsilon:=CM^2s_0^{1/2}(s\log(pn))^{3/4}/n^{1/4}}$ for some big enough universal constant $C>0$.
	\end{lemma}
	Note that for the case of $\alpha=0$,  the error term $\epsilon_i(\tilde{\btau})$ is bounded random variables, which satisfies the assumptions in Lemma \ref{lemma: key lemma for gaussian approximation} trivially. Hence, by Lemma \ref{lemma: negligible for alpha=0}, Lemma \ref{lemma: key lemma for gaussian approximation}, and using similar arguments of Steps 2 and 3  in Section \ref{sec: size control for alpha=1}, we finish the proof of Theorem \ref{theorem: size control for individual test}.
\end{proof}
\subsubsection{Gaussian approximation for $\alpha\in(0,1)$}\label{sec: size control for alpha}
\begin{proof}
	In this section, we give the size results for $\alpha\in(0,1)$. Recall $0<\tau_1<\cdots<\tau_K<1$ are user-specified $K$ quantile levels.  Let 
	$\tilde{\btau}:=(\tau_1,\ldots,\tau_K)^\top$ and $\bb=(b_1,\ldots,b_K)^\top$. For $\alpha\in(0,1)$, define the score function as:
	\begin{equation}
		z(\bx,y;\tilde{\btau},\bb,\bbeta):=(1-\alpha)\dfrac{1}{K}\sum_{k=1}^{K}\bx \big(\mathbf{1}\{y-b_k-\bx^\top\bbeta\leq 0\}-\tau_k\big)-\alpha \bx(y-\bx^\top\bbeta),
	\end{equation}
	and $Z(\bX_i,Y_i;\tilde{\btau},\bb,\bbeta)$ as its sample version. For $\alpha\in(0,1)$, we aim to prove:
	\begin{equation}\label{equation: proof of gaussian approximation for alpha}
		\sup_{z\in (0,\infty)}\big|\P(T_{\alpha}\leq z)-\P(T_{\alpha}^{b}\leq z|\mathcal{X})\big|=o_p(1), ~\text{as}~  n,p\rightarrow\infty.
	\end{equation}
	Note that for $\alpha\in(0,1)$, the score based CUSUM process reduces to:
	\begin{equation}\label{equation: C1 for alpha}
		\bC_\alpha(t)=\dfrac{1}{\sqrt{n}\hat{\sigma}(\alpha,\tilde{\btau})}\big(\sum\limits_{i=1}^{\lfloor nt\rfloor}\bZ(\bX_i,Y_i;\tilde{\btau},\hat{\bb},\hat{\bbeta})-\dfrac{\lfloor nt \rfloor }{n}\sum\limits_{i=1}^{n}\bZ(\bX_i,Y_i;\tilde{\btau},\hat{\bb},\hat{\bbeta})\big),
	\end{equation}
	where $\hat{\bb}$ and $\hat{\bbeta}$ are the lasso estimators defined in (\ref{equation: lasso estimator}), and $\hat{\sigma}(\alpha,\tilde{\btau})$ is the variance estimator defined in (\ref{equation: variance estimator}). By definition  of $T_\alpha$, we have $T_\alpha=\max\limits_{q_0\leq t\leq 1-q_0}\|\bC_\alpha(t)\|_{(s_0,2)}$. Recall $\bDelta={\bbeta}-\bbeta^{(0)}\in \RR^p$, $\bdelta={\bb}-\bb^{(0)}\in \RR^K$, $\delta_k=b_k-b_k^{(0)}\in \RR^1$, $\uwave{\bDelta_k}=(\bDelta^\top,\delta_k)^\top\in \RR^{p+1}$, and $\uwave{\bDelta}=(\bDelta^\top,\bdelta^\top)^\top\in \RR^{p+K}$. Accordingly, recall $\hat{\bDelta}$, $\hat{\bdelta}$, $\hat{\delta}_k$, $\uwave{\hat{\bDelta}_k}$, $\uwave{\hat{\bDelta}}$ by using the corresponding lasso estimators.
	Moreover, we define $\uwave{\bX_i}=(\bX_i^\top,1)^\top\in \RR^{p+1}$ or $\uwave{\bX_i}=(\bX_i^\top,\mathbf{1}_K)^\top\in \RR^{p+K}$ whenever it is used, where $\mathbf{1}_K$ is an $\RR^{K}$ dimensional vector with elements being 1s. Under $\Hb_0$, by the definition of $Y_i=\bX_i^\top \bbeta^{(0)}+\epsilon_i$, we have $Y_i\leq b_k+\bX_i^\top\bbeta$ which is equal to $\epsilon_i\leq \uwave{\bX_i}^\top\uwave{\bDelta_k}+b_k^{(0)}$. Hence, by replacing $\hat{\bbeta}$ by $\bbeta^{(0)}$ and $\hat{\bb}$ by $\bb^{(0)}$ in $\bC_\alpha(t)$ , under $\Hb_0$, we have the following decomposition:
	\begin{equation}\label{equation: C0t alpha}
		\bC_\alpha(t)=\bC^{I}_\alpha(t)+\bC^{II}_\alpha(t),
	\end{equation}
	where $	\bC^{I}_{\alpha}(t)$ and 	$\bC^{II}_{\alpha}(t)$ are defined as
	\begin{equation}\label{equation: C11+C12 alpha}
		\begin{array}{cc}
			\bC^{I}_{\alpha}(t)=\dfrac{1}{\sqrt{n}\hat{\sigma}(\alpha,\tilde{\btau})}\big(\sum\limits_{i=1}^{\floor{nt}}\bX_i((1-\alpha)e_i(\tilde{\btau})-\alpha\epsilon_i)-\dfrac{\floor{nt}}{n}\sum\limits_{i=1}^n\bX_i((1-\alpha)e_i(\tilde{\btau})-\alpha\epsilon_i)\big),\\
			~~\text{and}~~	\bC^{II}_{\alpha}(t)=(1-\alpha)	\bC^{II}_{0}(t)+\alpha\bC^{II}_{1}(t),
		\end{array}
	\end{equation}
	where $ e_i(\tilde{\btau}):=K^{-1}\sum\limits_{k=1}^{K}\big(\mathbf{1}\{\epsilon_i\leq b^{(0)}_k\}-\tau_k\big):=K^{-1}\sum\limits_{k=1}^{K}e_i(\tau_k)$, $\bC^{II}_{1}(t)$ is defined in (\ref{equation: C11+C12}), and $\bC^{II}_{0}(t)$ is defined in (\ref{equation: C11+C12 alpha=0}). Under this decomposition,  we can regard $\bC^{I}_{\alpha}(t)$ as the leading term of $\bC_\alpha(t)$ and $\bC^{II}_{\alpha}(t)$ as the residual term. Moreover, replacing $\hat{\sigma}(\alpha,\tilde{\btau})$ by $\sigma^2:=\text{Var}[ (1-\alpha)e_i(\tilde{\btau})-\alpha\epsilon_i]$ in $\bC^{I}_{\alpha}(t)$, we can define the oracle leading term as:
	\begin{equation}\label{equation: tilde-C-alpha-I}
		\tilde{\bC}^{I}_{\alpha}(t)=\dfrac{1}{\sqrt{n}\sigma}\big(\sum\limits_{i=1}^{\floor{nt}}\bX_i((1-\alpha)e_i(\tilde{\btau})-\alpha\epsilon_i)-\dfrac{\floor{nt}}{n}\sum\limits_{i=1}^n\bX_i((1-\alpha)e_i(\tilde{\btau})-\alpha\epsilon_i)\big).
	\end{equation}
	The following Lemma  \ref{lemma: negligible for alpha} shows that  we can approximate $T_\alpha$ by $\tilde{\bC}^{\alpha}_{1}(t)$ in terms of the $(s_0,2)$-norm. The proof of Lemma  \ref{lemma: negligible for alpha} is provided in Section \ref{sec: Proof of negligible alpha}.
	
	\begin{lemma}\label{lemma: negligible for alpha}
		Assume {\bf{Assumptions A, B, C.2, D, E.2 - E.4}} hold. Under $\Hb_0$, we have
		\begin{equation}\label{equation: Lemma B.1 alpha}
			\P\big(\max_{q_0\leq t\leq 1-q_0} \big\|\bC_\alpha(t)-\tilde{\bC}_\alpha^{I}(t)\big\|_{(s_0,2)}\geq \epsilon \big)=o(1),
		\end{equation}
		where ${\epsilon:=CM^2s_0^{1/2}{(s\log(pn))^{3/4}}/{n^{1/4}}}$ for some big enough constant $C>0$ and $C$ is a universal constant not depending on $n$ or $p$.
	\end{lemma}
	Note that for the case of $\alpha\in(0,1)$,  the error term $(1-\alpha)e_i(\tilde{\btau})-\alpha\epsilon_i$ is a combination of  a bounded random variable $e_i(\tilde{\btau})$ and $\epsilon_i$ , which can be proved to satisfy the assumptions in Lemma \ref{lemma: key lemma for gaussian approximation}. Hence, by Lemma \ref{lemma: negligible for alpha=0}, Lemma \ref{lemma: key lemma for gaussian approximation}, and using similar arguments of Steps 2 and 3 as in Section \ref{sec: size control for alpha=1}, we finish the proof of Theorem \ref{theorem: size control for individual test} with $\alpha\in(0,1)$. 
\end{proof}

\subsection{Proof of Theorem \ref{theorem: cpt estimation results}}
In this section, we give the change point estimation results for $\alpha=1$, $\alpha=0$ and $\alpha\in(0,1)$, respectively. Before the proof, we need some notations. Note that by {\bf{Assumption A}}, we have
$\|\bx\|_{(s_0,2)}\approx\|\bSigma\bx\|_{(s_0,2)}$ for any $\bx\in \RR^p$.
Hence, for simplicity, we assume $\bSigma=\Ib$. Moreover, to make a clear result, we assume  $s_0$ is fixed with $s_0\leq s:=|\cS^{(1)}|\vee |\cS^{(2)}|$.  Recall $\cM=\{j: \beta^{(1)}_j\neq \beta^{(2)}_j\}\subset\{1,\ldots,p\}$ as the set of coordinates having a change point. For any $\bx\in \RR^p$ and the subset $J\subset\{1,\ldots,p\}$, define the projection operator $\Pi_J \bx\in \RR^{|J|}$ being the sub-vector of $\bx$ with the same coordinates of $\bx$ on $J$, e.g.,  $\Pi_J \bx:=(x_1',\ldots,x'_{|J|})$ with $x'_j=x_j$ for $j\in J$. Based on the definition of $\|\bx\|_{(s_0,2)}$, we have $\|\bx\|_{(s_0,2)}=\max_{J\subset\{1,\ldots,p\},|J|=s_0}\|\Pi_J\bx\|_2$. In addition, for notational simplicity, we also assume $\floor{nt}=nt$ for any $t\in(0,1)$.

Throughout the following  Sections \ref{sec: proof of change point estimationfor alpha=1} - \ref{sec: proof of change point estimationfor alpha}, we assume 
\begin{equation}\label{equation: basic requirement for the signal}
	\|\bDelta\|_{(s_0,2)}\gg  C^* s_0^{1/2}M^2 \sqrt{\dfrac{\log(pn)}{n}} ~\text{and}~{s_0^{1/2}M^2 \sqrt{\dfrac{\log(pn)}{n}}=o(1)}
\end{equation}
for some big enough constant $C^*>0$.
\subsubsection{Change point estimation for $\alpha=1$}\label{sec: proof of change point estimationfor alpha=1}
\begin{proof}
	Recall $\bZ_i(\bX_i,Y_i;\bbeta)=\bX_i(Y_i-\bX_i^\top\bbeta)$ is the negative score for $\alpha=1$. For each $t\in[q_0,1-q_0]$, define $\tilde{\bC}_1(t)=(\tilde{C}_{11}(t),\ldots,\tilde{C}_{1p}(t))^\top$ with
	\begin{equation}\label{equation: C1 for alpha=0 without variance}
		\tilde{\bC}_1(t)=\dfrac{1}{\sqrt{n}}\big(\sum\limits_{i=1}^{\lfloor nt\rfloor}\bZ_i(\bX_i,Y_i;\hat{\bbeta})-\dfrac{\lfloor nt \rfloor }{n}\sum\limits_{i=1}^{n}\bZ_i(\bX_i,Y_i;\hat{\bbeta})\big).
	\end{equation}
	Note that there is no variance estimator in $\tilde{\bC}_1(t)$. 
	By definition, we have $$\hat{t}_{1}:=\argmax\limits_{t\in[q_0,1-q_0]}\|\tilde{\bC}_1(t)\|_{(s_0,2)}.$$ Let $\bDelta=\bbeta^{(1)}-\bbeta^{(2)}$ be the signal difference.  Moreover, define the estimation error $\epsilon_n$ as:
	\begin{equation}\label{equation: definition of epsilonn}
		\epsilon_{n}=C(s_0,M,q_0)\dfrac{\log(pn)}{n\|\bDelta\|^2_{(s_0,2)}}.
	\end{equation}
	To prove Theorem \ref{theorem: cpt estimation results} with $\alpha=1$, we need to prove that as $n,p\rightarrow\infty$, by choosing a large enough constant $C(s_0,M,q_0)$ in $\epsilon_{n}$, we have
	\begin{equation}\label{inequality: theorem 3.2}
		\P\big(|\hat{t}_{1}-t_1|\geq \epsilon_n\big) \rightarrow 0.
	\end{equation}
	To that end, we have to prove
	\begin{equation}\label{inequality: A1+A2}
		\begin{array}{ll}
			\P\big(|\hat{t}_{1}-t_1|\geq \epsilon_n\big)\\
			\leq \P\big(\hat{t}_{1}\geq t_1+ \epsilon_n\big)+\P\big(\hat{t}_{1}\leq t_1- \epsilon_n\big)\\
			\leq \P\big(\max\limits_{t\geq t_1+\epsilon_n}\|\tilde{\bC}_1(t)\|_{(s_0,2)}\geq \|\tilde{\bC}_1(t_1)\|_{(s_0,2)}\big)+\P\big(\max\limits_{t\leq t_1-\epsilon_n}\|\tilde{\bC}_1(t)\|_{(s_0,2)}\geq \|\tilde{\bC}_1(t_1)\|_{(s_0,2)}\big).\\
		\end{array}
	\end{equation}
	Hence, to prove $\P\big(|\hat{t}_{1}-t_1|\geq \epsilon_n\big)\rightarrow 0$, it is equivalent to prove 
	\begin{equation}\label{inequality: A1+A2}
		\begin{array}{ll}
			\P\big(\underbrace{\max\limits_{t\geq t_1+\epsilon_n}\|\tilde{\bC}_1(t)\|_{(s_0,2)}- \|\tilde{\bC}_1(t_1)\|_{(s_0,2)}\leq 0}_{A_1}\big)\\+\P\big(\underbrace{\max\limits_{t\leq t_1-\epsilon_n}\|\tilde{\bC}_1(t)\|_{(s_0,2)}- \| \tilde{\bC}_1(t_1)\|_{(s_0,2)}\leq 0}_{A_2}\big)\rightarrow 1.
		\end{array}
	\end{equation}
	Next, we prove $\P(A_1)\rightarrow 1$ and $\P(A_2)\rightarrow 1$. By the symmetry, we only consider $\P(A_1)\rightarrow 1$. Define the two events $\cH_1$ and $\cH_2$:
	\begin{equation}\label{equation: H2}
		\begin{array}{ll}
			\cH_1=\big\{\max\limits_{t\geq t_1+\epsilon_n}\|\tilde{\bC}_1(t)\|_{(s_0,2)}:=\max\limits_{t\geq t_1+\epsilon_n}\max\limits_{J\subset\{1,\ldots,p\}\atop |J|=s_0}\|\Pi_{J}\tilde{\bC}_1(t)\|_{2}=\max\limits_{t\geq t_1+\epsilon_n}\max\limits_{J\subset\cM\atop |J|=s_0}\|\Pi_{J}\tilde{\bC}_1(t)\|_{2}\big\},\\
			\cH_2=\big\{\|\tilde{\bC}_1(t_1)\|_{(s_0,2)}:=\max\limits_{J\subset\{1,\ldots,p\}\atop |J|=s_0}\|\Pi_{J}\tilde{\bC}_1(t_1)\|_{2}=\max\limits_{J\subset\cM\atop |J|=s_0}\|\Pi_{J}\tilde{\bC}_1(t_1)\|_{2}\big\}.
		\end{array}
	\end{equation}
	The following Lemma \ref{lemma: maximum at Pi} shows that $\cH_1$ and $\cH_2$ occur with high probability. The proof of Lemma \ref{lemma: maximum at Pi} is provided in Section \ref{sec: proof of maximum at Pi}.
	\begin{lemma}\label{lemma: maximum at Pi}
		Under {\bf{Assumptions A, B, C.2, E.2 - E.4}}, we have
		\begin{equation}
			\P(\cH_1)\rightarrow 1 ~~\text{and}~~	\P(\cH_2)\rightarrow 1.
		\end{equation}
	\end{lemma}
	\noindent Now, under $\cH_1\cap\cH_2$, we have: 
	\begin{equation*}
		\begin{array}{ll}
			\P(A_1)&= \P\big(\max\limits_{t\geq t_1+\epsilon_n}\|\tilde{\bC}_1(t)\|_{(s_0,2)}- \|\tilde{\bC}_1(t_1)\|_{(s_0,2)}\leq 0\big)\\
			&=\P\big(\max\limits_{t\geq t_1+\epsilon_n}\max\limits_{J\subset\cM\atop |J|=s_0}\|\Pi_{J}\tilde{\bC}_1(t)\|_{2}- \max\limits_{J\subset\cM\atop |J|=s_0}\|\Pi_{J}\tilde{\bC}_1(t_1)\|_{2}\leq 0\big)\\
			&=\P\big(\max\limits_{t\geq t_1+\epsilon_n}\max\limits_{J\subset\cM\atop |J|=s_0}\|\Pi_{J}\tilde{\bC}_1(t)\|^2- \max\limits_{J\subset\cM\atop |J|=s_0}\|\Pi_{J}\tilde{\bC}_1(t_1)\|^{2}\leq 0\big).\\
		\end{array}
	\end{equation*}
	Note that under $\Hb_1$, we have the following decomposition
	\begin{equation}\label{equation: tilde-C1 for estimation}
		\tilde{\bC}_1(t)={\bC}_1^{I}(t)+{\bdelta(t)}+{\bR(t)	},
	\end{equation}
	where ${\bC}^{I}_{1}(t)$, $\bdelta(t)$ and $\bR(t)$ are defined in (\ref{equation: delta-t alpha=1}) and (\ref{equation: C11+C12 under H1}), respectively. Similarly, we have 
	\begin{equation}\label{equation: tilde-Ct1}
		\tilde{\bC}_1(t_1)={\bC}_1^{I}(t_1)+{\bdelta(t_1)}+{\bR(t_1)	},
	\end{equation}
	by replacing $t$ by $t_1$. To prove $\P(A_1)\rightarrow 1$, we consider $\max\limits_{t\geq t_1+\epsilon_n}\max\limits_{J\subset\cM\atop |J|=s_0}\|\Pi_{J}\tilde{\bC}_1(t)\|^2- \max\limits_{J\subset\cM\atop |J|=s_0}\|\Pi_{J}\tilde{\bC}_1(t_1)\|^{2}\leq 0$. By the fact that $\max a_i-\max b_i \leq \max (a_i-b_i)$ for any $\{a_i\}$ and $\{b_i\}$, we have:
	\begin{equation*}
		\begin{array}{ll}
			\max\limits_{t\geq t_1+\epsilon_n}\max\limits_{J\subset\cM\atop |J|=s_0}\|\Pi_{J}\tilde{\bC}_1(t)\|^2- \max\limits_{J\subset\cM\atop |J|=s_0}\|\Pi_{J}\tilde{\bC}_1(t_1)\|^{2}\\
			\leq \max\limits_{t\geq t_1+\epsilon_n}\max\limits_{J\subset\cM\atop |J|=s_0}\big(\|\Pi_{J}\big({\bC}_1^{I}(t)+\bdelta(t)+\bR(t)\big)\|^2- \|\Pi_{J}\big({\bC}_1^{I}(t_1)+\bdelta(t_1)+\bR(t_1)\big)\|^2\big)\\
			\leq A_{1.1}+A_{1.2}+A_{1.3}+A_{1.4}+A_{1.5}+A_{1.6},
		\end{array}
	\end{equation*}
	where
	\begin{equation}
		\begin{array}{ll}
			A_{1.1}:=\max\limits_{t\geq t_1+\epsilon_n}\max\limits_{J\subset\cM\atop |J|=s_0}\{\|\Pi_{J}{\bC}_1^{I}(t)\|^2+\|\Pi_{J}{\bC}_1^{I}(t_1)\|^2\},\\
			A_{1.2}:=\dfrac{1}{3}\max\limits_{t\geq t_1+\epsilon_n}\max\limits_{J\subset\cM\atop |J|=s_0}\{\|\Pi_{J}\bdelta(t)\|^2-\|\Pi_{J}\bdelta(t_1)\|^2\},\\
			A_{1.3}:=\max\limits_{t\geq t_1+\epsilon_n}\max\limits_{J\subset\cM\atop |J|=s_0}\{\|\Pi_{J}\bR(t)\|^2+\|\Pi_{J}\bR(t_1)\|^2\}, \\ 
			A_{1.4}:=2\max\limits_{t\geq t_1+\epsilon_n}\max\limits_{J\subset\cM\atop |J|=s_0}\{\Pi_{J}{\bC}_1^{I}(t)^\top\Pi_{J}\bR(t)-\Pi_{J}{\bC}_1^{I}(t_1)^\top\Pi_{J}\bR(t_1)\},\\
			A_{1.5}\\
			:=\max\limits_{t\geq t_1+\epsilon_n}\max\limits_{J\subset\cM\atop |J|=s_0}\{2(\Pi_{J}{\bC}_1^{I}(t)^\top\Pi_{J}\bdelta(t)-\Pi_{J}{\bC}_1^{I}(t_1)^\top\Pi_{J}\bdelta(t_1))+\dfrac{1}{3}(\|\Pi_{J}\bdelta(t)\|^2-\|\Pi_{J}\bdelta(t_1)\|^2)\},\\
			A_{1.6}\\
			:=\max\limits_{t\geq t_1+\epsilon_n}\max\limits_{J\subset\cM\atop |J|=s_0}\{2(\Pi_{J}{\bdelta}(t)^\top\Pi_{J}\bR(t)-\Pi_{J}{\bdelta}(t_1)^\top\Pi_{J}\bR(t_1))+\dfrac{1}{3}(\|\Pi_{J}\bdelta(t)\|^2-\|\Pi_{J}\bdelta(t_1)\|^2)\}.
		\end{array}
	\end{equation}
	Our goal is to prove that $\P(A_{1.1}+A_{1.2}+A_{1.3}+A_{1.4}+A_{1.5}+A_{1.6}\leq 0)\rightarrow 1$. Next, we consider $A_{1.1}-A_{1.6}$, respectively. For $A_{1.1}$, we have:
	\begin{equation}\label{inequality: A.11}
		\begin{array}{ll}
			A_{1.1}&\leq 2\max\limits_{q_0\leq t\leq 1-q_0}\max\limits_{J\subset\cM\atop |J|=s_0}\|\Pi_{J}{\bC}_1^{I}(t)\|^2\\
			&\leq 2\max\limits_{q_0\leq t\leq 1-q_0}\max\limits_{J\subset\{1,\ldots,p\}\atop |J|=s_0}\|\Pi_{J}{\bC}_1^{I}(t)\|^2\\
			&\leq 2\max\limits_{q_0\leq t\leq 1-q_0}(\|{\bC}_1^{I}(t)\|^2_{(s_0,2)})\\
			&\leq 2\max\limits_{q_0\leq t\leq 1-q_0}(s_0^{1/2}\|{\bC}_1^{I}(t)\|_{\infty})^2\\
			&\leq C {s_0 M^2{\log(pn)}}:=C_1(s_0,M)\log(pn),
		\end{array}
	\end{equation}
	where the last inequality comes from Lemma \ref{lemma: exponential inequality for partial sum process}. Next, we consider $A_{1.2}$. 
	By the definition of $\bdelta(t)$ and $\bdelta(t_1)$ as defined in (\ref{equation: delta-t alpha=1}), for $t\geq t_1+\epsilon_n$ and $J\subset \cM$, we have:
	\begin{equation}\label{inequality: A.12}
		\begin{array}{ll}
			A_{1.2}&=_{(1)}\dfrac{1}{3}\max\limits_{t\geq t_1+\epsilon_n}\max\limits_{J\subset\cM\atop |J|=s_0}\{\|\Pi_{J}\bdelta(t)\|^2-\|\Pi_{J}\bdelta(t_1)\|^2\}\\
			\quad&=_{(2)}\dfrac{1}{3}\max\limits_{t\geq t_1+\epsilon_n}\big(nt_1^2(t_1-t)(2-t-t_1)\|\bDelta\|^2_{(s_0,2)}\big)\\
			\quad&=_{(3)}-\dfrac{1}{3}n\epsilon_nt_1^2(2-2t_1-\epsilon_n)\|\bDelta\|^2_{(s_0,2)}\\
			\quad &\leq_{(4)} {-\dfrac{1}{6}q_0n\epsilon_n\|\bDelta\|^2_{(s_0,2)}},
		\end{array}
	\end{equation}
	where the last inequality comes from $t_1\in[q_0,1-q_0]$, and $\epsilon_n=o(1)$. For $A_{1.3}$, by the definition of $\bR(t)$ and $\bR(t_1)$, and using Lemmas \ref{lemma: exponential inequality for partial sum process} and \ref{lemma: basic inequality for lasso with alpha=1}, we have: 
	\begin{equation}\label{inequality: A.13}
		A_{1.3}\leq {Cs_0s^2M^2\log(pn)\|\bDelta\|^2_{(s_0,2)}}:=C_3(s_0,M)s^2\log(pn)\|\bDelta\|^2_{(s_0,2)}.
	\end{equation}
	Next, we consider $A.14$. By the Cauchy-Swartz inequality, we have:
	\begin{equation}\label{inequality: A.14}
		\begin{array}{ll}
			A_{1.4}&=2\max\limits_{t\geq t_1+\epsilon_n}\max\limits_{J\subset\cM\atop |J|=s_0}\{\Pi_{J}\tilde{\bC}_1^{I}(t)^\top\Pi_{J}\bR_1(t)-\Pi_{J}\tilde{\bC}_1^{I}(t_1)^\top\Pi_{J}\bR_1(t_1)\}\\
			&\leq_{(1)} 4\max\limits_{t\in[q_0,1-q_0]}\max\limits_{J\subset\cM\atop |J|=s_0}|\{\Pi_{J}\tilde{\bC}_1^{I}(t)^\top\Pi_{J}\bR_1(t)|\\
			&\leq_{(2)} 4\max\limits_{t\in[q_0,1-q_0]}\max\limits_{J\subset\cM\atop |J|=s_0}\|\{\Pi_{J}\tilde{\bC}_1^{I}(t)\|_2\|\Pi_{J}\bR_1(t)\|_2\\
			&\leq_{(3)} 4\max\limits_{t\in[q_0,1-q_0]}\|\tilde{\bC}_1^{I}(t)\|_{(s_0,2)}\times\max\limits_{t\in[q_0,1-q_0]}\|\bR_1(t)\|_{(s_0,2)}\\
			&\leq_{(4)} Cs_0^{1/2}M\sqrt{\log(pn)}\times s_0^{1/2}\sqrt{n}M^2\sqrt{\dfrac{\log(pn)}{n}}s\|\bDelta\|_{(s_0,2)}\\
			&\leq_{(5)} {Cs_0sM^3{\log(pn)}\|\bDelta\|_{(s_0,2)}}:=C_4(s_0,M){s\log(pn)}\|\bDelta\|_{(s_0,2)},
		\end{array}
	\end{equation}
	where $(4)$ comes from Lemma \ref{lemma: exponential inequality for partial sum process} and Lemma \ref{lemma: concentration for covariance}. 
	Hence, combining (\ref{inequality: A.12}) - (\ref{inequality: A.13}), if  $\epsilon_n$ satisfies 
	\begin{equation}\label{equation: definition of epsilonn-1}
		\begin{array}{ll}
			\epsilon_n&=C\max\Big\{\underbrace{C_1(s_0,M)\dfrac{\log(pn)}{n\|\bDelta\|^2_{(s_0,2)}}}_{\rm by~ A_{1.1}},\underbrace{C_3(s_0,M)\dfrac{s^2\log(pn)}{n}}_{\rm by~A_{1.3}}, \underbrace{\dfrac{C_4(s_0,M)s\log(pn)}{n\|\bDelta\|_{(s_0,2)}}}_{\rm by~A_{1.4}}\Big\}\\
		\end{array}
	\end{equation}
	for some big enough constant $C>0$,  with probability tending to one, we have $A_{1.1}+A_{1.2}+A_{1.3}+A_{1.4}\leq 0$. 
	
	Next, we prove $A_{1.5}+A_{1.6}\leq 0$. For $A_{1.5}$, using the triangle inequality, we have:
	\begin{equation}\label{inequality: A.1.4}
		\begin{array}{ll}
			A_{1.5}&=\max\limits_{t\geq t_1+\epsilon_n}\max\limits_{J\subset\cM\atop |J|=s_0}\Big\{2(\Pi_{J}{\bC}_1^{I}(t)^\top\Pi_{J}\bdelta(t)-\Pi_{J}{\bC}_1^{I}(t_1)^\top\Pi_{J}\bdelta(t_1))\\&-\dfrac{1}{3}(\|\Pi_{J}\bdelta_(t_1)\|^2-\|\Pi_{J}\bdelta(t)\|^2)\Big\}\\
			&=A_{1.5.1}+A_{1.5.2},\\
		\end{array}
	\end{equation}
	where
	\begin{equation}\label{equation: A.1.4.1+A.1.4.2}
		\begin{array}{ll}
			A_{1.5.1}\\
			=\max\limits_{t\geq t_1+\epsilon_n}\max\limits_{J\subset\cM\atop |J|=s_0}\Big\{2\Pi_{J}{\bC}_1^{I}(t)^\top(\Pi_{J}\bdelta(t)-\Pi_{J}\bdelta(t_1))\}-\dfrac{1}{6}(\|\Pi_{J}\bdelta(t_1)\|^2-\|\Pi_{J}\bdelta(t)\|^2)\Big\},\\
			A_{1.5.2}\\
			=\max\limits_{t\geq t_1+\epsilon_n}\max\limits_{J\subset\cM\atop |J|=s_0}\Big\{2\Pi_{J}\bdelta(t_1)^\top(\Pi_{J}{\bC}_1^{I}(t)-\Pi_{J}{\bC}_1^{I}(t_1))-\dfrac{1}{6}(\|\Pi_{J}\bdelta(t_1)\|^2-\|\Pi_{J}\bdelta(t)\|^2)\Big\}.
		\end{array}
	\end{equation}
	To bound $A_{1.5}$, we prove $\P(A_{1.5.1}\leq 0)\rightarrow 1$ and $\P(A_{1.5.2}\leq 0)\rightarrow 1$, respectively. 
	To bound $A_{1.5.1}$, note that for any fixed $t\geq t_1+\epsilon_n$ and $J\subset\cM$ with $|J|=s_0$, we have:
	\begin{equation}\label{inequality: A.1.4.1}
		\begin{array}{ll}
			2\Pi_{J}{\bC}_1^{I}(t)^\top\Pi_{J}(\bdelta(t)-\bdelta_1(t_1))-\dfrac{1}{6}(\|\Pi_{J}\bdelta(t_1)\|^2-\|\Pi_{J}\bdelta(t)\|^2)\\
			\leq_{(1)} 2\|\Pi_{J}{\bC}_1^{I}(t)\|_2\|\Pi_{J}(\bdelta(t)-\bdelta(t_1))\|_2-\dfrac{1}{6}\big(nt_1^2(t-t_1)(2-t-t_1)\|\Pi_J\bDelta\|^2\big)\\
			\leq_{(2)} 2s_{0}^{1/2}\|\Pi_{J}{\bC}_1^{I}(t)\|_{\infty}\sqrt{n}t_1(t-t_1))\|\bDelta\|_{(s_0,2)}-\dfrac{1}{6}\big(nt_1^2(t-t_1)(1+q_0-t_1)\|\bDelta\|^2_{(s_0,2)}\big)\\
			\leq_{(3)} 2s_{0}^{1/2}\|{\bC}_1^{I}(t)\|_{\infty}\sqrt{n}t_1(t-t_1))\|\bDelta\|_{(s_0,2)}-\dfrac{1}{6}\big(nt_1^2(t-t_1)(1+q_0-t_1)\|\bDelta\|^2_{(s_0,2)}\big).\\
		\end{array}
	\end{equation}
	Hence, by (\ref{inequality: A.1.4.1}), to prove $\P(A_{1.5.1}\leq 0)\rightarrow 1$, it is sufficient to prove that :
	\begin{equation*}
		\begin{array}{ll}
			\P\Big(\max\limits_{t\geq t_1+\epsilon_n}\max\limits_{J\subset\cM\atop |J|=s_0}\{2s_{0}^{1/2}\|{\bC}_1^{I}(t)\|_{\infty}\sqrt{n}t_1(t-t_1))\|\bDelta\|_{(s_0,2)}\\
			\qquad-\dfrac{1}{6}\big(nt_1^2(t-t_1)(1+q_0-t_1)\|\bDelta\|^2_{(s_0,2)}\big)\}\leq 0\Big)\rightarrow 1.
		\end{array}
	\end{equation*}
	Equivalently, it is sufficient to prove that 
	\begin{equation*}
		\begin{array}{ll}
			\P\Big(\max\limits_{t\geq t_1+\epsilon_n}\max\limits_{J\subset\cM\atop |J|=s_0}\{2s_{0}^{1/2}\|{\bC}_1^{I}(t)\|_{\infty}\sqrt{n}t_1(t-t_1))\|\bDelta\|_{(s_0,2)}\\
			\qquad	-\dfrac{1}{6}\big(nt_1^2(t-t_1)(1+q_0-t_1)\|\bDelta\|^2_{(s_0,2)}\big)\}\leq 0\Big)\\
			\geq \P\Big(\max\limits_{q_0\leq t\leq 1-q_0}2s_{0}^{1/2}\|{\bC}_1^{I}(t)\|_{\infty}t_1-\dfrac{1}{6}\big(\sqrt{n}t_1^2(1+q_0-t_1)\|\bDelta\|_{(s_0,2)}\big)\leq 0\Big)\rightarrow 1.\\
		\end{array}
	\end{equation*}
	Note that by Lemma \ref{lemma: exponential inequality for partial sum process}, we have $\max\limits_{q_0\leq t\leq 1-q_0}\{2s_{0}^{1/2}\|\tilde{\bC}_1^{I}(t)\|_{\infty}=O_p(s_0^{1/2}M\sqrt{\log(pn)})$.  Moreover,  if we choose a big enough constant $C^*$ in by (\ref{equation: basic requirement for the signal}), it is easy to see that 
	\begin{equation*}
		\begin{array}{ll}
			\P\Big(\max\limits_{q_0\leq t\leq 1-q_0}2s_{0}^{1/2}\|\tilde{\bC}_1^{I}(t)\|_{\infty}t_1-\dfrac{1}{6}\big(\sqrt{n}t_1^2(1+q_0-t_1)\|\bDelta\|_{(s_0,2)}\big)\leq 0\Big)\rightarrow 1,\\
		\end{array}
	\end{equation*}
	which yields $\P(A_{1.5.1})\rightarrow 1$. After bounding $A_{1.5.1}$, we next consider $A_{1.5.2}$. Note that for any fixed $t\geq t_1+\epsilon_n$ and $J\subset\cM$ with $|J|=s_0$, we have:
	\begin{equation}\label{inequality: A.1.4.2}
		\begin{array}{ll}
			2\Pi_{J}\bdelta(t_1)^\top\Pi_{J}({\bC}_1^{I}(t)-{\bC}_1^{I}(t_1))-\dfrac{1}{6}(\|\Pi_{J}\bdelta(t_1)\|^2-\|\Pi_{J}\bdelta(t)\|^2)\\
			\leq_{(1)} 2\|\Pi_{J}\bdelta(t_1)\|_2\|\Pi_{J}({\bC}_1^{I}(t)-{\bC}_1^{I}(t_1))\|_2-\dfrac{1}{6}\big(nt_1^2(t-t_1)(2-t-t_1)\|\Pi_J\bDelta\|^2\\
			\leq 2\sqrt{n}t_1(1-t_1)\|\bDelta\|_{(s_0,2)}s_0^{1/2}\|\tilde{\bC}_1^{I}(t)-\tilde{\bC}_1^{I}(t_1)\|_{\infty}-
			\dfrac{1}{6}\big(nt_1^2(t-t_1)(1+q_0-t_1)\|\bDelta\|^2_{(s_0,2)}.
		\end{array}
	\end{equation}
	Note that by the definition of ${\bC}_1^{I}(t)$ and ${\bC}_1^{I}(t_1)$, we have:
	\begin{equation}\label{equation: partial sum local neighborhood}
		{\bC}_1^{I}(t)-{\bC}_1^{I}(t_1)=\dfrac{1}{\sqrt{n}}\big(\sum_{i=\floor{nt_1}+1}^{\floor{nt}}\bX_i\epsilon_i-\dfrac{\floor{nt}-\floor{nt_1}}{n}\sum_{i=1}^n\bX_i\epsilon_i\big).
	\end{equation} 
	Hence, combining (\ref{inequality: A.1.4.2}) and (\ref{equation: partial sum local neighborhood}), we have:
	\begin{equation*}
		\begin{array}{ll}
			2\Pi_{J}\bdelta(t_1)^\top\Pi_{J}({\bC}_1^{I}(t)-{\bC}_1^{I}(t_1))-\dfrac{1}{6}(\|\Pi_{J}\bdelta(t_1)\|^2-\|\Pi_{J}\bdelta(t)\|^2)\leq A_{1.5.2}^{I}+A_{1.5.2}^{II},
		\end{array}
	\end{equation*}
	where
	\begin{equation}\label{equation: A.1.4.2-1+A.1.4.2-2}
		\begin{array}{ll}
			A_{1.5.2}^{I}=2t_1(1-t_1)\|\bDelta\|_{(s_0,2)}s_0^{1/2}\|\sum\limits_{i=\floor{nt_1}+1}^{\floor{nt}}\bX_i\epsilon_i\|_{\infty}\\
			\qquad\qquad\qquad-\dfrac{1}{12}\big(nt_1^2(t-t_1)(1+q_0-t_1)\|\bDelta\|^2_{(s_0,2)},\\
			A_{1.5.2}^{II}=2t_1(1-t_1)\|\bDelta\|_{(s_0,2)}s_0^{1/2}\|\dfrac{\floor{nt}-\floor{nt_1}}{n}\sum\limits_{i=1}^n\bX_i\epsilon_i\|_{\infty}\\
			\qquad\qquad\qquad-\dfrac{1}{12}\big(nt_1^2(t-t_1)(1+q_0-t_1)\|\bDelta\|^2_{(s_0,2)}.
		\end{array}
	\end{equation}
	Considering (\ref{equation: A.1.4.1+A.1.4.2}), (\ref{inequality: A.1.4.2}), (\ref{equation: partial sum local neighborhood}), and (\ref{equation: A.1.4.2-1+A.1.4.2-2}), to prove $\P(A_{1.5.2})\rightarrow 1$, it is sufficient to prove $\P(\max_{t}\max_{J}A_{1.5.2}^{I}\leq 0)\rightarrow 1$ and $\P(\max_{t}\max_{J}A_{1.5.2}^{II}\leq 0)\rightarrow 1$. For $A_{1.5.2}^{I}$, we have to prove
	\begin{equation}\label{inequality: A.1.4.1-3}
		\begin{array}{ll}
			\P\Big(\max\limits_{t\geq t_1+\epsilon_n}\max\limits_{J\subset\cM\atop |J|=s_0}\Big\{2t_1(1-t_1)\|\bDelta\|_{(s_0,2)}s_0^{1/2}\|\sum\limits_{i=\floor{nt_1}+1}^{\floor{nt}}\bX_i\epsilon_i\|_{\infty}\\
			\qquad-\dfrac{1}{6}\big(nt_1^2(t-t_1)(1+q_0-t_1)\|\bDelta\|^2_{(s_0,2)}\Big\}\leq 0\Big)\\
			=\P\Big(\max\limits_{t\geq t_1+\epsilon_n}\|\dfrac{1}{\floor{nt}-\floor{nt_1}}\sum\limits_{i=\floor{nt_1}+1}^{\floor{nt}}\bX_i\epsilon_i\|_{\infty}\leq C(t_1,q_0)s_0^{-1/2}\|\bDelta\|_{(s_0,2)}\Big)\rightarrow 1.
		\end{array}
	\end{equation}
	Note that by Lemma \ref{lemma: exponential inequality for partial sum process}, we have 
	\begin{equation*}
		\max\limits_{t\geq t_1+\epsilon_n}\|\dfrac{1}{\floor{nt}-\floor{nt_1}}\sum\limits_{i=\floor{nt_1}+1}^{\floor{nt}}\bX_i\epsilon_i\|_{\infty}=O_p(Mn^{1/4}\dfrac{\log(pn)}{n\epsilon_{n}}).
	\end{equation*}
	Hence, if we choose 
	\begin{equation*}
		{\epsilon_{n}=C_5(s_0,M)\dfrac{\log(pn)}{n^{3/4}\|\bDelta\|_{(s_0,2)}}}
	\end{equation*}
	for some big enough constant $C_5(s_0,M)>0$, we have (\ref{inequality: A.1.4.1-3}) holds, which yields $\P(\max_{t}\max_{J}A_{1.5.2}^{I}\leq 0)\rightarrow 1$. Similarly, we can prove $\P(\max_{t}\max_{J}A_{1.5.2}^{II}\leq 0)\rightarrow 1$, which yields $\P(A_{1.5.2}\leq 0)\rightarrow 1$. 
	
	With a very similar proof technique, if we choose ${\epsilon_n=C_6(s_0,M)\dfrac{s^2\log(pn)}{n}}$ for some big enough constant $C_6>0$, we can prove
	\begin{equation*}
		\P(A_{1.6}\leq 0)\rightarrow 1.
	\end{equation*}
	Combining the previous results, if $\epsilon_{n}$ satisfies
	\begin{equation}\label{equation: definition of epsilonn-2}
		\begin{array}{ll}
			\epsilon_n=C\max\Big\{\underbrace{C_1(s_0,M)\dfrac{\log(pn)}{n\|\bDelta\|^2_{(s_0,2)}}}_{\rm by~A_{1.1}},\underbrace{\dfrac{C_3(s_0,M)s^2\log(pn)}{n}}_{\rm by~A_{1.3}}, \underbrace{\dfrac{C_4(s_0,M)s\log(pn)}{n\|\bDelta\|_{(s_0,2)}}}_{\rm by~A_{1.4}},\\ \qquad\qquad\underbrace{\dfrac{C_5(s_0,M)\log(pn)}{n^{3/4}\|\bDelta\|_{(s_0,2)}}}_{\rm by~A_{1.5}},\underbrace{C_6(s_0,M)\dfrac{s^2\log(pn)}{n}}_{\rm by~A_{1.6}}\Big\}\\
			=C(s_0,M)\dfrac{\log(pn)}{n\|\bDelta\|^2_{(s_0,2)}}\times \max\Big\{1,s^2\|\bDelta\|^2_{(s_0,2)},s\|\bDelta\|_{(s_0,2)},n^{1/4}\|\bDelta\|_{(s_0,2)}\Big\}.
		\end{array}
	\end{equation}
	we can prove $\P(A_1)\rightarrow 1$. By symmetry, we can prove $\P(A_2)\rightarrow 1$, which finishes the proof.
	
	Lastly, we need to discuss the five terms in (\ref{equation: definition of epsilonn-2}). Note that by {Assumption F} and the assumption that $\|\bbeta^{(1)}-\bbeta^{(2)}\|_1\leq C_{\bDelta}$, we have $s^2\|\bDelta\|^2_{(s_0,2)}=O(1)$ and $s\|\bDelta\|_{(s_0,2)}=O(1)$. Moreover, by the assumption that $n^{1/4}=o(s)$, we have $n^{1/4}\|\bDelta\|_{(s_0,2)}=o(1)$, which finishes the proof.
	
\end{proof}
\subsubsection{Change point estimation for $\alpha=0$}\label{sec: proof of change point estimationfor alpha=0}
\begin{proof}
	For $\alpha=0$, recall $\bZ(\bX_i,Y_i;\tilde{\btau},\hat{\bb},\hat{\bbeta}):=\dfrac{1}{K}\sum\limits_{k=1}^{K}\bX_i \big(\mathbf{1}\{Y_i-\hat{b}_k-\bX_i^\top\hat{\bbeta}\leq 0\}-\tau_k\big)$ as the score function.  For each $t\in[q_0,1-q_0]$, define $\tilde{\bC}_0(t)=(\tilde{C}_{01}(t),\ldots,\tilde{C}_{0p}(t))^\top$ with
	\begin{equation}
		\tilde{\bC}_0(t)=\dfrac{1}{\sqrt{n}}\big(\sum\limits_{i=1}^{\lfloor nt\rfloor}\bZ(\bX_i,Y_i;\tilde{\btau},\hat{\bb},\hat{\bbeta})-\dfrac{\lfloor nt \rfloor }{n}\sum\limits_{i=1}^{n}\bZ(\bX_i,Y_i;\tilde{\btau},\hat{\bb},\hat{\bbeta})\big).
	\end{equation}
	Note that there is no any variance estimator in $\tilde{\bC}_0(t)$.  Recall $\hat{t}_{0}:=\argmax\limits_{t\in[q_0,1-q_0]}\|\tilde{\bC}_0(t)\|_{(s_0,2)}$. To prove Theorem \ref{theorem: cpt estimation results} with $\alpha=0$, we need to prove that as $n,p\rightarrow\infty$, by choosing a large enough constant $C(s_0,M,\tilde{\btau})$ in $\epsilon_{n}$ (which will be given in (\ref{inequality: step-6 change point estimation for alpha=0})), we have
	\begin{equation}\label{inequality: theorem 3.2 alpha=0}
		\P\big(|\hat{t}_{0}-t_1|\geq \epsilon_n\big) \rightarrow 0.
	\end{equation}
	Similar to Section \ref{sec: proof of change point estimationfor alpha=1}, we have to prove $\P(A_1)\rightarrow 1$ and $\P(A_2)\rightarrow 1$, where
	\begin{equation}\label{inequality: A1+A2 alpha=0}
		\begin{array}{ll}
			A_1=\max\limits_{t\geq t_1+\epsilon_n}\|\tilde{\bC}_0(t)\|_{(s_0,2)}- \|\tilde{\bC}_0(t_1)\|_{(s_0,2)}\leq 0,\\
			A_2=\max\limits_{t\leq t_1-\epsilon_n}\|\tilde{\bC}_0(t)\|_{(s_0,2)}- \| \tilde{\bC}_0(t_1)\|_{(s_0,2)}\leq 0.
		\end{array}
	\end{equation}
	By the symmetry, we only consider $\P(A_1)\rightarrow 1$. Define the two events $\cH_1$ and $\cH_2$:
	\begin{equation}\label{equation: H2 alpha=0}
		\begin{array}{ll}
			\cH_1=\big\{\max\limits_{t\geq t_1+\epsilon_n}\|\tilde{\bC}_0(t)\|_{(s_0,2)}:=\max\limits_{t\geq t_1+\epsilon_n}\max\limits_{J\subset\{1,\ldots,p\}\atop |J|=s_0}\|\Pi_{J}\tilde{\bC}_0(t)\|_{2}=\max\limits_{t\geq t_1+\epsilon_n}\max\limits_{J\subset\cM\atop |J|=s_0}\|\Pi_{J}\tilde{\bC}_0(t)\|_{2}\big\},\\
			\cH_2=\big\{\|\tilde{\bC}_0(t_1)\|_{(s_0,2)}:=\max\limits_{J\subset\{1,\ldots,p\}\atop |J|=s_0}\|\Pi_{J}\tilde{\bC}_0(t_1)\|_{2}=\max\limits_{J\subset\cM\atop |J|=s_0}\|\Pi_{J}\tilde{\bC}_0(t_1)\|_{2}\big\}.
		\end{array}
	\end{equation}
	Similar to the proof of Lemma \ref{lemma: maximum at Pi}, we can prove  $\P(\cH_1\cap\cH_2)\rightarrow 1$. 
	\noindent Now, under $\cH_1\cap\cH_2$, we have: 
	\begin{equation*}
		\begin{array}{ll}
			\P(A_1)&= \P\big(\max\limits_{t\geq t_1+\epsilon_n}\|\tilde{\bC}_0(t)\|_{(s_0,2)}- \|\tilde{\bC}_0(t_1)\|_{(s_0,2)}\leq 0\big)\\
			&=\P\big(\max\limits_{t\geq t_1+\epsilon_n}\max\limits_{J\subset\cM\atop |J|=s_0}\|\Pi_{J}\tilde{\bC}_0(t)\|_{2}- \max\limits_{J\subset\cM\atop |J|=s_0}\|\Pi_{J}\tilde{\bC}_0(t_1)\|_{2}\leq 0\big)\\
			&=\P\big(\max\limits_{t\geq t_1+\epsilon_n}\max\limits_{J\subset\cM\atop |J|=s_0}\|\Pi_{J}\tilde{\bC}_0(t)\|^2- \max\limits_{J\subset\cM\atop |J|=s_0}\|\Pi_{J}\tilde{\bC}_0(t_1)\|^{2}\leq 0\big).\\
		\end{array}
	\end{equation*}
	Recall $\sigma^2(0,\tilde{\btau})=\sqrt{K^{-2}\sum_{k_1,k_2}\gamma_{k_1k_2}}$, with $\gamma_{k_1k_2}:=\min(\tau_{k_1},\tau_{k_2})-\tau_{k_1}\tau_{k_2}$ and $\bC_{0}(t)$ defined in (\ref{equation: C0(t)}). Then, under $\Hb_1$, we have the following decomposition
	\begin{equation}\label{equation: tilde-C1 for estimation alpha=0}
		\begin{array}{ll}
			\tilde{\bC}_0(t)&=\sigma(0,\tilde{\btau})\times\bC_{0}(t)\\
			&=\sigma(0,\tilde{\btau})\times \big(-SNR(0,\tilde{\btau})\times\bdelta(t)
			+\bC_{0}^{(1)}(t)
			-SNR(0,\tilde{\btau})\times\bR(t)+	\bC_{0}^{(3)}(t)+	\bC_{0}^{(4)}(t)\big),
		\end{array}
	\end{equation}
	where the second equation comes from the decomposition in (\ref{equation: decmposition for the score of composite quantile  under H1}).  By the fact that $\max a_i-\max b_i \leq \max (a_i-b_i)$ and $\max (a_i+b_i)\leq \max a_i +\max b_i$ for any $\{a_i\}$ and $\{b_i\}$, we have:
	\begin{equation*}
		\begin{array}{ll}
			\max\limits_{t\geq t_1+\epsilon_n}\max\limits_{J\subset\cM\atop |J|=s_0}\|\Pi_{J}\tilde{\bC}_0(t)\|^2- \max\limits_{J\subset\cM\atop |J|=s_0}\|\Pi_{J}\tilde{\bC}_0(t_1)\|^{2}\\
			\leq \sigma^2(0,\tilde{\btau})\times\max\limits_{t\geq t_1+\epsilon_n}\max\limits_{J\subset\cM\atop |J|=s_0}\big(\|\Pi_{J}\big(-SNR(0,\tilde{\btau})\times\bdelta(t)
			+\bC_{0}^{(1)}(t)\\
			\qquad\qquad	-SNR(0,\tilde{\btau})\times\bR(t)+	\bC_{0}^{(3)}(t)+	\bC_{0}^{(4)}(t)\big)\|^2\\
			\quad\quad-\|\Pi_{J}\big(-SNR(0,\tilde{\btau})\times\bdelta(t_1)
			+\bC_{0}^{(1)}(t_1)
			-SNR(0,\tilde{\btau})\times\bR(t_1)+	\bC_{0}^{(3)}(t_1)+	\bC_{0}^{(4)}(t_1)\big)\|^2\big)\\
			\leq \sigma^2(0,\tilde{\btau})\times(A_{1.1}+\cdots +A_{1.15}),
		\end{array}
	\end{equation*}
	where the fifteen parts $A_{1.1}\cdots A_{1.15}$ are defined as:
	\begin{equation*}
		\begin{array}{ll}
			A_{1.1}:=\max\limits_{t\geq t_1+\epsilon_n}\max\limits_{J\subset\cM\atop |J|=s_0}\{\|\Pi_{J}{\bC}_0^{(1)}(t)\|^2+\|\Pi_{J}{\bC}_0^{(1)}(t_1)\|^2\},\\
			A_{1.2}:=\dfrac{SNR^2(0,\tilde{\btau})}{5}\max\limits_{t\geq t_1+\epsilon_n}\max\limits_{J\subset\cM\atop |J|=s_0}\{\|\Pi_{J}\bdelta(t)\|^2-\|\Pi_{J}\bdelta(t_1)\|^2\},\\
			A_{1.3}:=SNR^2(0,\tilde{\btau})\max\limits_{t\geq t_1+\epsilon_n}\max\limits_{J\subset\cM\atop |J|=s_0}\{\|\Pi_{J}\bR(t)\|^2+\|\Pi_{J}\bR(t_1)\|^2\}, \\ 
			A_{1.4}:=2SNR(0,\tilde{\btau})\max\limits_{t\geq t_1+\epsilon_n}\max\limits_{J\subset\cM\atop |J|=s_0}\{-\Pi_{J}{\bC}_0^{(1)}(t)^\top\Pi_{J}\bR(t)+\Pi_{J}{\bC}_0^{(1)}(t_1)^\top\Pi_{J}\bR(t_1)\},\\
			A_{1.5}:=2SNR(0,\tilde{\btau})\max\limits_{t\geq t_1+\epsilon_n}\max\limits_{J\subset\cM\atop |J|=s_0}\Big\{-\Pi_{J}{\bC}_0^{(1)}(t)^\top\Pi_{J}\bdelta(t)+\Pi_{J}{\bC}_0^{(1)}(t_1)^\top\Pi_{J}\bdelta(t_1)\\
			\qquad\qquad\qquad		\qquad\qquad\qquad\qquad\qquad\qquad	+\dfrac{SNR^2(0,\tilde{\btau})}{5}(\|\Pi_{J}\bdelta(t)\|^2-\|\Pi_{J}\bdelta(t_1)\|^2)\Big\},\\
			A_{1.6}:=2SNR^2(0,\tilde{\btau})\max\limits_{t\geq t_1+\epsilon_n}\max\limits_{J\subset\cM\atop |J|=s_0}\Big\{\Pi_{J}{\bR}(t)^\top\Pi_{J}\bdelta(t)-\Pi_{J}{\bR}(t_1)^\top\Pi_{J}\bdelta(t_1)\\
			\qquad\qquad\qquad		\qquad\qquad\qquad\qquad\qquad\qquad+\dfrac{SNR^2(0,\tilde{\btau})}{5}(\|\Pi_{J}\bdelta(t)\|^2-\|\Pi_{J}\bdelta(t_1)\|^2)\Big\},\\
			A_{1.7}:=\max\limits_{t\geq t_1+\epsilon_n}\max\limits_{J\subset\cM\atop |J|=s_0}\{\|\Pi_{J}\bC_0^{(3)}(t)\|^2+\|\Pi_{J}\bC_0^{(3)}(t_1)\|^2\}, \\
			A_{1.8}:=\max\limits_{t\geq t_1+\epsilon_n}\max\limits_{J\subset\cM\atop |J|=s_0}\{\|\Pi_{J}\bC_0^{(4)}(t)\|^2+\|\Pi_{J}\bC_0^{(4)}(t_1)\|^2\}, \\
			A_{1.9}:=2\max\limits_{t\geq t_1+\epsilon_n}\max\limits_{J\subset\cM\atop |J|=s_0}\{\Pi_{J}{\bC}_0^{(1)}(t)^\top\Pi_{J}\bC_0^{(3)}(t)-\Pi_{J}{\bC}_0^{(1)}(t_1)^\top\Pi_{J}\bC_0^{(3)}(t_1)\}, \\
			A_{1.10}:=2\max\limits_{t\geq t_1+\epsilon_n}\max\limits_{J\subset\cM\atop |J|=s_0}\{\Pi_{J}{\bC}_0^{(1)}(t)^\top\Pi_{J}\bC_0^{(4)}(t)-\Pi_{J}{\bC}_0^{(1)}(t_1)^\top\Pi_{J}\bC_0^{(4)}(t_1)\},\\
			A_{1.11}:=2SNR(0,\tilde{\btau})\max\limits_{t\geq t_1+\epsilon_n}\max\limits_{J\subset\cM\atop |J|=s_0}\{-\Pi_{J}\bR(t)^\top\Pi_{J}\bC_0^{(3)}(t)+\Pi_{J}\bR(t_1)^\top\Pi_{J}\bC_0^{(3)}(t_1)\},\\
			A_{1.12}:=2SNR(0,\tilde{\btau})\max\limits_{t\geq t_1+\epsilon_n}\max\limits_{J\subset\cM\atop |J|=s_0}\{-\Pi_{J}\bR(t)^\top\Pi_{J}\bC_0^{(4)}(t)+\Pi_{J}\bR(t_1)^\top\Pi_{J}\bC_0^{(4)}(t_1)\},\\
			A_{1.13}:=2\max\limits_{t\geq t_1+\epsilon_n}\max\limits_{J\subset\cM\atop |J|=s_0}\{\Pi_{J}\bC_0^{(3)}(t)^\top\Pi_{J}\bC_0^{(4)}(t)-\Pi_{J}\bC_0^{(3)}(t_1)^\top\Pi_{J}\bC_0^{(4)}(t_1)\},\\
			A_{1.14}:=2SNR(0,\tilde{\btau})\max\limits_{t\geq t_1+\epsilon_n}\max\limits_{J\subset\cM\atop |J|=s_0}\Big\{-\Pi_{J}{\bC}_0^{(3)}(t)^\top\Pi_{J}\bdelta(t)+\Pi_{J}{\bC}_0^{(3)}(t_1)^\top\Pi_{J}\bdelta(t_1)\\
			\qquad\qquad\qquad		\qquad\qquad\qquad\qquad\qquad\qquad	+\dfrac{SNR^2(0,\tilde{\btau})}{5}(\|\Pi_{J}\bdelta(t)\|^2-\|\Pi_{J}\bdelta(t_1)\|^2)\Big\}\\
			A_{1.15}:=2SNR(0,\tilde{\btau})\max\limits_{t\geq t_1+\epsilon_n}\max\limits_{J\subset\cM\atop |J|=s_0}\Big\{-\Pi_{J}{\bC}_0^{(4)}(t)^\top\Pi_{J}\bdelta(t)+\Pi_{J}{\bC}_0^{(4)}(t_1)^\top\Pi_{J}\bdelta(t_1)\\
			\qquad\qquad\qquad		\qquad\qquad\qquad\qquad\qquad\qquad+\dfrac{SNR^2(0,\tilde{\btau})}{5}(\|\Pi_{J}\bdelta(t)\|^2-\|\Pi_{J}\bdelta(t_1)\|^2)\Big\}.\\
		\end{array}
	\end{equation*}
	Next, we aim to prove that $\P(A_{1.1}+\cdots+A_{1.15}\leq 0)\rightarrow 1$. The proof proceeds into five steps: 
	{\bf{Step~1:}} We aim to prove that, with probability tending to 1, 
	\begin{equation*}
		A_{1.1}+A_{1.2}+A_{1.3}+A_{1.4}+A_{1.7}+A_{1.8}+A_{1.9}+A_{1.10}+A_{1.11}+A_{1.12}+A_{1.13}\leq 0.
	\end{equation*}
	The main idea of step~1 is to obtain the upper bound for each item. Note that similar to the proofs in Section \ref{sec: proof of change point estimationfor alpha=1}, we can directly prove that:
	\begin{equation*}
		\begin{array}{ll}
			{A_{1.1}\leq C_1(s_0,M,\tilde{\btau})\log(pn)}, \\		
			A_{1.2}\leq -\dfrac{SNR^2(0,\tilde{\btau})}{10}q_0n\epsilon_n\|\bDelta\|^2_{(s_0,2)}, \\
			{A_{1.3}\leq C_3(s_0,M,\tilde{\btau})SNR^2(0,\tilde{\btau})s^2\log(pn)\|\bDelta\|^2_{(s_0,2)}},\\	
			{A_{1.4}\leq C_4(s_0,M,\tilde{\btau})SNR(0,\tilde{\btau}){s\log(pn)}\|\bDelta\|_{(s_0,2)}.}
		\end{array}
	\end{equation*}
	For $A_{1.7}$,  by (\ref{inequality: C0-3}), we have:
	\begin{equation*}
		\begin{array}{ll}
			A_{1.7}&:=\max\limits_{t\geq t_1+\epsilon_n}\max\limits_{J\subset\cM\atop |J|=s_0}\{\|\Pi_{J}\bC_0^{(3)}(t)\|^2+\|\Pi_{J}\bC_0^{(3)}(t_1)\|^2\}\\
			&\leq 2\max\limits_{q_0\leq t\leq 1-q_0}\|\bC_0^{(3)}(t)\|_{(s_0,2)}^2\\
			&\leq {C_7(s_0,M,\tilde{\btau})s^4\log(pn)\|\bDelta\|_{(s_0,2)}^4},
		\end{array}
	\end{equation*}
	For $A_{1.8}$,  by (\ref{inequality: C0-4}), we have:
	\begin{equation*}
		\begin{array}{ll}
			A_{1.8}&:=\max\limits_{t\geq t_1+\epsilon_n}\max\limits_{J\subset\cM\atop |J|=s_0}\{\|\Pi_{J}\bC_0^{(4)}(t)\|^2+\|\Pi_{J}\bC_0^{(4)}(t_1)\|^2\}\\
			&\leq 2\max\limits_{q_0\leq t\leq 1-q_0}\|\bC_0^{(4)}(t)\|_{(s_0,2)}^2\\
			&\leq {C_8(s_0,M,\tilde{\btau})s^2\log(pn)\|\bDelta\|_{(s_0,2)}^2}.
		\end{array}
	\end{equation*}
	For $A_{1.9}$,  by the Cauchy-Swartz inequality, Lemma \ref{lemma: exponential inequality for partial sum process}, and  (\ref{inequality: C0-3}),  we have:
	\begin{equation}
		\begin{array}{ll}
			A_{1.9}&=2\max\limits_{t\geq t_1+\epsilon_n}\max\limits_{J\subset\cM\atop |J|=s_0}\{\Pi_{J}{\bC}_0^{(1)}(t)^\top\Pi_{J}\bC_0^{(3)}(t)-\Pi_{J}\tilde{\bC}_0^{(1)}(t_1)^\top\Pi_{J}\bC_0^{(3)}(t_1)\}\\
			&\leq_{(1)} 4\max\limits_{t\in[q_0,1-q_0]}\max\limits_{J\subset\cM\atop |J|=s_0}|\{\Pi_{J}{\bC}_0^{(1)}(t)^\top\Pi_{J}\bC_0^{(3)}(t)|\\
			&\leq_{(2)} 4\max\limits_{t\in[q_0,1-q_0]}\max\limits_{J\subset\cM\atop |J|=s_0}\|\{\Pi_{J}{\bC}_0^{(1)}(t)\|_2\|\Pi_{J}\bC_0^{(3)}(t)\|_2\\
			&\leq_{(3)} 4\max\limits_{t\in[q_0,1-q_0]}\|{\bC}_0^{(1)}(t)\|_{(s_0,2)}\times\max\limits_{t\in[q_0,1-q_0]}\|\bC_0^{(3)}(t)\|_{(s_0,2)}\\
			&\leq_{(4)} Cs_0^{1/2}M\sqrt{\log(pn)}\times s_0^{1/2}\sqrt{\log(pn)}s^2\|\bDelta\|^2_{(s_0,2)}\\
			&\leq_{(5)} C_9(s_0,M,\tilde{\btau}){s^2\log(pn)}\|\bDelta\|^2_{(s_0,2)}.
		\end{array}
	\end{equation}
	Similarly, for $A_{1.10}-A_{1.13}$, by (\ref{inequality: bias term for alpha=0}), (\ref{inequality: C0-3}), (\ref{inequality: C0-4}), and the Cauchy-Swartz inequality, we can prove that:
	\begin{equation*}
		\begin{array}{ll}
			A_{1.10}\leq C_{10}(s_0,M,\tilde{\btau}){s\log(pn)}\|\bDelta\|_{(s_0,2)},\\
			A_{1.11}\leq C_{11}(s_0,M,\tilde{\btau})SNR(0,\tilde{\btau})s^3\log(pn)\|\bDelta\|^3_{(s_0,2)},\\
			A_{1.12}\leq C_{12}(s_0,M,\tilde{\btau})SNR(0,\tilde{\btau}){s^2\log(pn)}\|\bDelta\|^2_{(s_0,2)},\\
			A_{1.13}\leq C_{13}(s_0,M,\tilde{\btau})s^3\log(pn)\|\bDelta\|^3_{(s_0,2)}.
		\end{array}
	\end{equation*}
	Note that by {Assumption F} and the assumption that $\|\bDelta\|_1\leq C_{\bDelta}$, we have $s^2\|\bDelta\|^2_{(s_0,2)}=O(1)$, $s^3\|\bDelta\|^3_{(s_0,2)}=O(1)$, and $s\|\bDelta\|_{(s_0,2)}=O(1)$. Hence, for the above results, by Assumption E.2,  up to some constants, the upper bounds of $A_{1.1}$ dominates the others. Hence, if $\epsilon_{n}$ satisfies: 
	\begin{equation}\label{inequality: step-1 change point estimation for alpha=0}
		\epsilon_{n}\geq C(s_0,M,\tilde{\btau})\dfrac{\log(pn)}{nSNR^2(0,\tilde{\btau})\|\bDelta\|^2_{(s_0,2)}},
	\end{equation}
	for some big enough $C>0$, w.p.a.1, we have 
	\begin{equation*}
		A_{1.1}+A_{1.2}+A_{1.3}+A_{1.4}+A_{1.7}+A_{1.8}+A_{1.9}+A_{1.10}+A_{1.11}+A_{1.12}+A_{1.13}\leq 0.
	\end{equation*}
	{\bf{Step~2:}} We aim to prove that $\P(A_{1.5}\leq 0)\rightarrow 1$.
	With a very similar proof procedure as  (\ref{inequality: A.1.4}) - (\ref{inequality: A.1.4.1-3}) in Section \ref{sec: proof of change point estimationfor alpha=1}, if  $\epsilon_{n}$ satisfies:
	\begin{equation}\label{inequality: step-2 change point estimation for alpha=0}
		\epsilon_{n}\geq C_5(s_0,M,\tilde{\btau})\dfrac{\log(pn)}{nSNR^2(0,\tilde{\btau})\|\bDelta\|^2_{(s_0,2)}},
	\end{equation}
	for some big enough $C_5>0$, then, w.p.a.1, we have $A_{1.5}\leq 0$.\\
	{\bf{Step~3:}} We aim to prove that $\P(A_{1.6}\leq 0)\rightarrow 1$.
	Note that with a very similar proof procedure as  (\ref{inequality: A.1.4}) - (\ref{inequality: A.1.4.1-3}) in Section \ref{sec: proof of change point estimationfor alpha=1}, if  $\epsilon_{n}$ satisfies:
	\begin{equation}\label{inequality: step-3 change point estimation for alpha=0}
		{\epsilon_n=C_6(s_0,M,\tilde{\btau})\dfrac{s^2\log(pn)}{n}}:=C_6(s_0,M,\tilde{\btau})\dfrac{\log(pn)}{n\|\bDelta\|^2_{(s_0,2)}}s^2\|\bDelta\|^2_{(s_0,2)},
	\end{equation}
	for some big enough $C_6>0$, then, w.p.a.1, we have $A_{1.6}\leq 0$.\\
	{\bf{Step~4:}} We aim to prove that $\P(A_{1.14}\leq 0)\rightarrow 1$. Using similar analysis as in (\ref{equation: I alpha=0-1}) - (\ref{equation: I alpha=0-2}) and  with a very similar proof procedure but some tedious modifications of   (\ref{inequality: A.1.4}) - (\ref{inequality: A.1.4.1-3}) in Section \ref{sec: proof of change point estimationfor alpha=1}, if  $\epsilon_{n}$ satisfies:
	\begin{equation}\label{inequality: step-4 change point estimation for alpha=0}
		\epsilon_{n}\geq C_{14}(s_0,M,\tilde{\btau})\dfrac{\log(pn)}{nSNR^2(0,\tilde{\btau})\|\bDelta\|^2_{(s_0,2)}}s^4\|\bDelta\|^4_{(s_0,2)},
	\end{equation}
	for some big enough $C_{14}>0$,then, w.p.a.1, we have $A_{1.14}\leq 0$.\\
	{\bf{Step~5:}} We aim to prove $\P(A_{1.15}\leq 0)\rightarrow 1$.
	Using some tedious modifications of Lemma \ref{lemma: upper bound for the empirical process of alpha=0 under H0}, with a very similar proof procedure as  (\ref{inequality: A.1.4}) - (\ref{inequality: A.1.4.1-3}) in Section \ref{sec: proof of change point estimationfor alpha=1}, if $\epsilon_{n}$ satisfies:
	\begin{equation}\label{inequality: step-5 change point estimation for alpha=0}
		\epsilon_{n}\geq C_{15}(s_0,M,\tilde{\btau})\dfrac{\log(pn)}{nSNR^2(0,\tilde{\btau})\|\bDelta\|^2_{(s_0,2)}}s^2\|\bDelta\|^2_{(s_0,2)},
	\end{equation}
	for some big enough $C_{15}>0$, then w.p.a.1, we have $A_{1.15}\leq 0$.
	
	Lastly, considering (\ref{inequality: step-1 change point estimation for alpha=0}) - (\ref{inequality: step-5 change point estimation for alpha=0}), if $\epsilon_{n}$ satisfies:
	\begin{equation}\label{inequality: step-6 change point estimation for alpha=0}
		\epsilon_{n}\geq C^*(s_0,M,\tilde{\btau})\dfrac{\log(pn)}{nSNR^2(0,\tilde{\btau})\|\bDelta\|^2_{(s_0,2)}},
	\end{equation}
	for some big enough $C^*>0$, we have $\P(A_{1.1}+\cdots+A_{1.15}\leq 0)\rightarrow 1$,  which yields $\P(A_1)\rightarrow 1$. Similarly, we can prove $\P(A_2)\rightarrow 1$, which finishes the proof of Theorem 
	\ref{theorem: cpt estimation results} with $\alpha=0$.
\end{proof}
\subsubsection{Change point estimation for $\alpha\in(0,1)$}\label{sec: proof of change point estimationfor alpha}
\begin{proof}
	For $\alpha\in(0,1)$, recall $\bZ(\bX_i,Y_i;\tilde{\btau},\hat{\bb},\hat{\bbeta}):=(1-\alpha)\dfrac{1}{K}\sum\limits_{k=1}^{K}\bX_i \big(\mathbf{1}\{Y_i-\hat{b}_k-\bX_i^\top\hat{\bbeta}\leq 0\}-\tau_k\big)-\alpha\bX_i(Y_i-\bX_i^\top\hat{\bbeta})$ as the weighted score function.  For $\alpha\in(0,1)$ and  each $t\in[q_0,1-q_0]$, define $\tilde{\bC}_\alpha(t)=(\tilde{C}_{\alpha1}(t),\ldots,\tilde{C}_{\alpha p}(t))^\top$ with
	\begin{equation}
		\tilde{\bC}_\alpha(t)=\dfrac{1}{\sqrt{n}}\big(\sum\limits_{i=1}^{\lfloor nt\rfloor}\bZ(\bX_i,Y_i;\tilde{\btau},\hat{\bb},\hat{\bbeta})-\dfrac{\lfloor nt \rfloor }{n}\sum\limits_{i=1}^{n}\bZ(\bX_i,Y_i;\tilde{\btau},\hat{\bb},\hat{\bbeta})\big).
	\end{equation}
	Note that there is no any variance estimator in $\tilde{\bC}_\alpha(t)$. Recall $\hat{t}_{\alpha}:=\argmax\limits_{t\in[q_0,1-q_0]}\|\tilde{\bC}_\alpha(t)\|_{(s_0,2)}$. To prove Theorem \ref{theorem: cpt estimation results} with $\alpha\in(0,1)$, we need to prove that as $n,p\rightarrow\infty$, by choosing a large enough constant $C(s_0,M,q_0,\alpha)$ in $\epsilon_{n}$ (which will be given in (\ref{inequality: step-6 change point estimation for alpha})), we have
	\begin{equation}\label{inequality: theorem 3.2 alpha}
		\P\big(|\hat{t}_{\alpha}-t_1|\geq \epsilon_n\big) \rightarrow 0.
	\end{equation}
	Similar to Sections \ref{sec: proof of change point estimationfor alpha=1} and \ref{sec: proof of change point estimationfor alpha=0}, we have to prove $\P(A_1)\rightarrow 1$ and $\P(A_2)\rightarrow 1$, where
	\begin{equation}\label{inequality: A1+A2 alpha}
		\begin{array}{ll}
			A_1=\max\limits_{t\geq t_1+\epsilon_n}\|\tilde{\bC}_\alpha(t)\|_{(s_0,2)}- \|\tilde{\bC}_\alpha(t_1)\|_{(s_0,2)}\leq 0,\\
			A_2=\max\limits_{t\leq t_1-\epsilon_n}\|\tilde{\bC}_\alpha(t)\|_{(s_0,2)}- \| \tilde{\bC}_\alpha(t_1)\|_{(s_0,2)}\leq 0.
		\end{array}
	\end{equation}
	By the symmetry, we only consider $\P(A_1)\rightarrow 1$. Similar to the previous two sections, define the two events $\cH_1$ and $\cH_2$:
	\begin{equation}\label{equation: H2 alpha}
		\begin{array}{ll}
			\cH_1=\big\{\max\limits_{t\geq t_1+\epsilon_n}\|\tilde{\bC}_\alpha(t)\|_{(s_0,2)}:=\max\limits_{t\geq t_1+\epsilon_n}\max\limits_{J\subset\{1,\ldots,p\}\atop |J|=s_0}\|\Pi_{J}\tilde{\bC}_\alpha(t)\|_{2}=\max\limits_{t\geq t_1+\epsilon_n}\max\limits_{J\subset\cM\atop |J|=s_0}\|\Pi_{J}\tilde{\bC}_\alpha(t)\|_{2}\big\},\\
			\cH_2=\big\{\|\tilde{\bC}_\alpha(t_1)\|_{(s_0,2)}:=\max\limits_{J\subset\{1,\ldots,p\}\atop |J|=s_0}\|\Pi_{J}\tilde{\bC}_\alpha(t_1)\|_{2}=\max\limits_{J\subset\cM\atop |J|=s_0}\|\Pi_{J}\tilde{\bC}_\alpha(t_1)\|_{2}\big\}.
		\end{array}
	\end{equation}
	Similar to the proof of Lemma \ref{lemma: maximum at Pi}, we can prove  $\P(\cH_1\cap\cH_2)\rightarrow 1$. 
	\noindent Now, under $\cH_1\cap\cH_2$, we have: 
	\begin{equation*}
		\begin{array}{ll}
			\P(A_1)&= \P\big(\max\limits_{t\geq t_1+\epsilon_n}\|\tilde{\bC}_\alpha(t)\|_{(s_0,2)}- \|\tilde{\bC}_\alpha(t_1)\|_{(s_0,2)}\leq 0\big)\\
			&=\P\big(\max\limits_{t\geq t_1+\epsilon_n}\max\limits_{J\subset\cM\atop |J|=s_0}\|\Pi_{J}\tilde{\bC}_\alpha(t)\|_{2}- \max\limits_{J\subset\cM\atop |J|=s_0}\|\Pi_{J}\tilde{\bC}_\alpha(t_1)\|_{2}\leq 0\big)\\
			&=\P\big(\max\limits_{t\geq t_1+\epsilon_n}\max\limits_{J\subset\cM\atop |J|=s_0}\|\Pi_{J}\tilde{\bC}_\alpha(t)\|^2- \max\limits_{J\subset\cM\atop |J|=s_0}\|\Pi_{J}\tilde{\bC}_\alpha(t_1)\|^{2}\leq 0\big).\\
		\end{array}
	\end{equation*}
	Recall $\sigma^2(\alpha,\tilde{\btau}):=\text{Var}[ (1-\alpha)e_i(\tilde{\btau})-\alpha\epsilon_i]$ and $\bC_{\alpha}(t)$ defined in (\ref{equation: C-alpha-t}). Then, under $\Hb_1$, we have the following decomposition
	\begin{equation}\label{equation: tilde-C1 for estimation alpha}
		\begin{array}{ll}
			\tilde{\bC}_\alpha(t)&=\sigma(\alpha,\tilde{\btau})\times\bC_{\alpha}(t)\\
			&=\sigma(\alpha,\tilde{\btau})\times \big(\tilde{\bC}^{I}_{\alpha}(t)-SNR(\alpha,\tilde{\btau})\times\bdelta(t)-SNR(\alpha,\tilde{\btau})\times\bR(t)\\
			&+	(1-\alpha)\bC_{0}^{(3)}(t)+(1-\alpha)\bC_{0}^{(4)}(t)\big),
		\end{array}
	\end{equation}
	where the second equation comes from the decomposition in (\ref{equation: decomposition of C-alpha-t}).  By the fact that $\max a_i-\max b_i \leq \max (a_i-b_i)$ and $\max (a_i+b_i)\leq \max a_i +\max b_i$ for any $\{a_i\}$ and $\{b_i\}$, we have:
	\begin{equation*}
		\begin{array}{ll}
			\Big\{\max\limits_{t\geq t_1+\epsilon_n}\max\limits_{J\subset\cM\atop |J|=s_0}\|\Pi_{J}\tilde{\bC}_\alpha(t)\|^2- \max\limits_{J\subset\cM\atop |J|=s_0}\|\Pi_{J}\tilde{\bC}_\alpha(t_1)\|^{2}\leq 0\Big\}\\
			\subset \Big\{\sigma^2(\alpha,\tilde{\btau})\times\max\limits_{t\geq t_1+\epsilon_n}\max\limits_{J\subset\cM\atop |J|=s_0}\Big(\|\Pi_{J}\big(\tilde{\bC}^{I}_{\alpha}(t)-SNR(\alpha,\tilde{\btau})\times\bdelta(t)-SNR(\alpha,\tilde{\btau})\times\bR(t)\\
			+	(1-\alpha)\bC_{0}^{(3)}(t)+(1-\alpha)\bC_{0}^{(4)}(t)\big)\|^2\\
			\quad\quad\quad-\|\Pi_{J}\big(\tilde{\bC}^{I}_{\alpha}(t_1)-SNR(\alpha,\tilde{\btau})\times\bdelta(t_1)-SNR(\alpha,\tilde{\btau})\times\bR(t_1)\\
			+	(1-\alpha)\bC_{0}^{(3)}(t_1)+(1-\alpha)\bC_{0}^{(4)}(t_1)\big)\|^2\Big)\leq 0\Big\}.\\
		\end{array}
	\end{equation*}
	Note that similar to Section \ref{sec: proof of change point estimationfor alpha=0}, for the above inequality, we can decompose it into fifteen parts. Moreover,  using the obtained bounds in Sections \ref{sec: proof of change point estimationfor alpha=1} and \ref{sec: proof of change point estimationfor alpha=0},  if $\epsilon_{n}$ satisfies:
	\begin{equation}\label{inequality: step-6 change point estimation for alpha}
		\epsilon_{n}\geq C^*(s_0,M,\tilde{\btau},\alpha)\dfrac{\log(pn)}{nSNR^2(\alpha,\tilde{\btau})\|\bDelta\|^2_{(s_0,2)}},
	\end{equation}
	for some big enough $C^*>0$, it is not hard to prove that $\P(A_1)\rightarrow 1$ and $\P(A_2)\rightarrow 1$, which finishes the proof of Theorem \ref{theorem: cpt estimation results} with $\alpha\in(0,1)$.
	
\end{proof}

\subsection{Proof of Theorem \ref{theorem: variance estimator under H1}}
Let $r_{\alpha}(n)=\sqrt{s\log(pn)/n}$ if $\alpha=1$ and $r_{\alpha}(n)=s\sqrt{\dfrac{\log(pn)}{n}}\vee s^{\frac{1}{2}}(\dfrac{\log(pn)}{n})^{\frac{3}{8}}$ if $\alpha\in[0,1)$. In this section, we aim to prove the consistency of $\hat{\sigma}^2(\alpha,\tilde{\btau})$ in the sense that 
\begin{equation}\label{equation: proof variance estimation under H1}
	|\hat{\sigma}^2(\alpha,\tilde{\btau})-{\sigma}^2(\alpha,\tilde{\btau})|=O_p(r_{\alpha}(n)).
\end{equation}
We consider the proof in two cases:\\
{\bf Case 1: the signal jump satisfies $SNR(\alpha,\tilde{\btau})\|\bDelta\|_{(s_0,2)}\gg \sqrt{\log(pn)/n}$}. In this case, by Theorem \ref{theorem: cpt estimation results},  w.p.a.1, we have:
\begin{equation*}
	\big|n\hat{t}_{\alpha}-nt_1\big|=o(n).
\end{equation*}
Recall  $n_-:=\{i: i\leq nh\hat{t}_{\alpha}\}$ and $n_+:=\{i:   \hat{t}_{\alpha}n+(1-h)(1-\hat{t}_{\alpha})n   \leq i\leq n \}$ for some $0<h<1$. Hence, by Theorem \ref{theorem: cpt estimation results}, w.p.a.1, the samples in $n_-$ are before the true change point $t_1$ and those in $n_+$ are after $t_1$. Hence, we can use a very similar proof technique as in Section \ref{sec: proof of variance estimation under H0} to yield (\ref{equation: proof variance estimation under H1}).\\
{\bf Case 2: the signal jump satisfies $SNR(\alpha,\tilde{\btau})\|\bDelta\|_{(s_0,2)}=O(\sqrt{\log(pn)/n})$}. In this case, the change point estimator $\hat{t}_\alpha$ can be an arbitrary number which satisfies $\hat{t}_\alpha\in[q_0,1-q_0]$. Note that in this case, the signal jump $\bbeta^{(1)}-\bbeta^{(2)}$ is very small in the sense that:
\begin{equation*}
	\|\bbeta^{(1)}-\bbeta^{(2)}\|_1=O(s\sqrt{\dfrac{\log(pn)}{n}}),~~\|\bbeta^{(1)}-\bbeta^{(2)}\|_2=O(\sqrt{s\dfrac{\log(pn)}{n}}).
\end{equation*}
In this case, using some modifications of Theorem \ref{theorem: variance estimator under H0} in Section \ref{sec: proof of variance estimation under H0}, we can  still prove
\begin{equation}\label{equation: proof variance estimation under H1-2}
	|\hat{\sigma}^2(\alpha,\tilde{\btau})-{\sigma}^2(\alpha,\tilde{\btau})|=O_p(r_{\alpha}(n)).
\end{equation}
Since the modifications are lengthy, to save space, we omit the details.

\subsection{Proof of Theorem \ref{theorem: power control for individual test} }
Throughout the following proofs, we assume $\|\bbeta^{(2)}-\bbeta^{(1)}\|_{\infty}\geq \sqrt{\log(p)/n}$ and $\|\bbeta^{(2)}-\bbeta^{(1)}\|_{2}\geq \sqrt{s\log(p)/n}$, as well as $\|\bbeta^{(2)}-\bbeta^{(1)}\|_{1}\geq s\sqrt{\log(p)/n}$.
Next, we give the power results for $\alpha=1$, $\alpha=0$ and $\alpha\in(0,1)$, respectively. For simplicity, we will omit the subscript $\alpha$ whenever  needed.
\subsubsection{Power analysis for $\alpha=1$}\label{sec: power for alpha=1}
Firstly, we consider the oracle case that assumes the variance is known by letting $\hat{\sigma}^2(\alpha,\tilde{\btau})=\sigma^2$, where $\sigma^2:=\text{Var}[\epsilon]$. In addition, for the case of  $\alpha=1$, we have $SNR(\alpha,\tilde{\btau}):=SNR(1,\tilde{\btau})=1/\sigma$, where $\sigma^2=\text{Var}(\epsilon)$. Without loss of generality, we assume $\sigma^2=1$.
The proof of Theorem \ref{theorem: power control for individual test} proceeds in two steps. In Step 1, we obtain the upper bound of $c_{T_{1}^b}(1-\gamma)$, where $c_{T_{1}^b}(1-\gamma)$ is  the $1-\gamma$ th quantile of $T^{b}_{1}$, which is defined as
\begin{equation}
	c_{T_{1}^b}(1-\gamma):=\inf\big\{t:\P(T_{1}^b\leq t)\geq 1-\gamma  \big\}.
\end{equation}
In Step 2, using the obtained upper bound, we get the lower bound of $\P\big(T_{1}\geq c_{T_{1}^b}(1-\gamma)\big)$ and prove
\begin{equation}
	\P\big(T_{1}\geq c_{T_{1}^b}(1-\gamma)\big)\rightarrow1, ~\text{as}~n,p\rightarrow\infty.
\end{equation}
Note that $\{	\Psi_{\gamma,1}=1\}\Leftrightarrow\{T_{1}\geq \hat{c}_{T_{\alpha}^b}(1-\gamma) \}$, where 
\begin{equation}
	\hat{c}_{T_{1}^b}(1-\gamma):=\inf\big\{t:
	(B+1)^{-1}\sum_{b=1}^{B}\mathbf{1}\{T_{1}^{b}\leq t\}\geq 1-\gamma\big\}.
\end{equation}
Finally, using the fact that $\hat{c}_{T_{1}^b}(1-\gamma)$ is the estimation for $c_{T_{1}^b}(1-\gamma)$ based on the bootstrap samples, we complete the proof. Now, we consider the two steps in detail.  \\
{\bf{Step~1}}: By the definition of $T_1^b$, we have: $T_1^b=\max_{q_0\leq t\leq1-q_0}\|\bC_1^b(t)\|_{(s_0,2)}$, where
\begin{equation*}
	\begin{array}{ll}
		\bC_1^b(t):=\dfrac{1}{\sqrt{n}v(1,\tilde{\btau})}\Big(\sum\limits_{i=1}^{\floor{nt}}\bX_i\epsilon^b_i-\dfrac{\floor{nt}}{n}\sum\limits_{i=1}^n\bX_i\epsilon^b_i\Big),
	\end{array}
\end{equation*}
where $\epsilon_i^b$ are i.i.d $N(0,1)$,  $v(1,\tilde{\btau}):=\text{Var}[e_i^b]=1$. Our next goal is to obtain an upper bound of $c_{T_{1}^b}(1-\gamma)$. To this end, for any $1\leq i\leq n$, $1\leq j\leq p$, and $\floor{nq_0}\leq k\leq n-\floor{nq_0}$, we define $W^b_{ijk}=X_{ij}\epsilon^b_ia_{ik}$, where $a_{ik}:=\mathbf{1}\{i\leq k\}-k/n$. Using the above notations, for $T_1$, we have:
\begin{equation*}
	\begin{array}{ll}
		T_1=\max_{q_0\leq t\leq1-q_0}\|\bC_1^b(t)\|_{(s_0,2)}\leq s_0^{1/2}\max_{q_0\leq t\leq1-q_0}\|\bC_1^b(t)\|_{\infty}\\
		=s_0^{1/2}\dfrac{1}{\sqrt{n}}\underbrace{\max_{1\leq j\leq p,\atop\floor{nq_0}\leq k\leq n-\floor{nq_0}}|\sum_{i=1}^nW^b_{ijk}|}_{Z}.
	\end{array}
\end{equation*}
Hence, according to the above inequality, let $c_{Z}(1-\gamma)$ be the $1-\gamma$-th quantile of $Z$, then we have:
\begin{equation}\label{inequality: upper bound of 1-gamma quantile of T1-1}
	c_{T_{1}^b}(1-\gamma)\leq s_0^{1/2}\dfrac{1}{\sqrt{n}}c_{Z}(1-\gamma).
\end{equation} 
Next, we obtain an upper bound of $c_{Z}(1-\gamma)$. The main technique is to use Lemma \ref{lemma: maximum inequality} and Lemma \ref{lemma: concentration inequality for maximum}. Let
$M=\max_{i,j,k}|W_{ijk}|$ and $\sigma_*^2=\max_{jk}\sum_{i}\E[W_{ijk}^2]$. Then, we have
\begin{equation*}
	\sigma_*^2=\max_{jk}\sum_{i}\E[W_{ijk}^2]=\max_{jk}\E[X_{ij}\epsilon^b_ia_{ik}]^2\leq nM^2(1-q_0)^2\leq C_1(M,q_0)n
\end{equation*}
where the last inequality uses the fact that $|X_{ij}|\leq M$ and $|a_{ik}|\leq 1-q_0$. For $\E[M^2]$,  we have:
\begin{equation*}
	\E[M^2]=\E[\max_{ijk}|X_{ij}\epsilon^b_ia_{ik}|^2]\leq M^2(1-q_0)^2\E\big[\max_{i}|\epsilon^b_i|^2\big]\leq C_2(M,q_0)\log(n),
\end{equation*}
where the last inequality comes from Example 3.5.6 in \cite{embrechts2013modelling}. Let $n'=n-2\floor {nq_0}$. Using the above results, by Lemma \ref{lemma: maximum inequality}, we have:
\begin{equation*}
	\E[Z]\leq C(\sigma_*\sqrt{\log (pn')}+\sqrt{\E[M^2]}\log {pn'})\leq C_3(M,q_0)\sqrt{n\log (pn)}.
\end{equation*}
Note that $X_{ij}$ and $e_i^b$ are all sub-Gaussian random variables, which implies $\|X_{ij}\epsilon^b_{i}\|_{\psi_1}$ exists.  Hence, we have:
\begin{equation*}
	\|M\|_{\psi_{1}}:=\|\max_{i,j,k}|X_{ij}\epsilon^b_ia_{ik}|\|_{\psi_{1}}\leq C\log(pnn'+1)\max_{i,j,k}\|X_{ij}\epsilon^b_ia_{ik}\|_{\psi_1}\leq C_4(M,q_0)\log(pn).
\end{equation*}
By Lemma \ref{lemma: concentration inequality for maximum}, taking $\eta=1$ and $\beta=1$, we have:
\begin{equation*}
	\P(Z\geq 2\E [Z]+t)\leq \exp(-\dfrac{t^2}{3C_1(M,q_0)n})+ 3\exp\Big(-\dfrac{t}{C_4(M,q_0)\log(pn)}\Big).
\end{equation*}
Taking $t=2(t_1\vee t_2)$, where $t_1$ and $t_2$ satisfies:
\begin{equation*}
	-\dfrac{t_1^2}{3C_1(M,q_0)n}=\log(\gamma/2) ~~\text{and}~~-\dfrac{t_2}{C_4(M,q_0)\log(pn)}=\log(\gamma/6), 
\end{equation*}
we have:
\begin{equation*}
	\P(Z\geq 2\E [Z]+t)\leq \gamma.
\end{equation*}
By noting  that $t:=2(t_1\vee t_2)\leq C_{5}(M,q_0)\sqrt{n\log(1/\gamma)}$ and $\E[Z]\leq C_3(M,q_0)\sqrt{n\log (pn)}$, we have:
\begin{equation*}
	\begin{array}{ll}
		c_{Z}(1-\gamma)=2\E[Z]+t\\
		\leq 2C_3(M,q_0)\sqrt{n\log (pn)}+C_{5}(M,q_0)\sqrt{n\log(1/\gamma)}\leq C_{6}(\sqrt{n\log (pn)}+\sqrt{n\log(1/\gamma)}).
	\end{array}
\end{equation*}
Lastly, considering (\ref{inequality: upper bound of 1-gamma quantile of T1-1}), we have:
\begin{equation}
	c_{T_{1}^b}(1-\gamma)\leq C_{6}(M,q_0)s_0^{1/2}(\sqrt{\log (pn)}+\sqrt{\log(1/\gamma)}),
\end{equation}
where $C_{6}(M,q_0)$ is some universal constant not depending on $n$ or $p$. \\
\textbf{Step~2}: In this step, we aim to prove that $\P\big(T_{1}\geq c_{T_{1}^b}(1-\gamma)\big)\rightarrow1$ as $n,p\rightarrow\infty$. Note that in Step~1, we have obtained the upper bound of $c_{T_{1}^b}(1-\gamma)$. Hence, it is sufficient to prove that $H_1\rightarrow1$ as $n,p\rightarrow\infty$, where 
\begin{equation}\label{equation: H_1}
	H_1:=\P\big(T_{1}\geq C_{6}(M,q_0)s_0^{1/2}(\sqrt{\log (pn)}+\sqrt{\log(1/\gamma)})\big).
\end{equation}
To prove $H_1\rightarrow 1$, we need the decomposition of $T_1$ under $\Hb_1$. Recall the decomposition of $\bC_1(t)$ defined in (\ref{equation: C1 for alpha=0}). Let the signal jump be
\begin{equation}\label{equation: delta-t alpha=1}
	\bdelta(t):=\left\{
	\begin{array}{ll}
		\sqrt{n}\dfrac{\floor{nt}}{n}\dfrac{n-\floor{nt_1}}{n}\bSigma\big(\bbeta^{(1)}-\bbeta^{(2)}\big),~~\text{if}~~t\leq t_1,\\\\
		\sqrt{n}\dfrac{\floor{nt_1}}{n}\dfrac{n-\floor{nt}}{n}\bSigma\big(\bbeta^{(1)}-\bbeta^{(2)}\big),~~\text{if}~~t> t_1.
	\end{array} \right.
\end{equation}
Then, under $\Hb_1$, for $\alpha=1$, we have the following decomposition:
\begin{equation*}
	\bC_1(t)=\bC_1^{I}(t)+SNR(1,\tilde{\btau})\times{\bdelta(t)}+{SNR(1,\tilde{\btau})\times\bR(t)	},
\end{equation*}
where $\bC_1(t)$ and $\bR(t)$ are defined as 
\begin{equation}\label{equation: C11+C12 under H1}
	\small
	\begin{array}{ll}
		\bC^{I}_{1}(t):=\dfrac{1}{\sqrt{n}\hat{\sigma}(\alpha,\tilde{\btau})}\big(\sum\limits_{i=1}^{\floor{nt}}\bX_i\epsilon_i-\dfrac{\floor{nt}}{n}\sum\limits_{i=1}^n\bX_i\epsilon_i\big),
		~\bR(t):=\bR^{I}(t)\mathbf{1}\{i\leq \floor{nt_1}\}+\bR^{II}(t)\mathbf{1}\{i> \floor{nt_1}\},\\
	\end{array}
\end{equation}
with $\bR^{I}(t)$ and $\bR^{II}(t)$ being defined as
\begin{equation*}
	\begin{array}{ll}
		\bR^{I}(t):=	\dfrac{\floor {nt}(n-\floor{nt})}{n^{3/2}}
		\big(\hat{\bSigma}(0:t)-\bSigma\big)\big(\bbeta^{(1)}-\hat{\bbeta}\big)\\-\dfrac{\floor{nt}(\floor{nt_1}-\floor{nt})}{n^{3/2}}\big(\hat{\bSigma}(t:t_1)-\bSigma\big)\big(\bbeta^{(1)}-\hat{\bbeta}\big)\\
		\quad\quad\quad\quad\quad-\dfrac{\floor{nt}(n-\floor{nt_1})}{n^{3/2}}\big(\hat{\bSigma}(t_1:1)-\bSigma\big)\big(\bbeta^{(2)}-\hat{\bbeta}\big),
	\end{array}
\end{equation*}
and 
\begin{equation*}
	\begin{array}{ll}
		\bR^{II}(t):=	\dfrac{\floor {nt_1}(n-\floor{nt})}{n^{3/2}}
		\big(\hat{\bSigma}(0:t_1)-\bSigma\big)\big(\bbeta^{(1)}-\hat{\bbeta}\big)
		\\-\dfrac{(n-\floor{nt})(\floor{nt}-\floor{nt_1})}{n^{3/2}}\big(\hat{\bSigma}(t_1:t)-\bSigma\big)\big(\bbeta^{(2)}-\hat{\bbeta}\big)\\
		\qquad\qquad\qquad-\dfrac{\floor{nt}(n-\floor{nt})}{n^{3/2}}\big(\hat{\bSigma}(t:1)-\bSigma\big)\big(\bbeta^{(2)}-\hat{\bbeta}\big).
	\end{array}
\end{equation*}
To prove $H_1\rightarrow1$, we need the analysis of $\bC^{I}_{1}(t)$, $\bdelta(t)$, and $\bR(t)$, respectively. By definition, for $\bdelta(t)$, we have: $t_1=\argmax_{q_0\leq t\leq 1-q_0}\|\bdelta(t)\|_{(s_0,2)}$. In other words, $\|\bdelta(t)\|_{(s_0,2)}$ obtains its maximum value at the true change point location. For $\bC^{I}_1(t)$, by  Lemma \ref{lemma: exponential inequality for partial sum process} and the fact that $\|\bv\|_{(s_0,2)}\leq s_0^{1/2}\|\bv\|_{\infty}$ for any $\bv\in \RR^p$, we have $\max_{q_0\leq t\leq 1-q_0}\|\bC^{I}_1(t)\|_{(s_0,2)}\leq Cs_0^{1/2}M\sqrt{\log(pn)}$ for some constant $C>0$. As for $\bR(t)$, using the triangle inequality, we have:
\begin{equation*}
	\max_{q_0\leq t\leq 1-q_0}\|\bR(t)\|_{(s_0,2)}\leq \max_{q_0\leq t\leq t_1}\|\bR^{I}(t)\|_{(s_0,2)}+\max_{t_1\leq t\leq 1-q_0}\|\bR^{II}(t)\|_{(s_0,2)}.
\end{equation*}
For  $\max_{q_0\leq t\leq t_1}\|\bR^{I}(t)\|_{(s_0,2)}$, using Lemma \ref{lemma: concentration for covariance}, with probability at least $1-(pn)^{-C}$, we have
\begin{equation*}
	\max_{q_0\leq t\leq t_1}\|\bR^{I}(t)\|_{(s_0,2)}\leq C_1s_0^{1/2}\sqrt{\log p}\big(\big\|\bbeta^{(1)}-\hat{\bbeta}\big\|_1+\big\|\bbeta^{(2)}-\hat{\bbeta}\big\|_1\big).
\end{equation*}
Note that by Lemma \ref{lemma: basic inequality for lasso with alpha=0} and the fact that $\bbeta^*=t_1\bbeta^{(1)}+(1-t_1)\bbeta^{(2)}$, we have:
\begin{equation*}
	\begin{array}{ll}
		&\big\|\bbeta^{(1)}-\hat{\bbeta}\big\|_1\\
		&
		\quad\leq_{(1)} \big\|\bbeta^{(1)}-\bbeta^*\big\|_1+\big\|\bbeta^*-\hat{\bbeta}\big\|_1\\
		&\quad\leq_{(2)}  (1-t_1)\|\bbeta^{(1)}-\bbeta^{(2)}\|_1+C_1sM\sqrt{\dfrac{\log(pn)}{n}}\big(1+M\|\bbeta^{(2)}-\bbeta^{(1)}\|_1\big)\\
		&\quad\leq_{(3)} C_1\|\bbeta^{(1)}-\bbeta^{(2)}\|_1\\
		&\quad\leq_{(4)} C_1s\|\bbeta^{(1)}-\bbeta^{(2)}\|_{\infty}\\
		&\quad\leq_{(5)} C_1s\|\bSigma^{-1}\bSigma(\bbeta^{(1)}-\bbeta^{(2)})\|_{(s_0,2)}\\
		&\quad\leq_{(6)} C_2s\|\bSigma(\bbeta^{(1)}-\bbeta^{(2)})\|_{(s_0,2)},\\
	\end{array}
\end{equation*}
where $(2)$ comes from Lemma \ref{lemma: basic inequality for lasso with alpha=0}, $(3)$ comes from the assumption that {$sM^2\sqrt{\log(pn)/n}=o(1)$}, $(6)$ comes from {\bf{Assumption A}}. Similarly, we can prove $\big\|\bbeta^{(2)}-\hat{\bbeta}\big\|_1=O_p(s\|\bSigma(\bbeta^{(1)}-\bbeta^{(2)})\|_{(s_0,2)})$. Combining this result, we can prove that 
\begin{equation*}
	\begin{array}{ll}
		\max\limits_{q_0\leq t\leq 1-q_0}\|\bR(t)\|_{(s_0,2)}
		&\leq 2\max\big(\max\limits_{q_0\leq t\leq t_1}\|\bR^{I}(t)\|_{(s_0,2)},\max\limits_{t_1\leq t\leq 1-q_0}\|\bR^{II}(t)\|_{(s_0,2)}\big)\\
		&\leq Cs_0^{1/2}s\sqrt{\log p}\|\bSigma(\bbeta^{(1)}-\bbeta^{(2)})\|_{(s_0,2)}. \\
	\end{array}
\end{equation*}
Using the above bounds of $\bC^{I}_{1}(t)$, $\bdelta(t)$, and $\bR(t)$, and by the triangle inequality, we have:
\begin{equation}
	\begin{array}{ll}
		T_1&=\max\limits_{q_0\leq t\leq 1-q_0}\|\bC_1(t)\|_{(s_0,2)}\\
		&\geq SNR(1,\tilde{\btau})\times\max\limits_{q_0\leq t\leq 1-q_0}\|\bdelta(t)\|_{(s_0,2)}-\max\limits_{q_0\leq t\leq 1-q_0}\|\bC^{I}_1(t)\|_{(s_0,2)}-\max\limits_{q_0\leq t\leq 1-q_0}\|\bR(t)\|_{(s_0,2)}\\
		&\geq \sqrt{n}\times SNR(1,\tilde{\btau})\times t_1(1-t_1)\times\|\bSigma(\bbeta^{(1)}-\bbeta^{(2)})\|_{(s_0,2)}
		-C_1s_0^{1/2}M\sqrt{\log(pn)}\\
		&\quad\quad-C_2s_0^{1/2}s\sqrt{\log p}\|\bSigma(\bbeta^{(1)}-\bbeta^{(2)})\|_{(s_0,2)}, \\
		&\geq \sqrt{n}\times SNR(1,\tilde{\btau})\times t_1(1-t_1)\times\|\bSigma(\bbeta^{(1)}-\bbeta^{(2)})\|_{(s_0,2)}
		(1-\epsilon_n)-C_1s_0^{1/2}M\sqrt{\log(pn)},\\
	\end{array}
\end{equation}
where $\epsilon_n:=(SNR(1,\tilde{\btau})\times t_1(1-t_1))^{-1}s^{1/2}_0s\sqrt{\log(p)/n}={O(s^{1/2}_0s\sqrt{\log(p)/n})}$. 
Recall $H_1$ as  defined in (\ref{equation: H_1}).  Hence, to prove $H_1\rightarrow1$, it is sufficient to prove $H_1'\rightarrow 1$, where 
\begin{equation*}
	\begin{array}{ll}
		H_1'=\P\Big(
		\sqrt{n}\times SNR(1,\tilde{\btau})\times t_1(1-t_1)\times\|\bSigma(\bbeta^{(1)}-\bbeta^{(2)})\|_{(s_0,2)}\\
		\geq 
		\dfrac{Cs^{1/2}_0M\big(\sqrt{\log(pn)}+\sqrt{\log(1/\gamma)}\big)}{1-\epsilon_n}
		\Big).
	\end{array}
\end{equation*}
Hence, by (\ref{inequality: theoretical signal strengh}), it is straightforward to see that $H_1'\rightarrow 1$ as $n,p\rightarrow\infty$, which finishes the proof. 
\begin{remark}
	Note that for $\alpha=1$, if we replace $\sigma^2(1,\tilde{\btau})$ by an estimator $\hat{\sigma}^2(\alpha,\tilde{\btau})$ which satisfies: $|\hat{\sigma}^2(\alpha,\tilde{\btau})-\sigma^2(1,\tilde{\btau})|=o_p(1)$, then under condition (\ref{inequality: theoretical signal strengh}), the power still converges to 1.
\end{remark}

\subsubsection{Power analysis for $\alpha=0$}\label{sec: power for alpha=0}
\begin{proof}
	Firstly, we assume $\hat{\sigma}^2(0,\tilde{\btau})=\sigma^2(0,\tilde{\btau})$ by assuming the variance is unknown, where $\sigma^2(\alpha,\tilde{\btau})=\sqrt{K^{-2}\sum_{k_1,k_2}\gamma_{k_1k_2}}$, with $\gamma_{k_1k_2}:=\min(\tau_{k_1},\tau_{k_2})-\tau_{k_1}\tau_{k_2}$
	In addition, for the case of  $\alpha=0$, we have
	\begin{equation*}
		SNR(0,\tilde{\btau})=\dfrac{\sum\limits_{k=1}^{K}f_{\epsilon}(b_{k}^{(0)})}{\sqrt{\sum_{k_1=1}^{K}\sum_{k_2=1}^{K}\gamma_{k_1k_2}}}.
	\end{equation*}
	Similar to Section \ref{sec: power for alpha=1}, the proof of Theorem\ref{theorem: power control for individual test} proceeds in four steps. In Step 1, we obtain the upper bound of $c_{T_{0}^b}(1-\gamma)$, where $c_{T_{0}^b}(1-\gamma)$ is  the $1-\gamma$ th quantile of $T^{b}_{0}$, which is defined as
	\begin{equation}
		c_{T_{0}^b}(1-\gamma):=\inf\big\{t:\P(T_{0}^b\leq t)\geq 1-\gamma  \big\}.
	\end{equation}
	In Steps 2-4, using the  upper bound, we get the lower bound of $\P\big(T_{0}\geq c_{T_{0}^b}(1-\gamma)\big)$ and prove
	\begin{equation}
		\P\big(T_{0}\geq c_{T_{0}^b}(1-\gamma)\big)\rightarrow1, ~\text{as}~n,p\rightarrow\infty.
	\end{equation}
	{\bf{Step~1}}: By the definition of $T_0^b$, we have: $T_0^b=\max_{q_0\leq t\leq1-q_0}\|\bC_0^b(t)\|_{(s_0,2)}$, where
	\begin{equation*}
		\bC_0^b(t):=\dfrac{1}{\sqrt{n}v(0,\tilde{\btau})}\Big(\sum\limits_{i=1}^{\floor{nt}}\bX_ie_i^{b}(\tilde{\btau})-\dfrac{\floor{nt}}{n}\sum\limits_{i=1}^n\bX_ie_i^{b}(\tilde{\btau})\Big),
	\end{equation*}
	where $e_i^{b}(\tilde{\btau})=K^{-1}\sum\limits_{k=1}^{K}e_i^{b}(\tau_k)$ with $e_i^{b}(\tau_k):=\mathbf{1}\{\epsilon_i^b\leq \Phi^{-1}(\tau_k)\}-\tau_k$, $\epsilon_i^b~ \text{~is ~i.i.d} ~N(0,1)$, 
	$v(0,\tilde{\btau}):=\text{Var}[e^{b}(\tilde{\btau})]$, and  $\Phi(x)$ is the CDF for the standard normal distribution. Note that $|e_i^{b}(\tilde{\btau})|\leq 1$ by definition. Hence, we can use very similar proof procedure as in Step~1 in Section \ref{sec: power for alpha=1} to obtain 
	\begin{equation}\label{equation: upper bounds of critical value for alpha=0}
		c_{T_{0}^b}(1-\gamma)\leq C(M,q_0,\tilde{\btau})s_0^{1/2}(\sqrt{\log (pn)}+\sqrt{\log(1/\gamma)}),
	\end{equation}
	where $C(M,q_0,\tilde{\btau})$ is some universal constant only depending on $M,q_0$ and $\tilde{\btau}$. \\
	\textbf{Step~2 Decomposition of $\bC_0(t)$}. In this step, we aim to prove that $\P\big(T_{0}\geq {c}_{T_{0}^b}(1-\gamma)\big)\rightarrow1$ as $n,p\rightarrow\infty$. Note that in Step~1, we have obtained the upper bound of $c_{T_{\alpha}^b}(1-\gamma)$. Hence, it is sufficient to prove that $H_1\rightarrow1$ as $n,p\rightarrow\infty$, where 
	\begin{equation}\label{equation: H_1 alpha=0}
		H_1:=\P\big(T_{0}\geq C(M,q_0,\tilde{\btau})s_0^{1/2}(\sqrt{\log (pn)}+\sqrt{\log(1/\gamma)})\big).
	\end{equation}
	To prove $H_1\rightarrow 1$, we need the decomposition of $T_0$ under $\Hb_1$. Note that for $\alpha=0$, with known variance, the score based CUSUM process reduces to:
	\begin{equation}\label{equation: C0(t)}
		\bC_0(t)=\dfrac{1}{\sqrt{n}{\sigma}(0,\tilde{\btau})}\big(\sum\limits_{i=1}^{\lfloor nt\rfloor}\bZ(\bX_i,Y_i;\tilde{\btau},\hat{\bb},\hat{\bbeta})-\dfrac{\lfloor nt \rfloor }{n}\sum\limits_{i=1}^{n}\bZ(\bX_i,Y_i;\tilde{\btau},\hat{\bb},\hat{\bbeta})\big),
	\end{equation}
	where $\bZ(\bX_i,Y_i;\tilde{\btau},\hat{\bb},\hat{\bbeta}):=\dfrac{1}{K}\sum\limits_{k=1}^{K}\bX_i \big(\mathbf{1}\{Y_i-\hat{b}_k-\bX_i^\top\hat{\bbeta}\leq 0\}-\tau_k\big)$. Define
	\begin{equation}
		\begin{array}{ll}
			\hat{e}_{i}(\tilde{\btau})&:=\dfrac{1}{K}\sum\limits_{k=1}^{K}\Big(\mathbf{1}\{Y_i-\hat{b}_k-\bX_i^\top\hat{\bbeta}\leq 0\}-\tau_k\big):=\dfrac{1}{K}\sum\limits_{k=1}^{K}\hat{e}_{i}(\tau_k),
		\end{array}
	\end{equation}
	where $\hat{e}_{i}(\tau_k):=\mathbf{1}\{Y_i-\hat{b}_k-\bX_i^\top\hat{\bbeta}\leq 0\}-\tau_k$. For $\hat{e}_i(\tilde{\btau})$, we have the following decomposition:
	\begin{equation}\label{equation: decomposition for the error of composite quantile}
		\hat{e}_{i}(\tilde{\btau})=e_{i}(\tilde{\btau})+\E[\hat{e}_i(\tilde{\btau})-e_i(\tilde{\btau})]+\underbrace{\{\hat{e}_i(\tilde{\btau})-e_i(\tilde{\btau})-\E[\hat{e}_i(\tilde{\btau})-e_i(\tilde{\btau})]\}}_{V_i(\tilde{\btau})},
	\end{equation}
	where 
	\begin{equation}
		\begin{array}{ll}
			e_i(\tilde{\btau}):=\dfrac{1}{K}\sum\limits_{k=1}^{K}e_i(\tau_k),~\text{with}~e_i(\tau_k)=\mathbf{1}\{\epsilon_i\leq b^{(0)}_{k}\}-\tau_k,\\
			\E[\hat{e}_i(\tilde{\btau})-e_i(\tilde{\btau})]:=\dfrac{1}{K}\sum\limits_{k=1}^{K}\E\big[\hat{e}_i({\tau_k})-e_i({\tau_k})\big],&\\
			V_i(\tilde{\btau}):=\dfrac{1}{K}\sum\limits_{k=1}^{K}V_{i}(\tau_k),&
		\end{array}
	\end{equation}
	and 
	\begin{equation*}
		V_{i}(\tau_k)=[\mathbf{1}\{Y_i-\hat{b}_k-\bX_i^\top\hat{\bbeta}\leq 0\}-\mathbf{1}\{\epsilon_i\leq b^{(0)}_{k}\}]-\E[\mathbf{1}\{Y_i-\hat{b}_k-\bX_i^\top\hat{\bbeta}\leq 0\}-\mathbf{1}\{\epsilon_i\leq b^{(0)}_{k}\}].
	\end{equation*}
	Next, we analyze the three parts in (\ref{equation: decomposition for the error of composite quantile}). Note that the first term $e_{i}(\tilde{\btau})$ is a sum for simple Bernoulli  random variables. For the second term, by the Taylor's expansion, we have
	\begin{equation}\label{equation: taylor expansion for composite quantile}
		\begin{array}{ll}	
			\E[\hat{e}_i(\tilde{\btau})-e_i(\tilde{\btau})]
			&=\dfrac{1}{K}\sum\limits_{k=1}^{K}\E\big[\hat{e}_i({\tau_k})-e_i({\tau_k})\big]=\underbrace{\dfrac{1}{K}\sum\limits_{k=1}^{K}M_i^{(1)}(\tau_{k})}_{M_i^{(1)}(\tilde{\btau})}+\underbrace{\dfrac{1}{K}\sum\limits_{k=1}^{K}M_i^{(2)}(\tau_{k})}_{M_i^{(2)}(\tilde{\btau})},
		\end{array}
	\end{equation}
	where
	\begin{equation}
		\begin{array}{ll}
			M_i^{(1)}(\tau_{k})
			:=f_\epsilon(b_{k}^{(0)})\big(\hat{b}_k-b_{k}^{(0)}+\bX_i^\top(\hat{\bbeta}-\bbeta^{(1)})\big)\mathbf{1}\{i\leq \floor{nt_1}\}\\
			\qquad\quad\quad\quad\quad\quad\quad+f_\epsilon(b_{k}^{(0)})\big(\hat{b}_k-b_{k}^{(0)}+\bX_i^\top(\hat{\bbeta}-\bbeta^{(2)})\big)\mathbf{1}\{i>\floor{nt_1}\},
		\end{array}
	\end{equation}
	and
	\begin{equation}
		\begin{array}{ll}
			M_i^{(2)}(\tau_{k}):=\dfrac{1}{2}f'_{\epsilon}(\xi_{ik})\big(\hat{b}_k-b_{k}^{(0)}+\bX_i^\top(\hat{\bbeta}-\bbeta^{(1)})\big)^2\mathbf{1}\{i\leq \floor{nt_1}\}\\
			\qquad\quad\quad\quad\quad\quad\quad+\dfrac{1}{2}f'_{\epsilon}(\xi_{ik})\big(\hat{b}_k-b_{k}^{(0)}+\bX_i^\top(\hat{\bbeta}-\bbeta^{(2)})\big)^2\mathbf{1}\{i> \floor{nt_1}\},
		\end{array}
	\end{equation}
	with $\xi_{ik}$ being some constant that between ${b}_{k}^{(0)}$ and $\hat{b}_k+\bX_i^\top(\hat{\bbeta}-\bbeta^{(1)})$ (or $b_k+\bX_i^\top(\hat{\bbeta}-\bbeta^{(2)})$).
	
	Hence, based on the above decomposition, for the composite quantile based score function, its CUSUM process can be decomposed into four parts:
	\begin{equation}\label{equation: decomposition for the composite cusum}
		\bC_{0}(t)=	\bC_{0}^{(1)}(t)+	\bC^{(2)}_{0}(t)+\bC^{(3)}_{0}(t)+\bC^{(4)}_{0}(t),
	\end{equation}
	where $\bC_{0}^{(1)}(t),\ldots, \bC_{0}^{(4)}(t)$ are defined as:
	\begin{equation}\label{equation: C01-C04}
		\begin{array}{ll}
			\bC_{0}^{(1)}(t)=\dfrac{1}{\sqrt{n}\sigma(\alpha,\tilde{\btau})}\big(\sum\limits_{i=1}^{\floor{nt}}\bX_ie_i(\tilde{\btau})-\dfrac{\floor{nt}}{n}\sum\limits_{i=1}^n\bX_ie_i(\tilde{\btau})\big),\\
			\bC_{0}^{(2)}(t)=\dfrac{1}{\sqrt{n}\sigma(\alpha,\tilde{\btau})}\big(\sum\limits_{i=1}^{\floor{nt}}\bX_i M_i^{(1)}(\tilde{\btau})-\dfrac{\floor{nt}}{n}\sum\limits_{i=1}^n\bX_iM_i^{(1)}(\tilde{\btau})\big),\\
			\bC_{0}^{(3)}(t)=\dfrac{1}{\sqrt{n}\sigma(\alpha,\tilde{\btau})}\big(\sum\limits_{i=1}^{\floor{nt}}\bX_i M_i^{(2)}(\tilde{\btau})-\dfrac{\floor{nt}}{n}\sum\limits_{i=1}^n\bX_iM_i^{(2)}(\tilde{\btau})\big),\\
			\bC_{0}^{(4)}(t)=\dfrac{1}{\sqrt{n}\sigma(\alpha,\tilde{\btau})}\big(\sum\limits_{i=1}^{\floor{nt}}\bX_i V_i(\tilde{\btau})-\dfrac{\floor{nt}}{n}\sum\limits_{i=1}^n\bX_iV_i(\tilde{\btau})\big).
		\end{array}
	\end{equation}
	Note that 	$\bC_{0}^{(2)}(t)$ consists of the signal jump and is very important for detecting a change point. To see this, recall $M_i^{(1)}(\tilde{\btau})=\dfrac{1}{K}\sum\limits_{k=1}^{K}M_i^{(1)}(\tau_{k})$ defined in (\ref{equation: taylor expansion for composite quantile}). Then, we have
	\begin{equation}\label{equation: C02(t)}
		\begin{array}{ll}
			\bC_{0}^{(2)}(t)&=_{(1)}\dfrac{1}{\sqrt{n}\sigma(\alpha,\tilde{\btau})}\big(\sum\limits_{i=1}^{\floor{nt}}\bX_i \dfrac{1}{K}\sum\limits_{k=1}^{K}M_i^{(1)}(\tau_{k})-\dfrac{\floor{nt}}{n}\sum\limits_{i=1}^n\bX_i\dfrac{1}{K}\sum\limits_{k=1}^{K}M_i^{(1)}(\tau_{k})\big)\\
			&=_{(2)}\dfrac{1}{K}\sum\limits_{k=1}^{K}\big[\dfrac{1}{\sqrt{n}\sigma(\alpha,\tilde{\btau})}\big(\sum\limits_{i=1}^{\floor{nt}}\bX_i M_i^{(1)}(\tau_k)-\dfrac{\floor{nt}}{n}\sum\limits_{i=1}^n\bX_iM_i^{(1)}(\tau_k)\big)\big]\\
			&=_{(3)}\dfrac{1}{K}\sum\limits_{k=1}^{K}\big[\dfrac{-f_{\epsilon}(b_{k}^{(0)})}{\sigma(0,\tilde{\btau})}\big(\bdelta(t)+\bR(t)\big)\big],
		\end{array}
	\end{equation}
	where $\bdelta(t)$ is defined in (\ref{equation: delta-t alpha=1}), and $\bR(t)$ is defined in (\ref{equation: C11+C12 under H1}).
	Hence, combining  (\ref{equation: decomposition for the composite cusum}) -  (\ref{equation: C02(t)}), under $\Hb_1$, the score based CUSUM for the quantile loss can be decomposed into four terms:
	\begin{equation}\label{equation: decmposition for the score of composite quantile  under H1}
		\begin{array}{ll}
			\bC_{0}(t)=-SNR(0,\tilde{\btau})\times\bdelta(t)
			+\bC_{0}^{(1)}(t)
			-SNR(0,\tilde{\btau})\times\bR(t)+	\bC_{0}^{(3)}(t)+	\bC_{0}^{(4)}(t).
		\end{array}
	\end{equation}
	{\textbf{Step~3: obtain the upper bounds for the residuals and random noises in $\bC_0(t)$}.} \\
	We first consider $\max_{t}\|\bC_0^{(1)}\|_{(s_0,2)}$. By  definition, $\bC_0^{(1)}$ is a partial sum process based on $\bX_ie_i(\tilde{\btau})$. Hence, by Lemma \ref{lemma: exponential inequality for partial sum process}, we can prove that $\max_{q_0\leq t\leq 1-q_0}\|\bC^{(1)}_0(t)\|_{(s_0,2)}=O_p(s_0^{1/2}M\sqrt{\log(pn)})$. For $\max_{t}\|\bR(t)\|_{(s_0,2)}$, using Lemma \ref{lemma: basic inequality for lasso with alpha=0}, Remark \ref{remark: difference beta for alpha=0}, and using a similar proof procedure as in Step 2 of Section \ref{sec: power for alpha=1}, we can prove that 
	\begin{equation}\label{inequality: bias term for alpha=0}
		\max\limits_{q_0\leq t\leq 1-q_0}\|\bR(t)\|_{(s_0,2)}
		=O_p\big(s_0^{1/2}s\sqrt{\log p}\|\bSigma(\bbeta^{(1)}-\bbeta^{(2)})\|_{(s_0,2)}\big).
	\end{equation}
	Next, we consider $\max_{q_0\leq t\leq 1-q_0}\|\bC^{(3)}_0(t)\|_{(s_0,2)}$. To that end, we need some notations.  let $\uwave{\bbeta}^{(1)}:=((\bbeta^{(1)})^\top,(\bb^{(0)})^\top)^\top\in \RR^{p+K}$, $\uwave{\bbeta}^{(2)}:=((\bbeta^{(2)})^\top,(\bb^{(0)})^\top)^\top\in \RR^{p+K}$,  $\uwave{\bX}:=(\bX^\top,\mathbf{1}_{K})\in \RR^{p+K}$, and $\bS_k:=\diag(\mathbf{1}_p,\be_k)$, where $\be_k\in \RR^K$ is a vector with the $k$-th element being 1 and the others being zeros, and $\mathbf{1}_{K}$ is a $K$-dimensional vector with all elements being 1s. Moreover, recall $\bS$ as defined in (\ref{equation: bS}). Then, by the definition of $\bC^{(3)}_0(t)$, we have:
	\begin{equation*}
		\begin{array}{lc}
			\max\limits_{q_0\leq t\leq 1-q_0}\|\bC^{(3)}_0(t)\|_{(s_0,2)}\\
			\quad\leq_{(1)} s_0^{1/2}\max\limits_{q_0\leq t\leq 1-q_0}\|\bC^{(3)}_0(t)\|_{\infty}\\
			\quad\leq_{(2)} s_0^{1/2}\max\limits_{t}\dfrac{\floor{nt}}{\sqrt{n}\sigma(\alpha,\tilde{\btau})}\Big\|\dfrac{1}{\floor{nt}}\sum\limits_{i=1}^{\floor{nt}}\bX_i M_i^{(2)}(\tilde{\btau})\Big\|_{\infty}\\
			\qquad\qquad\qquad+s_0^{1/2}\max\limits_{t}\dfrac{\floor{nt}}{\sqrt{n}\sigma(\alpha,\tilde{\btau})}\Big\|\dfrac{1}{n}\sum\limits_{i=1}^{n}\bX_i M_i^{(2)}(\tilde{\btau})\Big\|_{\infty}\\
			\quad\leq_{(3)} C_1\sqrt{n}s_0^{1/2}\max\limits_{j}\max\limits_{t}\Big|\dfrac{1}{\floor{nt}}\sum\limits_{i=1}^{\floor{nt}} X_{ij}M_i^{(2)}(\tilde{\btau})\Big|+C_1\sqrt{n}s_0^{1/2}\max\limits_{j}\Big|\dfrac{1}{n}\sum\limits_{i=1}^{n} X_{ij}M_i^{(2)}(\tilde{\btau})\Big|\\
			\quad\leq_{(4)}  C_1\sqrt{n}s_0^{1/2}(I\vee II)+C_1M\sqrt{n}s_0^{1/2}(III\vee IV),
		\end{array}
	\end{equation*}
	where $I-IV$ are defined as
	\begin{equation}
		\begin{array}{cc}
			I:=\max\limits_{1\leq j\leq p}\max\limits_{q_0\leq t\leq 1-q_0}\big|\dfrac{1}{\floor{nt}}\sum\limits_{i=1}^{\floor{nt}} X_{ij}\dfrac{1}{K}\sum\limits_{k=1}^K\dfrac{1}{2}f'_{\epsilon}(\xi_{ik})\big(\hat{b}_k-b_{k}^{(0)}+\bX_i^\top(\hat{\bbeta}-\bbeta^{(1)})\big)^2\big|,\\
			II:=\max\limits_{1\leq j\leq p}\max\limits_{q_0\leq t\leq 1-q_0}\big|\dfrac{1}{\floor{nt}}\sum\limits_{i=1}^{\floor{nt}} X_{ij}\dfrac{1}{K}\sum\limits_{k=1}^K\dfrac{1}{2}f'_{\epsilon}(\xi_{ik})\big(\hat{b}_k-b_{k}^{(0)}+\bX_i^\top(\hat{\bbeta}-\bbeta^{(2)})\big)^2\big|,\\
			III:=\max\limits_{j}\big|\dfrac{1}{n}\sum\limits_{i=1}^{n}X_{ij} \dfrac{1}{K}\sum\limits_{k=1}^K\dfrac{1}{2}f'_{\epsilon}(\xi_{ik})\big(\hat{b}_k-b_{k}^{(0)}+\bX_i^\top(\hat{\bbeta}-\bbeta^{(1)})\big)^2\big|,\\
			IV:=\max\limits_{j}\big|\dfrac{1}{n}\sum\limits_{i=1}^{n}X_{ij} \dfrac{1}{K}\sum\limits_{k=1}^K\dfrac{1}{2}f'_{\epsilon}(\xi_{ik})\big(\hat{b}_k-b_{k}^{(0)}+\bX_i^\top(\hat{\bbeta}-\bbeta^{(1)})\big)^2\big|.\\
		\end{array}
	\end{equation}
	Next, we consider $I-IV$, respectively. To that end, define 
	\begin{equation*}
		\begin{array}{cc}
			M_i^{(2)}(\tilde{\btau};\bbeta^{(1)})=\dfrac{1}{K}\sum\limits_{k=1}^K\dfrac{1}{2}f'_{\epsilon}(\xi_{ik})\big(\hat{b}_k-b_{k}^{(0)}+\bX_i^\top(\hat{\bbeta}-\bbeta^{(1)})\big)^2,~\text{for}~i=1,\ldots,n,\\
			M_i^{(2)}(\tilde{\btau};\bbeta^{(2    })=\dfrac{1}{K}\sum\limits_{k=1}^K\dfrac{1}{2}f'_{\epsilon}(\xi_{ik})\big(\hat{b}_k-b_{k}^{(0)}+\bX_i^\top(\hat{\bbeta}-\bbeta^{(2)})\big)^2,~\text{for}~i=1,\ldots,n,\\
			\bM^{(2)}(\tilde{\btau};\bbeta^{(1)}):=(M_1^{(2)}(\tilde{\btau};\bbeta^{(1)}),\ldots,M_n^{(2)}(\tilde{\btau};\bbeta^{(1)}))^\top,\\
			\bM^{(2)}(\tilde{\btau};\bbeta^{(2)}):=(M_1^{(2)}(\tilde{\btau};\bbeta^{(2)}),\ldots,M_n^{(2)}(\tilde{\btau};\bbeta^{(2)}))^\top.\\
		\end{array}
	\end{equation*}
	For $I$, we then have:
	\begin{equation}\label{equation: I alpha=0-1}
		\begin{array}{ll}
			I&=_{(1)}\max\limits_{1\leq j\leq p}\max\limits_{q_0\leq t\leq 1-q_0}\Big|\dfrac{1}{\floor{nt}}\sum\limits_{i=1}^{\floor{nt}} X_{ij}\dfrac{M_i^{(2)}(\tilde{\btau};\bbeta^{(1)})}{\|\bM^{(2)}(\tilde{\btau};\bbeta^{(1)})\|}\Big|\big\|\bM^{(2)}(\tilde{\btau};\bbeta^{(1)})\big\|\\
			&\leq_{(2)} \max\limits_{\bw=(w_1,\ldots,w_n)^\top\atop \|\bw\|=1}\max\limits_{1\leq j\leq p}\max\limits_{q_0\leq t\leq 1-q_0}\Big|\dfrac{1}{\floor{nt}}\sum\limits_{i=1}^{\floor{nt}} X_{ij}w_i\Big|\big\|\bM^{(2)}(\tilde{\btau};\bbeta^{(1)})\big\|\\
			&\leq_{(3)} O_p(\dfrac{\sqrt{\log(pn)}}{n})\|\bM^{(2)}(\tilde{\btau};\bbeta^{(1)})\big\|,\\
		\end{array}
	\end{equation}
	where $(3)$ comes from Assumption (A.2) and the Hoeffding's inequality. Hence, to bound $I$, we need to consider $\|\bM^{(2)}(\tilde{\btau};\bbeta^{(1)})\big\|$.  In fact, we have:
	\begin{equation}\label{equation: I alpha=0-2}
		\begin{array}{ll}
			&\|\bM^{(2)}(\tilde{\btau};\bbeta^{(1)})\big\|^2\\
			&\quad=_{(1)}\sum\limits_{i=1}^n[M_i^{(2)}(\tilde{\btau};\bbeta^{(1)})]^2\\
			&\quad=_{(2)}\sum\limits_{i=1}^n\big[\dfrac{1}{K}\sum\limits_{k=1}^K\dfrac{1}{2}f'_{\epsilon}(\xi_{ik})\big(\hat{b}_k-b_{k}^{(0)}+\bX_i^\top(\hat{\bbeta}-\bbeta^{(1)})\big)^2\big]^2\\
			&\quad\leq_{(3)} \dfrac{C'^2_+n}{4K^2}\max\limits_{1\leq i \leq n}\big[\sum\limits_{k=1}^K\big(\hat{b}_k-b_{k}^{(0)}+\bX_i^\top(\hat{\bbeta}-\bbeta^{(1)})\big)^2\big]^2\\
			&\quad\leq_{(4)}  \dfrac{C'^2_+n}{4K^2}\big[\sum\limits_{k=1}^K( |\hat{b}_k-b_{k}^{(0)}|+M\|\hat{\bbeta}-\bbeta^{(1)}\|_1  )^2]^2\\
			&\quad\leq_{(5)}  \dfrac{C'^2_+n}{4K^2}\big[KM^2 \|\uwave{\hat{\bbeta}}-\uwave{\bbeta}^{(1)}\|^2_1\big]^2\\
			&\quad\leq_{(6)}  \dfrac{M^4C'^2_+n}{4}\big[ \|\uwave{\hat{\bbeta}}-\uwave{\bbeta}^{*}\|_1+\|\uwave{\bbeta}^{*}-\uwave{\bbeta}^{(1)}\|_1\big]^4\\
			&\quad\leq_{(7)}  \dfrac{M^4C'^2_+n}{4}\big[ O_p(s\sqrt{\dfrac{\log(pn)}{n}})+C_1\|{\bbeta}^{(2)}-{\bbeta}^{(1)}\|_1\big]^4\\
			&\quad\leq_{(8)}  O_p(ns^4 \|{\bbeta}^{(2)}-{\bbeta}^{(1)}\|^4_{(s_0,2)}). \\
		\end{array}
	\end{equation}
	where $(3)$ comes from {\bf Assumption D}, $(4)$ comes from {\bf Assumption A}, $(7)$ comes from Lemma \ref{lemma: basic inequality for lasso with alpha=0}, $(8)$ comes from  $\|{\bbeta}^{(2)}-{\bbeta}^{(1)}\|_1\leq s\|{\bbeta}^{(2)}-{\bbeta}^{(1)}\|_{(s_0,2)}$. Combining (\ref{equation: I alpha=0-1}) and (\ref{equation: I alpha=0-2}), we have 
	\begin{equation*}
		I=O_p(s^2\sqrt{\dfrac{\log(pn)}{n}} \|{\bbeta}^{(2)}-{\bbeta}^{(1)}\|^2_{(s_0,2)}).
	\end{equation*}
	With a similar proof procedure, we can prove  $II, III, IV=O_p(s^2\sqrt{\log(pn)/n} \|{\bbeta}^{(2)}-{\bbeta}^{(1)}\|^2_{(s_0,2)})$, which yields:
	\begin{equation}\label{inequality: C0-3}
		\max\limits_{q_0\leq t\leq 1-q_0}\|\bC^{(3)}_0(t)\|_{(s_0,2)}=O_p(s_0^{1/2}s^2\sqrt{\log(pn)} \|\bSigma({{\bbeta}^{(2)}}-{\bbeta}^{(1)})\|^2_{(s_0,2)}).
	\end{equation}
	
	Lastly, we consider the control of $\max_{q_0\leq t\leq 1-q_0}\|\bC^{(4)}_0(t)\|_{(s_0,2)}$. Similar to the proof of Lemma \ref{lemma: upper bound for the empirical process of alpha=0 under H0}, using some tedious modifications, it is not hard to prove that:
	\begin{equation}\label{inequality: C0-4}
		\max_{q_0\leq t\leq 1-q_0}\|\bC^{(4)}_0(t)\|_{(s_0,2)}=O_p\big(Ms_0^{1/2}s\sqrt{\log(pn)}\|\bSigma({{\bbeta}^{(2)}}-{\bbeta}^{(1)})\|_{(s_0,2)}\big).
	\end{equation} 
	{\textbf{Step~4: Combining the previous results}.} Recall (\ref{equation: upper bounds of critical value for alpha=0}), (\ref{equation: H_1 alpha=0}), (\ref{equation: decmposition for the score of composite quantile  under H1}). Using the above bounds of $\bC_{(0)}^{(1)}(t)$, $\bR(t)$, $\bC_{(0)}^{(3)}(t)$, $\bC_{(0)}^{(4)}(t)$, and by the triangle inequality, w.p.a.1, we have:
	\begin{equation}\label{inequality: lower bound of testing statistic alpha=0}
		\small
		\begin{array}{ll}
			T_0&=\max\limits_{q_0\leq t\leq 1-q_0}\|\bC_0(t)\|_{(s_0,2)}\\
			&\geq SNR(0,\tilde{\btau})\times\max\limits_{q_0\leq t\leq 1-q_0}\|\bdelta(t)\|_{(s_0,2)}-\max\limits_{q_0\leq t\leq 1-q_0}\|\bC^{(1)}_0(t)\|_{(s_0,2)}\\
			&-SNR(0,\tilde{\btau})\times\max\limits_{q_0\leq t\leq 1-q_0}\|\bR(t)\|_{(s_0,2)}-\max\limits_{q_0\leq t\leq 1-q_0}\|\bC^{(3)}_0(t)\|_{(s_0,2)}-\max\limits_{q_0\leq t\leq 1-q_0}\|\bC^{(4)}_0(t)\|_{(s_0,2)}\\
			&\geq \sqrt{n}\times SNR(0,\tilde{\btau})\times t_1(1-t_1)\times\|\bSigma(\bbeta^{(1)}-\bbeta^{(2)})\|_{(s_0,2)}
			-C_1s_0^{1/2}M\sqrt{\log(pn)}\\
			&\quad-C_2s_0^{1/2}s\sqrt{\log p}\|\bSigma(\bbeta^{(1)}-\bbeta^{(2)})\|_{(s_0,2)}-C_3s_0^{1/2}s^2\sqrt{\log(pn)} \|\bSigma({{\bbeta}^{(2)}}-{\bbeta}^{(1)})\|^2_{(s_0,2)}
			\\
			&\qquad\qquad-C_4Ms_0^{1/2}s\sqrt{\log(p)}\|\bSigma({{\bbeta}^{(2)}}-{\bbeta}^{(1)})\|_{(s_0,2)}\\
			&\geq \sqrt{n}\times SNR(0,\tilde{\btau})\times t_1(1-t_1)\times\|\bSigma(\bbeta^{(1)}-\bbeta^{(2)})\|_{(s_0,2)}
			(1-\epsilon_n)-C_1s_0^{1/2}M\sqrt{\log(pn)},\\
		\end{array}
	\end{equation}
	where 
	\begin{equation*}
		\epsilon_n:=O\big(s^{1/2}_0s\sqrt{\dfrac{\log p}{n}}\big) \vee O\big({s_0^{1/2}s^2\sqrt{\dfrac{\log p}{n}}\|\bbeta^{(2)}-\bbeta^{(1)}\|_{(s_0,2)}}\big).
	\end{equation*}
	Hence,  considering (\ref{inequality: lower bound of testing statistic alpha=0}), to prove (\ref{equation: H_1 alpha=0}), it is sufficient to prove $H_1'\rightarrow 1$, where 
	\begin{equation*}
		\begin{array}{ll}
			H_1'=\P\Big(
			\sqrt{n}\times SNR(0,\tilde{\btau})\times t_1(1-t_1)\times\|\bSigma(\bbeta^{(1)}-\bbeta^{(2)})\|_{(s_0,2)}\\
			\geq 
			\dfrac{Cs^{1/2}_0M\big(\sqrt{\log(pn)}+\sqrt{\log(1/\gamma)}\big)}{1-\epsilon_n}
			\Big).
		\end{array}
	\end{equation*}
	By (\ref{inequality: theoretical signal strengh}), it is straightforward to see that $H_1'\rightarrow 1$ as $n,p\rightarrow\infty$, which finishes the proof.

\end{proof}

\begin{remark}
	Note that for $\alpha=0$, if we replace $\sigma^2(0,\tilde{\btau})$ by an estimator $\hat{\sigma}^2(0,\tilde{\btau})$ which satisfies: $|\hat{\sigma}^2(\alpha,\tilde{\btau})-\sigma^2(0,\tilde{\btau})|=o_p(1)$, then under condition (\ref{inequality: theoretical signal strengh}), the change point test is still consistent.
\end{remark}	

\subsubsection{Power analysis for $\alpha\in(0,1)$}
\begin{proof}
	In what follows, we assume $\hat{\sigma}^2(\alpha,\tilde{\btau})=\sigma^2(\alpha,\tilde{\btau})$ by assuming the variance is unknown, where $\sigma^2(\alpha,\tilde{\btau}):=\text{Var}[ (1-\alpha)e_i(\tilde{\btau})-\alpha\epsilon_i]$.
	In addition, for the case of  $\alpha\in(0,1)$, we define
	\begin{equation*}
		SNR(\alpha,\tilde{\btau}):=\dfrac{(1-\alpha) \big(\dfrac{1}{K}\sum\limits_{k=1}^{K}f_{\epsilon}(b_{k}^{(0)})\big)+\alpha}{\sigma(\alpha,\tilde{\btau})}.
	\end{equation*}
	Similar to Sections \ref{sec: power for alpha=1} and \ref{sec: power for alpha=0} , the proof of Theorem\ref{theorem: power control for individual test} proceeds in four steps. In Step 1, we obtain the upper bound of $c_{T_{\alpha}^b}(1-\gamma)$, where $c_{T_{\alpha}^b}(1-\gamma)$ is  the $1-\gamma$ th quantile of $T^{b}_{\alpha}$, which is defined as
	\begin{equation}
		c_{T_{\alpha}^b}(1-\gamma):=\inf\big\{t:\P(T_{\alpha}^b\leq t)\geq 1-\gamma  \big\}.
	\end{equation}
	In Steps 2-4, using the  upper bound, we get the lower bound of $\P\big(T_{\alpha}\geq c_{T_{\alpha}^b}(1-\gamma)\big)$ and prove
	\begin{equation}
		\P\big(T_{\alpha}\geq c_{T_{\alpha}^b}(1-\gamma)\big)\rightarrow1, ~\text{as}~n,p\rightarrow\infty.
	\end{equation}
	{\bf{Step~1}}: By the definition of $T_\alpha^b$, we have: $T_\alpha^b=\max_{q_0\leq t\leq1-q_0}\|\bC_\alpha^b(t)\|_{(s_0,2)}$, where
	\begin{equation*}
		\bC_\alpha^b(t):=\dfrac{1}{\sqrt{n}v(\alpha,\tilde{\btau})}\Big(\sum\limits_{i=1}^{\floor{nt}}\bX_i((1-\alpha)e_i^{b}(\tilde{\btau})-\alpha e_i^b)-\dfrac{\floor{nt}}{n}\sum\limits_{i=1}^n\bX_i((1-\alpha)e_i^{b}(\tilde{\btau})-\alpha e_i^b)\Big),
	\end{equation*}
	where $e_i^{b}(\tilde{\btau})=K^{-1}\sum\limits_{k=1}^{K}e_i^{b}(\tau_k)$ with $e_i^{b}(\tau_k):=\mathbf{1}\{\epsilon_i^b\leq \Phi^{-1}(\tau_k)\}-\tau_k$, $e_i^b~ \text{~is ~i.i.d} ~N(0,1)$, and  $\Phi(x)$ is the CDF for the standard normal distribution, and $v^2(\alpha,\tilde{\btau}):=\text{Var}[(1-\alpha)e_i^{b}(\tilde{\btau})-\alpha e_i^b]$.
	
	Note that $(1-\alpha)e_i^{b}(\tilde{\btau})-\alpha e_i^b$ is just a linear combination of a bounded random variable $e_i^{b}(\tilde{\btau})$ and a standard normal distribution. Hence, using a  very similar proof procedure as in Step~1 in Section \ref{sec: power for alpha=1}, it is not hard to prove
	\begin{equation}\label{equation: upper bounds of critical value for alpha}
		c_{T_{\alpha}^b}(1-\gamma)\leq C(M,q_0,\tilde{\btau},\alpha)s_0^{1/2}(\sqrt{\log (pn)}+\sqrt{\log(1/\gamma)}),
	\end{equation}
	where $C(M,q_0,\tilde{\btau},\alpha)$ is some universal constant only depending on $M,q_0$, $\tilde{\btau}$ and $\alpha$. \\
	\textbf{Step~2 Decomposition of $\bC_\alpha(t)$}. In this step, we aim to prove that $\P\big(T_{\alpha}\geq c_{T_{\alpha}^b}(1-\gamma)\big)\rightarrow1$ as $n,p\rightarrow\infty$. Note that in Step~1, we have obtained the upper bound of $c_{T_{\alpha}^b}(1-\gamma)$. Hence, it is sufficient to prove that $H_1\rightarrow1$ as $n,p\rightarrow\infty$, where 
	\begin{equation}\label{equation: H_1 alpha in (0,1)}
		H_1:=\P\big(T_{\alpha}\geq C(M,q_0,\tilde{\btau},\alpha)s_0^{1/2}(\sqrt{\log (pn)}+\sqrt{\log(1/\gamma)})\big).
	\end{equation}
	To prove $H_1\rightarrow 1$, we need the decomposition of $T_\alpha$ under $\Hb_1$. Note that for $\alpha\in(0,1)$, with known variance, the score based CUSUM process reduces to:
	\begin{equation}\label{equation: C-alpha-t}
		\bC_\alpha(t)=\dfrac{1}{\sqrt{n}{\sigma}(\alpha,\tilde{\btau})}\Big(\sum\limits_{i=1}^{\lfloor nt\rfloor}\bZ(\bX_i,Y_i;\tilde{\btau},\hat{\bb},\hat{\bbeta})-\dfrac{\lfloor nt \rfloor }{n}\sum\limits_{i=1}^{n}\bZ(\bX_i,Y_i;\tilde{\btau},\hat{\bb},\hat{\bbeta})\Big),
	\end{equation}
	where $\bZ(\bX_i,Y_i;\tilde{\btau},\hat{\bb},\hat{\bbeta}):=(1-\alpha)\dfrac{1}{K}\sum\limits_{k=1}^{K}\bX_i \big(\mathbf{1}\{Y_i-\hat{b}_k-\bX_i^\top\hat{\bbeta}\leq 0\}-\tau_k\big)-\alpha\bX_i(Y_i-\bX_i^\top\hat{\bbeta})$. 
	
	Using the results obtained in Sections \ref{sec: power for alpha=1} and \ref{sec: power for alpha=0}, we have the following decomposition:
	\begin{equation}\label{equation: decomposition of C-alpha-t}
		\bC_\alpha(t)=\tilde{\bC}^{I}_{\alpha}(t)-SNR(\alpha,\tilde{\btau})\times\bdelta(t)-SNR(\alpha,\tilde{\btau})\times\bR(t)+	(1-\alpha)\bC_{0}^{(3)}(t)+(1-\alpha)\bC_{0}^{(4)}(t),
	\end{equation}
	where 	$\tilde{\bC}^{I}_{\alpha}(t)$ is the random noise based partial sum process defined in (\ref{equation: tilde-C-alpha-I}), $\bdelta(t)$ is the signal jump defined in (\ref{equation: delta-t alpha=1}), $\bR(t)$ is defined in (\ref{equation: C11+C12 under H1}), and $\bC_{0}^{(3)}(t)$ and $\bC_{0}^{(4)}(t)$ are defined in (\ref{equation: C01-C04}).\\
	{\textbf{Step~3: obtain the upper bounds for the residuals and random noises in $\bC_\alpha(t)$}.} We first bound $\max_t\|\tilde{\bC}^{I}_{\alpha}(t)\|_{(s_0,2)}$. Note that by its definition in (\ref{equation: tilde-C-alpha-I}), we have:
	\begin{equation*}
		\begin{array}{ll}
			&\max\limits_{t\in[q_0,1-q_0]}\|\tilde{\bC}^{I}_{\alpha}(t)\|_{(s_0,2)}\\
			&\quad=\max\limits_{t\in[q_0,1-q_0]}\|(1-\alpha)\tilde{\bC}^{I}_{0}(t)-\alpha \tilde{\bC}^{I}_{1}(t)\|_{(s_0,2)}\\
			&\quad\leq (1-\alpha)\max\limits_{t\in[q_0,1-q_0]}\|\tilde{\bC}^{I}_{0}(t)\|_{(s_0,2)}+\alpha\max\limits_{t\in[q_0,1-q_0]}\| \tilde{\bC}^{I}_{1}(t)\|_{(s_0,2)}\\
			&\quad\leq (1-\alpha)s_0^{1/2}\max\limits_{t\in[q_0,1-q_0]}\|\tilde{\bC}^{I}_{0}(t)\|_{\infty}+\alpha s_0^{1/2}\max\limits_{t\in[q_0,1-q_0]}\| \tilde{\bC}^{I}_{1}(t)\|_{\infty}\\
			&\quad=O_p(s_0^{1/2}M\sqrt{\log(pn)}),
		\end{array}
	\end{equation*}
	where the last equation comes from Lemma \ref{lemma: exponential inequality for partial sum process}. Next, we consider $\max_{t}\|\bR(t)\|_{(s_0,2)}$. Using Lemma \ref{lemma: basic inequality for lasso with alpha}, Remark \ref{remark: difference beta for alpha in (0,1)}, and using a similar proof procedure as Step 2 in Section \ref{sec: power for alpha=1}, we have
	\begin{equation*}
		\max\limits_{q_0\leq t\leq 1-q_0}\|\bR(t)\|_{(s_0,2)}
		=O_p\big(s_0^{1/2}s\sqrt{\log p}\|\bSigma(\bbeta^{(1)}-\bbeta^{(2)})\|_{(s_0,2)}\big).
	\end{equation*}
	For $\bC_{0}^{(3)}(t)$ and $\bC_{0}^{(4)}(t)$, using the obtained upper bounds in (\ref{inequality: C0-3}) and (\ref{inequality: C0-4}), we have:
	\begin{equation*}
		\begin{array}{ll}
			\max\limits_{q_0\leq t\leq 1-q_0}\|\bC^{(3)}_0(t)\|_{(s_0,2)}=O_p(s_0^{1/2}s^2\sqrt{\log(pn)} \|\bSigma({{\bbeta}^{(2)}}-{\bbeta}^{(1)})\|^2_{(s_0,2)}),~~\text{and}~~\\
			\max\limits_{q_0\leq t\leq 1-q_0}\|\bC^{(4)}_0(t)\|_{(s_0,2)}=O_p\big(Ms_0^{1/2}s\sqrt{\log(pn)}\|\bSigma({{\bbeta}^{(2)}}-{\bbeta}^{(1)})\|_{(s_0,2)}\big).
		\end{array}
	\end{equation*}
	{\textbf{Step~4: Combining the previous results}.} Recall (\ref{equation: upper bounds of critical value for alpha}), (\ref{equation: H_1 alpha in (0,1)}), (\ref{equation: decmposition for the score of composite quantile  under H1}). Using the above bounds of $\tilde{\bC}^{I}_{\alpha}(t)$, $\bR(t)$, $\bC_{(0)}^{(3)}(t)$, $\bC_{(0)}^{(4)}(t)$, and by the triangle inequality, w.p.a.1, we have:
	\begin{equation}\label{inequality: lower bound of testing statistic alpha}\\
		\begin{array}{ll}
			T_\alpha&=\max\limits_{q_0\leq t\leq 1-q_0}\|\bC_\alpha(t)\|_{(s_0,2)}\\
			&\geq SNR(\alpha,\tilde{\btau})\times\max\limits_{q_0\leq t\leq 1-q_0}\|\bdelta(t)\|_{(s_0,2)}-\max\limits_{q_0\leq t\leq 1-q_0}\|\tilde{\bC}^{I}_{\alpha}(t)\|_{(s_0,2)}\\
			&-SNR(\alpha,\tilde{\btau})\times\max\limits_{q_0\leq t\leq 1-q_0}\|\bR(t)\|_{(s_0,2)}
			-(1-\alpha
			)\max\limits_{q_0\leq t\leq 1-q_0}\|\bC^{(3)}_0(t)\|_{(s_0,2)}\\
			&	-(1-\alpha)\max\limits_{q_0\leq t\leq 1-q_0}\|\bC^{(4)}_0(t)\|_{(s_0,2)}\\
			&\geq \sqrt{n}\times SNR(\alpha,\tilde{\btau})\times t_1(1-t_1)\times\|\bSigma(\bbeta^{(1)}-\bbeta^{(2)})\|_{(s_0,2)}
			-C_1s_0^{1/2}M\sqrt{\log(pn)}\\
			&\quad-C_2s_0^{1/2}s\sqrt{\log p}\|\bSigma(\bbeta^{(1)}-\bbeta^{(2)})\|_{(s_0,2)}-C_3(1-\alpha)s_0^{1/2}s^2\sqrt{\log(pn)} \|\bSigma({{\bbeta}^{(2)}}-{\bbeta}^{(1)})\|^2_{(s_0,2)}\\
			&\qquad\qquad-C_4(1-\alpha)Ms_0^{1/2}s\sqrt{\log(p)}\|\bSigma({{\bbeta}^{(2)}}-{\bbeta}^{(1)})\|_{(s_0,2)}\\
			&\geq \sqrt{n}\times SNR(\alpha,\tilde{\btau})\times t_1(1-t_1)\times\|\bSigma(\bbeta^{(1)}-\bbeta^{(2)})\|_{(s_0,2)}
			(1-\epsilon_n)-C_1s_0^{1/2}M\sqrt{\log(pn)},\\
		\end{array}
	\end{equation}
	where 
	\begin{equation*}
		\epsilon_n:=O\big(s^{1/2}_0s\sqrt{\dfrac{\log p}{n}}\big) \vee O\big({s_0^{1/2}s^2\sqrt{\dfrac{\log p}{n}}\|\bbeta^{(2)}-\bbeta^{(1)}\|_{(s_0,2)}}\big).
	\end{equation*}
	Hence,  considering (\ref{inequality: lower bound of testing statistic alpha=0}), to prove (\ref{equation: H_1 alpha=0}), it is sufficient to prove $H_1'\rightarrow 1$, where 
	\begin{equation*}
		\begin{array}{ll}
			H_1'=\P\Big(
			\sqrt{n}\times SNR(\alpha,\tilde{\btau})\times t_1(1-t_1)\times\|\bSigma(\bbeta^{(1)}-\bbeta^{(2)})\|_{(s_0,2)}\\
			\geq 
			\dfrac{Cs^{1/2}_0M\big(\sqrt{\log(pn)}+\sqrt{\log(1/\gamma)}\big)}{1-\epsilon_n}
			\Big).
		\end{array}
	\end{equation*}
	By (\ref{inequality: theoretical signal strengh}), it is straightforward to see that $H_1'\rightarrow 1$ as $n,p\rightarrow\infty$, which finishes the proof. 
	\begin{remark}
		Note that for $\alpha\in(0,1)$, if we replace $\sigma^2(\alpha,\tilde{\btau})$ by an estimator $\hat{\sigma}^2(\alpha,\tilde{\btau})$ which satisfies: $|\hat{\sigma}^2(\alpha,\tilde{\btau})-\sigma^2(\alpha,\tilde{\btau})|=o_p(1)$, then under condition (\ref{inequality: theoretical signal strengh}), the power still converges to 1.
	\end{remark}

\end{proof}

\section{Proofs of lemmas in Section \ref{sec:proof of main results}}\label{sec: proofs of lemmas used in the main theory}
\subsection{Proof of Lemma \ref{lemma: negligible for alpha=1}}\label{sec: Proof of negligible alpha=1}
\begin{proof}
	In this section, we prove Lemma \ref{lemma: negligible for alpha=1}. In other words, we will prove 
	\begin{equation}\label{equation: proof of negligible alpha=1}
		\P\big(\max_{q_0\leq t\leq 1-q_0} \big\|\bC_1(t)-\tilde{\bC}_1^{I}(t)\big\|_{(s_0,2)}\geq \epsilon \big)=o(1).
	\end{equation}
	Using the triangle inequality, we have
	\begin{equation}\label{inequality: D1+D2}
		\P\big(\max_{q_0\leq t\leq 1-q_0} \big\|\bC_1(t)-\tilde{\bC}_1^{I}(t)\big\|_{(s_0,2)}\geq \epsilon \big)\leq D_1+D_2,
	\end{equation}
	where $D_1$ and $D_2$ are defined as 
	\begin{equation}
		\begin{array}{l}
			D_1:=\P\big(\max\limits_{q_0\leq t\leq 1-q_0}\big\|\bC_1(t)-\bC_1^{I}(t)\big\|_{(s_0,2)}\geq \epsilon/2 \big),\\
			D_2:=\P\big(\max\limits_{q_0\leq t\leq 1-q_0}\big\|\bC_1^{I}(t)-\tilde{\bC}_1^{I}(t)\big\|_{(s_0,2)}\geq \epsilon/2 \big).
		\end{array}
	\end{equation}
	By (\ref{inequality: D1+D2}), to prove (\ref{equation: proof of negligible alpha=1}), we need to bound $D_1$ and $D_2$, respectively. \\
	{\bf{Step~1: obtain the upper bound for $D_1$.}} We first consider $D_1$. To this end, we define 
	\begin{equation}\label{equation: set for variance estimation}
		\cE=\big\{c^2_{\epsilon}/4\leq \hat{\sigma}^2\leq 4C^2_{\epsilon}  \big\}
	\end{equation}
	where $c_\epsilon$ and $C_{\epsilon}$ are in {\bf{Assumption B}.}  By introducing $\cE$, we have
	\begin{equation}
		D_1\leq \P\big(\max\limits_{q_0\leq t\leq 1-q_0}\big\|\bC_1(t)-\bC_1^{I}(t)\big\|_{(s_0,2)}\geq \epsilon/2\cap\cE \big)+\P(\cE^c).
	\end{equation}
	By {Theorem \ref{theorem: variance estimator under H0}}, we have $\P(\cE^c)=o(1)$ as $n,p\rightarrow\infty$. Under the event $\cE$, we have
	\begin{equation}\label{inequality: basic inequality for D1}
		\begin{array}{ll}
			\P\big(\max\limits_{q_0\leq t\leq 1-q_0}\big\|\bC_1(t)-\bC_1^{I}(t)\big\|_{(s_0,2)}\geq \epsilon/2\cap\cE \big)\\
			= \P\big(\max\limits_{q_0\leq t\leq 1-q_0}\big\|\bC_1^{II}(t)\big\|_{(s_0,2)}\geq \epsilon/2\cap\cE \big),
		\end{array}
	\end{equation}
	where $\bC_1^{II}(t)$ is defined in (\ref{equation: C11+C12}). 
	Hence, under the event $\cE$, we have
	\begin{equation}\label{inequality: D1-2}
		\small
		\begin{array}{lc}
			\P\big(\max\limits_{q_0\leq t\leq 1-q_0}\big\|\bC_1^{II}(t)\big\|_{(s_0,2)}\geq \dfrac{\epsilon}{2}\cap\cE \big)
			\\
			=_{(1)}\P\Big(\max\limits_{t}\big\|\dfrac{1}{\sqrt{n}}\Big(\sum\limits_{i=1}^{\floor{nt}}\bX_i\bX_i^\top(\bbeta^{(0)}-\hat{\bbeta})-\dfrac{\floor{nt}}{n}\sum\limits_{i=1}^n\bX_i\bX_i^\top(\bbeta^{(0)}-\hat{\bbeta})\big\|_{(s_0,2)}\geq \dfrac{c_{\epsilon}\epsilon}{4}\cap\cE\Big)\\
			\leq_{(2)} \P\Big(\max\limits_{t}\big\|\dfrac{\floor{nt}}{\sqrt{n}}\big(\hat{\bSigma}(1:t)-\hat{\bSigma}(1:n)\big)(\bbeta^{(0)}-\hat{\bbeta})\big\|_{(s_0,2)}\geq \dfrac{c_{\epsilon}\epsilon}{4}\cap\cE\Big)\\
			\leq_{(3)} \P\Big(\max\limits_{t}\big\|\dfrac{\floor{nt}}{\sqrt{n}}\big(\hat{\bSigma}(1:t)-\hat{\bSigma}(1:n)\big)(\bbeta^{(0)}-\hat{\bbeta})\big\|_{\infty}\geq s_0^{-1/2}\dfrac{c_{\epsilon}\epsilon}{4}\cap\cE\Big)\\
			\leq_{(4)} \P\Big(\max\limits_{t}\big\|\dfrac{\floor{nt}}{\sqrt{n}}\big(\hat{\bSigma}(1:t)-\hat{\bSigma}(1:n)\big)\|_{\infty}\|(\bbeta^{(0)}-\hat{\bbeta})\big\|_{1}\geq s_0^{-1/2}\dfrac{c_{\epsilon}\epsilon}{4}\Big)\\
			\leq_{(5)} \P\Big(\max\limits_{t}\big\|\big(\hat{\bSigma}(1:t)-\hat{\bSigma}(1:n)\big)\|_{\infty}\|(\bbeta^{(0)}-\hat{\bbeta})\big\|_{1}\geq n^{-1/2}s_0^{-1/2}\dfrac{c_{\epsilon}\epsilon}{4}\Big),\\
		\end{array}
	\end{equation}
	where  $(3)$ comes from the fact that $\|\bv\|_{(s_0,2)}\leq s_0^{1/2}\|\bv\|_{\infty}$ for any $\bv\in \RR^p$, $(4)$ comes from the fact that $\|\Ab\bv\|_{\infty}\leq \|\Ab\|_{\infty}\|\bv\|_1$ for any matrix $\Ab$ and vector $\bv$. By  Lemma \ref{lemma: concentration for covariance}, we have $\max\limits_{t}\big\|\big(\hat{\bSigma}(1:t)-\hat{\bSigma}(1:n)\big)\|_{\infty}=O_p(M^2\sqrt{\log(p)/n})$. Moreover, under $\Hb_0$, for the lasso estimator $\hat{\bbeta}$, using Lemma \ref{lemma: basic inequality for lasso with alpha=0}, we have $\|\hat{\bbeta}-\bbeta^0\|_1=O_p(s\sqrt{\log(p/n)})$. Hence, combining (\ref{inequality: D1-2}) and letting $\epsilon:=Cs_0^{1/2}sM^2{\log(p)}/{\sqrt{n}}$ for some big enough constant $C>0$, we have:
	\begin{equation*}
		\P\big(\max\limits_{q_0\leq t\leq 1-q_0}\big\|\bC_1^{II}(t)\big\|_{(s_0,2)}\geq \dfrac{\epsilon}{2}\cap\cE \big)=o(1).
	\end{equation*}
	{\bf{Step~2: obtain the upper bound for $D_2$.}}  By definition, we have
	\begin{equation*}
		\begin{array}{ll}
			D_2&:=\P\Big(\max\limits_{q_0\leq t\leq 1-q_0}\big\|\bC_1^{I}(t)-\tilde{\bC}_1^{I}(t)\big\|_{(s_0,2)}\geq \epsilon/2 \Big)\\
			&=\P\Big(\big|\dfrac{1}{\hat{\sigma}}-\dfrac{1}{\sigma}\big|\max\limits_{q_0\leq t\leq 1-q_0}\big\|\dfrac{1}{\sqrt{n}}\big(\sum\limits_{i=1}^{\floor{nt}}\bX_i\epsilon_i-\dfrac{\floor{nt}}{n}\sum\limits_{i=1}^n\bX_i\epsilon_i\big)\big\|_{(s_0,2)}\geq \epsilon/2\Big)\\
			&=\P\Big(\underbrace{\big|\dfrac{\sigma}{\hat{\sigma}}-1\big|}_{I_1}\underbrace{\max\limits_{q_0\leq t\leq 1-q_0}\big\|\dfrac{1}{\sqrt{n}\sigma}\big(\sum\limits_{i=1}^{\floor{nt}}\bX_i\epsilon_i-\dfrac{\floor{nt}}{n}\sum\limits_{i=1}^n\bX_i\epsilon_i\big)\big\|_{(s_0,2)}}_{I_2}\geq \epsilon/2\Big).
		\end{array}
	\end{equation*}
	Hence, to bound $D_2$, we need to bound $I_1$ and $I_2$, respectively. To bound $I_1$, define $\tilde{I}_1=|1-\dfrac{\hat{\sigma}}{\sigma}|$. Using the fact that $a^2-b^2=(a-b)(a+b)$, we have:
	\begin{equation*}
		\tilde{I}_1=\Big|\dfrac{\hat{\sigma}^2-\sigma^2}{\sigma(\sigma+\hat{\sigma})}\Big|\leq_{(1)} \Big|\dfrac{\hat{\sigma}^2-\sigma^2}{\sigma^2}\Big|\leq_{(2)} C_1\Big|\hat{\sigma}^2-\sigma^2\Big|\leq_{(3)}\leq C_2\sqrt{s\dfrac{\log(pn)}{n}}, 
	\end{equation*}
	where $(2)$ comes from {\bf{Assumption B}}, $(3)$ comes from {Theorem \ref{theorem: variance estimator under H0}}. By Lemma C.1 in \cite{Zhou2017An}, we have: $I_1\leq C \tilde{I}_1$. Next, we consider $I_2$. Using Lemma \ref{lemma: exponential inequality for partial sum process},  and the fact that $\|\bv\|_{(s_0,2)}\leq s_0^{1/2}\|\bv\|_{\infty}$ for any $\bv\in \RR^p$, we have:
	\begin{equation*}
		I_2=O_p\big( Ms_0^{1/2}\sqrt{\log(pn)}\big).
	\end{equation*}
	Hence, we have $I_1I_2=O_p(s_0^{1/2}s^{1/2}M\dfrac{\log(pn)}{\sqrt{n}})$.
	
	Lastly, combining Steps 1 and 2, if we choose ${\epsilon:=Cs_0^{1/2}sM^2{\log(p)}/{\sqrt{n}}}$ for some big constant $C>0$, we have $D_1+D_2=o(1)$, which finishes the proof.
\end{proof}

\subsection{Proof of Lemma \ref{lemma: Gaussian approximation 2}}
\label{sec: proof of Gaussian approximation 2 with alpha=1}
\begin{proof}
	In this section, we aim to prove $\sup_{z>0}I_z=o_p(1)$, where
	\begin{equation*}
		I_z:= \big|\P(\max_{q_0\leq t\leq 1-q_0}\| \bC_1^{\bG}(t)\|_{(s_0,2)}> z\big)-\P(\max_{q_0\leq t\leq 1-q_0}\|\bC_1^b(t)\|_{(s_0,2)}> z|\mathcal{X}\big)\big|.
	\end{equation*}
	To that end,  let $R=Cs_0n$ and $L=\sup_{z\in(0,+\infty)}I_z$.  Then, we can write 
	$L$ as $$L=\max(L_1,L_2),$$ where $L_1=\sup_{z\in(0,R]}I_z$ and $L_2=\sup_{z\in(R,\infty)}I_z$. Therefore, to prove $L=o_p(1)$, we need to bound $L_1$ and $L_2$, respectively. We first bound $L_2=\sup_{z\in(R,\infty)}I_z$. Considering that for any $\bv\in \RR^p$, $\|\bv\|_{(s_0,2)}\leq s_0^{1/2}\|\bv\|_{\infty}\leq s_0\|\bv\|_{\infty}$ holds, we have 
	\begin{equation*}
		L_2=\sup_{z\in(R,\infty)}I_z\leq \P(\max_{q_0\leq t\leq 1-q_0}\|\bC_1^{\bG}(t)\|_{\infty}>Cn\big)+\P(\max_{q_0\leq t\leq 1-q_0}\|\bC_1^b(t)\|_{\infty}>Cn|\mathcal{X}\big).
	\end{equation*}
	By the exponential inequality and similar to the proof of Lemma \ref{lemma: exponential inequality for partial sum process}, we can prove that  
	\begin{equation*}
		\max_{q_0\leq t\leq 1-q_0}\|\bC_1^{\bG}(t)\|_{\infty}=O_p(M\sqrt{\log(pn)}), \text{and} \max_{q_0\leq t\leq 1-q_0}\|\bC_1^{b}(t)\|_{\infty}|\cX=O_p(M\sqrt{\log(pn)}),
	\end{equation*}
	which yields
	\begin{equation}\label{inequality: Iz: R to infinity}
		L_2=\sup_{z\in(R,\infty)}I_z=o_p(1).
	\end{equation}
	
	After bounding $L_2$ in (\ref{inequality: Iz: R to infinity}), we now bound $L_1:=\sup_{z\in(0,R]}I_z$. Let $\mathcal{E}^{R,p}=\{\bx\in \mathbb{R}^{p}: \|\bx\|\leq R\}$ and $V_{(s_0,2)}^{z,p}=\{\bx\in \mathbb{R}^{p}: \|\bx\|_{(s_0,2)}\leq z\}$. Considering $\|\bx\|\leq p^{1/2} \|\bx\|_{\infty}\leq p^{1/2} \|\bx\|_{(s_0,2)}$ for any $\bx\in\RR^p$, we have $V_{(s_0,2)}^{z,p}\subset \mathcal{E}^{Rp^{1/2},p}$ for $z\leq R$.  Therefore, considering Lemma \ref{lemma: convex approximation}, there is a $m$-generated convex set $A^{m}$ and a $\epsilon>0$ such that
	\begin{equation}\label{subset: A-m}
		A^{m}\subset V_{(s_0,2)}^{z,p}\subset A^{m,Rp^{1/2}\epsilon} ~~\text{and}~~ m\leq p^{s_0}(\frac{\gamma}{\sqrt{\epsilon}}\ln(\dfrac{1}{\epsilon}))^{s_0^2} ~~\big(\text{by} ~V_{(s_0,2)}^{z,p}\subset \mathcal{E}^{Rp^{1/2},p}~\text{for}~z\leq R\big).
	\end{equation}
	By setting $\epsilon=(pn)^{-3/2}$, we have $\epsilon'=Rp^{1/2}\epsilon=Cs_0p^{-1}n^{-1/2}$. By (\ref{subset: A-m}), for $z\in(0,R]$, we  have  $$I_z\leq I_{z,1}+I_{z,2},$$ where 
	\begin{equation*}
		\begin{array}{ll }
			I_{z,1}=&\max\Big(\P\big(\bigcap\limits_{q_0\leq t\leq 1-q_0}\bC_{1}^{\bG}(t)\in A^{m,\epsilon'}\setminus A^{m}\big),\P\big(\bigcap\limits_{q_0\leq t\leq 1-q_0}\bC_1^b(t)\in A^{m,\epsilon'}\setminus A^{m}|\cX\big)\Big),\\\\
			I_{z,2}=&\max\Big(\big|\P\big(\bigcap\limits_{q_0\leq t\leq 1-q_0}\bC_{1}^{\bG}(t)\in A^{m,\epsilon'})-\P\big(\bigcap\limits_{q_0\leq t\leq 1-q_0}\bC_1^b(t)\in A^{m,\epsilon'}|\cX\big)\big|,\\
			&\big|\P\big(\bigcap\limits_{q_0\leq t\leq 1-q_0}\bC_{1}^{\bG}(t)\in A^{m})-\P\big(\bigcap\limits_{q_0\leq t\leq 1-q_0}\bC_1^b(t)\in A^{m}|\cX\big)\big|\Big).
		\end{array}
	\end{equation*}
	\par Next, we consider $I_{z,1}$ and $I_{z,2}$, respectively. Recall $\epsilon'=Cs_0p^{-1}n^{-1/2}$. For $I_{z,1}$, by Lemma \ref{lemma:anti consentration inequality} and the definitions of $A^m$ and $ A^{m,\epsilon'}$ in (\ref{equation: Ame}) and (\ref{subset: A-m}), $\text{for~all}~z\in(0,R]$, we have
	\begin{equation}\label{inequality: Iz1}
		\begin{array}{ll}
			I_{z,1}&\leq Cs_0p^{-1}n^{-1/2}\sqrt{\log(m(n-2\floor{nq_0}))}\leq Cs_0^2p^{-1}n^{-1/2}\sqrt{\log(pn)}=o_p(1).
		\end{array}
	\end{equation}
	
	Recall $\mathcal{V}_{s_0}:=\{\bv\in \mathbb{S}^{q-1}: \|\bv\|=1, \|\bv\|_{0}\leq s_0\}$ and $\hat{\bSigma}(0:t)$ defined in (\ref{equation: big sigma}). We then have
	\begin{equation}\label{inequality: large deviation hyperrectangle1}
		\begin{array}{ll}
			&\sup\limits_{q_0\leq t_1,t_2\leq 1-q_0}\sup\limits_{\bv_1,\bv_2\in \mathcal{V}_{s_0}}\Big|\bv_1^\top\Big(\E\big[\bC_{1}^{\bG}(t_1)\bC_1^{\bG}(t_2)^\top\big]-\E\big[\bC_1^b(t_1)\bC^b(t_2)^\top\big|\cX]\Big)\bv_2\Big|\\
			&\quad\leq_{(1)} \sup\limits_{q_0\leq t_1,t_2\leq 1-q_0}\Big|\E\big[\bC_{1}^{\bG}(t_1)\bC_1^{\bG}(t_2)^\top\big]-\E\big[\bC_1^b(t_1)\bC^b(t_2)^\top\big|\cX]\Big\|_{\infty}\big\|\bv_1\big\|_{1}\big\|\bv_2\big\|_1\\
			&\quad\leq_{(2)} s_0\sup\limits_{q_0\leq t_1,t_2\leq 1-q_0}
			\big\|\min(t_1,t_2)(\hat{\bSigma}(0:\min(t_1,t_2))-\bSigma)\\
			&\quad\qquad\qquad-t_1t_2(\hat{\bSigma}(0:t_1)-\bSigma)-t_1t_2(\hat{\bSigma}(0:t_2)-\bSigma)+t_1t_2(\hat{\bSigma}(0:1)-\bSigma)\big\|_{\infty},
		\end{array}
	\end{equation}
	the last inequality in (\ref{inequality: large deviation hyperrectangle1}) comes from the Cauchy-Schwartz inequality, and the fact $\bv_1,\bv_2\in \mathcal{V}_{s_0}$. Therefore, based on (\ref{inequality: large deviation hyperrectangle1}), using Theorem 4.1 and Remark 4.1 in \cite{chernozhukov2017central} and Lemma \ref{lemma: concentration for covariance}, with probability tending to one, we have
	\begin{equation}\label{inequality: Iz2}
		I_{z,2}\leq C \Big(s_0 M^2\sqrt{\dfrac{\log(pn)}{n}}\Big)^{1/3}\log^{2/3}\big(m(n-2\floor{nq_0})\big)\leq C \Big(\dfrac{s_0^{10}\log^7(pn)}{n}\Big)^{1/6}.
	\end{equation}
	Considering (\ref{inequality: Iz2}), by {\bf{Assumptions A, E.1}}, we have $I_{z,2}=o_p(1)$ for all $z\in (0,R]$.
	
	Finally, combining (\ref{inequality: Iz: R to infinity}), (\ref{inequality: Iz1}), and (\ref{inequality: Iz2}), we have $I_z=o_p(1)$ uniformly holds for $z\geq 0$, which finishes the proof for Lemma \ref{lemma: Gaussian approximation 2}.

\end{proof}
\subsection{Proof of Lemma \ref{lemma: negligible for alpha=0}}\label{sec: Proof of negligible alpha=0}
\begin{proof}
	In this section, we prove Lemma \ref{lemma: negligible for alpha=0}. In other words, we aim to prove 
	\begin{equation}\label{equation: proof of negligible alpha=0}
		\P\Big(\max_{q_0\leq t\leq 1-q_0} \big\|\bC_0(t)-\tilde{\bC}_0^{I}(t)\big\|_{(s_0,2)}\geq \epsilon \Big)=o(1),
	\end{equation}
	where $\bC_{0}(t)$ is defined in (\ref{equation: C0t}), and $\tilde{\bC}_0^{I}(t)$ is defined in 
	(\ref{equation: tilde-C0-I}). Using the triangle inequality, we have
	\begin{equation}\label{inequality: D1+D2 alpha=0}
		\P\big(\max_{q_0\leq t\leq 1-q_0} \big\|\bC_0(t)-\tilde{\bC}_0^{I}(t)\big\|_{(s_0,2)}\geq \epsilon \big)\leq D_1+D_2,
	\end{equation}
	where $D_1$ and $D_2$ are defined as
	\begin{equation}
		\begin{array}{l}
			D_1:=\P\big(\max\limits_{q_0\leq t\leq 1-q_0}\big\|\bC_0(t)-\bC_0^{I}(t)\big\|_{(s_0,2)}\geq \epsilon/2 \big),\\
			D_2:=\P\big(\max\limits_{q_0\leq t\leq 1-q_0}\big\|\bC_0^{I}(t)-\tilde{\bC}_0^{I}(t)\big\|_{(s_0,2)}\geq \epsilon/2 \big).
		\end{array}
	\end{equation}
	By (\ref{inequality: D1+D2 alpha=0}), to prove (\ref{equation: proof of negligible alpha=0}), we need to bound $D_1$ and $D_2$, respectively. \\
	{\bf{Step~1: obtain the upper bound for $D_1$.}} We first consider $D_1$. To this end, we define 
	\begin{equation}\label{equation: set for variance estimation alpha=0}
		\cE=\big\{\sigma^2/2\leq \hat{\sigma}^2\leq 2\sigma^2  \big\},
	\end{equation}
	where $\sigma^2:=\text{Var}[e_i(\tilde{\btau})]$ is the true variance.  By introducing $\cE$, we have
	\begin{equation}
		D_1\leq \P\big(\max\limits_{q_0\leq t\leq 1-q_0}\big\|\bC_0(t)-\bC_0^{I}(t)\big\|_{(s_0,2)}\geq \epsilon/2\cap\cE \big)+\P(\cE^c).
	\end{equation}
	By {Theorem \ref{theorem: variance estimator under H0}}, we have $\P(\cE^c)=o(1)$ as $n,p\rightarrow\infty$. Under the event $\cE$, we have
	\begin{equation}\label{inequality: basic inequality for D1 alpha=0}
		\begin{array}{ll}
			\P\big(\max\limits_{q_0\leq t\leq 1-q_0}\big\|\bC_0(t)-\bC_0^{I}(t)\big\|_{(s_0,2)}\geq \epsilon/2\cap\cE \big)\\
			\quad= \P\big(\max\limits_{q_0\leq t\leq 1-q_0}\big\|\bC_0^{II}(t)\big\|_{(s_0,2)}\geq \epsilon/2\cap\cE \big),
		\end{array}
	\end{equation}
	where $\bC_0^{II}(t)$ is defined in (\ref{equation: C11+C12 alpha=0}). Before controlling $\bC_0^{II}(t)$, given $\cX=(\bX_1,\ldots,\bX_n)$, we need to decompose $\bC_0^{II}(t)|\cX$ into two terms: 
	\begin{equation}
		\bC_0^{II}(t)=\bC_0^{II,1}(t)+\bC_0^{II,2}(t),
	\end{equation}
	where $\bC_0^{II,1}(t)$ and $\bC_0^{II,2}(t)$ are defined as 
	\begin{equation}\label{equation: C02 alpha=0 two parts}
		\begin{array}{ll}
			\bC^{II,1}_{0}(t)=\dfrac{1}{\sqrt{n}\hat{\sigma}(0,\tilde{\btau})}\big(\sum\limits_{i=1}^{\floor{nt}}\dfrac{1}{K}\sum\limits_{k=1}^K\bX_i\big(\E [\mathbf{1}\{\epsilon_i\leq \uwave{\bX_i}^\top\uwave{\hat{\bDelta}_k}+b_k^{(0)}\}]-\E[\mathbf{1}\{\epsilon_i\leq b_k^{(0)} \}]\big)\\
			\qquad\qquad\qquad\qquad-\dfrac{\floor{nt}}{n}\sum\limits_{i=1}^n\dfrac{1}{K}\sum\limits_{k=1}^K\bX_i\big(\E[\mathbf{1}\{\epsilon_i\leq \uwave{\bX_i}^\top\uwave{\hat{\bDelta}_k}+b_k^{(0)}\}]-\E[\mathbf{1}\{\epsilon_i\leq b_k^{(0)} \}]\big)\big),\\
			\bC^{II,2}_{0}(t)=\dfrac{1}{\sqrt{n}\hat{\sigma}(0,\tilde{\btau})}\big(\sum\limits_{i=1}^{\floor{nt}}\dfrac{1}{K}\sum\limits_{k=1}^K\bX_i\big(g_{ik}(\uwave{\bX_i}^\top\uwave{\hat{\bDelta}_k})-g_{ik}(0)\big)\\
			\qquad\qquad\qquad\qquad-\dfrac{\floor{nt}}{n}\sum\limits_{i=1}^n\dfrac{1}{K}\sum\limits_{k=1}^K\bX_i\big(g_{ik}(\uwave{\bX_i}^\top\uwave{\hat{\bDelta}_k})-g_{ik}(0)\big)\big),
		\end{array}
	\end{equation}
	where $g_{ik}(t):=\mathbf{1}\{\epsilon_i\leq b_k^{(0)}+t\}-\P\{\epsilon_i\leq b_k^{(0)}+t\}$. Next, we control $\bC^{II,1}_{0}(t)$ and $\bC^{II,2}_{0}(t)$, respectively. 
	
	Let $F_{\epsilon}(t):=\P(\epsilon\leq t)$ be the CDF for $\epsilon$ and $f_{\epsilon}$ be its density function. For $\bC^{II,1}_{0}(t)$, by its definition, we have:
	\begin{equation}\label{equation: C02 alpha=0 two parts }
		\begin{array}{cc}
			\bC^{II,1}_{0}(t)=\dfrac{1}{\sqrt{n}\hat{\sigma}(0,\tilde{\btau})}\Big(\sum\limits_{i=1}^{\floor{nt}}\dfrac{1}{K}\sum\limits_{k=1}^K\bX_i\big(F_\epsilon( \uwave{\bX_i}^\top\uwave{\hat{\bDelta}_k}+b_k^{(0)})-F_{\epsilon}( b_k^{(0)})\big)\\
			\qquad\qquad\qquad\qquad\qquad\qquad-\dfrac{\floor{nt}}{n}\sum\limits_{i=1}^n\dfrac{1}{K}\sum\limits_{k=1}^K\bX_i\big(F_\epsilon( \uwave{\bX_i}^\top\uwave{\hat{\bDelta}_k}+b_k^{(0)})-F_{\epsilon}( b_k^{(0)})\big)\Big).\\
		\end{array}
	\end{equation}
	Using Taylor's expansion, we have:
	\begin{equation*}
		\begin{array}{ll}
			&F_\epsilon(\uwave{\bX_i}^\top\uwave{\hat{\bDelta}_k}+b_k^{(0)})-F_{\epsilon}( b_k^{(0)})\\
			&\quad=f_{\epsilon}(b_{k}^{(0)})\uwave{\bX_i}^\top\uwave{\hat{\bDelta}_k}+\dfrac{1}{2}f'_{\epsilon}(\xi_{ik})(\uwave{\hat{\bDelta}_k}^\top\uwave{\bX_i})^2\\
			&\quad=f_{\epsilon}(b_{k}^{(0)})\bX_i^\top\hat{{\bDelta}}+f_{\epsilon}(b_{k}^{(0)})(\hat{b_k}-b_{k}^{(0)})+\dfrac{1}{2}f'_{\epsilon}(\xi_{ik})(\uwave{\hat{\bDelta}_k}^\top\uwave{\bX_i})^2\\
			&\quad=f_{\epsilon}(b_{k}^{(0)})\bX_i^\top(\hat{\bbeta}-\bbeta^{(0)})+f_{\epsilon}(b_{k}^{(0)})(\hat{b_k}-b_{k}^{(0)})+\dfrac{1}{2}f'_{\epsilon}(\xi_{ik})(\uwave{\hat{\bDelta}_k}^\top\uwave{\bX_i})^2,
		\end{array}
	\end{equation*}
	where $\xi_{ik}$ is some random variable between $b_k^{(0)}$ and $\uwave{\bX_i}^\top\uwave{\hat{\bDelta}_k}+b_k^{(0)}$. Hence, by the above expansion, $\bC^{II,1}_{0}(t)$ can be decomposed into three terms:
	\begin{equation}\label{equation: decomposition of C_{II,1}}
		\bC^{II,1}_{0}(t)=\bC^{II,1,1}_{0}(t)+\bC^{II,1,2}_{0}(t)+\bC^{II,1,3}_{0}(t),
	\end{equation}
	where $\bC^{II,1,1}_{0}(t) - \bC^{II,1,3}_{0}(t)$ are defined as
	\begin{equation}\label{equation: CII11}
		\begin{array}{ll}
			\bC^{II,1,1}_{0}(t)
			=\dfrac{1}{\sqrt{n}\hat{\sigma}(0,\tilde{\btau})}\big(\sum\limits_{i=1}^{\floor{nt}}\bX_i\bX_i^\top\dfrac{1}{K}\sum\limits_{k=1}^Kf_{\epsilon}(b_{k}^{(0)})(\hat{\bbeta}-\bbeta^{(0)})\big)\\
			\qquad\qquad\qquad-\dfrac{\floor{nt}}{n}\sum\limits_{i=1}^n\bX_i\bX_i^\top\dfrac{1}{K}\sum\limits_{k=1}^Kf_{\epsilon}(b_{k}^{(0)})(\hat{\bbeta}-\bbeta^{(0)})\big),\\
			\bC^{II,1,2}_{0}(t)
			=\dfrac{1}{\sqrt{n}\hat{\sigma}(0,\tilde{\btau})}\big(\sum\limits_{i=1}^{\floor{nt}}\bX_i\dfrac{1}{K}\sum\limits_{k=1}^Kf_{\epsilon}(b_{k}^{(0)})(\hat{b}_k-b_k^{(0)})\big)\\
			\qquad\qquad\qquad-\dfrac{\floor{nt}}{n}\sum\limits_{i=1}^n\bX_i\dfrac{1}{K}\sum\limits_{k=1}^Kf_{\epsilon}(b_{k}^{(0)})(\hat{b}_k-b_k^{(0)})\big)\big),\\
			\bC^{II,1,3}_{0}(t)
			=\dfrac{1}{\sqrt{n}\hat{\sigma}(0,\tilde{\btau})}\big(\sum\limits_{i=1}^{\floor{nt}}\bX_i\dfrac{1}{K}\sum\limits_{k=1}^K\dfrac{1}{2}f'_{\epsilon}(\xi_{ik})(\uwave{\hat{\bDelta}_k}^\top\uwave{\bX_i})^2\big)\\
			\qquad\qquad\qquad-\dfrac{\floor{nt}}{n}\sum\limits_{i=1}^n\bX_i\dfrac{1}{K}\sum\limits_{k=1}^K\dfrac{1}{2}f'_{\epsilon}(\xi_{ik})(\uwave{\hat{\bDelta}_k}^\top\uwave{\bX_i})^2\big)\big).\\
		\end{array}
	\end{equation}
	Hence, to bound $\bC^{II,1}_{0}(t)$, we need to bound $\bC^{II,1,1}_{0}(t)-\bC^{II,1,3}_{0}(t)$ respectively. For $\bC^{II,1,1}_{0}(t)$, under the event $\cE$, with probability tending to 1, we have:
	\begin{equation}\label{inequality: upper bound for C{II,1,1}}
		\begin{array}{ll}
			&\max\limits_{t\in[q_0,1-q_0]}\|\bC^{II,1,1}_{0}(t)\|_{(s_0,2)}\\
			&\quad\leq_{(1)} Cs_0^{1/2} \max\limits_{t}\big\|\dfrac{\floor{nt}}{\sqrt{n}}\big(\hat{\bSigma}(1:t)-\hat{\bSigma}(1:n)\big)\dfrac{1}{K}\sum\limits_{k=1}^Kf_{\epsilon}(b_{k}^{(0)})(\hat{\bbeta}-\bbeta^{(0)})\big\|_{\infty}\\
			&\quad\leq_{(2)} Cs_0^{1/2}\max\limits_{t}\big\|\dfrac{\floor{nt}}{\sqrt{n}}\big(\hat{\bSigma}(1:t)-\hat{\bSigma}(1:n)\big)\|_{\infty}\times \|\dfrac{1}{K}\sum\limits_{k=1}^Kf_{\epsilon}(b_{k}^{(0)})(\hat{\bbeta}-\bbeta^{(0)})\big\|_{1}\\
			&\quad\leq_{(3)} Cs_0^{1/2}M^2\sqrt{\log(p)}\|\dfrac{1}{K}\sum\limits_{k=1}^Kf_{\epsilon}(b_{k}^{(0)})(\hat{\bbeta}-\bbeta^{(0)})\big\|_{1}\\
			&\quad\leq_{(4)} Cs_0^{1/2}M^2\sqrt{\log(p)}\|(\hat{\bbeta}-\bbeta^{(0)})\big\|_{1}\\
			&\quad\leq_{(5)} Cs_0^{1/2}M^2\sqrt{\log(p)}s\sqrt{\log(p)/n}\\
		\end{array}
	\end{equation}
	where $(1)$ comes from (\ref{equation: set for variance estimation alpha=0}), $(3)$ comes from Lemma \ref{lemma: concentration for covariance}, $(4)$ comes from Assumption D.2, and $(5)$ comes from Lemma \ref{lemma: basic inequality for lasso with alpha=0}. With a similar procedure,  we can prove that
	\begin{equation}\label{inequality: upper bound for C{II,1,2}}
		\max\limits_{t\in[q_0,1-q_0]}\|\bC^{II,1,2}_{0}(t)\|_{(s_0,2)}=O_p(s_0^{1/2}sM^2\log(p)/\sqrt{n}). 
	\end{equation}
	Next, we consider $\bC^{II,1,3}_{0}(t)$. Using $(a+b)^2\leq 2(a^2+b^2)$, we have
	\begin{equation}\label{inequality: talor's residual}
		\Big|\dfrac{1}{K}\sum\limits_{k=1}^K\dfrac{1}{2}f'_{\epsilon}(\xi_{ik})(\uwave{\hat{\bDelta}_k}^\top\uwave{\bX_i})^2\Big|\leq \dfrac{C'_+}{2K}\sum_{k=1}^K(\hat{\delta}_k+\bX_i^\top\hat{\bDelta})^2\leq C'_+\big(\dfrac{1}{K}\sum_{k=1}^K\hat{\delta}_k^2+{\hat{\bDelta}}^\top\bX_i\bX_i^\top{\hat{\bDelta}}\big),
	\end{equation}
	where $\hat{\delta}_k=\hat{b}_k-b^{(0)}$, $\hat{\bDelta}=\hat{\bbeta}-\bbeta^{(0)}$. Hence, using the above result, under $\cE$, we have
	\begin{equation}\label{inequality: upper bound for C{II,1,3}}
		\begin{array}{ll}
			\max\limits_{t\in[q_0,1-q_0]}\|\bC^{II,1,3}_{0}(t)\|_{(s_0,2)}\\
			\quad\leq_{(1)} Cs_0^{1/2}\Big\|\max\limits_{t}\dfrac{\floor{nt}}{\sqrt{n}}\big(\dfrac{1}{\floor{nt}}\sum\limits_{i=1}^{\floor{nt}}\bX_i \dfrac{1}{K}\sum\limits_{k=1}^K\dfrac{1}{2}f'_{\epsilon}(\xi_{ik})(\uwave{\hat{\bDelta}_k}^\top\uwave{\bX_i})^2\big) \Big\|_{\infty}\\
			\qquad\qquad\qquad+Cs_0^{1/2}\Big\|\max\limits_{t}\dfrac{\floor{nt}}{\sqrt{n}}\big(\dfrac{1}{n}\sum\limits_{i=1}^{n}\bX_i \dfrac{1}{K}\sum\limits_{k=1}^K\dfrac{1}{2}f'_{\epsilon}(\xi_{ik})(\uwave{\hat{\bDelta}_k}^\top\uwave{\bX_i})^2\big) \Big\|_{\infty}\\
			\quad\leq_{(2)} C\sqrt{n}s_0^{1/2}\max\limits_{t}\Big(\dfrac{1}{\floor{nt}}\sum\limits_{i=1}^{\floor{nt}}\|\bX_i\|_{\infty} \big|\dfrac{1}{K}\sum\limits_{k=1}^K\dfrac{1}{2}f'_{\epsilon}(\xi_{ik})(\uwave{\hat{\bDelta}_k}^\top\uwave{\bX_i})^2\big|\Big) \\
			\qquad\qquad\qquad +C\sqrt{n}s_0^{1/2}\Big(\dfrac{1}{n}\sum\limits_{i=1}^{n}\|\bX_i\|_{\infty} \big|\dfrac{1}{K}\sum\limits_{k=1}^K\dfrac{1}{2}f'_{\epsilon}(\xi_{ik})(\uwave{\hat{\bDelta}_k}^\top\uwave{\bX_i})^2\big|\Big) \\
			\quad\leq_{(3)} CC'_{+}\sqrt{n}Ms_0^{1/2}(\dfrac{1}{K}\|\hat{\bdelta}\|^2+\max\limits_{t}{\hat{\bDelta}}^\top\hat{\bSigma}(0:t){\hat{\bDelta}}) \\
			\qquad\quad\qquad +CC'_{+}\sqrt{n}Ms_0^{1/2}(\dfrac{1}{K}\|\hat{\bdelta}\|^2+{\hat{\bDelta}}^\top\hat{\bSigma}(0:1){\hat{\bDelta}}) \\
			\quad\leq_{(4)} CC'_{+}\sqrt{n}Ms_0^{1/2}(\dfrac{1}{K}\|\hat{\bdelta}\|^2+\max\limits_{t}|{\hat{\bDelta}}^\top(\hat{\bSigma}(0:t)-\bSigma){\hat{\bDelta}}|+\hat{\bDelta}^\top\bSigma\hat{\bDelta}\\
			\qquad\quad\qquad +CC'_{+}\sqrt{n}Ms_0^{1/2}(\dfrac{1}{K}\|\hat{\bdelta}\|^2+|{\hat{\bDelta}}^\top(\hat{\bSigma}(0:1)-\bSigma){\hat{\bDelta}})|+\hat{\bDelta}^\top\bSigma\hat{\bDelta} \\
			\quad \leq_{(5)}CC'_{+}\sqrt{n}Ms_0^{1/2}\Big(\dfrac{1}{K}\|\hat{\bdelta}\|^2+\lambda_{\rm max}(\bSigma)\|\hat{\bDelta}\|^2+M^2s\sqrt{\dfrac{\log(pn)}{n}}\|\hat{\bDelta}\|^2\Big)\\
			\quad \leq_{(6)} Cs_0^{1/2}sM^2\log(p)/\sqrt{n}.
		\end{array}
	\end{equation}
	where $(1)$ comes from (\ref{equation: set for variance estimation alpha=0}) and the triangle inequality, $(2)$ comes from $\max_{t}\floor{nt}/\sqrt{n}\leq \sqrt{n}$  ,$(3)$ comes from {\bf{Assumption D}} and (\ref{inequality: talor's residual}), and $(5)$ comes from the fact that $|{\hat{\bDelta}}^\top(\hat{\bSigma}(0:1)-\bSigma){\hat{\bDelta}})|\leq \|{\hat{\bDelta}}\|^2_1\|\hat{\bSigma}(0:1)-\bSigma\|_{\infty}\leq s\|{\hat{\bDelta}}\|^2_2\|\hat{\bSigma}(0:1)-\bSigma\|_{\infty} $ and Lemmas  \ref{lemma: concentration for covariance}, $(6)$ comes from {Lemma \ref{lemma: basic inequality for lasso with alpha=0}} and the fact that $sM^2\log(p)/\sqrt{n}=o(1)$. 
	
	Hence, combining (\ref{equation: CII11}), (\ref{inequality: upper bound for C{II,1,1}}), (\ref{inequality: upper bound for C{II,1,2}}), (\ref{inequality: upper bound for C{II,1,3}}), we obtain that
	\begin{equation}\label{inequality: final upper bound for C0II,1 with alpha=0}
		\max\limits_{t\in[q_0,1-q_0]}\|\bC^{II,1}_{0}(t)\|_{(s_0,2)}\leq C{s_0^{1/2}sM^2\log(p)/\sqrt{n}}.
	\end{equation}
	After bounding $\bC^{II,1,3}_{0}(t)$, we next consider $\bC^{II,2}_{0}(t)$. The following lemma provides the desired bound. The proof of Lemma \ref{lemma: upper bound for the empirical process of alpha=0 under H0} is given in Section  \ref{sec: proof of upper bound for the empirical process of alpha=0 under H0}.
	\begin{lemma}\label{lemma: upper bound for the empirical process of alpha=0 under H0}
		{\bf{Suppose Assumptions A, D, E.2 - E.4 hold.}} Then, with probability tending to 1, we have:
		\begin{equation*}
			\max\limits_{t\in[q_0,1-q_0]}\|\bC^{II,2}_{0}(t)\|_{(s_0,2)}\leq C{s_0^{1/2}(s\log(pn))^{3/4}/n^{1/4}},
		\end{equation*}
		for some big enough constant $C>0$.
	\end{lemma}
	\noindent Hence, combining (\ref{inequality: final upper bound for C0II,1 with alpha=0}) and Lemma \ref{lemma: upper bound for the empirical process of alpha=0 under H0}, we have:
	\begin{equation*}
		\max\limits_{t\in[q_0,1-q_0]}\|\bC^{II,2}_{0}(t)\|_{(s_0,2)}\leq C{s_0^{1/2}M^2\dfrac{(s\log(pn))^{3/4}}{n^{1/4}}}.
	\end{equation*}
	{\bf{Step~2: obtain the upper bound for $D_2$.}} By Theorem \ref{theorem: variance estimator under H0}, and similar to the proof of Step 2 in Section \ref{sec: Proof of negligible alpha=1}, we can prove that 
	\begin{equation*}
		\max\limits_{q_0\leq t\leq 1-q_0}\big\|\bC_0^{I}(t)-\tilde{\bC}_0^{I}(t)\big\|_{(s_0,2)}=r_0(n)\times O_p\big( Ms_0^{1/2}\sqrt{\log(pn)}\big)
	\end{equation*}
	where $r_{0}(n)=s\sqrt{\dfrac{\log(pn)}{n}}\vee s^{\frac{1}{2}}(\dfrac{\log(pn)}{n})^{\frac{3}{8}}$. 
	
	Lastly, combining Steps 1 and 2,  if we choose $\epsilon:=C{s_0^{1/2}(s\log(pn))^{3/4}/n^{1/4}}$ for some big constant $C>0$, we have $D_1+D_2=o(1)$, which finishes the proof.
\end{proof}

\subsection{Proof of Lemma \ref{lemma: negligible for alpha}}
\label{sec: Proof of negligible alpha}
\begin{proof}
	In this section, we prove Lemma \ref{lemma: negligible for alpha}. In other words, we aim to prove 
	\begin{equation}\label{equation: proof of negligible alpha}
		\P\big(\max_{q_0\leq t\leq 1-q_0} \big\|\bC_\alpha(t)-\tilde{\bC}_\alpha^{I}(t)\big\|_{(s_0,2)}\geq \epsilon \big)=o(1),
	\end{equation}
	where $\bC_{\alpha}(t)$  and $\tilde{\bC}_\alpha^{I}(t)$ are defined in (\ref{equation: C1 for alpha}) and (\ref{equation: tilde-C-alpha-I}). By the triangle inequality, we have
	\begin{equation}\label{inequality: D1+D2 alpha}
		\P\big(\max_{q_0\leq t\leq 1-q_0} \big\|\bC_\alpha(t)-\tilde{\bC}_\alpha^{I}(t)\big\|_{(s_0,2)}\geq \epsilon \big)\leq D_1+D_2,
	\end{equation}
	where $D_1$ and $D_2$ are defined as 
	\begin{equation}
		\begin{array}{l}
			D_1:=\P\big(\max\limits_{q_0\leq t\leq 1-q_0}\big\|\bC_\alpha(t)-\bC_\alpha^{I}(t)\big\|_{(s_0,2)}\geq \epsilon/2 \big),\\	D_2:=\P\big(\max\limits_{q_0\leq t\leq 1-q_0}\big\|\bC_\alpha^{I}(t)-\tilde{\bC}_\alpha^{I}(t)\big\|_{(s_0,2)}\geq \epsilon/2 \big).
		\end{array}
	\end{equation}
	By (\ref{inequality: D1+D2 alpha}), to prove (\ref{equation: proof of negligible alpha}), we need to bound $D_1$ and $D_2$, respectively. \\
	{\bf{Step~1: obtain the upper bound for $D_1$.}} We first consider $D_1$. To this end, we define 
	\begin{equation}\label{equation: set for variance estimation alpha}
		\cE=\big\{\sigma^2/2\leq \hat{\sigma}^2\leq 2\sigma^2  \big\},
	\end{equation}
	where $\sigma^2:=\text{Var}[(1-\alpha)e_i(\tilde{\btau})+\alpha \epsilon_i]$ is the true variance.  By introducing $\cE$, we have
	\begin{equation}
		D_1\leq \P\big(\max\limits_{q_0\leq t\leq 1-q_0}\big\|\bC_\alpha(t)-\bC_\alpha^{I}(t)\big\|_{(s_0,2)}\geq \epsilon/2\cap\cE \big)+\P(\cE^c).
	\end{equation}
	By {Theorem \ref{theorem: variance estimator under H0}}, we have $\P(\cE^c)=o(1)$ as $n,p\rightarrow\infty$. Under the event $\cE$, we have
	\begin{equation}\label{inequality: basic inequality for D1 alpha}
		\begin{array}{ll}
			\P\big(\max\limits_{q_0\leq t\leq 1-q_0}\big\|\bC_\alpha(t)-\bC_\alpha^{I}(t)\big\|_{(s_0,2)}\geq \epsilon/2\cap\cE \big)
			&= \P\big(\max\limits_{q_0\leq t\leq 1-q_0}\big\|\bC_\alpha^{II}(t)\big\|_{(s_0,2)}\geq \epsilon/2\cap\cE \big),
		\end{array}
	\end{equation}
	where $\bC_\alpha^{II}(t)$ is defined in (\ref{equation: C11+C12 alpha}), which is decomposed into two parts:
	\begin{equation}
		\bC^{II}_{\alpha}(t)=(1-\alpha)	\bC^{II}_{0}(t)+\alpha\bC^{II}_{1}(t),
	\end{equation}
	where $\bC^{II}_{1}(t)$ is defined in (\ref{equation: C11+C12}), and $\bC^{II}_{0}(t)$ is defined in (\ref{equation: C11+C12 alpha=0}). Note that by the proofs of Lemmas \ref{sec: Proof of negligible alpha=1} and \ref{sec: Proof of negligible alpha=0}, we have proved that:
	\begin{equation*}
		\begin{array}{ll}
			\max\limits_{q_0\leq t\leq 1-q_0}\big\|\bC_1^{II}(t)\big\|_{(s_0,2)}=O_p(M^2s_0^{1/2}s\dfrac{\log(pn)}{\sqrt{n}}),\\
			~~\max\limits_{q_0\leq t\leq 1-q_0}\big\|\bC_0^{II}(t)\big\|_{(s_0,2)}=O_p(M^2s_0^{1/2}\dfrac{(s\log(pn))^{3/4}}{n^{1/4}}),
		\end{array}
	\end{equation*}
	Moreover, by {\bf{Assumption E.2}}, we have  $s\dfrac{\log(pn)}{\sqrt{n}}\ll \dfrac{(s\log(pn))^{3/4}}{n^{1/4}} $, which implies that: 
	\begin{equation*}
		\max\limits_{q_0\leq t\leq 1-q_0}\big\|\bC_\alpha^{II}(t)\big\|_{(s_0,2)}=O_p\Big(M^2s_0^{1/2}\dfrac{(s\log(pn))^{3/4}}{n^{1/4}}\Big).
	\end{equation*}
	{\bf{Step~2: obtain the upper bound for $D_2$.}} By Theorem \ref{theorem: variance estimator under H0}, and similar to the proof of Step 2 in Sections \ref{sec: Proof of negligible alpha=1} and \ref{sec: Proof of negligible alpha=0}, we can prove that 
	\begin{equation*}
		\max\limits_{q_0\leq t\leq 1-q_0}\big\|\bC_\alpha^{I}(t)-\tilde{\bC}_\alpha^{I}(t)\big\|_{(s_0,2)}=r_\alpha(n)\times O_p\big( Ms_0^{1/2}\sqrt{\log(pn)}\big)
	\end{equation*}
	where $r_{\alpha}(n)=s\sqrt{\dfrac{\log(pn)}{n}}\vee s^{\frac{1}{2}}(\dfrac{\log(pn)}{n})^{\frac{3}{8}}$. 
	Note that by {\bf{Assumption E.2}}, we have 
	\begin{equation*}
		r_\alpha(n)\times Ms_0^{1/2}\sqrt{\log(pn)}\ll M^2s_0^{1/2}\dfrac{(s\log(pn))^{3/4}}{n^{1/4}}.
	\end{equation*}
	Hence, combining Steps 1 and 2,  if we choose $\epsilon:=C{s_0^{1/2}(s\log(pn))^{3/4}/n^{1/4}}$ for some big constant $C>0$, we have $D_1+D_2=o(1)$, which finishes the proof.
\end{proof}
\subsection{Proof of Lemma \ref{lemma: maximum at Pi}}\label{sec: proof of maximum at Pi}
\begin{proof}
	Note that the proof for $\cH_1$ and $\cH_2$ is similar. We only give the proof of $\cH_1$. The proof proceeds in two steps: In Step~1, we obtain the upper bounds of $\max\limits_{t\geq t_1+\epsilon_n}\max\limits_{J\subset\{1,\ldots,p\}\atop |J|=s_0}\|\Pi_{J}\tilde{\bC}_1(t)-\Pi_{J}\bdelta_1(t)\|_{2}$. In Step~2, using the upper bound and some regular inequalities, we finish the proof. \\
	{\bf{Step~1}}: By the decomposition of $\tilde{\bC}_1(t)$ as in (\ref{equation: tilde-C1 for estimation}), with probability tending to one,  we have:
	\begin{equation*}
		\begin{array}{ll}
			\max\limits_{t\geq t_1+\epsilon_n}\max\limits_{J\subset\{1,\ldots,p\}\atop |J|=s_0}\|\Pi_{J}\tilde{\bC}_1(t)-\Pi_{J}\bdelta(t)\|_{2}\\
			\quad\leq_{(1)} \max\limits_{t\geq t_1+\epsilon_n}\max\limits_{J\subset\{1,\ldots,p\}\atop |J|=s_0}\|\Pi_J\tilde{\bC}_1^{I}(t)+\Pi_J\bR(t)\|_2\\
			\quad\leq_{(2)} \max\limits_{t\geq t_1+\epsilon_n}\max\limits_{J\subset\{1,\ldots,p\}\atop |J|=s_0}\|\Pi_J\tilde{\bC}_1^{I}(t)\|_2+\max\limits_{t\geq t_1+\epsilon_n}\max\limits_{J\subset\{1,\ldots,p\}\atop |J|=s_0}\|\Pi_J\bR(t)\|_2\\
			\quad=_{(3)} \max\limits_{t\geq t_1+\epsilon_n}\|\tilde{\bC}_1^{I}(t)\|_{(s_0,2)}+\max\limits_{t\geq t_1+\epsilon_n}\|\bR(t)\|_{(s_0,2)}\\
			\quad \leq_{(4)}s_0^{1/2}\max\limits_{q_0\leq t\leq 1-q_0}\|\tilde{\bC}_1^{I}(t)\|_{\infty}+s_0^{1/2}\max\limits_{t\geq t_1+\epsilon_n}\|\bR(t)\|_{\infty}\\
			\quad \leq_{(5)}\underbrace{C^*(s_0^{1/2}M\sqrt{\log(pn)}+s_0^{1/2}s\sqrt{\log(pn)}\|\bDelta\|_{(s_0,2)})}_{t^*}.
		\end{array}
	\end{equation*}
	Recall $\cM=\{j: \beta^{(1)}_j\neq \beta^{(2)}_j\}$. Note that 
	
	\begin{equation*}
		\begin{array}{ll}
			\max\limits_{t\geq t_1+\epsilon_n}\max\limits_{J\subset\{1,\ldots,p\}\atop |J|=s_0}\|\Pi_{J}\tilde{\bC}_1(t)-\Pi_{J}\bdelta(t)\|_{2}\\
			\quad=\max\limits_{t\geq t_1+\epsilon_n}\max\limits_{J\subset\cM\atop |J|=s_0}\|\Pi_{J}\tilde{\bC}_1(t)-\Pi_{J}\bdelta(t)\|_{2}+\max\limits_{t\geq t_1+\epsilon_n}\max\limits_{J\subset\cM^c\atop |J|=s_0}\|\Pi_{J}\tilde{\bC}_1(t)-\Pi_{J}\bdelta(t)\|_{2}.
		\end{array}
	\end{equation*}
	Using the fact that $|\max_i \|\ba_i\|_2-\max_{i}\|\bb_i\|_2|\leq \max_i|\|\ba_i\|_2-\|\bb_i\|_2|\leq \max_i|\|\ba_i-\bb_i\|_2|$ for any vectors $\ba_i$ and $\bb_i$,  we have:
	\begin{equation}\label{inequality: maximum within and without Pi}
		\begin{array}{ll}
			\P\Big(\Big|\max\limits_{t\geq t_1+\epsilon_n}\max\limits_{J\subset\cM\atop |J|=s_0}\|\Pi_{J}\tilde{\bC}_1(t)\|_{2}-\max\limits_{t\geq t_1+\epsilon_n}\max\limits_{J\subset\cM\atop |J|=s_0}\|\Pi_{J}\bdelta(t)\|_{2}\Big|\leq t^*\Big)\rightarrow 1\\
			~\text{and}~~\P\Big(\max\limits_{t\geq t_1+\epsilon_n}\max\limits_{J\subset\cM^c\atop |J|=s_0}\|\Pi_{J}\tilde{\bC}_1(t)\|_{2}\leq t^*\Big)\rightarrow 1.
		\end{array}
	\end{equation}
	{\bf{Step~2}}: Note that $\max\limits_{t\geq t_1+\epsilon_n}\max\limits_{J\subset\cM\atop |J|=s_0}\|\Pi_{J}\bdelta(t)\|_{2}\Big|=\sqrt{n}t_1(1-t_1-\epsilon_n)\|\bDelta\|_{(s_0,2)}$. By chooing a big enough constant in (\ref{equation: basic requirement for the signal}), we  have $\max\limits_{t\geq t_1+\epsilon_n}\max\limits_{J\subset\cM\atop |J|=s_0}\|\Pi_{J}\bdelta(t)\|_{2}\Big|\geq 2t^*$.
	Moreover, by (\ref{inequality: maximum within and without Pi}), we see that:
	\begin{equation*}
		\begin{array}{ll}
			\P\Big(\max\limits_{t\geq t_1+\epsilon_n}\max\limits_{J\subset\cM\atop |J|=s_0}\|\Pi_{J}\tilde{\bC}_1(t)\|_{2}-\max\limits_{t\geq t_1+\epsilon_n}\max\limits_{J\subset\cM^c\atop |J|=s_0}\|\Pi_{J}\tilde{\bC}_1(t)\|_{2}\leq 0\Big)\\
			\leq_{(1)} \P\Big(\max\limits_{t\geq t_1+\epsilon_n}\max\limits_{J\subset\cM\atop |J|=s_0}\|\Pi_{J}\tilde{\bC}_1(t)\|_{2}-\max\limits_{t\geq t_1+\epsilon_n}\max\limits_{J\subset\cM^c\atop |J|=s_0}\|\Pi_{J}\tilde{\bC}_1(t)\|_{2}\\
			\qquad	\qquad\qquad\leq \max\limits_{t\geq t_1+\epsilon_n}\max\limits_{J\subset\cM\atop |J|=s_0}\|\Pi_{J}\bdelta(t)\|_{2}-t^*-t^*\Big)\\
			\leq_{(2)}\P\Big(\max\limits_{t\geq t_1+\epsilon_n}\max\limits_{J\subset\cM\atop |J|=s_0}\|\Pi_{J}\tilde{\bC}_1(t)\|_{2}\leq \max\limits_{t\geq t_1+\epsilon_n}\max\limits_{J\subset\cM\atop |J|=s_0}\|\Pi_{J}\bdelta(t)\|_{2}-t^*\Big)\\
			\qquad+\P\Big(-\max\limits_{t\geq t_1+\epsilon_n}\max\limits_{J\subset\cM^c\atop |J|=s_0}\|\Pi_{J}\tilde{\bC}_1(t)\|_{2}\leq -t^*\Big)
			\rightarrow 0,~~\text{as}~~(n,p)\rightarrow\infty,
		\end{array}
	\end{equation*}
	which finishes the proof.
\end{proof}

\section{Proofs of useful lemmas in Section \ref{section: useful lemmas}}\label{sec: proof of useful lemmas}
\subsection{Proof of Lemma \ref{lemma: key lemma for gaussian approximation}}
\begin{proof}
	In this section, we aim to prove (\ref{equation: key lemma for gaussian approximation1}). Firstly, we 
	define $\mathcal{E}^{R}=\{\bx\in \mathbb{R}^{p}: \|\bx\|\leq R\}$ and $V_{(s_0,2)}^{z}=\{\bx\in \mathbb{R}^{p}: \|\bx\|_{(s_0,2)}\leq z\}$. Then, by the definition of $V_{(s_0,2)}^{z}$, we have
	\begin{equation}\label{equation: proof of lemma A.1}
		\begin{array}{ll}
			&\sup\limits_{z\in(0,\infty)} \big|\P(\max\limits_{k_0\leq k\leq n-k_0}\|\bS^{\bZ}(k)\|_{(s_0,2)}\leq z\big)-\P(\max\limits_{k_0\leq k\leq n-k_0}\|\bS^{\bG}(k)\|_{(s_0,2)}\leq z\big)\big|\\
			=&\sup\limits_{z\in(0,\infty)} \underbrace{\big|\P\big(\bigcap\limits_{k_0\leq k\leq n-k_0}\big\{\bS^{\bZ}(k)\in V_{(s_0,2)}^{z} \big)\big\}-\P\big(\bigcap\limits_{k_0\leq k\leq n-k_0}\big\{\bS^{\bG}(k)\in V_{(s_0,2)}^{z}\big)\big\}\big|}_{A_z}.
		\end{array}
	\end{equation}
	By interting $\cE^{R}$ and $(\cE^{R})^c$ in $A_z$, we have $A_z\leq A_z^{(1)}+A_z^{(2)}$, where 
	\begin{equation}\label{equation: Az1+Az2}
		\small
		\begin{array}{l}
			A^{(1)}_z:=\big|\P\big(\bigcap\limits_{k_0\leq k\leq n-k_0}\big\{\bS^{\bZ}(k)\in (\mathcal{E}^{R})^c\cap V_{(s_0,2)}^{z} \big\}\big)-\P\big(\bigcap\limits_{k_0\leq k\leq n-k_0}\big\{\bS^{\bG}(k)\in (\mathcal{E}^{R})^c\cap V_{(s_0,2)}^{z}\big\}\big)\big|,\\\\
			A^{(2)}_z:=\big|\P\big(\bigcap\limits_{k_0\leq k\leq n-k_0}\big\{\bS^{\bZ}(k)\in V_{(s_0,2)}^{z}\cap\mathcal{E}^{R} \big\}\big)-\P\big(\bigcap\limits_{k_0\leq k\leq n-k_0}\big\{\bS^{\bG}(k)\in V_{(s_0,2)}^{z}\cap\mathcal{E}^{R}\big\}\big)\big|.
		\end{array}
	\end{equation}
	Next, we bound $A_z^{(1)}$ and $A_z^{(2)}$ respectively. For $A_z^{(1)}$, using the triangle inequality, we have
	\begin{equation*}
		A_{z}^{(1)}\leq \P\big(\bigcap\limits_{k_0\leq k\leq n-k_0}\big\{\bS^{\bZ}(k)\in (\mathcal{E}^{R})^c\cap V_{(s_0,2)}^{z} \big\}\Big)+\P\Big(\bigcap\limits_{k_0\leq k\leq n-k_0}\big\{\bS^{\bG}(k)\in (\mathcal{E}^{R})^c\cap V_{(s_0,2)}^{z}\big\}\big).
	\end{equation*}
	Recall  $\bS^{\bZ}$ and $\bS^{\bG}$ in (\ref{equation: SZ+SG}). Let $a_{ik}=\mathbf{1}\{i\leq k\}-k/n$ for $i=1\ldots,n$ and $k_0\leq k\leq n-k_0$. We then have $\bS^{\bZ}(k)=n^{-1/2}\sum_{i=1}^n\bZ_ia_{ik}$ and $\bS^{\bG}(k)=n^{-1/2}\sum_{i=1}^n\bG_ia_{ik}$. Moreover, by the definition of  $k_0=\floor {nq_0}$, we have $q_0\leq |a_{ik}|\leq 1-q_0$ for $i=1,\ldots,n$ and $k_0\leq k\leq n-k_0$.    Hence, we have
	\begin{equation}\label{inequality: Az(1) first-1}
		\begin{array}{ll}
			\P\Big(\bigcap\limits_{k_0\leq k\leq n-k_0}\big\{\bS^{\bZ}(k)\in (\mathcal{E}^{R})^c\cap V_{(s_0,2)}^{z} \big\}\Big)\\
			\quad\leq \P\Big(\bigcap\limits_{k_0\leq k\leq n-k_0}\big\{\bS^{\bZ}(k)\in (\mathcal{E}^{R})^c \big\}\Big)\\
			\quad\leq  \P\Big(\bigcap\limits_{k_0\leq k\leq n-k_0}\big\|n^{-1/2}\sum\limits_{i=1}^n\bZ_ia_{ik}\big\|_2\geq R\Big)\\
			\quad\leq \P\Big(\bigcap\limits_{k_0\leq k\leq n-k_0}n^{-1/2}\sum\limits_{i=1}^n\|\bZ_i\|_2\geq R\Big)\\
			\quad\leq \sum\limits_{i=1}^n\P\Big(\sum\limits_{j=1}^pZ_{ij}^2\geq \dfrac{R^2}{n}\Big).
		\end{array}
	\end{equation}
	By Assumption {\bf (M2)} and Markov's inequality, we further have:
	\begin{equation}\label{inequality: Az(1) first-2}
		\begin{array}{ll}
			\sum\limits_{i=1}^n\P\Big(\sum\limits_{j=1}^pZ_{ij}^2\geq \dfrac{R^2}{n}\Big)\leq \dfrac{n\sum\limits_{i=1}^n\sum\limits_{j=1}^p\E Z_{ij}^2}{R^2}\leq  \dfrac{np\max_{1\leq j\leq p}\sum\limits_{i=1}^n\E Z_{ij}^2}{R^2}\leq \dfrac{n^2pK^2}{R^2}.
		\end{array}
	\end{equation}
	Hence, taking $R^2=n^{5/2}p$ and combining (\ref{inequality: Az(1) first-1}) and (\ref{inequality: Az(1) first-2}), we have 
	\begin{equation*}
		\P\Big(\bigcap\limits_{k_0\leq k\leq n-k_0}\big\{\bS^{\bZ}(k)\in (\mathcal{E}^{R})^c\cap V_{(s_0,2)}^{z} \big\}\Big)\leq C\dfrac{1}{\sqrt{n}}.
	\end{equation*}
	Similarly, for $\bS^{\bG}(k)$, we have $\P\big(\bigcap\limits_{k_0\leq k\leq n-k_0}\big\{\bS^{\bG}(k)\in (\mathcal{E}^{R})^c\cap V_{(s_0,2)}^{z} \big\}\big)\leq C\dfrac{1}{\sqrt{n}}$. The above results yield that
	$A_{z}^{(1)}\leq  C\dfrac{1}{\sqrt{n}}$.
	
	About bounding  $A_{z}^{(1)}$, we next consider $A_{z}^{(2)}$. By Lemma \ref{lemma: convex approximation}, there exists an $m$-generated convex set $A^m$ such that 
	\begin{equation*}
		A^{m}\subset \mathcal{E}^{R,p}\cap V_{(s_0,2)}^{z,p}\subset A^{m,R\epsilon} ~~\text{and}~~ m\leq p^{s_0}\Big(\frac{\gamma}{\sqrt{\epsilon}}\ln(\dfrac{1}{\epsilon})\Big)^{s_0^2}.
	\end{equation*}
	By letting 
	\begin{equation*}
		\begin{array}{ll}
			&\bar{\rho}_1:=\big|\P\big(\bigcap\limits_{k_0\leq k\leq n-k_0}\big(\bS^{\bZ}(k)\in A^{m}\big)\big)-\P\big(\bigcap\limits_{k_0\leq k\leq n-k_0}\big(\bS^{\bG}(k)\in A^{m}\big)\big)\big|,\\
			&\bar{\rho}_2:=\big|\P\big(\bigcap\limits_{k_0\leq k\leq n-k_0}\big(\bS^{\bZ}(k)\in A^{m,R\epsilon}\big)\big)-\P\big(\bigcap\limits_{k_0\leq k\leq n-k_0}\big(\bS^{\bG}(k)\in A^{m,R\epsilon}\big)\big)\big|,
		\end{array}
	\end{equation*}
	we have 
	\begin{equation}\label{inequality: upper bound lemma A.1-1}
		\begin{array}{ll}
			&\P\big(\bigcap\limits_{k_0\leq k\leq n-k_0}\big\{\bS^{\bZ}(k)\in\mathcal{E}^{R}\cap V_{(s_0,2)}^{z}\big\}\big)
			\\
			&\quad\leq \P\big(\bigcap\limits_{k_0\leq k\leq n-k_0}\big(\bS^{\bZ}(k)\in A^{m,R\epsilon}\big)\big)~\big(\text{by}~~\mathcal{E}^{R}\cap V_{(s_0,2)}^{z}\subset A^{m,R\epsilon}\big)\\
			&\quad\leq \underbrace{\P\big(\bigcap\limits_{k_0\leq k\leq n-k_0}\big(\bS^{\bG}(k)\in A^{m,R\epsilon}\big)\big)}_{P_z}+\max(\bar{\rho}_1,\bar{\rho}_2).
		\end{array}
	\end{equation}
	Using  Assumption {$\bf (M1)$}, by the definition of $A^{m,R\epsilon}$ in (\ref{equation: Ame}) and  Lemma \ref{lemma:anti consentration inequality}, we have
	\begin{equation}\label{inequality: upper bound lemma A.1-2}
		\begin{array}{ll}
			P_z
			&= \P\Big(\bigcap\limits_{k_0\leq k\leq n-k_0}\bigcap\limits_{\bv\in \mathcal{V}(A^{m})}\big(\bS^{\bG}(k)^\top\bv\leq S_{A^m}(\bv)+R\epsilon\big)\Big)\\
			&\leq \P\Big(\bigcap\limits_{k_0\leq k\leq n-k_0\atop \bv\in \mathcal{V}(A^{m})}\big(\bS^{\bG}(k)^\top\bv\leq S_{A^m}(\bv)\big)\Big)\\		
			&+\P\Big(\bigcap\limits_{k_0\leq k\leq n-k_0\atop \bv\in \mathcal{V}(A^{m})}\big(S_{A^m}(\bv)\leq \bS^{\bG}(k)^\top\bv\leq S_{A^m}(\bv)+R\epsilon\big)\Big)\\
			&\leq \P\Big(\bigcap\limits_{k_0\leq k\leq n-k_0}\big(\bS^{\bG}(k)\in \mathcal{E}^{R}\cap V_{(s_0,2)}^{z}\big)\Big)+CR\epsilon\sqrt{\log nm} \big(\text{by}~A^m\subset\mathcal{E}^{R,d}\cap V_{(s_0,2)}^{z}\Big).
		\end{array}
	\end{equation}
	Therefore, by (\ref{inequality: upper bound lemma A.1-1}) and (\ref{inequality: upper bound lemma A.1-2}),  we have 
	\begin{equation}\label{equation:upper}
		\begin{array}{ll}
			&\P\big(\bigcap\limits_{k_0\leq k\leq n-k_0}\big(\bS^{\bZ}(k)\in\mathcal{E}^{R}\cap V_{(s_0,2)}^{z}\big)\big)\\
			&\quad\leq \P\big(\bigcap\limits_{k_0\leq k\leq n-k_0}\big(\bS^{\bG}(k)\in \mathcal{E}^{R}\cap V_{(s_0,2)}^{z}\big)\big)+CR\epsilon\sqrt{\log nm}+\max(\bar{\rho}_1,\bar{\rho}_2).
		\end{array}
	\end{equation}
	Similar to the procedures in (\ref{inequality: upper bound lemma A.1-1}), (\ref{inequality: upper bound lemma A.1-2}), and (\ref{equation:upper}), we also have
	\begin{equation}\label{equation:lower}
		\begin{array}{ll}
			&\P\big(\bigcap\limits_{k_0\leq k\leq n-k_0}\big(\bS^{\bZ}(k)\in\mathcal{E}^{R}\cap V_{(s_0,2)}^{z}\big)\big)\\
			&\quad\geq \P\big(\bigcap\limits_{k_0\leq k\leq n-k_0}\big(\bS^{\bG}(k)\in \mathcal{E}^{R,d}\cap V_{(s_0,p)}^{z,d}\big)\big)-CR\epsilon\sqrt{\log nm}-\max(\bar{\rho}_1,\bar{\rho}_2).
		\end{array}
	\end{equation}
	Therefore, by (\ref{equation: Az1+Az2}), (\ref{equation:upper}), and (\ref{equation:lower}), we obtain
	\begin{equation} \label{inequality: Az2}
		A^{(2)}_z\leq \max(\bar{\rho}_1,\bar{\rho}_2)+CR\epsilon\sqrt{\log nm}.
	\end{equation}
	Next, we consider $\bar{\rho}_1$ and $\bar{\rho}_2$. For $\bar{\rho}_1$, we have 
	\begin{equation*}
		\begin{array}{ll}
			\bar{\rho}_1:=\big|\P\big(\bigcap\limits_{k_0\leq k\leq n-k_0}\bigcap\limits_{\bv\in \cV(A^m)}\bS^{\bZ}(k)^\top\bv\leq S_{A^m}(\bv)\big)\\
			\qquad\qquad-\P\big(\bigcap\limits_{k_0\leq k\leq n-k_0}\bigcap\limits_{\bv\in \cV(A^m)}\bS^{\bG}(k)^\top\bv\leq S_{A^m}(\bv)\big)\big|.
		\end{array}
	\end{equation*}
	Define $\tilde{Z}_{i}(k,\bv)=\bv^\top\bZ_ia_{ik}$ and  $\tilde{G}_{i}(k,\bv)=\bv^\top\bG_i a_{ik}$ for $i=1,\ldots,n$, $k=k_0,\ldots,n-k_0$ and $\bv\in \cV(A^m)$. By letting
	\begin{equation*}
		S^{\tilde{Z}_{i}(k,v)}=\dfrac{1}{\sqrt{n}}\sum_{i=1}^n\tilde{Z}_{i}(k,v), ~\text{and}~S^{\tilde{G}_{i}(k,v)}=\dfrac{1}{\sqrt{n}}\sum_{i=1}^n\tilde{G}_{i}(k,v), 
	\end{equation*}
	we have 
	\begin{equation*}
		\begin{array}{ll}
			\bar{\rho}_1:=\big|\P\big(S^{\tilde{Z}_{i}(k,v)}\leq S_{A^m}(\bv),k_0\leq k\leq n-k_0, \bv\in \cV(A^m) \big)\\
			\qquad\qquad-\P\big(S^{\tilde{G}_{i}(k,v)}\leq S_{A^m}(\bv),k_0\leq k\leq n-k_0, \bv\in \cV(A^m) \big)\big|,
		\end{array}
	\end{equation*}
	which is high dimensional Gaussian approximation for hyperrectangle in terms of $\{\tilde{Z}_{i}(k,v)\}$. To use Proposition 2.1 in \cite{chernozhukov2017central}, we need to verify that under Assumptions ({\bf{M1}})-({\bf{M3}}), $\tilde{Z}_{i}(k,v)=\bv^\top\bZ_ia_{ik}$ satisfies Conditions (M.1), (M.2) and (E.2) in 
	\cite{chernozhukov2017central}. In fact, by {Assumption (\bf{M1})}, we have $\inf_{k,\bv}\E\tilde{Z}_{i}(k,\bv)^2\geq b $ holds for $i=1,\ldots,n$, which implies Condition (M.1). Moreover, for $\bv\in \cV(A^m) $, let $J(\bv)$ be the set of non-zero coordinates of $\bv$ with $|J(\bv)|\leq s_0$. Using H{\"o}lder's inequality, for any vector $\ba=(a_1,\ldots,a_p)^\top $, we have $(\sum_{j\in J(\bv)}|a_j|)^{2+\ell}\leq s_0^{1+\ell}\sum_{j\in J(\bv)}|a_j|^{2+\ell}$.  This implies that
	\begin{equation*}
		\begin{array}{ll}
			&\dfrac{1}{n}\sum\limits_{i=1}^{n}\E|\tilde{Z}_{i}(k,\bv)|^{2+\ell}\\
			&\quad\leq \dfrac{1}{n}\sum\limits_{i=1}^{n}\E|\bv^\top \bZ_i|^{2+\ell}~~(\text{by}~ q_0\leq |a_{ik}|\leq 1-q_0)\\
			&\quad=\dfrac{1}{n}\sum\limits_{i=1}^{n}\E\sum\limits_{j\in J(\bv)}|Z_{ij}|^{2+\ell}~~(|J(\bv)|\leq s_0 ~\text{and}~\|\bv\|=1)\\
			&\quad\leq s_0^{1+\ell}\dfrac{1}{n}\sum\limits_{i=1}^{n}\sum\limits_{j\in J(\bv)}\E|Z_{ij}|^{2+\ell}\\
			&\quad\leq s_0^{2+\ell}\max\limits_{1\leq j\leq p}\dfrac{1}{n}\sum\limits_{i=1}^{n}\E|Z_{ij}|^{2+\ell}\\
			&\quad\leq s_0^{2+\ell}K^\ell:=(B_n)^\ell,~~(\text{by~Assumption}~{\bf{(M2)}}),
		\end{array}
	\end{equation*}
	where $B_n:=Ks_0^{(2+\ell)/\ell}$. Hence, Condition (M.2) holds by taking $B_n:=Ks_0^{(2+\ell)/\ell}$. Lastly, we verify Condition (E.2). In fact, we have
	\begin{equation*}
		\begin{array}{ll}
			&\E \Big((\max\limits_{k_0\leq k\leq n-k_0 \atop\bv\in \cV(A^m) } |\tilde{Z}_{i}(k,\bv)|)^q\Big)\\
			&\quad\leq \E \Big((\max\limits_{k_0\leq k\leq n-k_0 \atop\bv\in \cV(A^m) } |\tilde{Z}_{i}(k,\bv)|)^q\Big)\\
			&\quad\leq \E \Big((\max\limits_{\bv\in \cV(A^m) } |\bv^\top\bZ_i|)^q\Big)~~(\text{by}~ q_0\leq |a_{ik}|\leq 1-q_0)\\
			&\quad\leq\E \Big((\max\limits_{\bv\in \cV(A^m) } |\sum_{j\in J(\bv)}Z_{ij}|)^q\Big)\\
			&\quad\leq s_0^q\E \Big((\max\limits_{1\leq j\leq p} |Z_{ij}|)^q\Big):=(B_n')^q,
		\end{array}
	\end{equation*}
	where $B_n':=s_0K$. Hence, Condition (E.2) in \cite{chernozhukov2017central} holds by taking $B_n':=s_0K$. Lastly, taking $\tilde{B}_n=s_0^3K$, we have 
	\begin{equation*}
		\begin{array}{ll}
			\max\limits_{k_0\leq k\leq n-k_0 \atop\bv\in \cV(A^m)}\dfrac{1}{n}\sum\limits_{i=1}^{n}\E|\tilde{Z}_{i}(k,v)|^{2+\ell}\leq (\tilde{B}_n)^\ell~~\text{for}~\ell=1,2;\\
			~~\text{and}~~\max_{1\leq i\leq n}\E \Big((\max\limits_{k_0\leq k\leq n-k_0 \atop\bv\in \cV(A^m) } |\tilde{Z}_{i}(k,v)|)^q\Big)\leq (\tilde{B}_n)^q.
		\end{array}
	\end{equation*}
	Let $D_{n}^{(1)}=\Big(\dfrac{s_0^6K^2\log^7(mn^2)}{n}\Big)^{1/6}$ and $D_n^{(2)}=\Big(\dfrac{s_0^6K^2\log^3(mn^2)}{n^{1-2/q}}\Big)^{1/3}$. Using Proposition 2.1 in \cite{chernozhukov2017central}, for $\bar{\rho}_1$ and $\bar{\rho}_2$, we have
	\begin{equation}\label{inequality: max rho1+rho2}
		\max(\bar{\rho}_1,\bar{\rho}_2)\leq C\Big(D_{n}^{(1)}+D_{n}^{(2)}\Big),
	\end{equation}
	where $C$ is some universal constant not depending on $n$ or $p$. Combining (\ref{equation: proof of lemma A.1}), (\ref{equation: Az1+Az2}), (\ref{inequality: Az2}), and (\ref{inequality: max rho1+rho2}), we have
	\begin{equation}\label{inequality: three parts-1}
		\sup\limits_{z\in(0,\infty)}A_z\leq C_1\dfrac{1}{\sqrt{n}}+C_2R\epsilon\sqrt{\log nm}+C_3\Big(D_{n}^{(1)}+D_{n}^{(2)}\Big).
	\end{equation}
	Recall $R:=n^{5/4}p^{1/2}$ and $m\leq p^{s_0}\Big(\frac{\gamma}{\sqrt{\epsilon}}\ln(\dfrac{1}{\epsilon})\Big)^{s_0^2}$. By letting $\epsilon=(pn^2)^{-1}$, we have 
	\begin{equation}\label{inequality: three parts-2}
		R\epsilon\sqrt{\log mn}\preceq \Big(\dfrac{s_0^6K^2\log^7(mn^2)}{n}\Big)^{1/6},~~\text{and}~~R\epsilon\sqrt{\log mn}\preceq \Big(\dfrac{s_0^6K^2\log^3(mn^2)}{n^{1-2/q}}\Big)^{1/3}.
	\end{equation}
	Moreover, using the Assumption that $s_0^3K^{2/7}\log(pn)=O(n^{\xi_1})$ for some $0<\xi_1<1/7$ and $s_0^4K^{2/3}\log(pn)=O(n^{\xi_2})$ for some $0<\xi_2<\frac{1}{3}(1-2/q)$, we have 
	\begin{equation}\label{inequality: Dn1+Dn2}
		D_{n}^{(1)}+D_{n}^{(2)}\leq n^{-\xi_0},~~\text{for~some~}\xi_0>0.
	\end{equation}
	Lastly, combining (\ref{inequality: three parts-1}), (\ref{inequality: three parts-2}) and (\ref{inequality: Dn1+Dn2}), we finish the proof of Lemma \ref{lemma: key lemma for gaussian approximation}.
\end{proof}
\subsection{Proof of Lemma \ref{lemma: exponential inequality for partial sum process}}

\begin{proof}
	Let $a_{ik}=\mathbf{1}\{i\leq k\}-k/n$ for $i=1\ldots,n$ and $\underline{k}_n\leq k\leq n-\overline{k}_n$ with $\underline{k}_n:=\floor{na_n}$ and $\overline{k}_n:=\floor{nb_n}$. Define $Z_{ij}(k)=X_{ij}\epsilon_ia_{ik}$ for $i=1,\ldots,n$, $j=1,\ldots,p$ and $k=\underline{k}_n,\ldots,n-\overline{k}_n$. By definition, we have:
	\begin{equation}
		\begin{array}{ll}
			\max\limits_{t\in[a_n,1-b_n]}\max\limits_{1\leq j\leq p}\Big|\dfrac{1}{\sqrt{n}}\Big( \sum\limits_{i=1}^{\lfloor nt \rfloor}X_{ij}\epsilon_i-\dfrac{\lfloor nt \rfloor}{n} \sum\limits_{i=1}^{n}X_{ij}\epsilon_i\Big)\Big|\\
			\quad=\max\limits_{\underline{k}_n\leq k\leq n-\overline{k}_n}\max\limits_{1\leq j\leq p}\Big|\dfrac{1}{\sqrt{n}}\sum\limits_{i=1}^nZ_{ij}(k)\Big|.
		\end{array}	
	\end{equation}
	Note that by {\bf{Assumption A, C}}, we have $\E|{Z_{ij}(k)}|^{2+\ell}\leq a_{ik}^{2+\ell}M^{2+\ell}K^\ell$ for $\ell=1,2$. Let $M=\max_{i,j,k}|Z_{ij}(k)|$ and $\sigma^2=\max_{j,k}\sum_{i}\E[Z_{ij}^2]$. Then, by Lemma \ref{lemma: concentration inequality for maximum}, we have:
	\begin{equation*}
		\begin{array}{ll}
			\E \Big[\max\limits_{\underline{k}_n\leq k\leq n-\overline{k}_n}\max\limits_{1\leq j\leq p}\Big|\dfrac{1}{\sqrt{n}}\sum\limits_{i=1}^nZ_{ij}(k)\Big|\Big]\\
			\leq \dfrac{C}{\sqrt{n}}\big(\sigma\sqrt{\log p(n-\underline{k}_n-\overline{k}_n)}+\sqrt{\E[M^2]}\log p(n-\underline{k}_n-\overline{k}_n)\big).
		\end{array}
	\end{equation*}
	For $\sigma^2$, using H{\"o}lder's inequality, we have $\sigma^2\leq_{(1)} C\sum\limits_{i=1}^na_{ik}^2M_{n}^2\leq_{(2)} CnM_{n}^2$, where $(2)$ comes from $a_n\leq |a_{ik}|\leq 1-b_n$. For $\E[M^2]$, by definition, we have:
	\begin{equation*}
		\E[M^2]=\E[\max_{i,j,k}|X_{ij}\epsilon_ia_{ik}|^2]\leq_{(1)}M^2\E[\max_{i}|\epsilon_i|^2]\leq_{(2)} CM^2n^{1/2},
	\end{equation*}
	where $(1)$ comes from {\bf{Assumption A}}, $(2)$ comes from {\bf{Assumption C}} and Theorem 3 in \cite{1990Distribution}. Hence, we have 
	\begin{equation}
		\begin{array}{ll}
			\E \Big[\max\limits_{\underline{k}_n\leq k\leq n-\overline{k}_n}\max\limits_{1\leq j\leq p}\Big|\dfrac{1}{\sqrt{n}}\sum\limits_{i=1}^nZ_{ij}(k)\Big|\Big]\leq  CM\sqrt{\log p(n-\underline{k}_n-\overline{k}_n)}.
		\end{array}	
	\end{equation}
	Hence, using Lemma 	\ref{lemma: concentration inequality for maximum}, taking $\eta=1$, $s=2$ and $t=C^*M\sqrt{{\log(p(n-\underline{k}_n-\overline{k}_n))}}$ for some big enough constant $C^*>0$, we have:
	\begin{equation*}
		\P\Big(\max\limits_{\underline{k}_n\leq k\leq n-\overline{k}_n}\max\limits_{1\leq j\leq p}\Big|\dfrac{1}{\sqrt{n}}\sum\limits_{i=1}^nZ_{ij}(k)\Big|\geq C_1M\sqrt{\log(p(n-\underline{k}_n-\overline{k}_n))}\Big)\leq C_2n^{-1/2},
	\end{equation*}
	which completes the proof.
\end{proof}

\subsection{Proof of Lemma \ref{lemma: basic inequality for lasso with alpha=1}}
\label{sec: Proof of lasso property with alpha=1}
\begin{proof}
	In this section, we prove Lemma \ref{lemma: basic inequality for lasso with alpha=1}. Note that Lemma \ref{lemma: basic inequality for lasso with alpha=1} applies to both $\Hb_0$ and $\Hb_1$. To cover the above two cases in a unified way, we prove the results by assuming there is a change point $t_1$ such that  $\bbeta=\bbeta^{(1)}$ if $i\leq \floor{nt_1}$ and $\bbeta=\bbeta^{(2)}$ if $i> \floor{nt_1}$. Note that under $\Hb_0$, we can always set $\bbeta^{(1)}=\bbeta^{(2)}$ even though $t_1$  is not identifiable. Now, we are ready to prove Lemma \ref{lemma: basic inequality for lasso with alpha=1}. 
	
	Recall $\bbeta^*$ is the minimizer under the population level which is defined as:
	\begin{equation*}
		\bbeta^*=\argmin_{\bbeta}\dfrac{1}{n}\sum_{i=1}^n\E\Big[(Y_i-\bX_i^\top\bbeta)^2\Big].
	\end{equation*}
	By the first-order condition, it is easy to see that $\bbeta^*$ satisfies: 
	\begin{equation}\label{inequality: first order of lasso alpha=1}
		\dfrac{1}{n}\sum_{i=1}^n\E\Big[\bX_i(Y_i-\bX_i^\top\bbeta^* )\Big]=\mathbf{0}_{p}.
	\end{equation}
	Moreover, since the model is linear,  it is not hard to see that $\bbeta^*\in \RR^p$ has the following explicit form, which is a linear combination of $\bbeta^{(1)}$ and $\bbeta^{(2)}$:
	\begin{equation*}
		\bbeta^{*}=t_1\bbeta^{(1)}+(1-t_1)\bbeta^{(2)}.
	\end{equation*}
	Note that under $\Hb_0$ with $\bbeta^{(1)}=\bbeta^{(2)}$, we have $\bbeta^{*}=\bbeta^{(1)}$. In this case, $\bbeta^{*}$ is the true parameter for the linear model. Recall $\hat{\bbeta}$ is the minimizer of the empirical loss defined in (\ref{equation: lasso estimator}). Hence, we have:
	\begin{equation}\label{equation: difference of quadratic loss}
		\begin{array}{ll}
			\dfrac{1}{2n}\sum\limits_{i=1}^{n}(Y_i-\bX_i^\top\hat{\bbeta})^2-\dfrac{1}{2n}\sum\limits_{i=1}^{n}(Y_i-\bX_i^\top{\bbeta^*})^2\\
			\quad=_{(1)}\dfrac{1}{2n}\sum\limits_{i=1}^{n}\big(Y_i-\bX_i^\top\bbeta^*-(\bX_i^\top(\hat{\bbeta}-\bbeta^*)\big)^2-\dfrac{1}{2n}\sum\limits_{i=1}^{n}(Y_i-\bX_i^\top{\bbeta^*})^2\\
			\quad =_{(2)}\dfrac{1}{2}(\hat{\bbeta}-\bbeta^*)^\top \hat{\bSigma}(0:1)(\hat{\bbeta}-\bbeta^*)-(\hat{\bbeta}-\bbeta^*)^\top\dfrac{1}{n}\sum\limits_{i=1}^n\bX_i(Y_i-\bX_i^\top\bbeta^*)\\
			\quad =_{(3)}\dfrac{1}{2}(\hat{\bbeta}-\bbeta^*)^\top \hat{\bSigma}(0:1)(\hat{\bbeta}-\bbeta^*)\\
			\qquad-(\hat{\bbeta}-\bbeta^*)^\top\dfrac{1}{n}\sum\limits_{i=1}^n\Big(\bX_i(Y_i-\bX_i^\top\bbeta^*)-\E\bX_i(Y_i-\bX_i^\top\bbeta^*)\Big),\\
		\end{array}
	\end{equation}
	where $\hat{\bSigma}(0:1):=\dfrac{1}{n}\sum\limits_{i=1}^n\bX_i\bX_i^\top$, and 
	$(3)$ comes from the first order condition in (\ref{inequality: first order of lasso alpha=1}). Hence, by the fact that $\hat{\bbeta}$ is the minimizer of (\ref{equation: lasso estimator}), we have:
	\begin{equation*}
		\dfrac{1}{2n}\sum\limits_{i=1}^{n}(Y_i-\bX_i^\top\hat{\bbeta})^2-\dfrac{1}{2n}\sum\limits_{i=1}^{n}(Y_i-\bX_i^\top{\bbeta^*})^2+\lambda\big(\|\hat{\bbeta}\|_1-\|\bbeta^*\|_1\big)\leq 0,
	\end{equation*}
	where $\text{MSE}(\hat{\bbeta}):=\big\|\Xb(\hat{\bbeta}-\bbeta^*)\big\|^2$. Moreover, by (\ref{equation: difference of quadratic loss}), we have:
	\begin{equation}\label{inequality: basic lasso alpha=1-1}
		\begin{array}{ll}
			\dfrac{1}{2}(\hat{\bbeta}-\bbeta^*)^\top \hat{\bSigma}(0:1)(\hat{\bbeta}-\bbeta^*)\\-(\hat{\bbeta}-\bbeta^*)^\top\dfrac{1}{n}\sum\limits_{i=1}^n\Big(\bX_i(Y_i-\bX_i^\top\bbeta^*)-\E\bX_i(Y_i-\bX_i^\top\bbeta^*)\Big)+\lambda\big(\|\hat{\bbeta}\|_1-\|\bbeta^*\|_1\big)\\
			\quad\leq \dfrac{1}{2}\text{MSE}(\hat{\bbeta})-\|\dfrac{1}{n}\sum\limits_{i=1}^n\Big(\bX_i(Y_i-\bX_i^\top\bbeta^*)-\E\bX_i(Y_i-\bX_i^\top\bbeta^*)\Big)\|_{\infty}\|\hat{\bbeta}-\bbeta^*\|_1\\
			\qquad\qquad\qquad+\lambda\big(\|\hat{\bbeta}\|_1-\|\bbeta^*\|_1\big)\leq 0.
		\end{array}
	\end{equation}
	Moreover, by the fact that $Y_i=\epsilon_i+\bbeta^{(1)}\mathbf{1}\{i\leq \floor{nt_1}\}+\bbeta^{(2)}\mathbf{1}\{i> \floor{nt_1}\}$, we have:
	\begin{equation}\label{inequality: basic lasso alpha=1-2}
		\begin{array}{ll}
			\big\|\dfrac{1}{n}\sum\limits_{i=1}^n\Big(\bX_i(Y_i-\bX_i^\top\bbeta^*)-\E\bX_i(Y_i-\bX_i^\top\bbeta^*)\Big)\big\|_{\infty}\\ =_{(1)}\Big\|\dfrac{1}{n}\sum\limits_{i=1}^n\bX_i\epsilon_i+t_1(1-t_1)(\hat{\bSigma}(0:t_1)-\bSigma)(\bbeta^{(1)}-\bbeta^{(2)})\\
			\qquad\qquad\qquad+t_1(1-t_1)(\hat{\bSigma}(t_1:1)-\bSigma)(\bbeta^{(2)}-\bbeta^{(1)})\Big\|_{\infty}\\
			\leq_{(2)} \|\dfrac{1}{n}\sum\limits_{i=1}^n\bX_i\epsilon_i\|_{\infty}+t_1(1-t_1)\|\hat{\bSigma}(0:t_1)-\bSigma\|_{\infty}\|\bbeta^{(1)}-\bbeta^{(2)}\|_{1}\\
			\qquad\qquad\qquad+t_1(1-t_1)\|\hat{\bSigma}(t_1:1)-\bSigma\|_{\infty}\|\bbeta^{(1)}-\bbeta^{(2)}\|_{1}\\
			\leq_{(3)}C_1M\sqrt{\dfrac{\log(pn)}{n}}+C_2M^2\sqrt{\dfrac{\log(pn)}{n}}\|\bbeta^{(1)}-\bbeta^{(2)}\|_{1}+C_3M^2\sqrt{\dfrac{\log(pn)}{n}}\|\bbeta^{(1)}-\bbeta^{(2)}\|_{1}\\
			\leq_{(4)} C_4M^2\sqrt{\dfrac{\log(pn)}{n}}\big(1+\|\bbeta^{(1)}-\bbeta^{(2)}\|_{1}\big)
			\leq_{(5)} C_5M^2\sqrt{\dfrac{\log(pn)}{n}}\\
		\end{array}
	\end{equation}
	where $(3)$ comes from Lemmas \ref{lemma: exponential inequality for partial sum process} and \ref{lemma: concentration for covariance}, and $(5)$ comes from the assumption that $\|\bbeta^{(1)}-\bbeta^{(2)}\|=O(1)$. Hence, by letting $\lambda\geq 2C_5M^2\sqrt{\dfrac{\log(pn)}{n}}$, and combining 
	(\ref{inequality: basic lasso alpha=1-1}) and (\ref{inequality: basic lasso alpha=1-2}), we have:
	\begin{equation*}
		\dfrac{1}{2}\text{MSE}(\hat{\bbeta})-\dfrac{\lambda}{2}\|\hat{\bbeta}-\bbeta^*\|_1+\lambda\big(\|\hat{\bbeta}\|_1-\|\bbeta^*\|_1\big)\leq 0.
	\end{equation*}
	Adding $\lambda\|\hat{\bbeta}-\bbeta^*\|_1$  on both sides of the above inequality, we have:
	\begin{equation}\label{inequality: basic inequality lasso alpha=1 final 1}
		\dfrac{1}{2}\text{MSE}(\hat{\bbeta})+\dfrac{\lambda}{2}\|\hat{\bbeta}-\bbeta^*\|_1\leq \lambda(\|\hat{\bbeta}-\bbeta^*\|_1-\|\hat{\bbeta}\|_1+\|\bbeta^*\|_1).
	\end{equation}
	Hence, by (\ref{inequality: basic inequality lasso alpha=1 final 1}) and the fact that $\|(\hat{\bbeta}-\bbeta^*)_{J^c(\bbeta^*)}\|_1-\|(\hat{\bbeta})_{J^c(\bbeta^*)}\|_1+\|(\bbeta^*)_{J^c(\bbeta^*)}\|_1=0$, we have
	\begin{equation*}
		\begin{array}{ll}
			\dfrac{1}{2}\|\hat{\bbeta}-\bbeta^*\|_1
			&\leq\|(\hat{\bbeta}-\bbeta^*)_{J(\bbeta^*)}\|_1-\|(\hat{\bbeta})_{J(\bbeta^*)}\|_1+\|(\bbeta^*)_{J(\bbeta^*)}\|_1\\
			&\leq 2\|(\hat{\bbeta}-\bbeta^*)_{J(\bbeta^*)}\|_1,
		\end{array}
	\end{equation*}
	which implies $\|(\hat{\bbeta}-\bbeta^*)_{J^c(\bbeta^*)}\|_1\leq 3\|(\hat{\bbeta}-\bbeta^*)_{J(\bbeta^*)}\|_1$. Combining this result and (\ref{inequality: basic inequality lasso alpha=1 final 1}), we have:
	\begin{equation*}
		\dfrac{1}{2}\text{MSE}(\hat{\bbeta})+\lambda\big\|(\hat{\bbeta}-\bbeta^*)_{J(\bbeta^*)^c}\big\|_1\leq 3\lambda\big\|(\hat{\bbeta}-\bbeta^*)_{J(\bbeta^*)}\big\|_1,
	\end{equation*}
	
	Note that by {\bf{Assumptions A, E.2}}, the restricted eigenvalue condition holds for $\hat{\bbeta}-\bbeta^*$. Hence, using similar proof techniques as in \cite{bickel2009simultaneous}, it is not hard to derive  (\ref{inequality: lasso estimator inequality}). To save space, we omit the details here.
\end{proof}

\subsection{Proof of Lemma \ref{lemma: basic inequality for lasso with alpha=0}}
\label{sec: Proof of lasso property with alpha=0}
\begin{proof}
	In this section, we prove Lemma \ref{lemma: basic inequality for lasso with alpha=0}. Similar to Lemma  \ref{lemma: basic inequality for lasso with alpha=1}, we prove the results by assuming there is a change point $t_1$ such that  $\bbeta=\bbeta^{(1)}$ if $i\leq \floor{nt_1}$ and $\bbeta=\bbeta^{(2)}$ if $i> \floor{nt_1}$. Recall $\uwave{\bbeta}^*=((\bbeta^*)^\top,(\bb^*)^\top)^\top\in \RR^{p+K}$ defined in  (\ref{equation: parameters under population level}). By the first order condition, for $\alpha=0$, it is not hard to see that $\uwave{\bbeta}^*=((\bbeta^*)^\top,(\bb^*)^\top)^\top\in \RR^{p+K}$ satisfies the following equation:
	\begin{equation}\label{equation: first order condition alpha=0}
		\begin{array}{ll}
			\E \Big[\sum_{i=1}^n\sum_{k=1}^K\bX_i(\mathbf{1}\{Y_i\leq \bX_i^\top \bbeta^*+b_k^*\}-\tau_k)\Big]=\mathbf{0}_p, \\
			\E \Big[\sum_{i=1}^n(\mathbf{1}\{Y_i\leq \bX_i^\top \bbeta^*+b_k^*\}-\tau_k)\Big]={0}, ~\text{for}~k=1,\ldots,K.
		\end{array}
	\end{equation}
	By the fact that $Y_i=\epsilon_i+\bbeta^{(1)}\mathbf{1}\{i\leq \floor{nt_1}\}+\bbeta^{(2)}\mathbf{1}\{i> \floor{nt_1}\}$, for the above equation, we have:
	\begin{equation*}
		\begin{array}{ll}
			t_1\E \Big[\sum_{k=1}^K\bX \big(F_{\epsilon}( \bX^\top (\bbeta^*-\bbeta^{(1)})+b_k^*)-F_{\epsilon}(b_k^{(0)}\big))\Big]\\
			+t_2\E \Big[\sum_{k=1}^K\bX \big(F_{\epsilon}( \bX^\top (\bbeta^*-\bbeta^{(2)})+b_k^*)-F_{\epsilon}(b_k^{(0)}\big))\Big]=\mathbf{0}_p,
		\end{array}
	\end{equation*}
	and $\text{for}~k=1,\ldots,K$, 
	\begin{equation*}
		t_1\E \Big[ \big(F_{\epsilon}( \bX^\top (\bbeta^*-\bbeta^{(1)})+b_k^*)-F_{\epsilon}(b_k^{(0)}\big))\Big]+t_2\E \Big[ \big(F_{\epsilon}( \bX^\top (\bbeta^*-\bbeta^{(2)})+b_k^*)-F_{\epsilon}(b_k^{(0)}\big))\Big]=0,
	\end{equation*}
	where $t_2:=1-t_1$. Moreover, let $\uwave{\bbeta}^{(1)}:=((\bbeta^{(1)})^\top,(\bb^{(0)})^\top)^\top\in \RR^{p+K}$, $\uwave{\bbeta}^{(2)}:=((\bbeta^{(2)})^\top,$\\$(\bb^{(0)})^\top)^\top$$\in \RR^{p+K}$,  $\uwave{\bX}:=(\bX^\top,\mathbf{1}_{K})\in \RR^{p+K}$, and $\bS_k:=\diag(\mathbf{1}_p,\be_k)$, where $\be_k\in \RR^K$ is a vector with the $k$-th element being 1 and the others being zeros, and $\mathbf{1}_{K}$ is a $K$-dimensional vector with all elements being 1s. With the above notations, for the above equations, we have:
	\begin{equation*}
		\begin{array}{ll}
			t_1\E \Big[\sum\limits_{k=1}^K\bS_k\uwave{\bX} \big(F_{\epsilon}( (\bS_k\uwave{\bX})^\top (\uwave{\bbeta^*}-\uwave{\bbeta}^{(1)})+b_k^{(0)})-F_{\epsilon}(b_k^{(0)}\big))\Big]\\
			\quad+t_2\E \Big[\sum\limits_{k=1}^K\bS_k\uwave{\bX} \big(F_{\epsilon}( (\bS_k\uwave{\bX})^\top (\uwave{\bbeta^*}-\uwave{\bbeta}^{(2)})+b_k^{(0)})-F_{\epsilon}(b_k^{(0)}\big))\Big]=\mathbf{0}_{p+K}.
		\end{array}
	\end{equation*}
	Furthermore, by the Taylor's expansion, we have:
	\begin{equation}\label{equation: tilde-sigma-1+tilde-sigma-2}
		\begin{array}{ll}
			t_1\underbrace{\Big\{\sum\limits_{k=1}^{K}\E \Big[\int_{0}^{1}(\bS_k\uwave{\bX})(\bS_k\uwave{\bX})^\top f_{\epsilon}\big(b_k^{(0)}+t((\bS_k\uwave{\bX})^\top (\uwave{\bbeta^*}-\uwave{\bbeta}^{(1)})\big)dt\Big]\Big\}}_{\tilde{\bSigma}^{(1)}\in \RR^{(p+K)\times (p+K)}}(\uwave{\bbeta^*}-\uwave{\bbeta}^{(1)})\\
			+t_2\underbrace{\Big\{\sum\limits_{k=1}^{K}\E \Big[\int_{0}^{1}(\bS_k\uwave{\bX})(\bS_k\uwave{\bX})^\top f_{\epsilon}\big(b_k^{(0)}+t((\bS_k\uwave{\bX})^\top (\uwave{\bbeta^*}-\uwave{\bbeta}^{(2)})\big)dt\Big]\Big\}}_{\tilde{\bSigma}^{(2)}\in \RR^{(p+K)\times (p+K)}}(\uwave{\bbeta^*}-\uwave{\bbeta}^{(2)})=\mathbf{0}_{p+K}.
		\end{array}
	\end{equation}
	Hence, for $\uwave{\bbeta}^*$, by defining  $\tilde{\bSigma}^{(1)}$ and $\tilde{\bSigma}^{(2)}$,  it has the following explicit form:
	\begin{equation*}
		\uwave{\bbeta}^*=(t_1\tilde{\bSigma}^{(1)}+t_2\tilde{\bSigma}^{(2)})^{-1}(t_1\tilde{\bSigma}^{(1)}\bbeta^{(1)}+t_2\tilde{\bSigma}^{(2)}\bbeta^{(2)}).
	\end{equation*}
	Moreover, using some calculations, we have:
	\begin{equation}\label{equation: difference between pooled parameter and the true parameter}
		\begin{array}{ll}
			\uwave{\bbeta}^*-\uwave{\bbeta}^{(1)}=(t_1\tilde{\bSigma}^{(1)}+t_2\tilde{\bSigma}^{(2)})^{-1}t_2\tilde{\bSigma}^{(2)}(\uwave{\bbeta}^{(2)}-\uwave{\bbeta}^{(1)})\\
			\uwave{\bbeta}^*-\uwave{\bbeta}^{(2)}=(t_1\tilde{\bSigma}^{(1)}+t_2\tilde{\bSigma}^{(2)})^{-1}t_1\tilde{\bSigma}^{(1)}(\uwave{\bbeta}^{(1)}-\uwave{\bbeta}^{(2)}).
		\end{array}
	\end{equation}
	\begin{remark}\label{remark: difference beta for alpha=0}
		Note that for any matrix $\Ab\in \RR^{p\times p}$ and $\bx\in \RR^p$, we have $\|\Ab\bx\|_1\leq \|\Ab\|_{1,1}\|\bx\|_1$. Hence, if we assume that
		\begin{equation*}
			\Big\|(t_1\tilde{\bSigma}^{(1)}+t_2\tilde{\bSigma}^{(2)})^{-1}t_2\tilde{\bSigma}^{(2)}\Big\|_{1,1}\leq C_1~\text{and}~	\Big\|(t_1\tilde{\bSigma}^{(1)}+t_2\tilde{\bSigma}^{(2)})^{-1}t_1\tilde{\bSigma}^{(1)}\Big\|_{1,1}\leq C_2,
		\end{equation*}
		we can prove that $\|\uwave{\bbeta}^*-\uwave{\bbeta}^{(1)}\|_{1}\leq C_1\|\uwave{\bbeta}^{(1)}-\uwave{\bbeta}^{(2)}\|_1= C_1\|{\bbeta}^{(1)}-{\bbeta}^{(2)}\|_1$ and $\|\uwave{\bbeta}^*-\uwave{\bbeta}^{(2)}\|_{1}\leq C_2\|\uwave{{\bbeta}}^{(1)}-\uwave{{\bbeta}}^{(2)}\|_1=C_2\|{\bbeta}^{(1)}-{\bbeta}^{(2)}\|_1$ for the above positive constants $C_1,C_2>0$.
	\end{remark}

	So far, we have derived the explicit form for $\uwave{\bbeta}^*$ and the difference between $\uwave{\bbeta}^*-\uwave{\bbeta}^{(1)}$ or $\uwave{\bbeta}^*-\uwave{\bbeta}^{(2)}$, which is very important for proving Lemma \ref{lemma: basic inequality for lasso with alpha=0}. Now, we are ready to give the detailed proof. To that end, we define the following parameter space. Let $\uwave{\bDelta}=(\bDelta^\top,\bdelta^\top)^\top\in \RR^{p+K} $ with $\bDelta\in \RR^p$ and $\bdelta\in \RR^K$, we define
	\begin{equation}\label{equation: parameter space}
		\cA=\big\{\  (\bDelta^\top,\bdelta^\top)^\top: \|\bDelta_{J^c(\bbeta^*)}\|_1\leq 3\|\bDelta_{J(\bbeta^*)}\|_1+ \|\bdelta\|_1 \big\}.
	\end{equation}
	For any $\uwave{\bbeta}=((\bbeta)^\top,(\bb)^\top)^\top\in \RR^{p+K}$, let $\uwave{\bDelta}=\uwave{\bbeta}-\uwave{\bbeta}^*$ and $\uwave{\hat{\bDelta}}=\hat{\uwave{\bbeta}}-\uwave{\bbeta}^*$, where $\hat{\bbeta}$ is the minimizer of the empirical loss defined in (\ref{equation: lasso estimator}) with $\alpha=0$. Define the empirical loss and its expectation: 
	\begin{equation*}
		\begin{array}{ll}
			L_{n,K}(\uwave{\bbeta}):=\dfrac{1}{n}\sum_{i=1}^n\dfrac{1}{K}\sum_{k=1}^{K}\rho_{\tau_k}(Y_i-\bX_{i}^\top\bbeta-b_k),\\
			~\text{and}~L_{K}(\uwave{\bbeta}):=\E[L_{n,K}(\uwave{\bbeta})]=\dfrac{1}{n}\sum_{i=1}^n\dfrac{1}{K}\sum_{k=1}^{K}\E[\rho_{\tau_k}(Y_i-\bX_{i}^\top\bbeta-b_k)].
		\end{array}
	\end{equation*}
	Then, we can further define the excess risk as:
	\begin{equation*}
		H(\uwave{{\bDelta}})=L_{K}(\uwave{\bbeta}^*+\uwave{{\bDelta}})-L_{K}(\uwave{\bbeta}^*).
	\end{equation*}
	The proof of Lemma \ref{lemma: basic inequality for lasso with alpha=0} relies on the following three lemmas.
	Lemma \ref{lemma: basic inequality for quantile regression} shows that $\uwave{\hat{\bbeta}}-\uwave{\bbeta}$ belongs to $\cA$ with a large probability. The proof of Lemma \ref{lemma: basic inequality for quantile regression} is given in Section \ref{lemma: proof of basic inequality for quantile regression}.
	\begin{lemma}\label{lemma: basic inequality for quantile regression}
		Assume	{\bf{Asssumptions A, D, E.2 - E.4}} hold. Then, with probability tending to one, we have
		\begin{equation*}
			\uwave{\hat{\bbeta}}-\uwave{\bbeta}^*\in \cA
		\end{equation*}
	\end{lemma}
	Next, Lemma \ref{lemma: lower bound for the excess risk with alpha=0} shows that the excess risk $H(\uwave{{\bDelta}})$ can be bounded by the quadratic form of $\uwave{{\bDelta}}$. To show this, define 
	\begin{equation}\label{equation: bS}
		\bS=\sum_{k=1}^{K}
		\Big(\begin{array}{lc}
			\bSigma,\mathbf{0}\\
			\mathbf{0},\diag(\be_k)
		\end{array}\Big)\in \RR^{(p+K)\times (p+K)},~~~ \|\uwave{{\bDelta}}\|^2_{\bS}=\uwave{{\bDelta}}^\top\bS \uwave{{\bDelta}}=\sum_{k=1}^K(\bDelta^\top\bSigma\bDelta+\delta_k^2).
	\end{equation}
	
	\begin{lemma}\label{lemma: lower bound for the excess risk with alpha=0}
		Assume	{\bf{Assumptions A, D, E.2 - E.4}} hold. For any $\uwave{{\bDelta}}\in \cA$, with probability tending to one, we have
		\begin{equation*}
			H(\uwave{{\bDelta}})\geq c_*\min\Big(\dfrac{ \|\uwave{{\bDelta}}\|^2_{\bS}}{4},\dfrac{\|\uwave{{\bDelta}}\|_{\bS}}{4} \Big),
		\end{equation*}
		where $c_*>0$ is some universal constant not depending on $n$ or $p$. 
	\end{lemma}
	Lastly, Lemma \ref{lemma: large deviation for excess risk with alpha=0} shows that we can uniformly control the difference between the excess risk and its empirical version.
	\begin{lemma}\label{lemma: large deviation for excess risk with alpha=0}
		Assume	{\bf{Asssumptions A, D, E.2 - E.4}} hold. With probability tending to one, we have:
		\begin{equation*}
			\sup_{\uwave{{\bDelta}}\in \cA\atop \|\uwave{{\bDelta}}\|_{\bS}\leq \xi }\big|(L_{n,K}(\uwave{\bbeta}^*+\uwave{{\bDelta}})-L_{n,K}(\uwave{\bbeta}^*))-(L_K(\uwave{\bbeta}^*+\uwave{{\bDelta}})-L_K(\uwave{\bbeta}^*))\big|\leq C^*M\xi \sqrt{s\dfrac{\log(pn)}{n}},
		\end{equation*}
		where $C^*>0$ is some universal constant not depending on $n$ or $p$ and $s:=|J({\bbeta}^*)|$.
		
		With the above lemmas, we are ready to prove Lemma \ref{lemma: basic inequality for lasso with alpha=0}. Define two events $\cE_1$ and $\cE_2$ as:
		\begin{equation*}
			\begin{array}{cc}
				\cE_1=\{ \uwave{\hat{\bbeta}}-\uwave{\bbeta}^*\in \cA\}\\
				\cE_2=\Big\{ \sup\limits_{\uwave{{\bDelta}}\in \cA\atop \|\uwave{{\bDelta}}\|_{\bS}\leq \xi }\big|(L_{n,K}(\uwave{\bbeta}^*+\uwave{{\bDelta}})-L_{n,K}(\uwave{\bbeta}^*))-(L_K(\uwave{\bbeta}^*+\uwave{{\bDelta}})-L_K(\uwave{\bbeta}^*))\big|\\
				\leq C^*M\xi \sqrt{s\dfrac{\log(pn)}{n}}\Big\}.
			\end{array}
		\end{equation*}
		By Lemmas \ref{lemma: basic inequality for quantile regression} and \ref{lemma: large deviation for excess risk with alpha=0}, we have $\P(\cE_1\cap \cE_2)\rightarrow 1$. Hence, in what follows, we give the proof under the event $\cE_1\cap \cE_2$. Let $ \|\uwave{\hat{\bbeta}}-\uwave{\bbeta}^*\|_{\bS}=\xi$. By the optimality of $\uwave{\hat{\bbeta}}$, we have:
		\begin{equation*}
			L_{n,K}(\uwave{\hat{\bbeta}})-L_{n,K}(\uwave{\bbeta}^*)+\lambda(\|\uwave{\hat{\bbeta}}\|_1-\|\uwave{{\bbeta}^*}\|_1)\leq 0.
		\end{equation*}
		Moreover, using the above inequality, under $\cE_2$, we have:
		\begin{equation}\label{inequality: basic inequality alpha=0 final-1}
			\begin{array}{ll}
				&\lambda(\|\uwave{\hat{\bbeta}}-\uwave{{\bbeta}^*}\|_1)\\&
				\quad\geq_{(1)} L_{n,K}(\uwave{\hat{\bbeta}})-L_{n,K}(\uwave{\bbeta}^*)\\
				&\quad\geq_{(2)} L_{K}(\uwave{\hat{\bbeta}})-L_{K}(\uwave{\bbeta}^*)-C^*M\xi \sqrt{s\dfrac{\log(pn)}{n}}\\
				&\quad\geq _{(3)}c_*\min\Big(\dfrac{ \xi^2}{4},\dfrac{\xi}{4} \Big)-C^*M\xi \sqrt{s\dfrac{\log(pn)}{n}}.
			\end{array}
		\end{equation}
		Note that under $\cE_1$, we have:
		\begin{equation}\label{inequality: basic inequality alpha=0 final-2}
			\begin{array}{ll}
				&\|\uwave{\hat{\bbeta}}-\uwave{{\bbeta}^*}\|_1\\
				&\quad\leq 4\|(\hat{\bbeta}-\bbeta^*)_{J(\bbeta^*)}\|_1+\|\hat{\bb}-\bb^*\|_1\\
				&\quad\leq 4\sqrt{s}\|(\hat{\bbeta}-\bbeta^*)_{J(\bbeta^*)}\|_2+\sqrt{K}\|\hat{\bb}-\bb^*\|_2\\
				&\quad\leq 4\sqrt{s}\|(\hat{\bbeta}-\bbeta^*)\|_2+\sqrt{K}\|\hat{\bb}-\bb^*\|_2\\
				&\quad\leq C_{E}\sqrt{s}\|\uwave{\hat{\bbeta}}-\uwave{\bbeta}^*\|_{\bS},
			\end{array}
		\end{equation}
		where $C_{E}>0$ is some universal constant. Note that we can choose $\lambda=C_{\lambda}\sqrt{\log(p)/n}$ for some big enough constant $C_{\lambda}>0$. Combining (\ref{inequality: basic inequality alpha=0 final-1}) and (\ref{inequality: basic inequality alpha=0 final-2}), we have:
		\begin{equation*}
			c_*\min\Big(\dfrac{ \xi^2}{4},\dfrac{\xi}{4} \Big)-C^*M\xi \sqrt{s\dfrac{\log(pn)}{n}}-C_{\lambda}\xi\sqrt{s\dfrac{\log(pn)}{n}}\leq 0,
		\end{equation*}
		which implies 
		\begin{equation*}
			c_*\dfrac{\xi}{4} -C^*M\xi\sqrt{s\dfrac{\log(pn)}{n}}-C_{\lambda}\xi\sqrt{s\dfrac{\log(pn)}{n}}\leq 0,
		\end{equation*}
		or
		\begin{equation*}
			c_*\dfrac{\xi^2}{4} -C^*M\xi\sqrt{s\dfrac{\log(pn)}{n}}-C_{\lambda}\xi\sqrt{s\dfrac{\log(pn)}{n}}\leq 0,
		\end{equation*}
		Note that $\sqrt{s\dfrac{\log(pn)}{n}}=o(1)$. Hence, only the second case applies. As a result, we have:
		\begin{equation}\label{equation: xi for alpha=0}
			\xi=\|\uwave{\hat{\bbeta}}-\uwave{\bbeta}^*\|_{\bS}\leq CM\sqrt{s\dfrac{\log(pn)}{n}}.
		\end{equation}
		Lastly, by (\ref{equation: xi for alpha=0}) and some trivial calculations, we can directly derive (\ref{inequality: lasso estimator inequality with alpha=0}).
	\end{lemma}
\end{proof}
\subsection{Proof of Lemma \ref{lemma: basic inequality for lasso with alpha}}
\label{sec: Proof of lasso property with alpha}

\begin{proof}
	In this section, we prove Lemma \ref{lemma: basic inequality for lasso with alpha}. Similar to Lemmas  \ref{lemma: basic inequality for lasso with alpha=1} and \ref{lemma: basic inequality for lasso with alpha=0}, we givethe results by assuming there is a change point $t_1$ such that  $\bbeta=\bbeta^{(1)}$ if $i\leq \floor{nt_1}$ and $\bbeta=\bbeta^{(2)}$ if $i> \floor{nt_1}$. Note that the results are still applicable even though there is no change point.

	Before the proof, we need some discussion about  $\uwave{\bbeta}^*=((\bbeta^*)^\top,(\bb^*)^\top)^\top\in \RR^{p+K}$, which is defined in  (\ref{equation: parameters under population level}). By the first order condition, for $\alpha\in(0,1)$, it is not hard to see that $\uwave{\bbeta}^*=((\bbeta^*)^\top,(\bb^*)^\top)^\top\in \RR^{p+K}$ satisfies the following equation:
	\begin{equation}\label{equation: first order condition alpha}
		\begin{array}{ll}	
			(1-\alpha)\E \Big[\sum\limits_{i=1}^n\sum\limits_{k=1}^K\bX_i(\mathbf{1}\{Y_i\leq \bX_i^\top \bbeta^*+b_k^*\}-\tau_k)\Big]-\alpha\sum\limits_{i=1}^n\E\Big[\bX_i(Y_i-\bX_i^\top\bbeta^* )\Big]=\mathbf{0}_p,\\
			(1-\alpha) \E \Big[\sum\limits_{i=1}^n(\mathbf{1}\{Y_i\leq \bX_i^\top \bbeta^*+b_k^*\}-\tau_k)\Big]={0}, ~\text{for}~k=1,\ldots,K.
		\end{array}
	\end{equation}
	Note that  $Y_i=\epsilon_i+\bbeta^{(1)}\mathbf{1}\{i\leq \floor{nt_1}\}+\bbeta^{(2)}\mathbf{1}\{i> \floor{nt_1}\}$. Similar to the analysis in 
	Section \ref{sec: Proof of lasso property with alpha=0}, for the above equation, we have:
	\begin{equation*}
		\begin{array}{ll}
			t_1\Big\{(1-\alpha)\E \Big[\sum\limits_{k=1}^K\bX \big(F_{\epsilon}( \bX^\top (\bbeta^*-\bbeta^{(1)})+b_k^*)-F_{\epsilon}(b_k^{(0)}\big))\Big]+\alpha\E\Big[\bX\bX^\top (\bbeta^*-\bbeta^{(1)})\Big]\Big\}\\
			+t_2\Big\{(1-\alpha)\E \Big[\sum\limits_{k=1}^K\bX \big(F_{\epsilon}( \bX^\top (\bbeta^*-\bbeta^{(2)})+b_k^*)-F_{\epsilon}(b_k^{(0)}\big))\Big]+\alpha\E\Big[\bX\bX^\top (\bbeta^*-\bbeta^{(2)})\Big]\Big\}\\=\mathbf{0}_{p}
		\end{array}
	\end{equation*}
	and $\text{for}~k=1,\ldots,K$, 
	\begin{equation*}
		\begin{array}{ll}
			t_1\E \Big[\sum_{k=1}^K \big(F_{\epsilon}( \bX^\top (\bbeta^*-\bbeta^{(1)})+b_k^*)-F_{\epsilon}(b_k^{(0)}\big)\big)\Big]\\
			+t_2\E \Big[\sum_{k=1}^K \big(F_{\epsilon}( \bX^\top (\bbeta^*-\bbeta^{(2)})+b_k^*)-F_{\epsilon}(b_k^{(0)}\big)\big)\Big]=0,
		\end{array}
	\end{equation*}
	where $t_2:=1-t_1$. Moreover, let
	\begin{equation}\label{equation: sigma-tilde}
		\tilde{\bSigma}=
		\Big(\begin{array}{lc}
			\bSigma,\mathbf{0}\\
			\mathbf{0},\mathbf{0}
		\end{array}\Big)\in \RR^{(p+K)\times (p+K)}.
	\end{equation}
	Then, using similar analysis as in Section \ref{sec: Proof of lasso property with alpha=0}, we have:
	\begin{equation*}
		t_1\underbrace{\Big[(1-\alpha)\tilde{\bSigma}^{(1)}+\alpha\tilde{\bSigma}\Big]}_{\breve{\bSigma}^{(1)}}(\uwave{\bbeta^*}-\uwave{\bbeta}^{(1)})+t_2\underbrace{\Big[(1-\alpha)\tilde{\bSigma}^{(2)}+\alpha\tilde{\bSigma}\Big]}_{\breve{\bSigma}^{(2)}}(\uwave{\bbeta^*}-\uwave{\bbeta}^{(2)})=\mathbf{0}_{p+K},
	\end{equation*}
	where $\tilde{\bSigma}^{(1)}$ and $\tilde{\bSigma}^{(2)}$ are defined in	(\ref{equation: tilde-sigma-1+tilde-sigma-2}). Hence, for $\uwave{\bbeta}^*$,  it has the following explicit form:
	\begin{equation*}
		\uwave{\bbeta}^*=(t_1\breve{\bSigma}^{(1)}+t_2\breve{\bSigma}^{(2)})^{-1}(t_1\breve{\bSigma}^{(1)}\bbeta^{(1)}+t_2\breve{\bSigma}^{(2)}\bbeta^{(2)}).
	\end{equation*}
	Moreover, using some calculations, we have:
	\begin{equation}\label{equation: difference between pooled parameter and the true parameter alpha=0}
		\begin{array}{ll}
			\uwave{\bbeta}^*-\uwave{\bbeta}^{(1)}=(t_1\breve{\bSigma}^{(1)}+t_2\breve{\bSigma}^{(2)})^{-1}t_2\breve{\bSigma}^{(2)}(\uwave{\bbeta}^{(2)}-\uwave{\bbeta}^{(1)})\\
			\uwave{\bbeta}^*-\uwave{\bbeta}^{(2)}=(t_1\breve{\bSigma}^{(1)}+t_2\breve{\bSigma}^{(2)})^{-1}t_1\breve{\bSigma}^{(1)}(\uwave{\bbeta}^{(1)}-\uwave{\bbeta}^{(2)}).
		\end{array}
	\end{equation}
	
	\begin{remark}\label{remark: difference beta for alpha in (0,1)}
		If we assume that
		\begin{equation*}
			\Big\|(t_1\breve{\bSigma}^{(1)}+t_2\breve{\bSigma}^{(2)})^{-1}t_2\breve{\bSigma}^{(2)}\Big\|_{1,1}\leq C_1~\text{and}~	\Big\|(t_1\breve{\bSigma}^{(1)}+t_2\breve{\bSigma}^{(2)})^{-1}t_1\breve{\bSigma}^{(1)}\Big\|_{1,1}\leq C_2,
		\end{equation*}
		we can prove that $\|\uwave{\bbeta}^*-\uwave{\bbeta}^{(1)}\|_{1}\leq C_1\|\uwave{\bbeta}^{(1)}-\uwave{\bbeta}^{(2)}\|_1= C_1\|{\bbeta}^{(1)}-{\bbeta}^{(2)}\|_1$ and $\|\uwave{\bbeta}^*-\uwave{\bbeta}^{(2)}\|_{1}\leq C_2\|\uwave{{\bbeta}}^{(1)}-\uwave{{\bbeta}}^{(2)}\|_1=C_2\|{\bbeta}^{(1)}-{\bbeta}^{(2)}\|_1$ for the above positive constants $C_1,C_2>0$.
	\end{remark}
	
	For $\alpha\in(0,1)$, we have derived the explicit form for $\uwave{\bbeta}^*$ and the difference between $\uwave{\bbeta}^*-\uwave{\bbeta}^{(1)}$ or $\uwave{\bbeta}^*-\uwave{\bbeta}^{(2)}$, which is very important for proving Lemma \ref{lemma: basic inequality for lasso with alpha}. Now, we are ready to give the proof. 
	
	Recall the parameter space $\cA$ defined in (\ref{equation: parameter space}). For any $\uwave{\bbeta}=((\bbeta)^\top,(\bb)^\top)^\top\in \RR^{p+K}$, let $\uwave{\bDelta}=\uwave{\bbeta}-\uwave{\bbeta}^*$ and $\uwave{\hat{\bDelta}}=\hat{\uwave{\bbeta}}-\uwave{\bbeta}^*$, where $\hat{\bbeta}$ is the minimizer of the empirical loss defined in (\ref{equation: lasso estimator}) with $\alpha\in(0,1)$. Define the empirical loss and its expectation: 
	\begin{equation}\label{equation: Lnk-alpha}
		\begin{array}{cc}
			L^{\alpha}_{n,K}(\uwave{\bbeta}):=(1-\alpha)\dfrac{1}{n}\sum\limits_{i=1}^{n}\dfrac{1}{K}\sum\limits_{k=1}^{K}\rho_{\tau_k}(Y_i-b_i-\bX_i^\top\bbeta)+ \dfrac{\alpha}{2n}\sum\limits_{i=1}^{n}(Y_i-\bX_i^\top \bbeta)^2,\\
			~~\text{and}~L^\alpha_{K}(\uwave{\bbeta}):=\E[L^\alpha_{n,K}(\uwave{\bbeta})].
		\end{array}
	\end{equation}
	Then, for each $\alpha\in(0,1)$, we can further define the excess risk as:
	\begin{equation*}
		H^{\alpha}(\uwave{{\bDelta}})=L^{\alpha}_{K}(\uwave{\bbeta}^*+\uwave{{\bDelta}})-L^{\alpha}_{K}(\uwave{\bbeta}^*).
	\end{equation*}
	Similar to Section \ref{sec: Proof of lasso property with alpha=0}, the proof of Lemma \ref{lemma: basic inequality for lasso with alpha} relies on the following three lemmas. In detail, 
	Lemma \ref{lemma: basic inequality for quantile regression alpha} shows that $\uwave{\hat{\bbeta}}-\uwave{\bbeta}$ belongs to $\cA$ with a large probability.
	Lemma \ref{lemma: lower bound for the excess risk with alpha} shows that the excess risk $H^\alpha(\uwave{{\bDelta}})$ can be bounded by the quadratic form of $\uwave{{\bDelta}}$.  Lastly, Lemma \ref{lemma: large deviation for excess risk with alpha} shows that we can uniformly control the difference between the excess risk and its empirical version. The proofs of those lemmas are given in Sections \ref{sec: proof of basic inequality for quantile regression alpha} - \ref{sec: proof of large deviation for excess risk with alpha}. 
	\begin{lemma}\label{lemma: basic inequality for quantile regression alpha}
		Assume	{\bf{Assumptions A, B, C.2, D, E.2 - E.4}} hold.Then, with probability tending to one, we have
		\begin{equation*}
			\uwave{\hat{\bbeta}}-\uwave{\bbeta}^*\in \cA
		\end{equation*}
	\end{lemma}
	
	\begin{lemma}\label{lemma: lower bound for the excess risk with alpha}
		Assume	{\bf{Assumptions A, B, C.2, D, E.2 - E.4}} hold. For any $\uwave{{\bDelta}}\in \cA$, with probability tending to one,
		\begin{equation*}
			H^\alpha(\uwave{{\bDelta}})\geq c_*\min\Big(\dfrac{ \|\uwave{{\bDelta}}\|^2_{\bS}}{4},\dfrac{\|\uwave{{\bDelta}}\|_{\bS}}{4} \Big),
		\end{equation*}
		where $c_*>0$ is some universal constant not depending on $n$ or $p$. 
	\end{lemma}
	
	\begin{lemma}\label{lemma: large deviation for excess risk with alpha}
		Assume	{\bf{Assumptions A, B, C.2, D, E.2 - E.4}} hold. With probability tending to one, we have:
		\begin{equation*}
			\sup_{\uwave{{\bDelta}}\in \cA\atop \|\uwave{{\bDelta}}\|_{\bS}\leq \xi }\big|(L^\alpha_{n,K}(\uwave{\bbeta}^*+\uwave{{\bDelta}})-L^\alpha_{n,K}(\uwave{\bbeta}^*))-(L^\alpha_K(\uwave{\bbeta}^*+\uwave{{\bDelta}})-L^\alpha_K(\uwave{\bbeta}^*))\big|\leq C^*M\xi \sqrt{s\dfrac{\log(pn)}{n}},
		\end{equation*}
		where $C^*>0$ is some universal constant not depending on $n$ or $p$ and $s:=|J({\bbeta}^*)|$.
		
		With the above three lemmas, using similar proof procedures as in Section \ref{sec: Proof of lasso property with alpha=0}, we can directly prove Lemma \ref{lemma: basic inequality for lasso with alpha}. To save space, we omit the details  here.
	\end{lemma}

\end{proof}

\section{Additional lemmas}\label{sec: additional lemmas}
\subsection{Proof of Lemma \ref{lemma: upper bound for the empirical process of alpha=0 under H0}}\label{sec: proof of upper bound for the empirical process of alpha=0 under H0}
\begin{proof}
	Recall 
	\begin{equation*}
		\begin{array}{ll}
			\bC^{II,2}_{0}(t)=\dfrac{1}{\sqrt{n}\hat{\sigma}(\alpha,\tilde{\btau})}\big(\sum\limits_{i=1}^{\floor{nt}}\dfrac{1}{K}\sum\limits_{k=1}^K\bX_i\big(g_{ik}(\uwave{\bX_i}^\top\uwave{\hat{\bDelta}_k})-g_{ik}(0)\big)\\
			-\dfrac{\floor{nt}}{n}\sum\limits_{i=1}^n\dfrac{1}{K}\sum\limits_{k=1}^K\bX_i\big(g_{ik}(\uwave{\bX_i}^\top\uwave{\hat{\bDelta}_k})-g_{ik}(0)\big)
		\end{array}
	\end{equation*}
	where $g_{ik}(t):=\mathbf{1}\{\epsilon_i\leq b_k^{(0)}+t\}-\P\{\epsilon_i\leq b_k^{(0)}+t\}$. Note that under $\cE$, we have:
	\begin{equation*}
		\begin{array}{ll}
			\max\limits_{t\in[q_0,1-q_0]}\|\bC^{II,2}_{0}(t)\|_{(s_0,2)}\\
			\quad\leq C\sqrt{n}s_0^{1/2}\max\limits_{t\in[q_0,1-q_0]}\big\|\big(\dfrac{1}{\floor{nt}}\sum\limits_{i=1}^{\floor{nt}}\dfrac{1}{K}\sum\limits_{k=1}^K\bX_i\big(g_{ik}(\uwave{\bX_i}^\top\uwave{\hat{\bDelta}_k})-g_{ik}(0)\big)\big\|_{\infty}\\
			\qquad\quad+C\sqrt{n}s_0^{1/2}\big\|\big(\dfrac{1}{n}\sum\limits_{i=1}^{n}\dfrac{1}{K}\sum\limits_{k=1}^K\bX_i\big(g_{ik}(\uwave{\bX_i}^\top\uwave{\hat{\bDelta}_k})-g_{ik}(0)\big)\big\|_{\infty}\\
			\quad\leq
			C\sqrt{n}s_0^{1/2}\max\limits_{t}\max\limits_{j}\big|\big(\dfrac{1}{\floor{nt}}\sum\limits_{i=1}^{\floor{nt}}\dfrac{1}{K}\sum\limits_{k=1}^KX_{ij}\big(g_{ik}(\uwave{\bX_i}^\top\uwave{\hat{\bDelta}_k})-g_{ik}(0)\big)\big|\\
			\qquad\quad+C\sqrt{n}s_0^{1/2}\max\limits_{j}\big|\big(\dfrac{1}{n}\sum\limits_{i=1}^{n}\dfrac{1}{K}\sum\limits_{k=1}^KX_{ij}\big(g_{ik}(\uwave{\bX_i}^\top\uwave{\hat{\bDelta}_k})-g_{ik}(0)\big)\big|.\\
		\end{array}
	\end{equation*}
	Define 
	\begin{equation*}
		\begin{array}{ll}
			\psi_{j}(\epsilon_i,\bX_i;\uwave{{\bDelta}_k})&:=X_{ij}(\mathbf{1}\{\epsilon_i\leq \uwave{\bX_i}^\top\uwave{{\bDelta}_k}+b_k^{(0)}\}-\mathbf{1}\{\epsilon_i\leq b_k^{(0)}\})\\
			&=X_{ij}(\mathbf{1}\{\epsilon_i\leq {\bX_i}^\top{{\bDelta}}+\delta_k+b_k^{(0)}\}-\mathbf{1}\{\epsilon_i\leq b_k^{(0)}\}),
		\end{array}
	\end{equation*}
	where $\bDelta:=\bbeta-\bbeta^{(0)}$ and $\delta_k:=b_k-b_k^{(0)}$ for $1\leq k\leq K$. Hence, by definition, conditional on $\cX:=(\bX_1,\ldots,\bX_n)$, we have:
	\begin{equation}\label{inequality: tail I+II}
		\begin{array}{ll}
			\max\limits_{t\in[q_0,1-q_0]}\|\bC^{II,2}_{0}(t)\|_{(s_0,2)}|\cX\\
			\quad\leq
			C\sqrt{n}s_0^{1/2}\underbrace{\max\limits_{t}\max\limits_{j}\big|\big(\dfrac{1}{\floor{nt}}\sum\limits_{i=1}^{\floor{nt}}\dfrac{1}{K}\sum\limits_{k=1}^K(\psi_{j}(\epsilon_i,\bX_i;\uwave{\hat{\bDelta}_k})-\E[\psi_{j}(\epsilon_i,\bX_i;\uwave{\hat{\bDelta}_k})])\big)}_{I}\big|\\
			\qquad\quad+C\sqrt{n}s_0^{1/2}\underbrace{\max\limits_{j}\big|\big(\dfrac{1}{n}\sum\limits_{i=1}^{n}\dfrac{1}{K}\sum\limits_{k=1}^K(\psi_{j}(\epsilon_i,\bX_i;\uwave{\hat{\bDelta}_k})-\E[\psi_{j}(\epsilon_i,\bX_i;\uwave{\hat{\bDelta}_k})])\big)\big|}_{II}.\\
		\end{array}
	\end{equation}
	Hence, to bound $\max\limits_{t\in[q_0,1-q_0]}\|\bC^{II,2}_{0}(t)\|_{(s_0,2)}|\cX$, we need to consider $I$ and $II$, respectively. We first consider $I$. To that end, conditional on $\cX:=(\bX_1,\ldots,\bX_n)$, define the function:
	\begin{equation*}
		\begin{array}{ll}
			G_{t,j}(\uwave{{\bDelta}})& =\dfrac{1}{\floor{nt} }\sum\limits_{i=1}^{\floor{nt}}\dfrac{1}{K}\sum\limits_{k=1}^K(\psi_{j}(\epsilon_i,\bX_i;\uwave{{\bDelta}_k})-\E[\psi_{j}(\epsilon_i,\bX_i;\uwave{{\bDelta}_k})])\\
			& :=\dfrac{1}{n' }\sum\limits_{i=1}^{n'}\dfrac{1}{K}\sum\limits_{k=1}^K(\psi_{j}(\epsilon_i,\bX_i;\uwave{{\bDelta}_k})-\E[\psi_{j}(\epsilon_i,\bX_i;\uwave{{\bDelta}_k})])\\
			&:=G_{n',j}(\uwave{{\bDelta}}),
		\end{array}
	\end{equation*}
	where $n':=\floor{nt}$. Moreover, for the  sparsity parameter $s$ of $\bbeta^{(0)}$, and some big enough real numbers $\xi_1,\xi_2,\xi_3>0$, define the parameter space:
	\begin{equation*}
		\begin{array}{ll}
			\cR(\xi_1,\xi_2,\xi_3)\\
			:=\Big\{\uwave{\bDelta}=(\bDelta^\top,\bdelta^\top)^\top: \|\bDelta\|_0\leq \xi_1 s, \|\bDelta\|_2\leq \xi_2\sqrt{s\dfrac{\log(pn)}{n}},\|\bdelta\|_2\leq \xi_3\sqrt{s\dfrac{\log(pn)}{n}}    \Big\}.
		\end{array}
	\end{equation*}
	By Lemma \ref{lemma: basic inequality for lasso with alpha=0}, with probability tending to 1, we have $\uwave{\hat{\bDelta}}\in \cR(\xi_1,\xi_2,\xi_3)$ for some large enough constants $\xi_1,\xi_2,\xi_3>0$. Hence, to bound $I$, it is sufficient to bound:
	\begin{equation*}
		\max\limits_{1\leq j\leq p}\max\limits_{t\in[q_0,1-q_0]}\sup_{\uwave{{\bDelta}}\in \cR(\xi_1,\xi_2,\xi_3)}|G_{t,j}(\uwave{{\bDelta}})||\cX=\max\limits_{1\leq j\leq p}\max\limits_{n'\in[\floor{nq_0},n-\floor{nq_0}]}\sup_{\uwave{{\bDelta}}\in \cR(\xi_1,\xi_2,\xi_3)}|G_{n',j}(\uwave{{\bDelta}})||\cX.
	\end{equation*}
	Thoughout the following proofs, we assume $K$ is fixed which does not grow with $n$. To obtain the desired bound, we define the functional class:
	\begin{equation}\label{equation: final functional class with alpha=0}
		\cF=\Big\{f_{\uwave{{\bDelta}}}(\epsilon,\bX)=\dfrac{1}{K}\sum\limits_{k=1}^K(\psi_{j}(\epsilon,\bX;\uwave{{\bDelta}_k})|\uwave{{\bDelta}}\in \cR(\xi_1,\xi_2,\xi_3)\Big\}.
	\end{equation}
	Firstly, we obtain the upper bound for each fixed $n'\in[\floor{nq_0},n-\floor{nq_0}]$ and $1\leq j\leq p$. The main idea is to use Theorem 3.11 in Koltchinskii (2011) (Lemma A.1 in \cite{2014A}) and the Bousquet inequality (Corollary 14.2 in \cite{Peter2011}) to obtain the tail probability of $\sup_{\uwave{{\bDelta}}\in \cR(\xi_1,\xi_2,\xi_3)}|G_{n',j}(\uwave{{\bDelta}})||\cX$. The proofs proceed into five steps.\\
	{\bf{Step~1: obtain the envelope for $f_{\uwave{{\bDelta}}}(\epsilon,\bX)$. }} In fact, by {\bf{Assumption A}}, we have: 
	\begin{equation*}
		\begin{array}{ll}
			\sup_{\uwave{{\bDelta}}\in \cR(\xi_1,\xi_2,\xi_3)}|f_{\uwave{{\bDelta}}}(\epsilon,\bX)|\\
			\quad=\sup_{\uwave{{\bDelta}}\in \cR(\xi_1,\xi_2,\xi_3)}\dfrac{1}{K}|\sum\limits_{k=1}^K(\psi_{j}(\epsilon,\bX;\uwave{{\bDelta}_k})|\\
			\quad=\sup_{\uwave{{\bDelta}}\in \cR(\xi_1,\xi_2,\xi_3)}\dfrac{1}{K}|\sum\limits_{k=1}^K(X_{j}\mathbf{1}\{\epsilon\leq {\bX}^\top{{\bDelta}}+\delta_k+b_k^{(0)}\}-\mathbf{1}\{\epsilon\leq b_k^{(0)}\})|\\
			\quad \leq M,
		\end{array}
	\end{equation*}
	where the last inequality comes from the assumption that $|X_{j}|\leq M$ for $1\leq j\leq p$. \\
	{\bf{Step~2: obtain the upper bound for  $\sigma_{n'}^2:=\sup_{\uwave{{\bDelta}}}\dfrac{1}{n'}\sum\limits_{i=1}^{n'}\text{Var}[f_{\uwave{{\bDelta}}}(\epsilon_i,\bX_i)|\cX]$, }} where $\cX:=\{\bX_1,\ldots,\bX_n\}$. In fact, similar to the proof of Lemma 6.1 in \cite{2014A}, we have:
	\begin{equation}\label{inequality: sigma-n}
		\begin{array}{ll}
			&\dfrac{1}{n'}\sum\limits_{i=1}^{n'}\text{Var}[f_{\uwave{{\bDelta}}}(\epsilon_i,\bX_i)|\cX]\\
			&\quad\leq_{(1)} \dfrac{1}{n'}\sum\limits_{i=1}^{n'}\dfrac{6C_+M^2}{K}\sum\limits_{k=1}^K|\tilde{\bX}_i^\top\uwave{{\bDelta}_k}|\\
			&\quad\leq_{(2)}\dfrac{6C_+M^2}{K} \Big(\dfrac{1}{n'}\sum\limits_{i=1}^{n'}K|{\bX}_i^\top{{\bDelta}}|+\|\bdelta\|_1\Big)\\
			&\quad\leq_{(3)}\dfrac{6C_+M^2}{K} \Big(K\sqrt{{{\bDelta}}^\top(\dfrac{1}{n'}\sum\limits_{i=1}^{n'} \bX_i\bX_i^\top){{\bDelta}}}+\sqrt{K}\|\bdelta\|_2\Big)\\
			&\quad\leq_{(4)}\dfrac{6C_+M^2}{K} \Big(K\sqrt{{{\bDelta}}^\top(\dfrac{1}{n'}\sum\limits_{i=1}^{n'} \bX_i\bX_i^\top-\bSigma){{\bDelta}}+{{\bDelta}}^\top \bSigma{{\bDelta}}}+\sqrt{K}\|\bdelta\|_2\Big)\\
			&\quad\leq_{(5)} CM^2(\|\bDelta\|_2+\|\bdelta\|_2)\\
			&\quad\leq_{(6)} CM^2\sqrt{s\dfrac{\log(pn)}{n}}:=\sigma_{n'}^2,
		\end{array}
	\end{equation}
	where $(3)$ comes from the Cauchy-Swartz inequality, $(5)$ comes from Lemma \ref{lemma: concentration for covariance}. \\
	{\bf{Step~3: } obtain the covering number of the functional class $\cF$ as defined in (\ref{equation: final functional class with alpha=0}).} Let $\cT\subset\{1,\ldots,p\}$  with $|\cT|=\xi_1s$. Moreover, define the following functional classes:
	\begin{equation*}
		\small
		\begin{array}{ll}
			\cF_{k}:=\Big\{f_k(\epsilon,\bX)= \mathbf{1}\{\epsilon\leq {\bX}^\top{{\bDelta}}+\delta_k+b_k^{(0)}\}-\mathbf{1}\{\epsilon\leq b_k^{(0)}\} | \uwave{{\bDelta}}\in \cR(\xi_1,\xi_2,\xi_3), \text{supp}(\bDelta) \subset \cT \Big\},\\
			\cF_{K}:=\Big\{f_K(\epsilon,\bX)= \sum\limits_{k=1}^K\big(\mathbf{1}\{\epsilon\leq {\bX}^\top{{\bDelta}}+\delta_k+b_k^{(0)}\}-\mathbf{1}\{\epsilon\leq b_k^{(0)}\}\big) | \uwave{{\bDelta}}\in \cR(\xi_1,\xi_2,\xi_3), \text{supp}(\bDelta) \subset \cT \Big\}\\
			\cF_{0}:=\Big\{f_0(\epsilon,\bX)= \dfrac{1}{K}X_j\}\\
			\cF_{\cT}=\cF_{K}\cF_0=\Big\{f_{\cT}(\epsilon,\bX)= f_K(\epsilon,\bX)f_0(\epsilon,\bX)|\uwave{{\bDelta}}\in \cR(\xi_1,\xi_2,\xi_3), \text{supp}(\bDelta) \subset \cT \Big\},
		\end{array}
	\end{equation*}
	where $\text{supp}(\bDelta)$ denotes the support set for $\bDelta$. Note that $\cF_{k}$ is a VC-class with VC index smaller than $\xi_1s+2$, and $|f_k(\epsilon,\bX)|\leq  1$, $|f_K(\epsilon,\bX)|\leq  K$, and $|f_0(\epsilon,\bX)|\leq  M/K$.
	Let $N(\epsilon,\cF,L_2(Q))$ be the covering number for some functional class $\cF$ under the $L_2(Q)$ distance. Then, by Lemma 24 (ii) in \cite{2016Quantile} and the definition of $\cF_{\cT}$, we have:
	\begin{equation}\label{inequality: relation bewteen covering num with alpha=0}
		\begin{array}{ll}
			N(K\epsilon,\cF_{K},L_2(Q))\leq [N(\dfrac{\epsilon}{K},\cF_{k},L_2(Q))]^K,\\
			N(\epsilon K\dfrac{M}{K},\cF_{\cT},L_2(Q))=N(\epsilon M,\cF_{\cT},L_2(Q))\leq N(\dfrac{\epsilon K}{2},\cF_{K},L_2(Q)),\\
			N(\epsilon M,\cF,L_2(Q))\leq C_p^{\xi_1s}N(\epsilon M,\cF_{\cT},L_2(Q)).
		\end{array}
	\end{equation}
	Hence, by (\ref{inequality: relation bewteen covering num with alpha=0}), we have:
	\begin{equation}\label{equality: covering num for final-1}
		N(\epsilon M,\cF,L_2(Q))\leq C_p^{\xi_1s}[N(\dfrac{\epsilon}{2K},\cF_{k},L_2(Q))]^K.
	\end{equation}
	Furthermore, by Lemma 2.6.7 in \cite{1996Weak}, we have 
	\begin{equation}\label{equality: covering num for final-2}
		N(\dfrac{\epsilon}{2K},\cF_{k},L_2(Q))]\leq C(\xi_1s+2)(16e)^{\xi_1s+2}(\dfrac{2K}{\epsilon})^{2(\xi_1s+1)},
	\end{equation}
	where $C$ is some universal constant. Combining (\ref{equality: covering num for final-1}) and (\ref{equality: covering num for final-2}), for any probability measure $Q$, we have:
	\begin{equation*}
		\begin{array}{ll}
			N(\epsilon M,\cF,L_2(Q))
			&\leq C_p^{\xi_1s}[N(\dfrac{\epsilon}{2K},\cF_{k},L_2(Q))]^K\\
			&\leq C\Big(\dfrac{pe}{\xi_1s}\Big)^{\xi_1s}(\xi_1s+2)^{K}(16e)^{K(\xi_1s+2)}(\dfrac{2K}{\epsilon})^{2K(\xi_1s+1)}\\
			&\leq C\Big(\dfrac{pe}{\xi_1s}\Big)^{\xi_1s}\Big(\dfrac{32eK}{\epsilon}\Big)^{c\xi_1s},
		\end{array}
	\end{equation*} 
	where $c$ and $C$ are some big enough positive constants. \\
	{\bf{Step~4: obtain the upper bound of $\E\Big[\sup_{\uwave{{\bDelta}}\in \cR(\xi_1,\xi_2,\xi_3)}|G_{n',j}(\uwave{{\bDelta}})|\big|\cX\Big]$. }} Recall $\sigma_{n'}$ defined in (\ref{inequality: sigma-n}). By Lemma A.1 in \cite{2014A}, and using some basic calculations, we have:
	\begin{equation*}
		\begin{array}{ll}
			&\E\Big[\sup_{\uwave{{\bDelta}}\in \cR(\xi_1,\xi_2,\xi_3)}|G_{n',j}(\uwave{{\bDelta}})|\big|\cX\Big]\\
			&\quad\leq \dfrac{C}{\sqrt{n'}}\E\Big[\displaystyle{\Large\int_{0}^{2\sigma_{n'}} \sqrt{\log N(\epsilon,\cF,L_{2}(\PP_n|\cX))} d\epsilon }        \Big]\\
			&\quad\leq \dfrac{C}{\sqrt{n'}}\E\Big[\displaystyle{\Large\int_{0}^{2\sigma_{n'}} \sqrt{\sup_{Q}\log N(\epsilon,\cF,L_{2}(Q))}d\epsilon  }        \Big]\\
			&\quad\leq \dfrac{C}{\sqrt{n'}}\displaystyle{\Large\int_{0}^{2\sigma_{n'}} \sqrt{s\log(\dfrac{p}{\epsilon}) }}  d\epsilon      \\
			&\quad\leq C\sigma_{n'}\sqrt{\dfrac{s\log(p\vee n')}{n'}}:=r_{n'}.
		\end{array}
	\end{equation*}	
	{\bf{Step~5: obtain the tail bound of  $\sup_{\uwave{{\bDelta}}\in \cR(\xi_1,\xi_2,\xi_3)}|G_{n',j}(\uwave{{\bDelta}})|\cX$. }} In fact, by the Bousquet inequality (Corollary 14.2 in \cite{Peter2011}), we have:
	\begin{equation*}
		\begin{array}{ll}
			\P\Big(\sup_{\uwave{{\bDelta}}\in \cR(\xi_1,\xi_2,\xi_3)}|G_{n',j}(\uwave{{\bDelta}})|\geq r_{n'}+t\sqrt{2(\sigma_{n'}^2+2Mr_{n'})}+\dfrac{2t^2M}{3}|\cX\Big)\\
			\leq_{(1)} \exp(-n't^2)\leq_{(2)}\exp(-q_0nt^2),
		\end{array}
	\end{equation*}
	where $(2)$ comes from $n'=\floor{nt}$ with $t\in[q_0,1-q_0]$. It is straightforward to see that if we take $t=C^*\sqrt{\log(pn)/n}$ for some big enough constant $C^*>0$, we have:
	\begin{equation*}
		\P\Big(\sup_{\uwave{{\bDelta}}\in \cR(\xi_1,\xi_2,\xi_3)}|G_{n',j}(\uwave{{\bDelta}})|\geq C_1(\dfrac{s\log(pn)}{n})^{\frac{3}{4}}|\cX\Big)\leq (pn)^{-C_2}.
	\end{equation*}
	The above result yieds that:
	\begin{equation*}
		\P\Big(\max_{n',j}\sup_{\uwave{{\bDelta}}\in \cR(\xi_1,\xi_2,\xi_3)}|G_{n',j}(\uwave{{\bDelta}})|\geq C_1(\dfrac{s\log(pn)}{n})^{\frac{3}{4}}|\cX\Big)\leq (pn)^{-C_3},
	\end{equation*}
	which proves that $I=O_p((\dfrac{s\log(pn)}{n})^{\frac{3}{4}})$, where $I$ is defined in (\ref{inequality: tail I+II}). With a similar proof technique, we can also prove $II=O_p((\dfrac{s\log(pn)}{n})^{\frac{3}{4}})$. Combining with (\ref{inequality: tail I+II}), we have proved that:
	\begin{equation*}
		\max\limits_{t\in[q_0,1-q_0]}\|\bC^{II,2}_{0}(t)\|_{(s_0,2)}|\cX=O_p(s_0^{1/2}(s\log(pn))^{3/4}/n^{1/4}),
	\end{equation*}
	which finishes the proof of Lemma \ref{lemma: upper bound for the empirical process of alpha=0 under H0}.
\end{proof}
\subsection{Proof of Lemma \ref{lemma: basic inequality for quantile regression}}\label{lemma: proof of basic inequality for quantile regression}
\begin{proof}
	Recall $\uwave{\bbeta}=((\bbeta)^\top,(\bb)^\top)^\top\in \RR^{p+K}$ and  $L_{n,K}(\uwave{\bbeta}):=\dfrac{1}{n}\sum\limits_{i=1}^n\dfrac{1}{K}\sum\limits_{k=1}^{K}\rho_{\tau_k}(Y_i-\bX_{i}^\top\bbeta-b_k)$. Define 
	\begin{equation*}
		\begin{array}{ll}
			\nabla L_{n,K}(\uwave{\bbeta})=\dfrac{\partial L_{n,K}(\uwave{\bbeta})}{\partial \uwave{\bbeta}}\in \RR^{p+K},\\
			~~\nabla_1 L_{n,K}(\uwave{\bbeta})=\dfrac{\partial L_{n,K}(\uwave{\bbeta})}{\partial {\bbeta}}\in \RR^{p}, \\
			\nabla_2 L_{n,K}(\uwave{\bbeta})=\dfrac{\partial L_{n,K}(\uwave{\bbeta})}{\partial {\bb}}\in \RR^{K}.
		\end{array}
	\end{equation*}
	Hence, if we define $a^*_{i,k}=\mathbf{1}\{Y_i-\bX_i^\top\bbeta^*-b^*_k\leq 0\}-\tau_k$ for $i=1,\ldots,n$ and $k=1,\ldots,K$, we have:
	\begin{equation*}
		\nabla_1 L_{n,K}(\uwave{\bbeta}^*)=\dfrac{1}{n}\sum_{i=1}^n\dfrac{1}{K}\sum_{k=1}^K\bX_ia_{i,k}^*,~~\text{and}~~\nabla_2L_{n,K}(\uwave{\bbeta}^*)=\dfrac{1}{n}\sum_{i=1}^n\ba_i^*,
	\end{equation*}
	where $\ba_i^*=(a_{i,1},\ldots,a_{i,K})^\top\in \RR^K$. The proof of  Lemma  \ref{lemma: basic inequality for quantile regression} proceeds into two steps.
	
	{\bf{Step~1:}} obtain the upper bounds of $\|\nabla_1 L_{n,K}(\uwave{\bbeta}^*)\|_{\infty}$ and $\|\nabla_2 L_{n,K}(\uwave{\bbeta}^*)\|_{\infty}$. We first consider $\|\nabla_2 L_{n,K}(\uwave{\bbeta}^*)\|_{\infty}$. In fact, we have:
	\begin{equation*}
		\begin{array}{ll}
			&\|\nabla_2 L_{n,K}(\uwave{\bbeta}^*)\|_{\infty}\\&\quad=_{(1)}\max\limits_{1\leq k\leq K}|\dfrac{1}{n}\sum\limits_{i=1}^na_{i,k}^*|\\
			&\quad=_{(2)}\max\limits_{1\leq k\leq K}\big|\dfrac{1}{n}\sum\limits_{i=1}^n(a_{i,k}^*-\E[a^*_{i,k}])\big|\\
			&\quad\leq _{(3)}\max\limits_{1\leq k\leq K}t_1\underbrace{\big|\dfrac{1}{nt_1}\sum\limits_{i=1}^{nt_1}\big(\mathbf{1}\{\epsilon_i\leq \bX_{i}^\top(\bbeta^*-\bbeta^{(1)})+b_k^{*}\}-\E (F_{\epsilon}(\bX_{i}^\top(\bbeta^*-\bbeta^{(1)})+b_k^{*})\big)\big|}_{I}\\
			&\quad+\max\limits_{1\leq k\leq K}t_2\underbrace{\big|\dfrac{1}{nt_2}\sum\limits_{i=nt_1+1}^{n}\big(\mathbf{1}\{\epsilon_i\leq \bX_{i}^\top(\bbeta^*-\bbeta^{(2)})+b_k^{*}\}-\E (F_{\epsilon}(\bX_{i}^\top(\bbeta^*-\bbeta^{(2)})+b_k^{*})\big)\big|}_{II},
		\end{array}
	\end{equation*}
	where $(2)$ comes from the first order condition in (\ref{equation: first order condition alpha=0}). Hence, to control $\|\nabla_2 L_{n,K}(\uwave{\bbeta}^*)\|_{\infty}$, we need to consider $I$ and $II$. Let $Z_{i,k}:=\mathbf{1}\{\epsilon_i\leq \bX_{i}^\top(\bbeta^*-\bbeta^{(1)})+b_k^{*}\}-\E (F_{\epsilon}(\bX_{i}^\top(\bbeta^*-\bbeta^{(1)})+b_k^{*})$. Note that $\E[Z_{i,k}]=0$ and $-1\leq Z_{i,k}\leq 1$. Hence, by the Hoeffding's inequality, we can prove that $(I\vee II) \leq C_1\sqrt{\log(p)/n}$ w.p.a.1. Hence, we prove $\|\nabla_2L_{n,K}(\uwave{\bbeta}^*)\|_{\infty}\leq C_1\sqrt{\log(p)/n}$ w.p.a.1. for some $C_1>0$. Next, we consider $\|\nabla_1 L_{n,K}(\uwave{\bbeta}^*)\|_{\infty}$. In fact, we have:
	\begin{equation*}
		\begin{array}{ll}
			&\|\nabla_1 L_{n,K}(\uwave{\bbeta}^*)\|_{\infty}\\
			&\quad=_{(1)}\max\limits_{1\leq j\leq p}\big|\dfrac{1}{n}\sum\limits_{i=1}^n\dfrac{1}{K}\sum\limits_{k=1}^K\bX_ia_{i,k}^*\big|\\
			&\quad=_{(2)}\max\limits_{1\leq j\leq p}\big|\dfrac{1}{n}\sum\limits_{i=1}^n\dfrac{1}{K}\sum\limits_{k=1}^K(X_{ij}a_{i,k}^*-\E[X_{ij}a_{i,k}^*])\big|\\
			&\quad\leq_{(3)} \max\limits_{1\leq j\leq p}\max\limits_{1\leq k\leq K}\big|\dfrac{1}{n}\sum\limits_{i=1}^n(X_{ij}a_{i,k}^*-\E[X_{ij}a_{i,k}^*])\big|\\
			&\quad\leq_{(4)} \underbrace{\max\limits_{1\leq j\leq p}\max\limits_{1\leq k\leq K}t_1\big|\dfrac{1}{nt_1}\sum\limits_{i=1}^{nt_1}(X_{ij}a_{i,k}^*-\E[X_{ij}a_{i,k}^*])\big|}_{III}\\
			&\qquad\quad+\underbrace{\max\limits_{1\leq j\leq p}\max\limits_{1\leq k\leq K}t_2\big|\dfrac{1}{nt_2}\sum\limits_{i=nt_1+1}^{n}(X_{ij}a_{i,k}^*-\E[X_{ij}a_{i,k}^*])\big|}_{IV},\\
		\end{array}
	\end{equation*}
	where $(2)$ comes from the first order condition in (\ref{equation: first order condition alpha=0}).
	Let $W_{ijk}=X_{ij}a_{i,k}^*-\E[X_{ij}a_{i,k}^*]$. Conditional on $\Xb$, for fixed $j,k$, we have $ -M\leq -|X_{ij}|\leq W_{ijk}\leq |X_{ij}|\leq M$ and $\E[W_{ijk}]=0$. Hence, by the Hoeffding's inequality, it is not hard to see that $(III\vee IV)\leq C_2M\sqrt{\log(p)/n}$ w.p.a.1, which yields
	$\|\nabla_1 L_{n,K}(\uwave{\bbeta}^*)\|_{\infty}\leq C_2M\sqrt{\log(p)/n}$ w.p.a.1. 
	
	{\bf{Step~2:}} Let $\lambda\geq 2M(C_1\vee C_2)\sqrt{\log(p)/n}$, where $C_1$ and $C_2$ are defined in Step 1. Hence, we have $\|\nabla_1 L_{n,K}(\uwave{\bbeta}^*)\|_{\infty}\leq \lambda/2$ and $\|\nabla_2 L_{n,K}(\uwave{\bbeta}^*)\|_{\infty}\leq \lambda/2$ w.p.a.1. By the convexity of $L_{n,K}(\uwave{\bbeta})$, we have:
	\begin{equation*}
		L_{n,K}(\uwave{\hat{\bbeta}})-L_{n,K}(\uwave{\bbeta}^*)\geq \nabla L_{n,K}^\top({{\bbeta}^*})(\uwave{\hat{\bbeta}}-\uwave{{\bbeta}}^*)=\nabla_1 L_{n,K}^\top(\uwave{{\bbeta}^*})({\hat{\bbeta}}-{{\bbeta}}^*)+\nabla_2 L_{n,K}^\top(\uwave{{\bbeta}^*})({\hat{\bb}}-{{\bb}}^*).
	\end{equation*}
	Combining the above inequality and by the optimality of $\uwave{\hat{\bbeta}}$, we have:
	\begin{equation*}
		\begin{array}{ll}
			0&\leq L_{n,K}(\uwave{\bbeta}^*)-L_{n,K}(\uwave{\hat{\bbeta}})+\lambda(\|\bbeta^*\|_1-\|\hat{\bbeta}\|_1)\\
			&\leq \|\nabla_1 L_{n,K}(\uwave{{\bbeta}}^*)\|_{\infty}\|{\hat{\bbeta}}-{{\bbeta}}^*\|_{1}+\|\nabla_2 L_{n,K}(\uwave{\bbeta}^*)\|_{\infty}\|{\hat{\bb}}-{{\bb}}^*\|_{1}+\lambda(\|\bbeta^*\|_1-\|\hat{\bbeta}\|_1)\\
			&\leq \dfrac{\lambda}{2}\|{\hat{\bbeta}}-{{\bbeta}}^*\|_{1}+\dfrac{\lambda}{2}\|{\hat{\bb}}-{{\bb}}^*\|_{1}+\lambda(\|\bbeta^*\|_1-\|\hat{\bbeta}\|_1).
		\end{array}
	\end{equation*}
	Adding  $\dfrac{\lambda}{2}\|\hat{\bbeta}-\bbeta^*\|_1$  on both sides of the above inequality, and using the same proof as in Section \ref{sec: Proof of lasso property with alpha=1}, we can derive that:
	\begin{equation*}
		\| ({\hat{\bbeta}}-{\bbeta}^*)_{J^c(\bbeta^*)}\|_1\leq 3\| ({\hat{\bbeta}}-{\bbeta}^*)_{J(\bbeta^*)}\|_1+\|\hat{\bb}-\bb^*\|,
	\end{equation*}
	which finishes the proof.
	
\end{proof}
\subsection{Proof of Lemma \ref{lemma: lower bound for the excess risk with alpha=0}}
\label{lemma: proof of ower bound for the excess risk with alpha=0}
\begin{proof}
	By the well-known Knight's equation that: $\rho_{\tau}(x-y)-\rho_\tau(x)=-y(\tau-\mathbf{1}\{x\leq 0\})+\int_{0}^y\mathbf{1}\{x\leq s\}-\mathbf{1}\{x\leq 0\}ds$, and the	definition of $H(\uwave{{\bDelta}})$, we have:
	\begin{equation*}
		\begin{array}{ll}
			H(\uwave{{\bDelta}})&=L_{K}(\uwave{\bbeta}^*+\uwave{{\bDelta}})-L_{K}(\uwave{\bbeta}^*)\\
			&=\dfrac{1}{n}\sum\limits_{i=1}^n\dfrac{1}{K}\sum\limits_{k=1}^{K}\E\big[\rho_{\tau_k}(Y_i-\bX_{i}^\top\bbeta^*-b^*_k-(\bX_{i}^\top\bDelta+\delta_k))-\rho_{\tau_k}(Y_i-\bX_{i}^\top\bbeta^*-b^*_k)\big]\\
			&=I+II,
		\end{array}
	\end{equation*}
	where 
	\begin{equation*}
		\begin{array}{ll}
			I&=\dfrac{1}{n}\sum\limits_{i=1}^n\dfrac{1}{K}\sum\limits_{k=1}^{K}\E[(\bX_{i}^\top\bDelta+b_k)(\mathbf{1}\{Y_i-\bX_{i}^\top\bbeta^*-b^*_k\leq 0\}-\tau_k)]\\
			&=\bDelta^\top\dfrac{1}{n}\sum\limits_{i=1}^n\dfrac{1}{K}\sum\limits_{k=1}^{K}\E[\bX_{i}(\mathbf{1}\{Y_i-\bX_{i}^\top\bbeta^*-b^*_k\leq 0\}-\tau_k)]\\
			&\qquad\qquad\qquad+\dfrac{1}{K}\sum\limits_{k=1}^{K}b_k\dfrac{1}{n}\sum\limits_{i=1}^n\E[(\mathbf{1}\{Y_i-\bX_{i}^\top\bbeta^*-b^*_k\leq 0\}-\tau_k)]
		\end{array}
	\end{equation*}	
	and 
	\begin{equation*}
		\begin{array}{ll}
			II&=\dfrac{1}{n}\sum\limits_{i=1}^n\dfrac{1}{K}\sum\limits_{k=1}^{K}\E\int_{0}^{(\bX_{i}^\top\bDelta+\delta_k)}(\mathbf{1}\{Y_i-\bX_{i}^\top\bbeta^*-b^*_k\leq s\}-\mathbf{1}\{Y_i-\bX_{i}^\top\bbeta^*-b^*_k\leq 0\})\\
			&=\underbrace{t_1\dfrac{1}{K}\sum\limits_{k=1}^{K}\E\int_{0}^{(\bX^\top\bDelta+\delta_k)}(F_{\epsilon}(\bX^\top(\bbeta^*-\bbeta^{(1)})+b_k^*+s)-F_{\epsilon}(\bX^\top(\bbeta^*-\bbeta^{(1)})+b_k^*))}_{III}\\
			&\quad+\underbrace{t_2\dfrac{1}{K}\sum\limits_{k=1}^{K}\E\int_{0}^{(\bX^\top\bDelta+\delta_k)}(F_{\epsilon}(\bX^\top(\bbeta^*-\bbeta^{(2)})+b_k^*+s)-F_{\epsilon}(\bX^\top(\bbeta^*-\bbeta^{(2)})+b_k^*))}_{IV}.
		\end{array}
	\end{equation*}
	Note that by the first order condition of $\uwave{\bbeta}^*$ in (\ref{equation: first order condition alpha=0}), we have $I=0$. Recall $\bS_k:=\diag(\mathbf{1}_p,\be_k)$, $\uwave{\bX}:=(\bX^\top,\mathbf{1}_{K})\in \RR^{p+K}$, and $\bS:=\sum_{k=1}^K\E[(\bS_k\uwave{\bX})(\bS_k\uwave{\bX})^\top] $ defined in (\ref{equation: bS}). For $III$, by the Taylor's expansion, we have:
	\begin{equation*}
		\begin{array}{ll}
			III&=_{(1)}t_1\dfrac{1}{K}\sum\limits_{k=1}^{K}\E\Big[\int_{0}^{(\bS_k\uwave{\bX})^\top\uwave{\bDelta}}f_{\epsilon}(\bX^\top(\bbeta^*-\bbeta^{(1)})+b_k^*)s+\dfrac{s^2}{2}f'_{\epsilon}(W)ds\Big]\\
			&\geq_{(2)} t_1\dfrac{1}{K}\dfrac{C_-}{2}\sum\limits_{k=1}^{K}\E[|(\bS_k\uwave{\bX})^\top\uwave{\bDelta}|^2]-t_1\dfrac{1}{K}\dfrac{C'_+}{6}\sum\limits_{k=1}^{K}\E[|(\bS_k\uwave{\bX})^\top\uwave{\bDelta}|^3]\\
			&\geq_{(3)} t_1\dfrac{1}{K}\dfrac{C_-}{2}\|\uwave{\bDelta}\|^2_{\bS}-t_1\dfrac{1}{K}\dfrac{C'_+m_0}{6}\|\uwave{\bDelta}\|^3_{\bS},
		\end{array}
	\end{equation*}
	where $W$ in $(1)$ is some random variable between $\bX^\top(\bbeta^*-\bbeta^{(1)})+b_k^*+s$ and $\bX^\top(\bbeta^*-\bbeta^{(1)})+b_k^*$, $(2)$ follows from the assumption that $\inf_{1\leq k\leq K} f_{\epsilon}(\bX^\top(\bbeta^*-\bbeta^{(1)})+b_k^*)\geq C_-$ and $|f'_{\epsilon}(t)|\leq C'_+$, $(3)$ follows from the assumption that $\sum\limits_{k=1}^{K}\E[|(\bS_k\uwave{\bX})^\top\uwave{\bDelta}|^3]\leq m_0\|\uwave{\bDelta}\|^3_{\bS}$ for some $m_0>0$. Similarly, for $IV$, we have $IV\geq t_2\dfrac{1}{K}\dfrac{C_-}{2}\|\uwave{\bDelta}\|^2_{\bS}-t_2\dfrac{1}{K}\dfrac{C'_+m_0}{6}\|\uwave{\bDelta}\|^3_{\bS}$, which implies the final result:
	\begin{equation*}
		\begin{array}{ll}
			H(\uwave{{\bDelta}})&=_{(1)}I+II\\
			&=_{(2)}III+IV\\
			&\geq_{(3)} \dfrac{1}{K}\dfrac{C_-}{2}\|\uwave{\bDelta}\|^2_{\bS}-\dfrac{1}{K}\dfrac{C'_+m_0}{6}\|\uwave{\bDelta}\|^3_{\bS}\\
			&\geq_{(4)}c_*\min\Big(\dfrac{ \|\uwave{{\bDelta}}\|^2_{\bS}}{4},\dfrac{\|\uwave{{\bDelta}}\|_{\bS}}{4} \Big),
		\end{array}
	\end{equation*}
	where $(4)$ is very similar to the proof of Lemma C.1 in \cite{2014A}, which is omitted.
\end{proof}
\subsection{Proof of Lemma \ref{lemma: large deviation for excess risk with alpha=0}}
\label{sec: proof of large deviation for excess risk with alpha=0}
\begin{proof}
	Let $r_{i,k}=Y_i-\bX_{i}^\top\bbeta^*-b^*_k$ and 
	\begin{equation}\label{equation: Ui}
		\begin{array}{ll}
			U_{i}(\bDelta,\bdelta)&=\dfrac{1}{K}\sum\limits_{k=1}^{K}\big[\rho_{\tau_k}(Y_i-\bX_{i}^\top\bbeta^*-b^*_k-(\bX_{i}^\top\bDelta+\delta_k))-\rho_{\tau_k}(Y_i-\bX_{i}^\top\bbeta^*-b^*_k)\big]\\
			&=\dfrac{1}{K}\sum\limits_{k=1}^{K}\big[\rho_{\tau_k}(r_{i,k}-(\bX_{i}^\top\bDelta+\delta_k))-\rho_{\tau_k}(r_{i,k})\big].
		\end{array}	
	\end{equation}
	Hence, using the above notations, we have:
	\begin{equation*}
		\begin{array}{ll}
			\big(L_{n,K}(\uwave{\bbeta}^*+\uwave{{\bDelta}})-L_{n,K}(\uwave{\bbeta}^*))-(L_K(\uwave{\bbeta}^*+\uwave{{\bDelta}})-L_K(\uwave{\bbeta}^*)\big)=	\dfrac{1}{n}\sum\limits_{i=1}^n\Big[U_{i}(\bDelta,\bdelta)-\E[	U_{i}(\bDelta,\bdelta)]\Big].
		\end{array}
	\end{equation*}
	By the lipschitz continuity of $|\rho_{\tau}(t)-\rho_{\tau}(s)|\leq |s-t|$, we have
	\begin{equation}
		\begin{array}{ll}
			U_{i}(\bDelta,\bdelta)-\E [U_{i}(\bDelta,\bdelta)] \leq \dfrac{2}{K}\sum\limits_{k=1}^{K}|\bX_{i}^\top\bDelta+\delta_k|:=C_i(\bDelta,\bdelta).\\
		\end{array}
	\end{equation}
	Let $Z=\sup_{\uwave{{\bDelta}}\in \cA, \|\uwave{{\bDelta}}\|_{\bS}\leq \xi } \Big|\dfrac{1}{n}\sum\limits_{i=1}^n\Big[U_{i}(\bDelta,\bdelta)-\E[	U_{i}(\bDelta,\bdelta)]\Big|$.
	In what follows, we will use the Massart’s inequality (Theorem 14.2 in \cite{Peter2011}) to obtain the tail bound:
	\begin{equation}\label{inequality: marssart's inequality}
		\P\Big(Z>\E Z +t\Big)\leq \exp\Big(-\dfrac{nt^2}{8\sigma^2}\Big),
	\end{equation}
	where  $\sup\limits_{\uwave{{\bDelta}}\in \cA, \|\uwave{{\bDelta}}\|_{\bS}\leq \xi }\dfrac{1}{n}\sum\limits_{i=1}^nC^2_i(\bDelta,\bdelta)\leq \sigma^2$. Hence, to use Massart’s inequality, we need two steps.\\
	{\bf{Step~1:}} obtain the upper bound for $\sigma^2$. Recall $\bS_k:=\diag(\mathbf{1}_p,\be_k)$, $\uwave{\bX_i}:=(\bX_i^\top,\mathbf{1}_{K})\in \RR^{p+K}$, and $\bS:=\sum_{k=1}^K\E[(\bS_k\uwave{\bX})(\bS_k\uwave{\bX})^\top] $. With probability tending to 1, we have:
	\begin{equation}\label{inequality: upper bound for the variance of marssart}
		\begin{array}{ll}
			\dfrac{1}{n}\sum\limits_{i=1}^nC^2_i(\bDelta,\bdelta)&=_{(1)}\dfrac{1}{n}\sum\limits_{i=1}^n(\dfrac{2}{K}\sum\limits_{k=1}^{K}|\bX_{i}^\top\bDelta+\delta_k|)^2\\
			&\leq_{(2)} \dfrac{4}{n}\sum\limits_{i=1}^n\dfrac{1}{K}\sum\limits_{k=1}^K(\bX_{i}^\top\bDelta+\delta_k)^2\\
			&\leq_{(3)} \dfrac{8}{n}\sum\limits_{i=1}^n\dfrac{1}{K}\sum\limits_{k=1}^K((\bX_{i}^\top\bDelta)^2+\delta^2_k)\\
			&=_{(4)}\dfrac{8}{n}\sum\limits_{i=1}^n\dfrac{1}{K}\sum\limits_{k=1}^K((\bS_k\uwave{\bX_i})^\top \uwave{\bDelta})^2\\
			&=_{(5)}\dfrac{8}{K}\uwave{\bDelta}^\top \Big[\dfrac{1}{n}\sum\limits_{i=1}^n\sum\limits_{k=1}^K((\bS_k\uwave{\bX_i})(\bS_k\uwave{\bX_i})^\top\Big] \uwave{\bDelta}\\
			&=_{(6)} \dfrac{8}{K}\uwave{\bDelta}^\top \bS \uwave{\bDelta}+\dfrac{8}{K}\uwave{\bDelta}^\top \Big[\dfrac{1}{n}\sum\limits_{i=1}^n\sum\limits_{k=1}^K((\bS_k\uwave{\bX_i})(\bS_k\uwave{\bX_i})^\top-\bS\Big] \uwave{\bDelta}\\
			&\leq_{(7)} \dfrac{8}{K}\| \uwave{\bDelta}\|^2_{\bS}+\dfrac{8}{K}\|\uwave{\bDelta}\|_1^2\Big\|\dfrac{1}{n}\sum\limits_{i=1}^n\sum\limits_{k=1}^K((\bS_k\uwave{\bX_i})(\bS_k\uwave{\bX_i})^\top-\bS\Big\|_{\infty}\\
			&\leq_{(8)} \dfrac{8}{K}\| \uwave{\bDelta}\|^2_{\bS}+\dfrac{8}{K}M\sqrt{\log(p)/n}\| \uwave{\bDelta}\|^2_1\\
			&\leq_{(9)} \dfrac{8}{K}\| \uwave{\bDelta}\|^2_{\bS}+O(Ms\sqrt{\log(p)/n})\| \uwave{\bDelta}\|^2_{\bS}\\
			&\leq_{(10)} \dfrac{9}{K}\xi^2,
		\end{array}
	\end{equation}
	where $(2)$ follows from the Cauchy-Swarchz inequality, $(3)$ follows from $(a+b)^2\leq 2a^2+2b^2$, $(8)$ follows from the large deviation for $\Big\|\dfrac{1}{n}\sum\limits_{i=1}^n\sum\limits_{k=1}^K((\bS_k\uwave{\bX_i})(\bS_k\uwave{\bX_i})^\top-\bS\Big\|_{\infty}$, $(9)$ follows from the fact that $ \|\uwave{\bDelta}\|_1\leq 4\|\bDelta_{J(\bbeta^*)}\|_1+ \|\bdelta\|_1$ and the  Cauchy-Swarchz inequality, and $(10)$ comes from the assumption that $Ms\sqrt{\log(p)/n}=o(1)$.\\
	{\bf{Step~2:}} obtain the upper bound for $\E [Z]$. Let $e_1,\ldots,e_n$ be i.i.d Rademacher random variables with $\P(e_i=1)=\P(e_i=-1)=1/2$. In fact, by the symmetrization procedure (Theorem 14.3 in \cite{Peter2011}) and 
	the contraction principle (Theorem 14.4 in \cite{Peter2011}), we have:
	\begin{equation}\label{inequality: upper bound for maximum expectation of marssart}
		\begin{array}{ll}
			\E[Z]&=_{(1)}\E\Big[\sup\limits_{\uwave{{\bDelta}}\in \cA, \|\uwave{{\bDelta}}\|_{\bS}\leq \xi } \Big|\dfrac{1}{n}\sum\limits_{i=1}^n(U_{i}(\bDelta,\bdelta)-\E U_{i}(\bDelta,\bdelta))\Big|\Big]\\
			&\leq_{(2)} 2\E\Big[\sup\limits_{\uwave{{\bDelta}}\in \cA, \|\uwave{{\bDelta}}\|_{\bS}\leq \xi } \Big|\dfrac{1}{n}\sum\limits_{i=1}^n(e_iU_{i}(\bDelta,\bdelta))\Big|\Big]\\
			&=_{(3)}2\E\Big[\sup\limits_{\uwave{{\bDelta}}\in \cA, \|\uwave{{\bDelta}}\|_{\bS}\leq \xi } \Big|\dfrac{1}{n}\sum\limits_{i=1}^n\dfrac{1}{K}\sum\limits_{k=1}^{K}e_i[\rho_{\tau_k}(r_{i,k}-(\bX_{i}^\top\bDelta+\delta_k))-\rho_{\tau_k}(r_{i,k})\big]\Big|\Big]\\
			&\leq_{(4)} 2\max\limits_{1\leq k\leq K}\E\Big[\sup\limits_{\uwave{{\bDelta}}\in \cA, \|\uwave{{\bDelta}}\|_{\bS}\leq \xi } \Big|\dfrac{1}{n}\sum\limits_{i=1}^ne_i[\rho_{\tau_k}(r_{i,k}-(\bX_{i}^\top\bDelta+\delta_k))-\rho_{\tau_k}(r_{i,k})\big]\Big|\Big]\\
			&\leq_{(5)} 4\max\limits_{1\leq k\leq K}\E\Big[\sup\limits_{\uwave{{\bDelta}}\in \cA, \|\uwave{{\bDelta}}\|_{\bS}\leq \xi } \Big|\dfrac{1}{n}\sum\limits_{i=1}^ne_i[\bX_{i}^\top\bDelta+\delta_k\big]\Big|\Big]\\
			&=_{(6)}4\max\limits_{1\leq k\leq K}\E\Big[\sup\limits_{\uwave{{\bDelta}}\in \cA, \|\uwave{{\bDelta}}\|_{\bS}\leq \xi } \Big|\dfrac{1}{n}\sum\limits_{i=1}^ne_i(\bS_k \uwave{\bX_i})^\top\uwave{\bDelta}\Big|\Big]\\
			&\leq_{(7)} 4\underbrace{\Big[\sup\limits_{\uwave{{\bDelta}}\in \cA, \|\uwave{{\bDelta}}\|_{\bS}\leq \xi }\|\uwave{\bDelta}\|_1\Big]}_{I}\times\underbrace{\max\limits_{1\leq k\leq K}\E \Big\|\dfrac{1}{n}\sum\limits_{i=1}^ne_i(\bS_k \uwave{\bX_i})\Big\|_\infty}_{II}.
		\end{array}
	\end{equation}
	Hence, to control $\E[Z]$, it is sufficient to consider $I$ and $II$, respectively. For $I$, using the fact that $ \|\uwave{\bDelta}\|_1\leq 4\|\bDelta_{J(\bbeta^*)}\|_1+ \|\bdelta\|_1$ and the  Cauchy-Swarchz inequality, we can prove that: $I\leq C_1\sqrt{s}\xi$ for some $C_1>0$. For $II$, it is not hard to prove that $II\leq C_2M\sqrt{\log(p)/n}$ for some $C_2>0$. Hence, combining (\ref{inequality: marssart's inequality}), (\ref{inequality: upper bound for the variance of marssart}), and (\ref{inequality: upper bound for maximum expectation of marssart}), if we take $t=C_3\xi\sqrt{\log(p)/n}$ for some large enough $C_3>0$, with w.p.a.1, we have $Z\leq C_3M\xi\sqrt{s\log(p)/n}$, which finishes the proof.
\end{proof}

\subsection{Proof of Lemma \ref{lemma: basic inequality for quantile regression alpha}}
\label{sec: proof of basic inequality for quantile regression alpha}

\begin{proof}
	Recall $\uwave{\bbeta}=((\bbeta)^\top,(\bb)^\top)^\top\in \RR^{p+K}$ and  $L^\alpha_{n,K}(\uwave{\bbeta})$ defined in (\ref{equation: Lnk-alpha}). Recall $a^*_{i,k}:=\mathbf{1}\{Y_i-\bX_i^\top\bbeta^*-b^*_k\leq 0\}-\tau_k$ for $i=1,\ldots,n$ and $k=1,\ldots,K$. Define 
	\begin{equation*}
		\begin{array}{ll}
			\nabla L^\alpha_{n,K}(\uwave{\bbeta})=\dfrac{\partial L^\alpha_{n,K}(\uwave{\bbeta})}{\partial \uwave{\bbeta}}\in \RR^{p+K},\\
			~~\nabla_1 L^\alpha_{n,K}(\uwave{\bbeta})=\dfrac{\partial L^\alpha_{n,K}(\uwave{\bbeta})}{\partial {\bbeta}}\in \RR^{p},  \\
			\nabla_2 L^\alpha_{n,K}(\uwave{\bbeta})=\dfrac{\partial L^\alpha_{n,K}(\uwave{\bbeta})}{\partial {\bb}}\in \RR^{K}.
		\end{array}
	\end{equation*}
	Then, we have:
	\begin{equation*}
		\begin{array}{ll}
			\nabla_1 L^\alpha_{n,K}(\uwave{\bbeta}^*)=(1-\alpha)\dfrac{1}{n}\sum_{i=1}^n\dfrac{1}{K}\sum_{k=1}^K\bX_ia_{i,k}^*-\alpha\dfrac{1}{n}\sum_{i=1}^n\bX_i(Y_i-\bX_i^\top\bbeta^*),\\
			~~\text{and}~~\nabla_2L^\alpha_{n,K}(\uwave{\bbeta}^*)=\dfrac{1}{n}\sum_{i=1}^n\ba_i^*,
		\end{array}
	\end{equation*}
	where $\ba_i^*=(a_{i,1},\ldots,a_{i,K})^\top\in \RR^K$. The proof of  Lemma  \ref{lemma: basic inequality for quantile regression} proceeds into two steps.
	
	{\bf{Step~1:}} obtain the upper bounds of $\|\nabla_1 L^\alpha_{n,K}(\uwave{\bbeta}^*)\|_{\infty}$ and $\|\nabla_2 L^\alpha_{n,K}(\uwave{\bbeta}^*)\|_{\infty}$. We first consider $\|\nabla_2 L^\alpha_{n,K}(\uwave{\bbeta}^*)\|_{\infty}$. In fact, by Step~1 in Section \ref{lemma: proof of basic inequality for quantile regression}, we have $\|\nabla_2L^\alpha_{n,K}(\uwave{\bbeta}^*)\|_{\infty}\leq C_1\sqrt{\log(p)/n}$ w.p.a.1. for some $C_1>0$.  Next, we consider $\|\nabla_1 L^\alpha_{n,K}(\uwave{\bbeta}^*)\|_{\infty}$. In fact, we have:
	\begin{equation*}
		\begin{array}{ll}
			\|\nabla_1 L^\alpha_{n,K}(\uwave{\bbeta}^*)\|_{\infty}\\
			\quad=_{(1)}\|(1-\alpha)\dfrac{1}{n}\sum\limits_{i=1}^n\dfrac{1}{K}\sum\limits_{k=1}^K\bX_ia_{i,k}^*-\alpha\dfrac{1}{n}\sum\limits_{i=1}^n\bX_i(Y_i-\bX_i^\top\bbeta^*)\|_{\infty}\\
			\quad=_{(2)}\|(1-\alpha)\dfrac{1}{n}\sum\limits_{i=1}^n\dfrac{1}{K}\sum\limits_{k=1}^K\big(\bX_ia_{i,k}^*-\E(\bX_ia_{i,k}^*)\big)\\
			\qquad\qquad\qquad-\alpha\dfrac{1}{n}\sum\limits_{i=1}^n\big(\bX_i(Y_i-\bX_i^\top\bbeta^*)-\E(\bX_i(Y_i-\bX_i^\top\bbeta^*))\big)\|_{\infty}\\
			\quad \leq_{(3)}(1-\alpha)\underbrace{\|\dfrac{1}{n}\sum\limits_{i=1}^n\dfrac{1}{K}\sum\limits_{k=1}^K\big(\bX_ia_{i,k}^*-\E(\bX_ia_{i,k}^*)\big)\|_{\infty}}_{I}\\
			\qquad\qquad\qquad+\alpha\underbrace{\|\dfrac{1}{n}\sum\limits_{i=1}^n\big(\bX_i(Y_i-\bX_i^\top\bbeta^*)-\E(\bX_i(Y_i-\bX_i^\top\bbeta^*))\big)\|_{\infty}}_{II},	
		\end{array}
	\end{equation*}
	where $(1)$ comes from the first order condition in (\ref{equation: first order condition alpha}). By Step 2 in Section \ref{lemma: proof of basic inequality for quantile regression}, we have $I\leq C_2M\sqrt{\log(p)/n}$ w.p.a.1. Next, we consider $II$. In fact, by noting that $Y_i=\epsilon_i+\bbeta^{(1)}\mathbf{1}\{i\leq \floor{nt_1}\}+\bbeta^{(2)}\mathbf{1}\{i> \floor{nt_1}\}$, we have:
	\begin{equation*}
		\small
		\begin{array}{ll}
			II&=_{(1)}\big\|\dfrac{1}{n}\sum\limits_{i=1}^n\bX_i\epsilon_i+t_1(\hat{\bSigma}(0:t_1)-\bSigma)(\bbeta^{(1)}-\bbeta^*)+t_2(\hat{\bSigma}(t_1:1)-\bSigma)(\bbeta^{(2)}-\bbeta^*)\big\|_{\infty}\\
			&\leq_{(2)} \big\|\dfrac{1}{n}\sum\limits_{i=1}^n\bX_i\epsilon_i\|_{\infty}+t_1\|\hat{\bSigma}(0:t_1)-\bSigma\|_{\infty}\|\bbeta^{(1)}-\bbeta^*\|_1+t_2\|\hat{\bSigma}(t_1:1)-\bSigma\|_{\infty}\|\bbeta^{(2)}-\bbeta^*\|_1\\
			&=_{(3)}O_p\big( M\sqrt{\dfrac{\log(pn)}{n}}\big)+O_p\big( M^2\sqrt{\dfrac{\log(pn)}{n}}\big)\|\bbeta^{(1)}-\bbeta^*\|_1+O_p\big( M^2\sqrt{\dfrac{\log(pn)}{n}}\big)\|\bbeta^{(2)}-\bbeta^*\|_1\\
			&=_{(4)}O_p\big( M\sqrt{\dfrac{\log(pn)}{n}}\big)+O_p\big( M^2\sqrt{\dfrac{\log(pn)}{n}}\big)\|\bbeta^{(1)}-\bbeta^{(2)}\|_1
			=_{(5)}O_p\big( M^2\sqrt{\dfrac{\log(pn)}{n}}\big),
		\end{array}
	\end{equation*}
	where $(3)$ comes from Lemmas \ref{lemma: exponential inequality for partial sum process} and \ref{lemma: concentration for covariance}, $(4)$ and $(5)$ come from Remark \ref{remark: difference beta for alpha in (0,1)} and the assumption that $\|\bbeta^{(1)}-\bbeta^{(2)}\|_1\leq C_{\bDelta}$. Hence, combining the above bounds, w.p.a.1, we have: 
	\begin{equation*}
		\|\nabla_1 L^\alpha_{n,K}(\uwave{\bbeta}^*)\|_{\infty}\leq C_1M^2\sqrt{\dfrac{\log(pn)}{n}},\|\nabla_2L^\alpha_{n,K}(\uwave{\bbeta}^*)\|_{\infty}\leq C_2M^2\sqrt{\dfrac{\log(pn)}{n}}.
	\end{equation*}
	
	{\bf{Step~2:}} Let $\lambda\geq 2M^2(C_1\vee C_2)\sqrt{\log(p)/n}$, where $C_1$ and $C_2$ are defined in Step 1. Using similar proof procedure as in Step 2 of Section \ref{lemma: proof of basic inequality for quantile regression}, we can derive that:
	\begin{equation*}
		\| ({\hat{\bbeta}}-{\bbeta}^*)_{J^c(\bbeta^*)}\|_1\leq 3\| ({\hat{\bbeta}}-{\bbeta}^*)_{J(\bbeta^*)}\|_1+\|\hat{\bb}-\bb^*\|,
	\end{equation*}
	which finishes the proof.
	
\end{proof}

\subsection{Proof of Lemma \ref{lemma: lower bound for the excess risk with alpha}}
\label{sec: proof of lower bound for the excess risk with alpha}
\begin{proof}
	Recall 
	\begin{equation*}
		\begin{array}{cc}
			L^{\alpha}_{K}(\uwave{\bbeta}):=(1-\alpha)\dfrac{1}{n}\sum\limits_{i=1}^{n}\dfrac{1}{K}\sum\limits_{k=1}^{K}\E[\rho_{\tau_k}(Y_i-b_i-\bX_i^\top\bbeta)]+ \dfrac{\alpha}{2n}\sum\limits_{i=1}^{n}\E(Y_i-\bX_i^\top \bbeta)^2.
		\end{array}
	\end{equation*}
	Note that $L^{\alpha}_{K}(\uwave{\bbeta})$ is a combination of the composite quantile loss and the squared loss. Moreover, it is easy to see that the excess risk for the squared loss is lower bounded by a squared form. Hence, combining the results in Section \ref{lemma: proof of ower bound for the excess risk with alpha=0}, we can prove that there exists some $c_*>0$ such that 
	\begin{equation*}
		H^\alpha(\uwave{{\bDelta}})\geq c_*\min\Big(\dfrac{ \|\uwave{{\bDelta}}\|^2_{\bS}}{4},\dfrac{\|\uwave{{\bDelta}}\|_{\bS}}{4} \Big).
	\end{equation*}
	To save space, we omit the details.
\end{proof}

\subsection{Proof of Lemma \ref{lemma: large deviation for excess risk with alpha}}
\label{sec: proof of large deviation for excess risk with alpha}
Define
\begin{equation}
	\begin{array}{cl}
		U^{(1)}_{i}(\bDelta,\bdelta)&=\dfrac{1}{K}\sum\limits_{k=1}^{K}\big[\rho_{\tau_k}(Y_i-\bX_{i}^\top\bbeta^*-b^*_k-(\bX_{i}^\top\bDelta+\delta_k))-\rho_{\tau_k}(Y_i-\bX_{i}^\top\bbeta^*-b^*_k)\big]\\
		U^{(2)}_{i}(\bDelta,\bdelta)&=(Y_i-\bX_i^\top\bbeta^*-b^*_k-(\bX_{i}^\top\bDelta+\delta_k))^2-(Y_i-\bX_{i}^\top\bbeta^*-b^*_k)^2.\\
	\end{array}	
\end{equation}
Hence, using the above notations, we have:
\begin{equation*}
	\begin{array}{ll}
		\big(L^\alpha_{n,K}(\uwave{\bbeta}^*+\uwave{{\bDelta}})-L^\alpha_{n,K}(\uwave{\bbeta}^*))-(L^\alpha_K(\uwave{\bbeta}^*+\uwave{{\bDelta}})-L^\alpha_K(\uwave{\bbeta}^*)\big)\\
		\quad=	(1-\alpha)\dfrac{1}{n}\sum\limits_{i=1}^n\Big[ U^{(1)}_{i}(\bDelta,\bdelta)-\E[U^{(1)}_{i}(\bDelta,\bdelta)]    \Big]+\dfrac{\alpha}{2}\dfrac{1}{n}\sum\limits_{i=1}^n\Big[ U^{(2)}_{i}(\bDelta,\bdelta)-\E[U^{(2)}_{i}(\bDelta,\bdelta)]    \Big].
	\end{array}
\end{equation*}
Hence, to prove Lemma \ref{lemma: large deviation for excess risk with alpha}, it is sufficient to bound $I$ and $II$, where:
\begin{equation}\label{inequality: empirical process for I+II alpha=0}
	\begin{array}{ll}
		I=\sup\limits_{\uwave{{\bDelta}}\in \cA, \|\uwave{{\bDelta}}\|_{\bS}\leq \xi }\Big|\dfrac{1}{n}\sum\limits_{i=1}^n U^{(1)}_{i}(\bDelta,\bdelta)-\E[U^{(1)}_{i}(\bDelta,\bdelta)]    \Big|,\\
		II=\sup\limits_{\uwave{{\bDelta}}\in \cA, \|\uwave{{\bDelta}}\|_{\bS}\leq \xi }\Big|\dfrac{1}{n}\sum\limits_{i=1}^n U^{(2)}_{i}(\bDelta,\bdelta)-\E[U^{(2)}_{i}(\bDelta,\bdelta)]    \Big|.
	\end{array}
\end{equation}
Note that in Section \ref{sec: proof of large deviation for excess risk with alpha=0}, we have proved that $I=O_p(\xi \sqrt{s\log(p)/n})$. Hence, it only remains to consider $II$. Let $  \uwave{\bbeta}=\uwave{\bbeta}^*+\uwave{{\bDelta}}$. Then, it is equivalent to consider :
\begin{equation}\label{inequality: empirical process for II.1-II.3 alpha=0}
	\begin{array}{ll}
		II=\sup\limits_{\uwave{\bbeta}-\uwave{\bbeta}^*\in \cA, \|\uwave{\bbeta}-\uwave{\bbeta}^*\|_{\bS}\leq \xi }\Big|\dfrac{1}{n}\sum\limits_{i=1}^n [(Y_i-\bX_i^\top\bbeta)^2- (Y_i-\bX_i^\top\bbeta^*)^2 ]\\
		\qquad\qquad\qquad- \E[(Y_i-\bX_i^\top\bbeta)^2- (Y_i-\bX_i^\top\bbeta^*)^2]\Big|\\
		\leq II.1+II.2+II.3,
	\end{array}
\end{equation}
where 
\begin{equation*}
	\begin{array}{ll}
		II.1:=\sup\limits_{\uwave{\bbeta}-\uwave{\bbeta}^*\in \cA\atop\|\uwave{\bbeta}-\uwave{\bbeta}^*\|_{\bS}\leq \xi }\Big|\dfrac{1}{n}\sum\limits_{i=1}^n\epsilon_i\bX_i^\top(\bbeta^*-\bbeta)\Big|,\\
		II.2:=\sup\limits_{\uwave{\bbeta}-\uwave{\bbeta}^*\in \cA\atop\|\uwave{\bbeta}-\uwave{\bbeta}^*\|_{\bS}\leq \xi }t_1\Big|\dfrac{1}{nt_1}\sum\limits_{i=1}^{nt_1}\big[(\bX_i^\top\bbeta-\bX_i^\top\bbeta^{(1)})^2-(\bX_i^\top\bbeta^*-\bX_i^\top\bbeta^{(1)})^2\big]\\
		\qquad\qquad\qquad\qquad-\E\big[(\bX_i^\top\bbeta-\bX_i^\top\bbeta^{(1)})^2-(\bX_i^\top\bbeta^*-\bX_i^\top\bbeta^{(1)})^2\big]\Big|,\\
		II.3:=\sup\limits_{\uwave{\bbeta}-\uwave{\bbeta}^*\in \cA\atop\|\uwave{\bbeta}-\uwave{\bbeta}^*\|_{\bS}\leq \xi }t_2\Big|\dfrac{1}{nt_2}\sum\limits_{i=nt_1+1}^{n}\big[(\bX_i^\top\bbeta-\bX_i^\top\bbeta^{(2)})^2-(\bX_i^\top\bbeta^*-\bX_i^\top\bbeta^{(2)})^2\big]\\
		\qquad\qquad\qquad\qquad-\E\big[(\bX_i^\top\bbeta-\bX_i^\top\bbeta^{(2)})^2-(\bX_i^\top\bbeta^*-\bX_i^\top\bbeta^{(2)})^2\big]\Big|.\\
	\end{array}
\end{equation*}
Next, we consider $II.1-II.3$, respectively. For $II.1$, we have
\begin{equation*}
	\begin{array}{ll}
		II.1&\leq_{(1)} \|\dfrac{1}{n}\sum\limits_{i=1}^n\epsilon_i\bX_i^\top\|_{\infty}\sup\limits_{\uwave{\bbeta}-\uwave{\bbeta}^*\in \cA,\|\uwave{\bbeta}-\uwave{\bbeta}^*\|_{\bS}\leq \xi }\|\bbeta^*-\bbeta\|_1\\
		&\leq_{(2)} CM\sqrt{\dfrac{\log(pn)}{n}}\sup\limits_{\uwave{\bbeta}-\uwave{\bbeta}^*\in \cA,\|\uwave{\bbeta}-\uwave{\bbeta}^*\|_{\bS}\leq \xi }\|\bbeta^*-\bbeta\|_1\\
		&\leq_{(3)} CM\sqrt{\dfrac{\log(pn)}{n}}\sup\limits_{\uwave{\bbeta}-\uwave{\bbeta}^*\in \cA,\|\uwave{\bbeta}-\uwave{\bbeta}^*\|_{\bS}\leq \xi }(4\|(\bbeta^*-\bbeta)_{J(\bbeta^*)}\|_1+\|\bb-\bb^*\|_1)\\
		&\leq_{(4)} CM\sqrt{\dfrac{\log(pn)}{n}}\sup\limits_{\uwave{\bbeta}-\uwave{\bbeta}^*\in \cA,\|\uwave{\bbeta}-\uwave{\bbeta}^*\|_{\bS}\leq \xi }(4\sqrt{s}\|(\bbeta^*-\bbeta)_{J(\bbeta^*)}\|_2+\sqrt{K}\|\bb-\bb^*\|_2)\\
		&\leq_{(5)} CM\xi\sqrt{s\dfrac{\log(pn)}{n}}
	\end{array}
\end{equation*}
where $(2)$ comes from Lemma \ref{lemma: exponential inequality for partial sum process}, $(3)$ follows from the definition of $\cA$, $(5)$ follows from the definiteness of $\bS$. 

Our next goal is to bound $II.2$. To that end, we suppose that there exists some universal constant $\eta>0$ such that for all $\uwave{\bbeta}$ satisfying  $\|\uwave{\bbeta}-\uwave{\bbeta}^*\|_{\bS}\leq \xi$, we have:
\begin{equation*}
	|\bX^\top\bbeta-\bX^\top\bbeta^{(1)}|\leq \eta.
\end{equation*}
Note that this is a very common assumption for proving the concentration inequality for squared error loss (see \cite{Peter2011}). Define the functional class:
\begin{equation*}
	\gamma(\bX^\top\bbeta)=\dfrac{(\bX^\top\bbeta-\bX^\top\bbeta^{(1)})^2-(\bX^\top\bbeta^*-\bX^\top\bbeta^{(1)})^2}{2\eta}.
\end{equation*}
By definition, we can see that $|\gamma(\bX^\top\bbeta)-\gamma(\bX^\top\bbeta')|\leq |\bX^\top\bbeta-\bX^\top\bbeta'|$, which is 1-Lipschitz continous. Moreover, by defining $\gamma(\bX^\top\bbeta)$, $II.2$ reduces to:
\begin{equation*}
	II.2=2\eta t_1\underbrace{\sup\limits_{\uwave{\bbeta}-\uwave{\bbeta}^*\in \cA\atop\|\uwave{\bbeta}-\uwave{\bbeta}^*\|_{\bS}\leq \xi }\Big|\dfrac{1}{nt_1}\sum\limits_{i=1}^{nt_1}[\gamma(\bX_i^\top\bbeta)-\gamma(\bX_i^\top\bbeta^*)]-\E[\gamma(\bX_i^\top\bbeta)-\gamma(\bX_i^\top\bbeta^*)]}_{Z}.
\end{equation*}
Note that 
\begin{equation*}
	\begin{array}{ll}
		&\big|\gamma(\bX_i^\top\bbeta)-\gamma(\bX_i^\top\bbeta^*)-\E[\gamma(\bX_i^\top\bbeta)-\gamma(\bX_i^\top\bbeta^*)]\big|\leq  2|\bX_i^\top(\bbeta-\bbeta^*)|:=C_{i}(\bbeta).
	\end{array}
\end{equation*}
In what follows, we will use the Massart’s inequality (Theorem 14.2 in \cite{Peter2011}) to obtain the tail bound:
\begin{equation}\label{inequality: marssart's inequality-2}
	\P\Big(Z>\E Z +t\Big)\leq \exp\Big(-\dfrac{nt^2}{8\sigma^2}\Big),
\end{equation}
where  $\sup\limits_{\uwave{\bbeta}-\uwave{\bbeta}^*\in \cA,\|\uwave{\bbeta}-\uwave{\bbeta}^*\|_{\bS}\leq \xi }\dfrac{1}{n}\sum\limits_{i=1}^nC^2_i(\bbeta)\leq \sigma^2$. Hence, to use Massart’s inequality, we need two steps.\\
{\bf{Step~1:}} obtain the upper bound for $\sigma^2$. In fact, w.p.a.1, we have:
\begin{equation*}
	\begin{array}{ll}
		\sup\limits_{\uwave{\bbeta}-\uwave{\bbeta}^*\in \cA,\|\uwave{\bbeta}-\uwave{\bbeta}^*\|_{\bS}\leq \xi }\dfrac{1}{n}\sum\limits_{i=1}^nC^2_i(\bbeta)\\
		=_{(1)}4\sup\limits_{\uwave{\bbeta}-\uwave{\bbeta}^*\in \cA,\|\uwave{\bbeta}-\uwave{\bbeta}^*\|_{\bS}\leq \xi }(\bbeta-\bbeta^*)^\top\hat{\bSigma}(0:1)(\bbeta-\bbeta^*)\\
		=_{(2)}4\sup\limits_{\uwave{\bbeta}-\uwave{\bbeta}^*\in \cA,\|\uwave{\bbeta}-\uwave{\bbeta}^*\|_{\bS}\leq \xi }|(\bbeta-\bbeta^*)^\top(\hat{\bSigma}(0:1)-\bSigma)(\bbeta-\bbeta^*)|\\
		\qquad\qquad\qquad	+4\sup\limits_{\uwave{\bbeta}-\uwave{\bbeta}^*\in \cA,\|\uwave{\bbeta}-\uwave{\bbeta}^*\|_{\bS}\leq \xi }|(\bbeta-\bbeta^*)^\top\bSigma(\bbeta-\bbeta^*)|\\
		\leq_{(3)} 4\sup\limits_{\uwave{\bbeta}-\uwave{\bbeta}^*\in \cA,\|\uwave{\bbeta}-\uwave{\bbeta}^*\|_{\bS}\leq \xi }\|\hat{\bSigma}(0:1)-\bSigma)\|_{\infty}\|\bbeta-\bbeta^*\|_{1}^2\\
		\qquad\qquad\qquad+4\lambda_{\rm max}\sup\limits_{\uwave{\bbeta}-\uwave{\bbeta}^*\in \cA,\|\uwave{\bbeta}-\uwave{\bbeta}^*\|_{\bS}\leq \xi }\|\bbeta-\bbeta^*\|^2\\
		\leq_{(4)} C_1M^2\sqrt{\dfrac{\log(pn)}{n}}s\xi^2+C_2\xi^2\leq_{(5)}C_3\xi^2:=\sigma^2.
	\end{array}
\end{equation*}
{\bf{Step~2:}} obtain the upper bound for $\E [Z]$. Let $e_1,\ldots,e_n$ be i.i.d Rademacher random variables with $\P(e_i=1)=\P(e_i=-1)=1/2$. In fact, by the symmetrization procedure (Theorem 14.3 in \cite{Peter2011}) and 
the contraction principle (Theorem 14.4 in \cite{Peter2011}), we have:
\begin{equation*}
	\begin{array}{ll}
		\E[Z]&\leq 2\E\Big[\sup\limits_{\uwave{\bbeta}-\uwave{\bbeta}^*\in \cA,\|\uwave{\bbeta}-\uwave{\bbeta}^*\|_{\bS}\leq \xi }\Big|\dfrac{1}{nt_1}\sum\limits_{i=1}^{nt_1}e_i(\gamma(\bX_i^\top\bbeta)-\gamma(\bX_i^\top\bbeta^*))\Big|\Big]\\
		&\leq 4\E\Big[\sup\limits_{\uwave{\bbeta}-\uwave{\bbeta}^*\in \cA,\|\uwave{\bbeta}-\uwave{\bbeta}^*\|_{\bS}\leq \xi }\Big|\dfrac{1}{nt_1}\sum\limits_{i=1}^{nt_1}e_i(\bX_i^\top\bbeta-\bX_i^\top\bbeta^*)\Big|\Big]\\
		&\leq 4\sup\limits_{\uwave{\bbeta}-\uwave{\bbeta}^*\in \cA,\|\uwave{\bbeta}-\uwave{\bbeta}^*\|_{\bS}\leq \xi }\|\bbeta-\bbeta^*\|_1\E \Big\|\dfrac{1}{nt_1}\sum\limits_{i=1}^{nt_1}e_i\bX_i\|_{\infty}
		\leq C\xi\sqrt{s}M\sqrt{\dfrac{\log(pn)}{n}},
	\end{array}
\end{equation*}
where the last inequality comes from the Hoeffding's inequality. Hence, combining Steps 1 and 2, taking $t=C\xi\sqrt{\log(p)/n}$ for some big enough constant $C>0$, we have, w.p.a.1, $Z=O(\xi\sqrt{s\log(p)/n})$, which implies $II.2=O_p(\xi\sqrt{s\log(p)/n})$. 

Similarly, we can prove $II.3=O_p(\xi\sqrt{s\log(p)/n})$. Considering (\ref{inequality: empirical process for I+II alpha=0}) and (\ref{inequality: empirical process for II.1-II.3 alpha=0}), we proved:
\begin{equation*}
	\sup_{\uwave{{\bDelta}}\in \cA\atop \|\uwave{{\bDelta}}\|_{\bS}\leq \xi }\big|(L^\alpha_{n,K}(\uwave{\bbeta}^*+\uwave{{\bDelta}})-L^\alpha_{n,K}(\uwave{\bbeta}^*))-(L^\alpha_K(\uwave{\bbeta}^*+\uwave{{\bDelta}})-L^\alpha_K(\uwave{\bbeta}^*))\big|=O_p\big(M\xi \sqrt{s\dfrac{\log(pn)}{n}}\big),
\end{equation*}
which finishes the proof.

\end{document}